\documentclass[prx,aps,twocolumn,superscriptaddress]{revtex4-1}
\usepackage{graphicx}
\usepackage{amsmath}
\usepackage{amssymb}
\usepackage{bm}
\usepackage{dsfont}
\usepackage{multirow}
\usepackage{color}
\usepackage{placeins}

\usepackage{float}

\newlength{\bibitemsep}
\newlength{\bibparskip}
\let\oldthebibliography\thebibliography
\renewcommand\thebibliography[1]{%
  \oldthebibliography{#1}%
  \setlength{\parskip}{\bibitemsep}%
  \setlength{\itemsep}{\bibparskip}%
}

\usepackage{xr-hyper}
\usepackage[colorlinks, citecolor=blue]{hyperref}
\usepackage{xr}
\newcommand{\bs}[1]{\boldsymbol{#1}}

\begin{document}
\title{Topological Zero-Dimensional Defect and Flux States\\ in Three-Dimensional Insulators}

\author{Frank Schindler}
\thanks{Corresponding author: \url{schindler@princeton.edu}}
\affiliation{Princeton Center for Theoretical Science, Princeton University, Princeton, NJ 08544, USA}

\author{Stepan S. Tsirkin}
\affiliation{Department of Physics, University of Zurich, Winterthurerstrasse 190, 8057 Zurich, Switzerland}
            
\author{Titus Neupert}
\affiliation{Department of Physics, University of Zurich, Winterthurerstrasse 190, 8057 Zurich, Switzerland}
            
\author{B. Andrei Bernevig$^\ddag$}
\affiliation{Department of Physics,
Princeton University,
Princeton, NJ 08544, USA}
\affiliation{Donostia International Physics Center, P. Manuel de Lardizabal 4, 20018 Donostia-San Sebastian, Spain}
\affiliation{IKERBASQUE, Basque Foundation for Science, Bilbao, Spain}

\author{Benjamin J. Wieder}
\thanks{Corresponding author: \url{bwieder@mit.edu}}
\affiliation{Department of Physics, Massachusetts Institute of Technology, Cambridge, MA 02139, USA}
\affiliation{Department of Physics, Northeastern University, Boston, MA 02115, USA}
\affiliation{Department of Physics,
Princeton University,
Princeton, NJ 08544, USA}

\begin{abstract}
In insulating crystals, it was previously shown that defects with two fewer dimensions than the bulk can bind topological electronic states.  We here further extend the classification of topological defect states by demonstrating that the corners of crystalline defects with integer Burgers vectors can bind 0D higher-order end (HEND) states with anomalous charge and spin.  We demonstrate that HEND states are intrinsic topological consequences of the bulk electronic structure and introduce new bulk topological invariants that are predictive of HEND dislocation states in solid-state materials.  We demonstrate the presence of first-order 0D defect states in PbTe monolayers and HEND states in 3D SnTe crystals.  We relate our analysis to magnetic flux insertion in insulating crystals.  We find that $\pi$-flux tubes in inversion- and time-reversal-symmetric (helical) higher-order topological insulators bind Kramers pairs of spin-charge-separated HEND states, which represent observable signatures of anomalous surface half quantum spin Hall states.
\end{abstract}

\maketitle

{
\vspace{0.08in}
\centerline{\bf Introduction}
\vspace{0.08in}
}

In crystalline solids, there are numerous sources of disorder and defects.  One type of crystal defect -- integer dislocations -- can manifest as edge dislocations, in which planes of atoms are missing within a region of the sample.  Integer dislocations can also manifest as screw dislocations, in which planes of atoms in a portion of the crystal are successively shifted by an integer linear combination of lattice vectors~\cite{MerminReview}.  Screw and edge dislocations -- which locally represent 1D line defects in 3D crystals -- are each characterized by a gauge-invariant Burgers vector $\bs{B}$.

\begin{table*}[t]
\begin{small}
\begin{tabular}{|c|c|c|c|}
\hline
\multicolumn{4}{|c|}{Summary of Higher-Order Dislocation- and Flux-State Responses in Inversion-Symmetric 3D Insulators} \\
\hline
 & Fragile Topological Insulators & Magnetic Axion & Helical Higher-Order  \\
 & and Obstructed Atomic Limits & Insulators (AXIs) &  Topological Insulators (HOTIs) \\
\hline
\hline
Integer & Nontrivial if and only if: & Nontrivial if and only if: & Nontrivial if and only if: \\
Dislocation & $\bs{B}\cdot \bs{M}_{\nu}^{\mathrm{F}}\text{ mod }2\pi = \pi$ & $\bs{B}\cdot \bs{M}_{\nu}^{\mathrm{F}}\text{ mod }2\pi = \pi$ & $\bs{B}\cdot \bs{M}_{\nu}^{\mathrm{F}}\text{ mod }2\pi = \pi$ \\
\hline
$\pi$-Flux  &  & Nontrivial, signature of: & Nontrivial, signature of: \\
Tube  & Trivial & surface half quantum Hall state, & surface half quantum spin Hall state \\ 
 & & bulk magnetoelectric polarizability & bulk magneto-spinon polarizability (MSP)\\
\hline
\end{tabular}
\end{small}
\caption{{\bf Summary of dislocation- and flux-state responses derived in this work.} We have uncovered a new bulk weak topological index $\bs{M}_{\nu}^{\mathrm{F}}$ [detailed in the Methods section and in Supplementary Notes (SN)~\ref{sec:kpEdge},~\ref{sec:kpScrew},~\ref{sec:defecttopology},~\ref{sec:FrankScrew}, and~\ref{sec:weakFragile}] that indicates whether integer dislocations with Burgers vector $\bs{B}$ in an inversion-symmetric 3D insulator bind anomalous 0D states on their ends and corners, which we term higher-order end (HEND) states.  While $\bs{M}_{\nu}^{\mathrm{F}}$ can be nontrivial in stable topological crystalline insulators (TCIs) with 1D hinge modes [\emph{e.g.} magnetic axion insulators (AXIs)~\cite{WilczekAxion,QHZ,VDBAxion,FuKaneInversion,AndreiInversion,AshvinAxion2,FanHOTI,VDBHOTI,WiederAxion,MTQC} and helical higher-order topological insulators (HOTIs)~\cite{multipole,WladTheory,HOTIBernevig,TMDHOTI,ChenTCI,AshvinIndicators,AshvinTCI}], $\bs{M}_{\nu}^{\mathrm{F}}$ can also be nontrivial in insulators with less robust forms of topology, such as fragile topological insulators~\cite{AshvinFragile,YoungkukMonopole,FragileKoreanInversion,ZhidaFragileAffine} and obstructed atomic limits~\cite{QuantumChemistry,HingeSM,WladCorners}.  Integer dislocations therefore do not provide an unambiguous probe of bulk higher-order topology.  Through first-principles and tight-binding calculations, we further demonstrate a nontrivial HEND-state dislocation response in the 3D TCI~\cite{HsiehTCI} and HOTI SnTe~\cite{HOTIBernevig} (see the Methods section and SN~\ref{sec:DFTSnTe} for calculation details).  We have also studied the related problem of magnetic $\pi$-flux insertion in 3D insulators.  For AXIs, $\pi$-flux tubes are known to provide probes of the anomalous half quantum Hall states on gapped 2D surfaces, as well as the bulk topological magnetoelectric response~\cite{QHZ,FranzWormhole}.  For helical HOTIs, we find that $\pi$-flux tubes reveal previously unrecognized bulk and surface topological features similar to those of AXIs, including surface halves of time-reversal-symmetric quantum spin Hall states, and a spin-charge-separated variant of the bulk axionic magnetoelectric effect.}
\label{tb:ResultsMain}
\end{table*}

In pristine crystals -- defined by the absence of disorder and defects -- the electronic states form bands, which may be classified by their topological properties~\cite{AndreiTI,KaneMeleZ2,FuKaneInversion,AndreiInversion,QHZ,LiangTCI,HsiehTCI,HourglassInsulator,DiracInsulator,ChenRotation,SYBiBr,BiBrFanHOTI,ZhijunRotationHOTI,WladTheory,HOTIBernevig,WiederAxion}.  When a crystal exhibits unitary symmetries beyond translation -- such as spatial inversion ($\mathcal{I}$), then the band topology may conveniently be diagnosed by symmetry eigenvalues through elementary band representations, which give rise to symmetry-based indicators~\cite{QuantumChemistry,MTQC}.  Well-established symmetry-based indicators of insulating band topology include the Fu-Kane parity criterion~\cite{FuKaneInversion}, and the strong 3D $\mathbb{Z}_{4}$ and weak 2D $\mathbb{Z}_{2}$ invariants of $\mathcal{I}$- and time-reversal- ($\mathcal{T}$-) symmetric 3D insulators~\cite{AshvinIndicators,ChenTCI,AshvinTCI,TMDHOTI}.

Over the past decade, numerous proposals have been introduced to link the seemingly disparate limits of pristine crystalline solids with nontrivial electronic band topology and the more realistic setting of crystals hosting defects~\cite{AshvinScrewTI,TeoKaneDefect,QiDefect2,Vlad2D,TanakaDefect,VladScrewTI,JenDefect}.  This has led to the identification of electronic defect states in both topological insulators (TIs)~\cite{AndreiTI,KaneMeleZ2,FuKaneInversion,AndreiInversion,QHZ} and topological crystalline insulators (TCIs)~\cite{LiangTCI,HsiehTCI,HourglassInsulator,DiracInsulator,ChenRotation,SYBiBr,BiBrFanHOTI,ZhijunRotationHOTI}.  In particular, it has been extensively demonstrated~\cite{AshvinScrewTI,TeoKaneDefect,QiDefect2,Vlad2D,TanakaDefect,VladScrewTI} that screw and edge dislocations in $\mathcal{T}$-symmetric 3D insulators can bind helical pairs of 1D states if the defect Burgers vector aligns with the weak-index vector $\bs{M}_{\nu}=(\nu_{x},\nu_{y},\nu_{z})$:
\begin{equation}
\bs{B}\cdot \bs{M}_{\nu}\text{ mod }2\pi = \pi,
\label{eq:Burgers}
\end{equation}
where $\nu_{i}$ is the $\mathbb{Z}_{2}$-valued weak index in the $k_{i}=\pi$ plane~\cite{FuKaneInversion}.  For a $\mathcal{T}$-symmetric 3D insulator with vanishing strong indices~\cite{FuKaneInversion,WladTheory,AshvinIndicators,DiracInsulator,AshvinTCI,ChenTCI,TMDHOTI}, $\bs{M}_{\nu}\neq\bs{0}$ further indicates that the insulator can be adiabatically deformed without breaking a symmetry or closing a gap into a decoupled stack of 2D TIs -- known as a weak TI~\cite{FuKaneInversion}.  In weak TIs hosting defects with $\bs{B}$ directed along the stacking direction, $(\bs{B}\cdot \bs{M}_{\nu})/\pi$ indicates the number of decoupled 2D layers connecting the crystal defects.  Hence intuitively, if $(\bs{B}\cdot \bs{M}_{\nu})/\pi$ is odd [\emph{i.e.} Eq.~(\ref{eq:Burgers}) is satisfied], then the defects carry robust helical modes.  In terms of momentum-space band topology, Eq.~(\ref{eq:Burgers}) and its $\mathcal{T}$-broken variant~\cite{QiDefect2} predict defect bound states. They respectively diagnose which of the Brillouin-zone- \mbox{(BZ-)} boundary planes have Hamiltonians that are topologically equivalent to 2D TIs and magnetic Chern insulators.  In the above discussion of Eq.~(\ref{eq:Burgers}) and throughout the remainder of this work, we have defined the BZ boundary as the set of momentum-space surfaces for which ${\bs k}\cdot {\bs b}_{i}=\pi$, where ${\bs b}_{i}$ is a primitive reciprocal lattice vector.

In addition to crystal defects, static magnetic flux has also been proposed as a probe of bulk topology~\cite{QHZ,QiFlux,AshvinFlux,FranzWormhole,Vlad2D}.  For example, static $\pi$-flux cores in Chern insulators (2D TIs) have been shown to bind 0D solitons with $e/2$ charge (spin-charge separation), where we have defined all charges with respect to the point of charge neutrality.  In 3D TIs and magnetic axion insulators (AXIs)~\cite{FuKaneInversion,AndreiInversion,QHZ,WilczekAxion,FanHOTI,VDBAxion,VDBHOTI,WiederAxion,WladTheory,HOTIBernevig,TMDHOTI,MTQC,AshvinAxion2}, $\pi$-flux tubes provide a means of probing the topologically-quantized bulk magnetoelectric polarizability.  Specifically in 3D TIs, a pair of $\pi$-flux tubes will bind a pair of ``wormhole-like'' helical modes (subdivided into one pair of helical modes per tube)~\cite{FranzWormhole}.  If $\mathcal{T}$ is relaxed in a manner that preserves the quantized bulk axion angle $\theta=\pi$, the 3D TI is converted into a magnetic AXI, and the flux-tube helical modes will become gapped and leave behind anomalous $\pm e/2$ end charges, one at one end of each flux tube, in a manifestation of the axionic magnetoelectric effect.  Specifically, the topological axion angle $\theta=\pi$ is the coefficient of the magnetoelectric response ${\bs E}_{e}\cdot{\bs B}_{e}$, where ${\bs E}_{e}$ and ${\bs B}_{e}$ are the electric and magnetic fields, respectively.  Hence, the $e/2$ end charges bound to $\pi$-flux tubes in an AXI represents signatures of the quantized bulk magnetoelectric polarizability (nontrivial axion angle), because the external magnetic field has induced a quantized electric polarization aligned with the magnetic field.

In recent years, the set of topologically nontrivial 2D and 3D insulating phases has been greatly extended beyond TIs, Chern insulators, and AXIs by incorporating the constraints imposed by crystalline symmetry on electronic band structures~\cite{QuantumChemistry,AshvinIndicators}.  Recently introduced symmetry-protected 2D topological insulating phases include 2D TCIs with mirror-protected edge states~\cite{HsiehTCI,HingeSM,LiangTCIMonolayer,PbTeMonolayer1}, as well as fragile TIs (FTIs)~\cite{AshvinFragile,HingeSM,YoungkukMonopole,TMDHOTI,WiederAxion,FragileKoreanInversion,ZhidaFragileAffine} and 2D obstructed atomic limits (OALs)~\cite{multipole,WladTheory,HingeSM,QuantumChemistry,TMDHOTI,WiederAxion,WladCorners} with 0D fractionally charged or spin-charge-separated corner states.  In 3D, TCI phases with gapped 2D surfaces and gapless 1D hinges have recently been discovered, and have become known as higher-order TIs (HOTIs)~\cite{multipole,WladTheory,HOTIBernevig,FanHOTI,TMDHOTI,ChenTCI,AshvinIndicators,AshvinTCI,VDBHOTI,WiederAxion}.  After the discovery of higher-order topology, earlier examples of magnetic AXIs were recognized to in fact be magnetic chiral HOTIs~\cite{WiederAxion,VDBHOTI}.  In an AXI, each surface exhibits an odd number of massive or massless twofold Dirac cones corresponding to an anomalous half-integer surface Hall conductivity, and domain walls between gapped surfaces with differing half-integer Hall conductivities bind chiral hinge modes~\cite{FanHOTI,VDBAxion,VDBHOTI,WiederAxion,MTQC}.

$\mathcal{T}$-symmetric HOTI phases with helical hinge modes have also been predicted in rhombohedral bismuth crystals~\cite{HOTIBismuth}, the transition metal dichalcogenides MoTe$_2$ and WTe$_2$~\cite{TMDHOTI,AshvinMaterials1}, and BiBr~\cite{SYBiBr,BiBrFanHOTI,AshvinMaterials1}.  Through scanning tunneling microscopy (STM) and quantum oscillation experiments, incipient support for the existence of helical hinge states was subsequently reported in the aforementioned candidate HOTIs bismuth~\cite{HOTIBismuth}, MoTe$_2$~\cite{DavidMoTe2Exp,PhuanOngMoTe2Hinge}, WTe$_2$~\cite{MazWTe2Exp}, and BiBr~\cite{BiBrHOTIExp,FanZahidRoomTempBiBrExp}.   However, the experimental data attributed to helical higher-order topology has also attracted alternative explanations~\cite{BismuthScrewProposal,JenDefect,BismuthSurfaceBallisticExp}.  Unlike AXIs, $\mathcal{T}$-symmetric helical HOTIs exhibit trivial axion angles $\theta\text{ mod }2\pi=0$ and are therefore non-axionic.  To date, there does not yet exist a $\theta$-like bulk topological field theory for non-axionic HOTIs to provide clarity for the experimental data~\cite{BarryBenCDW,WiederAxion,MTQC}.

In this work, we present novel defect and static flux response effects in 3D insulators, which provide experimentally observable signatures of fragile and non-axionic higher-order topology in solid-state materials (see Table~\ref{tb:ResultsMain}).  We begin below by reviewing spin-charge separation in non-interacting electronic materials.  We then introduce a more general formulation of Eq.~(\ref{eq:Burgers}) that captures the dislocation bound states of all possible topologically nontrivial insulating phases, including FTIs and OALs; this formulation is based on a mapping from $(d-1)$-dimensional [$(d-1)$-D] subspaces of the BZ to $(d-1)$-D real-space surfaces in $d$-D crystals with $(d-2)$-D defects.  Next, we show that our extended formulation of topological defect response captures all previously identified topological electronic crystal dislocation states, and reveals the existence of higher-order end (HEND) states bound to the surface and corner terminations of screw and edge dislocations in FTIs, OALs, and HOTIs [see Supplementary Note (SN)~\ref{sec:numerics} for numerical defect-state calculation details].  We analytically and numerically demonstrate that 0D HEND states are equivalent to the fractionally charged or spin-charge-separated corner states of 2D FTIs and OALs, and are anomalous, intrinsic consequences of the bulk electronic structure.  Using tight-binding calculations (detailed in SN~\ref{sec:numerics}), we specifically demonstrate the presence of topological HEND states in 3D HOTIs and weak FTIs driven by double band inversion~\cite{YoungkukMonopole,HOTIBismuth,TMDHOTI} on the BZ boundary.  Lastly, we use density functional theory (DFT) to demonstrate the presence of intrinsic HEND corner states on edge dislocation networks in the 3D TCI~\cite{HsiehTCI} and HOTI~\cite{HOTIBernevig} SnTe (SN~\ref{sec:DFTSnTe}).  Following our crystal-defect calculations, we next extend the TI and TCI magnetic flux-threading analyses in Refs.~\onlinecite{QHZ,QiFlux,AshvinFlux,FranzWormhole,Vlad2D} to $\mathcal{T}$-symmetric helical HOTIs.  Below and in SN~\ref{sec:kpFlux},~\ref{sec:fluxfluxtubetopologymapping}, and~\ref{sec:numerics2}, we first reproduce the earlier results of Refs.~\onlinecite{QHZ,QiFlux,AshvinFlux,FranzWormhole,Vlad2D} by analytically and numerically demonstrating the static $\pi$-flux response of 2D TIs and Chern insulators, as well as 3D AXIs.  We then demonstrate the existence of a novel quantized $\pi$-flux response in $\mathcal{I}$- and $\mathcal{T}$-symmetric HOTIs.  Specifically, we show that a pair of static $\pi$-flux tubes in an $\mathcal{I}$- and $\mathcal{T}$-symmetric HOTI together binds an odd (anomalous) number of chargeless spinons per surface at a half system filling, suggesting that the bulk exhibits a novel form of quantized ``magneto-spinon polarizability'' (MSP).  Because a half-filled pair of fluxes in an isolated 2D TI binds an even number of chargeless spinons, then our results further imply that each gapped surface of an $\mathcal{I}$- and $\mathcal{T}$-symmetric HOTI is topologically equivalent to ``half'' of a 2D TI.  We conclude by discussing experimental venues for observing the HEND states and response effects introduced in this work.

\begin{figure}[t]
\includegraphics[width=\columnwidth]{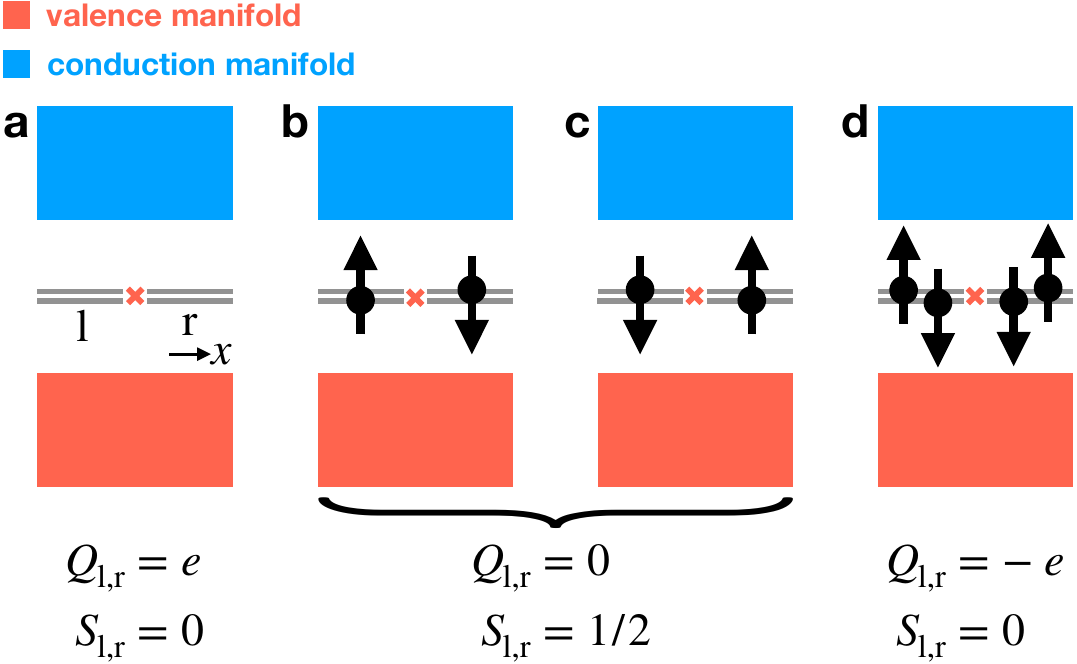}
\caption{{\bf Spin-charge-separated Kramers pairs of defect or flux states.} $\bs{a-d}$~An inversion- ($\mathcal{I}$-) related pair of Kramers pairs of 0D defect or flux states in a spinful, time-reversal- ($\mathcal{T}$-) symmetric insulator (where the $\mathcal{I}$ center is represented with a red $\times$ symbol in~$\bs{a-d}$).  $\bs{b,c}$~When the Fermi level lies at charge neutrality, each Kramers pair is filled by only a single electron, and therefore carries an excess chargeless spin-1/2 moment ($Q=0$, $S=1/2$).  Hence at half filling, and taking the spins of the electrons occupying each pair of states to point in opposite directions, $\mathcal{I}$ (which relates the positions of the Kramers pairs) and $\mathcal{T}$ symmetries are ``softly'' broken~\cite{WilczekAxion,GoldstoneWilczek,WiederAxion,HingeSM}, and each half-filled Kramers pair of states forms an effective spinon quasiparticle with a free-angle spin-1/2 moment (depicted in~$\bs{b,c}$ in configurations that preserve $\mathcal{I}\times\mathcal{T}$ symmetry).  By~$\bs{a}$ removing or~$\bs{d}$ adding two electrons to the system (one electron per Kramers pair), we may realize a system configuration in which each Kramers pair respectively carries a net charge of $\pm e$ (taking electrons to carry a charge $-e$), but carries a net-zero spin ($Q=\pm e$, $S=0$).  Hence, each Kramers pair of states either carries chargeless spin or spinless charge, and therefore exhibits the same reversed spin-charge relations as the solitons in polyacetylene~\cite{HeegerReview}.}
\label{fig:spinCharge_main}
\end{figure}

{
\vspace{0.15in}
\centerline{\bf Results}
\vspace{0.08in}
}

\textbf{Review of Spin-Charge Separation without Interactions} -- Throughout this work, we will demonstrate the existence of 0D defect and flux bound states with spin-charge separation in non-interacting insulating crystals.  Hence, before discussing defect and flux states in 2D and 3D insulators, we will briefly review spin-charge separation in non-interacting $\mathcal{T}$- and spin-rotation- [SU(2)-] invariant systems as a generalization of the familiar charge (fermion number) fractionalization previously discussed by Jackiw, Rebbi, Goldstone, and Wilczek~\cite{JackiwRebbi,WilczekAxion,GoldstoneWilczek,WiederAxion,HingeSM}.

\begin{figure*}[t]
\includegraphics[width=0.85\textwidth]{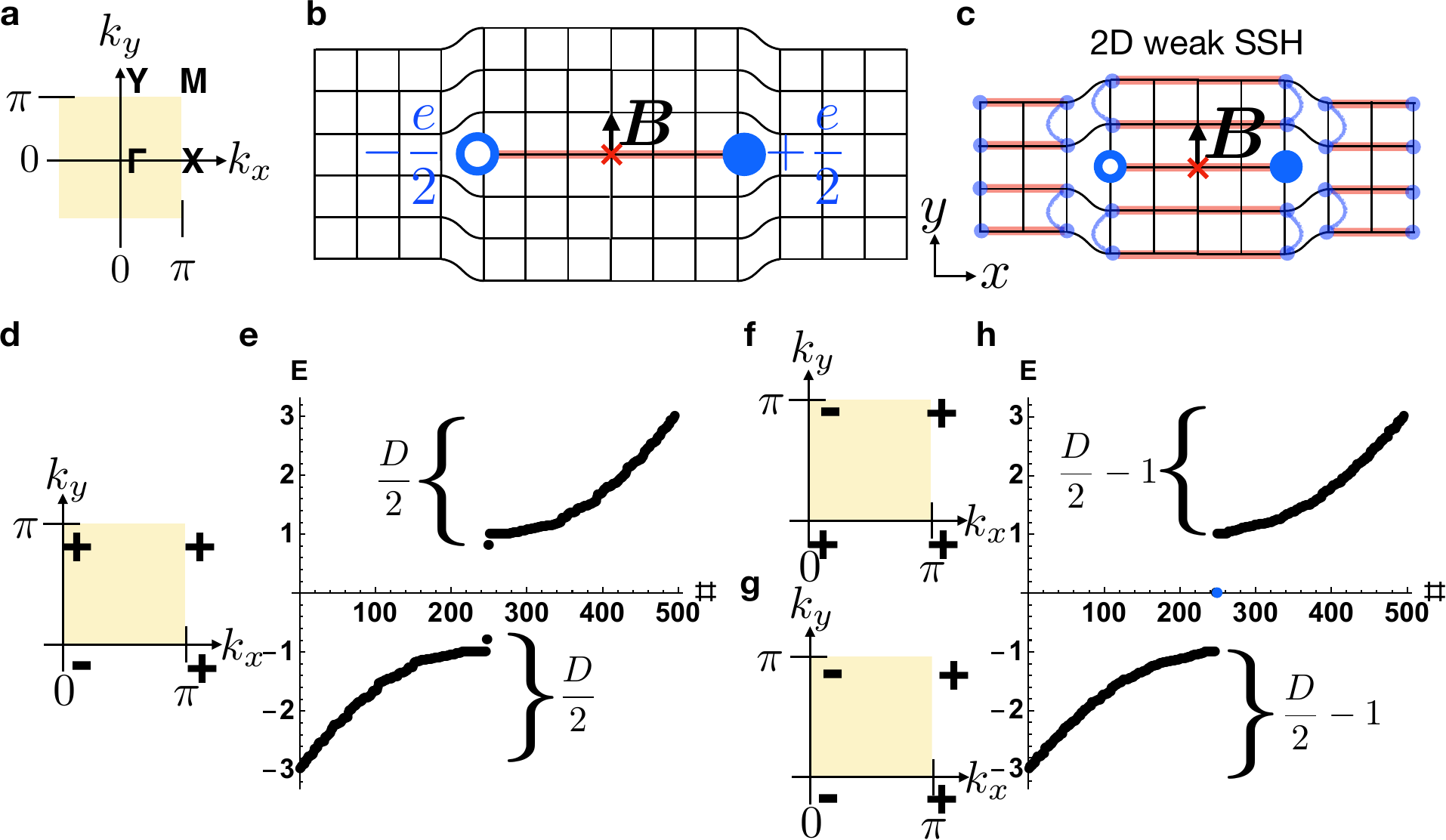}
\caption{{\bf First-order 0D dislocation states in 2D crystals from 1D polarization topology.} $\bs{a}$~The bulk Brillouin zone (BZ) of a 2D rectangular magnetic crystal with only $\mathcal{I}$ symmetry.  $\bs{b}$~An $\mathcal{I}$-related pair of 0D dislocations with Burgers vector $\bs{B} = \hat{y}$ in an $\mathcal{I}$-symmetric crystal, where the global $\mathcal{I}$ center is represented with a red $\times$ symbol.  $\bs{d}$-$\bs{h}$ Bulk parity ($\mathcal{I}$) eigenvalues and periodic-boundary-condition (PBC) energy spectra for the defect in $\bs{b}$ when the bulk is equivalent to $\bs{d}$ a $|C|=1$ Chern insulator with band inversion at $\Gamma$, $\bs{f}$ a $|C|=1$ Chern insulator with band inversion at $Y$, $\bs{g}$ a weak $y$-directed array $\bs{c}$ of $x$-directed Su-Schrieffer-Heeger (SSH) chains~\cite{HeegerReview}.  Anomalous 0D defect states $\bs{h}$ with charge $\pm e/2$ are present in cases $\bs{f}$ and $\bs{g}$, but not $\bs{d}$, which instead exhibits the trivial PBC spectrum in $\bs{e}$ [Eq.~(\ref{eq:SSH})].  Specifically, the spectrum in~$\bs{e}$ may be deformed to that of a trivial insulator (\emph{i.e.} a finite-sized insulator without midgap 0D states or without an imbalance in the number of states above or below the gap) without breaking $\mathcal{I}$ symmetry or closing the bulk gap, whereas the spectrum in~$\bs{h}$ cannot.  Hence, as defined in Refs.~\onlinecite{WiederAxion,HingeSM,WladCorners}, the midgap dislocation states in~$\bs{h}$ are filling-anomalous.  Next, by considering the limit in~$\bs{c}$ in which the bulk is equivalent to a decoupled array of SSH chains, we find that the two dislocations correspond to the ends of a ``leftover'' SSH chain that is decoupled from the bulk.  This implies that the $\pm e/2$-charged defect states are equivalent to the end states of an $\mathcal{I}$-symmetric SSH chain~\cite{HeegerReview} (red line in $\bs{b}$), and thus persist under the relaxation of particle-hole symmetry~\cite{WilczekAxion,GoldstoneWilczek,WiederAxion,HingeSM}. The explicit details of the numerical calculations shown in this figure are provided in SN~\ref{sec:2Dtoysubsec}.}
\label{fig:2D}
\end{figure*}

We begin by considering two $\mathcal{I}$-related pairs of topological defects or flux tubes in a 2D or 3D insulator that each bind a pair of 0D states (four degenerate single-particle states in total), taking each pair of states to be half-filled at charge neutrality (Fig.~\ref{fig:spinCharge_main}~$\bs{b,c}$).  The arguments below do not depend on whether the twofold degeneracy of each pair of states is enforced by spinful $\mathcal{T}$ or SU(2) symmetry, and therefore for simplicity we will focus on the case in which the two states within each pair are time-reversal (Kramers) pairs.  Enforcing $\mathcal{I} \times \mathcal{T}$ symmetry (where we have denoted a global $\mathcal{I}$ center with a red $\times$ symbol in Fig.~\ref{fig:spinCharge_main}), there is one filled state per Kramers pair.  Hence, each Kramers pair carries a balanced (net-zero) charge with respect to charge neutrality, but necessarily ``softly'' breaks $\mathcal{T}$ symmetry, because each pair of states is filled with an unpaired spin-1/2 degree of freedom.  We emphasize that without a spin conservation symmetry such as $s^{z}$, however, each unpaired electron is not required to exhibit a quantized spin projection along a particular high-symmetry axis.

Next, if the system is doped away from charge neutrality by adding two more electrons, $\mathcal{T}$ and global $\mathcal{I}$ symmetries can conversely be satisfied individually (Fig.~\ref{fig:spinCharge_main}~$\bs{d}$).  In the system configuration with two extra electrons, each fully-filled Kramers pair of states carries a charge $-e$ (taking electrons to have charge $-e$).  Unlike in the previous system configuration with chargeless spin-1/2 0D states at zero doping depicted in Fig.~\ref{fig:spinCharge_main}~$\bs{b,c}$, at a system doping of $-2e$, each Kramers pair of states is charged, but exhibits a net-zero spin, because $\mathcal{T}$ [or SU(2)] symmetry pairs electrons with reversed spins.  Similarly, if we remove one electron from each Kramers pair of states in Fig.~\ref{fig:spinCharge_main}~$\bs{b,c}$, then we realize a system configuration in which there is a total charge of $+2e$, implying that each fully empty pair of states carries a charge $+e$ and does not carry an electron spin (Fig.~\ref{fig:spinCharge_main}~$\bs{a})$.  Hence, the 0D Kramers pairs of states exhibit the same well-established spin-charge separation and reversed spin-charge relations as the solitons in polyacetylene~\cite{HeegerReview}.

\begin{figure*}
\includegraphics[width=0.9\textwidth]{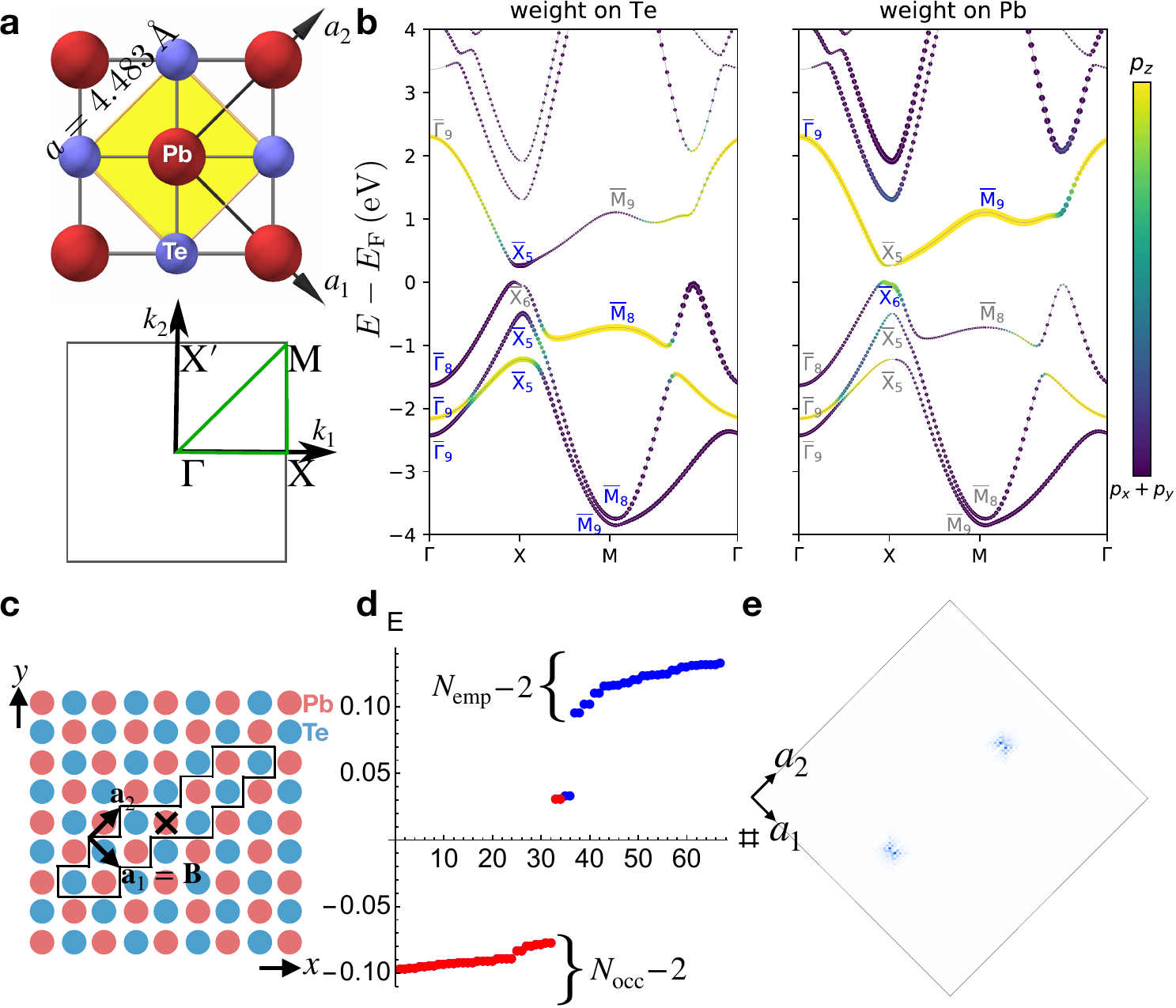}
\caption{{\bf First-order dislocation states in 2D PbTe monolayers.} $\bs{a}$~The crystal structure of monolayer PbTe and the bulk BZ.  The yellow diamond in~$\bs{a}$ indicates the primitive cell.  A PbTe monolayer~\cite{LiangTCIMonolayer,PbTeMonolayer1} has fourfold rotation, $\mathcal{I}$, and mirror symmetries (layer group $p4/mmm1'$~\cite{HingeSM}).  $\bs{b}$~Band structure of a PbTe monolayer along the high-symmetry lines of the 2D BZ in $\bs{a}$.  The size of the circle at each plotted point in ${\bs b}$ indicates the spectral weight on the Te (left panel) and Pb (right panel) atoms, and the color bars indicate the orbital character of each Bloch state on a scale of $p_z$ or $p_{x,y}$ orbitals.  We have additionally labeled the irreducible small corepresentations at the symmetry-independent TRIM points in the BZ in $\bs{a}$ (see SN~\ref{sec:DFTPbTe} for details).  The bands in ${\bs b}$ are inverted at $X$ and $X'$, driving the bulk into a 2D mirror TCI phase~\cite{LiangTCIMonolayer,PbTeMonolayer1} with mirror Chern number $C_{M_{z}}=2$ and nontrivial weak (partial) SSH indices $\bs{M}^{\mathrm{SSH}}_{\nu}=({\bs b}_{1} + {\bs b}_{2})/2$ (see SN~\ref{sec:weakSSH} and~\ref{sec:DFTPbTe}).  $\bs{c}$~Schematic of our real-space implementation of an $\mathcal{I}$-related pair of $\bs{B} = \bs{a}_1$ point dislocations in a Wannier-based tight-binding model of a PbTe monolayer obtained from first-principles calculations (details provided in SN~\ref{sec:DFTPbTe}), where the $\mathcal{I}$ center is marked with a black $\times$ symbol.  In~$\bs{c}$, the sites enclosed within the black line have been removed to implement the pair of point dislocations.  $\bs{d}$~PBC energy spectrum of a tight-binding model of PbTe with the $\mathcal{I}$-related pair of $\bs{B} = \bs{a}_1$ point dislocations shown in $\bs{c}$; there are four midgap, filling-anomalous~\cite{TMDHOTI,WiederAxion,WladCorners,HingeSM} dislocation states, consistent with Eq.~(\ref{eq:SSH}) [$\bs{B}\cdot \bs{M}^{\mathrm{SSH}}_{\nu}\text{ mod }2\pi = \pi$].  $\bs{e}$~The real-space localization of the four midgap states in $\bs{d}$, which subdivide into two $\mathcal{I}$-related Kramers pairs.  One Kramers pair of states is localized on each dislocation core, and corresponds when half-filled to a chargeless, spin-1/2 quasiparticle (\emph{i.e.} a spinon) that is equivalent to the end state of a spinful SSH chain.}
\label{fig:monolayer} 
\end{figure*}

\textbf{Defect Response of Inversion-Symmetric 2D Insulators} -- In this work, we rigorously establish a prescription for identifying insulators that bind anomalous 0D defect states as a consequence of the bulk topology.  We will first here numerically demonstrate that $\mathcal{I}$-symmetric 2D insulators with band inversion at high-symmetry points on the 2D BZ boundary exhibit a nontrivial dislocation response.  We will then bolster the numerical results through first-principles and tight-binding calculations demonstrating a nontrivial first-order defect response in PbTe monolayers (see the Methods section and SN~\ref{sec:DFTPbTe} for calculation details).

We begin by considering a simple magnetic 2D insulator with only rectangular lattice translations $T_{x,y}$ and $\mathcal{I}$ symmetry, such that the system respects the symmetries of magnetic layer group $p\bar{1}$~\cite{MTQC,HingeSM} (Fig.~\ref{fig:2D}, numerical details provided in SN~\ref{sec:2Dtoysubsec}).  We consider the case in which the pristine crystal is initially furnished with a single occupied, uncoupled, spinful $s$ orbital and a single unoccupied, uncoupled, spinful $p$ orbital -- both at the origin of each unit cell.  This implies that initially, the electronic structure at each $\mathcal{I}$-invariant crystal momentum (TRIM point) consists of one occupied state with a positive parity ($\mathcal{I}$) eigenvalue and one unoccupied state with a negative parity eigenvalue~\cite{QuantumChemistry,MTQC}.

Next, by tuning model parameters to invert the bands at different TRIM points (Fig.~\ref{fig:2D}~$\bs{a}$), we may realize several different insulating phases.  When only one of the parity ($\mathcal{I}$) eigenvalues of the occupied band is negative, the bulk is a symmetry-indicated Chern insulator with Chern number $C \text{ mod }2=1$~\cite{MTQC}.  In Fig.~\ref{fig:2D}~$\bs{d}$ (Fig.~\ref{fig:2D}~$\bs{f}$), we show the occupied parity eigenvalues of a $|C|=1$ Chern insulator driven by band inversion at $\Gamma$ ($Y$).  Inserting a pair of dislocations with Burgers vector $\bs{B} =\hat{y}$ that preserves global $\mathcal{I}$ symmetry (Fig.~\ref{fig:2D}~$\bs{b}$) and calculating the energy spectrum of the corresponding tight-binding model with periodic boundary conditions (PBC), we observe a pair of anomalous midgap states with charges $\pm e/2$~\cite{HeegerReview,WilczekAxion,GoldstoneWilczek,WiederAxion,HingeSM} for the parity eigenvalue in Fig.~\ref{fig:2D}~$\bs{f}$, but not for the parity eigenvalues in Fig.~\ref{fig:2D}~$\bs{d}$, reproducing the conclusions of Refs.~\onlinecite{QiDefect2,Vlad2D}.  Specifically, the spectrum in Fig.~\ref{fig:2D}~$\bs{e}$ is the same as that of a trivial (uninverted) insulator with two $\mathcal{I}$-related point dislocations.  On the other hand, the spectrum in Fig.~\ref{fig:2D}~$\bs{h}$ cannot be symmetrically deformed into the spectrum of an $\mathcal{I}$-symmetric trivial insulator with two point dislocations.  Hence, as defined in Refs.~\onlinecite{WiederAxion,HingeSM,WladCorners}, the midgap dislocation states in~$\bs{h}$ are filling-anomalous.  Throughout this work, we will use PBC and filling anomalies to numerically identify topologically nontrivial 0D defect- and flux-state responses in insulating crystals with $\mathcal{I}$ or $\mathcal{I}$ and $\mathcal{T}$ symmetries.

To understand the pattern of dislocation responses for the Chern insulators in Fig.~\ref{fig:2D}~$\bs{d},\bs{f}$, we next form a new insulator that is equivalent to a weak, $y$-directed array of $x$-directed, $\mathcal{I}$-symmetric Su-Schrieffer-Heeger (SSH) chains~\cite{HeegerReview} (Fig.~\ref{fig:2D}~$\bs{c,g}$); we observe that $\bs{B} = \hat{y}$ dislocations in this Wannierizable~\cite{QuantumChemistry,MTQC} ($C=0$) insulator also bind $\pm e/2$ charges.  By analogy to the weak TI discussion in Ref.~\onlinecite{AshvinScrewTI}, the center red line in the weak SSH array in Fig.~\ref{fig:2D}~$\bs{c}$ represents a ``leftover'' SSH chain that may be adiabatically decoupled from the bulk crystal and binds $\pm e/2$ charges on its ends, the dislocations.  The results of Fig.~\ref{fig:2D}~$\bs{d}$-$\bs{h}$ can be summarized by defining a weak polarization invariant $\bs{M}_{\nu}^{\mathrm{SSH}}=\pi(n_{XM},n_{YM})$, where $n_{ab}$ is the $\mathbb{Z}_{2}$ SSH polarization invariant of the occupied bands along the BZ-edge line $ab$, such that for example, the index $n_{XM}$ is nontrivial for $y$-directed SSH chains (see SN~\ref{sec:weakSSH}).  Analogously to the weak-index vectors of $\mathcal{T}$-symmetric 3D insulators~\cite{FuKaneInversion}, $\bs{M}^{\mathrm{SSH}}_{\nu}$ can only realize values equal to half-integer linear combinations of 2D reciprocal lattice vectors.  For the insulators in Fig.~\ref{fig:2D}~$\bs{d},\bs{f},\bs{g}$, $\bs{M}_{\nu}^{\mathrm{SSH}} = (0,0),$ $(0,\pi)$, and $(0,\pi)$, respectively.  Hence, for magnetic 2D insulators with $\mathcal{I}$ symmetry and integer Burgers vectors $\bs{B}$, we conclude that dislocations bind anomalous $\pm e/2$ charges if and only if:
\begin{equation}
\bs{B}\cdot\bs{M}^{\mathrm{SSH}}_{\nu}\text{ mod }{ 2\pi }=\pi,
\label{eq:SSH}
\end{equation}
in direct analogy to Eq.~(\ref{eq:Burgers}).

In SN~\ref{sec:weakSSH}, we additionally extend Eq.~(\ref{eq:SSH}) to $\mathcal{I}$- and $\mathcal{T}$-symmetric 2D insulators by instead computing the BZ-boundary weak time-reversal (partial) polarization indices, which reduce to the polarization per spin sector in the limit of $s^{z}$-spin conservation symmetry~\cite{TRPolarization}.  For $\mathcal{I}$- and $\mathcal{T}$-symmetric 2D insulators with nontrivial $\bs{M}_{\nu}^{\mathrm{SSH}}$ vectors, we show in SN~\ref{sec:weakSSH} and~\ref{sec:2DtoysubsecTRS} that dislocations satisfying Eq.~(\ref{eq:SSH}) bind spin-charge-separated 0D solitons, rather than $\pm e/2$ charges.

To further confirm Eq.~(\ref{eq:SSH}) and its $\mathcal{T}$-invariant extension, we have performed first-principles calculations of the electronic structure of a PbTe monolayer~\cite{LiangTCIMonolayer,PbTeMonolayer1} (layer group $p4/mmm1'$) [Fig.~\ref{fig:monolayer}~$\bs{a}$].  The lattice vectors of a PbTe monolayer are given by:
\begin{equation}
\bs{a}_1 = (1/2,-1/2),\ \bs{a}_2 = (1/2,1/2),
\end{equation} 
and the reciprocal lattice vectors are given by:
\begin{equation}
\bs{b}_1 = 2\pi(1,-1),\ \bs{b}_2 = 2\pi(1,1). 
\label{eq:monolayerGvecsMain}
\end{equation}
Previous works~\cite{LiangTCIMonolayer,PbTeMonolayer1} have demonstrated that PbTe monolayers are mirror-Chern $C_{M_{z}}=2$ TCIs driven by band inversions at the $X$ [${\bs k}_{X}={\bs b}_{1}/2$] and $X'$ [${\bs k}_{X'} = {\bs b}_{2}/2$] TRIM points [Fig.~\ref{fig:monolayer}~$\bs{b}$].  Computing the weak partial polarization indices along $XM$ and $X'M$, we determine that PbTe monolayers carry a nontrivial dislocation response vector:
\begin{equation}
\bs{M}^{\mathrm{SSH}}_{\nu}=({\bs b}_{1} + {\bs b}_{2})/2,
\label{eq:PbTe2DMain}
\end{equation}
where the details of our calculation are provided in SN~\ref{sec:DFTPbTe}.

To probe the dislocation response, we next construct a Wannier-based tight-binding model of a PbTe monolayer and insert an $\mathcal{I}$-related pair of $\bs{B} = \bs{a}_1$ point dislocations, as shown in Fig.~\ref{fig:monolayer}~$\bs{c}$.  In the dislocation geometry with PBC, the energy spectrum is filling-anomalous (Fig.~\ref{fig:monolayer}~$\bs{d}$), with each dislocation binding a Kramers pair of states (Fig.~\ref{fig:monolayer}~$\bs{e}$) where, at half filling, each pair carries a net-zero charge and a free-angle $|{\bs S}|=1/2$ spin moment (\emph{i.e.} a spinon).  Hence, the Kramers pairs of dislocation bound states in PbTe monolayers are equivalent to the spin-charge-separated end states of a spinful SSH chain~\cite{HeegerReview,TRPolarization}.  In summary, the appearance of filling-anomalous dislocation bound states in an $\mathcal{I}$-symmetric defect geometry in a PbTe monolayer provides further evidence for a first-order dislocation response in 2D insulators whose pristine electronic structure and dislocation Burgers vectors satisfy Eq.~(\ref{eq:SSH}).

\textbf{Defect Response from Momentum-Space Band Topology} -- We will next describe proofs -- summarized in the Methods section and provided in complete detail in SN~\ref{sec:kpEdge},~\ref{sec:kpScrew},~\ref{sec:defecttopology}, and~\ref{sec:FrankScrew} -- explicitly linking the topology of pristine, insulating crystals to the electronic states bound to dislocations.  In this work, we specifically show that dislocations with integer Burgers vectors~\cite{MerminReview} bind edge and corner modes deriving from the momentum-space topology of lower-dimensional surfaces of the BZs of pristine crystals.  In the 3D case -- which is most relevant to solid-state materials -- we use this mapping to analytically demonstrate that the corners and ends of 1D edge and screw dislocations in 3D insulators can bind anomalous 0D HEND states as an intrinsic consequence of nontrivial bulk topology.

A central result of this work is the recognition that Eqs.~(\ref{eq:Burgers}) and~(\ref{eq:SSH}) represent specific cases of a more general statement, which we will summarize below.  First, for a $d$-D crystal hosting $(d-2)$-D dislocations with integer-valued Burgers vectors~\cite{TeoKaneDefect}, the exact location of the real-space $(d-1)$-D surface spanning the dislocations is a gauge-dependent quantity~\cite{MerminReview} (it can be moved at zero energy cost and changed by redefinition), while the locations of the $(d-2)$-D dislocations are gauge-invariant, as they carry quantized and measurable Burgers vectors.  Specifically, $\bs{B}$ is defined by measuring the total displacement along a loop around a dislocation; though the amount of displacement assigned to a given $(d-1)$-D surface between a pair of dislocations represents a numerical choice of gauge, the location of each dislocation and the value of the total displacement $\bs{B}$ are conversely gauge-independent.  In the momentum-space $d$-D Hamiltonians of pristine insulators with the same bulk topology as the crystal with dislocations, we next consider the topology in the $(d-1)$-D BZ-boundary surface defined by the normal momentum vector ${\bs M}$ [\emph{e.g.}, in the $k_{x}=\pi$ plane of a 3D insulator, ${\bs M}=(\pi,0,0)$].  In this work, we find that the $(d-1)$-D position-space surface spanning a pair or closed loop of dislocations -- regardless of its gauge-dependent shape -- hosts the same topological boundary states as a $(d-1)$-D crystal whose bulk topology is equivalent to that of the $(d-1)$-D BZ-boundary surface defined by ${\bs M}$, provided that two conditions are satisfied:
\begin{enumerate}
\item{${\bs B}\cdot{\bs M}\text{ mod }2\pi = \pi$.}
\item{The position-space system with dislocations preserves the same symmetries that enforce the momentum-space bulk $(d-1)$-D topology in the $(d-1)$-D BZ surface defined by ${\bs M}$.}  
\end{enumerate}
In a weak TI~\cite{FuKaneInversion,AshvinScrewTI}, the necessary symmetry is $\mathcal{T}$; however as shown in this work, the required symmetry may also be spatial (\emph{e.g.} $\mathcal{I}$).

To reconcile our results with previous works, we have formulated two alternative and equivalent sets of proofs demonstrating the aforementioned dislocation topological mapping from momentum space to position space.  Our proofs reproduce the results of all previous studies of crystal dislocation bound states with integer $\bs{B}$~\cite{AshvinScrewTI,TeoKaneDefect,QiDefect2,Vlad2D,TanakaDefect,VladScrewTI}.  First, building upon the ``cutting'' and ``gluing'' construction of topological defect states developed in Ref.~\onlinecite{AshvinScrewTI} to predict helical dislocation modes in weak TIs~\cite{FuKaneInversion}, we have employed $k\cdot p$ theory to predict HEND states in 3D crystals (see SN~\ref{sec:kpEdge} and~\ref{sec:kpScrew}).  Next, we use more general arguments to demonstrate that $(d-2)$-D dislocations in $d$-D crystals can map $(d-1)$-D BZ surfaces to $(d-1)$-D real-space surfaces, leading in 3D crystals to the presence of 1D and 0D topological defect states (see SN~\ref{sec:defecttopology} and~\ref{sec:FrankScrew}).

Through both sets of proofs, we deduce that given an $\mathcal{I}$-symmetric, $\mathcal{T}$-broken 3D insulator with vanishing weak Chern numbers~\cite{QiDefect2,TeoKaneDefect,MTQC,BarryBenCDW}, $\mathcal{I}$-symmetric dislocations with Burgers vector $\bs{B}$ will bind anomalous 0D states at $\mathcal{I}$-related locations along the set of dislocations if and only if:
\begin{equation}
\bs{B}\cdot \bs{M}_{\nu}^{\mathrm{F}}\text{ mod }2\pi = \pi,
\label{eq:HEND}
\end{equation}
where $\bs{M}_{\nu}^{\mathrm{F}}=\pi(\nu_{x}^{\mathrm{F}},\nu_{y}^{\mathrm{F}},\nu_{z}^{\mathrm{F}})$ is a new weak index vector characterizing which of the BZ-boundary planes host Hamiltonians that are topologically equivalent to the $\mathcal{I}$-symmetric 2D FTI introduced in Refs.~\onlinecite{TMDHOTI,WiederAxion}, or the OAL that results from adding trivial bands without anomalous corner charges to the $\mathcal{I}$-symmetric 2D FTI.  Like the weak-index vectors of $\mathcal{T}$-symmetric 3D insulators~\cite{FuKaneInversion}, $\bs{M}_{\nu}^{\mathrm{F}}$ can only realize values equal to half-integer linear combinations of 3D reciprocal lattice vectors.  In SN~\ref{sec:weakFragile}, we rigorously define $\bs{M}_{\nu}^{\mathrm{F}}$ using elementary band representations~\cite{QuantumChemistry,MTQC}.  Heuristically, $\nu_{i}^{\mathrm{F}}$ is nontrivial when the Hamiltonian in the $k_{i}=\pi$ BZ boundary plane differs by $2+4n$ band inversions from an $\mathcal{I}$-symmetric 2D trivial atomic limit (counting each state individually, as opposed to Kramers pairs of states).

Analogously to $\bs{M}^{\mathrm{SSH}}_{\nu}$ [defined in the text preceding Eq.~(\ref{eq:SSH})], $\bs{M}_{\nu}^{\mathrm{F}}$ can also be adapted to $\mathcal{I}$- and $\mathcal{T}$-symmetric 3D systems by analyzing nonmagnetic insulators with four band inversions (two Kramers pairs) in a BZ boundary plane.  In SN~\ref{sec:FTIwithT} and~\ref{sec:nested}, we respectively define the $\mathcal{T}$-symmetric invariant $\bs{M}_{\nu}^{\mathrm{F}}$, using elementary band representations and by introducing a nested Wilson loop formulation~\cite{multipole,WladTheory,HOTIBernevig,HingeSM,WiederAxion,TMDHOTI} of partial nested Berry phase (which reduces to the nested Berry phase per spin sector in the limit of $s^{z}$-spin conservation symmetry).  As with the $\mathcal{T}$-symmetric generalization of $\bs{M}^{\mathrm{SSH}}_{\nu}$ discussed earlier in the context of PbTe monolayers [see Eq.~(\ref{eq:PbTe2DMain}) and the surrounding text], for $\mathcal{I}$- and $\mathcal{T}$-symmetric 3D insulators with nontrivial $\bs{M}_{\nu}^{\mathrm{F}}$ vectors, the corners of edge dislocations and the ends of screw dislocations satisfying Eq.~(\ref{eq:HEND}) bind spin-charge-separated 0D solitons, rather than $\pm e/2$ charges.

\begin{figure*}[t]
\includegraphics[width=\textwidth]{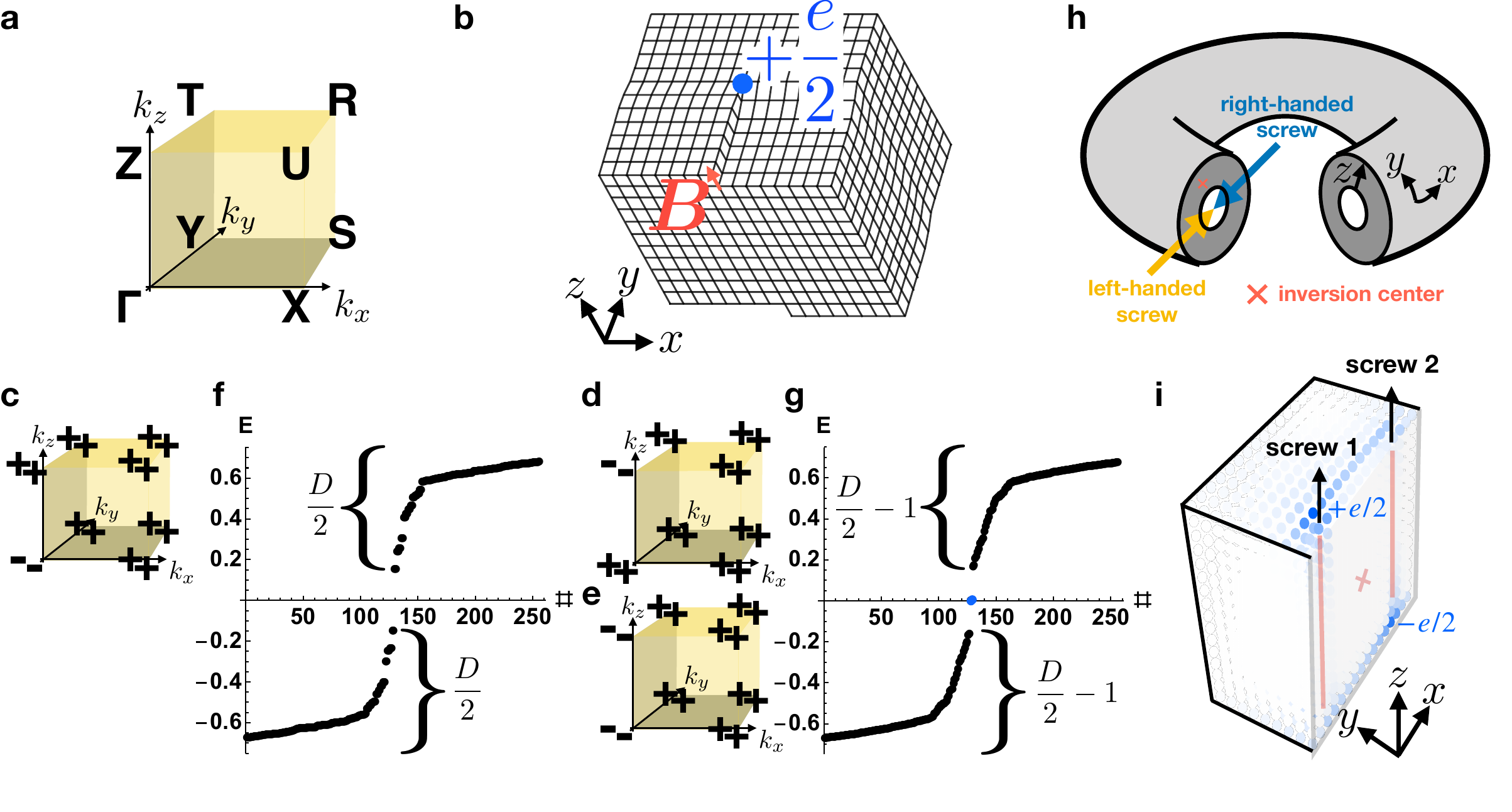}
\caption{{\bf Higher-order end dislocation states in 3D crystals from 2D fragile topology.} $\bs{a}$~The bulk BZ of a 3D orthorhombic magnetic crystal with only $\mathcal{I}$ symmetry.  $\bs{b}$~A screw dislocation in an $\mathcal{I}$-symmetric crystal with Burgers vectors $\bs{B} =\hat{z}$. $\bs{c}$-$\bs{g}$ Bulk parity ($\mathcal{I}$) eigenvalues and hollow-doughnut-boundary-condition (HDBC) energy spectra for the defects in $\bs{b,h}$ when the bulk is topologically equivalent to $\bs{c}$ an $\mathcal{I}$-symmetric axion insulator (AXI)~\cite{HOTIBernevig,WladTheory,FanHOTI,TMDHOTI,MTQC,VDBHOTI,WiederAxion} with double band inversion at $\Gamma$, $\bs{d}$ an AXI with double band inversion at $Z$, and $\bs{e}$ a weak stack of $\mathcal{I}$-symmetric 2D fragile TIs (FTIs) with $\pm e/2$ corner charges~\cite{TMDHOTI,WiederAxion}.  $\bs{h,i}$~The HDBC geometry is defined by imposing periodic boundary conditions in two directions (here $x$ and $y$), and open boundary conditions in the remaining direction (here $z$).  The screw dislocations in~$\bs{h,i}$ are related by global $\mathcal{I}$ symmetry (red $\times$ symbol in $\bs{h,i}$).  Filling-anomalous higher-order end (HEND) states with charge $\pm e/2$ (the midgap states in $\bs{g}$) are present at two of the four $\mathcal{I}$-related ends of the two screw dislocations in $\bs{d}$ and $\bs{e}$ (the top end of screw 1 and the bottom end of screw 2 in $\bs{i}$), but are absent in $\bs{c}$ [Eq.~(\ref{eq:HEND})], which instead displays the trivial HDBC spectrum in $\bs{f}$ (see SN~\ref{sec:3Dscrew} for calculation details).  The $\pm e/2$-charged HEND states of the insulators in $\bs{d,e}$ are equivalent to the corner modes of the 2D FTI stacked to form $\bs{e}$, and thus persist under the relaxation of particle-hole symmetry~\cite{TMDHOTI,WiederAxion,HingeSM} (see SN~\ref{sec:kpScrew},~\ref{sec:FrankScrew}, and~\ref{subsec:weakFTIstack}).  Each gapped dislocation therefore carries an anomalous half of the $e/2$ polarization of an isolated SSH chain in the case in which global $\mathcal{I}$ symmetry is enforced, as it is in our numerics for the purpose of detecting filling anomalies.}
\label{fig:AXI}
\end{figure*}

\begin{figure}
\includegraphics[width=\columnwidth]{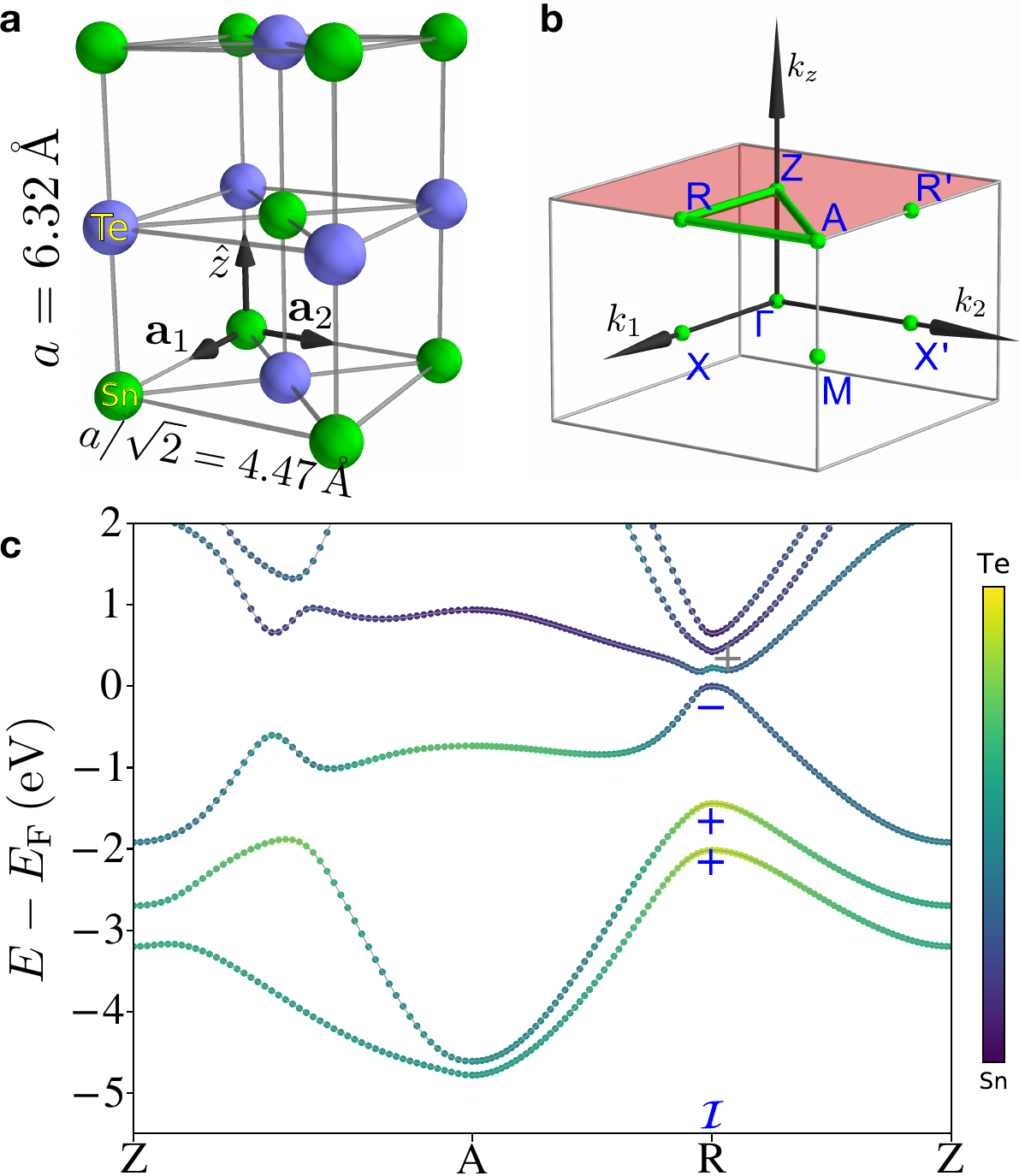}
\caption{{\bf Nontrivial weak partial fragile indices in 3D SnTe.} $\bs{a}$ Crystal structure of 3D SnTe~\cite{HsiehTCI,HOTIBernevig} in a tetragonal supercell that contains four atoms and respects the symmetries of space group 123 $P4/mmm1'$.  $\bs{b}$ The BZ of the tetragonal supercell in the left panel of~$\bs{a}$.  $\bs{c}$~The first-principles electronic structure of SnTe plotted along the path indicated in~$\bs{b}$ with a green line (see SN~\ref{sec:SnTeDFTSPECIFIC} for calculation details).  The bands in $\bs{c}$ exhibit a fourfold degeneracy at all ${\bs k}$ points due to the combined effects of spinful $\mathcal{I} \times \mathcal{T}$ symmetry and supercell BZ folding.  We have specifically employed a supercell geometry that preserves the primitive lattice translation symmetries of SnTe in order to simplify the system geometry when dislocations are inserted.  Bands in the $k_{z}=\pi$ plane are hence fourfold degenerate due to band backfolding and $\mathcal{I}\times\mathcal{T}$ symmetry.  However, the tetragonal supercell only represents a choice of convention, and does not affect the generalization of our results to real SnTe crystals, which are face-centered cubic~\cite{HsiehTCI,HOTIBernevig}.  The $\pm$ signs in $\bs{c}$ denote the parity eigenvalues per Kramers pair of the Bloch states at the TRIM point $R$ [${\bs k}_{R} = {\bs b}_{1}/2$, see Eq.~(\ref{eq:3DSnTeReciprocalVectorMain}) for the definitions of ${\bs b}_{1,2,3}$].  The band structure in $\bs{c}$ indicates that SnTe differs from an unobstructed atomic limit [that is topologically equivalent to 3D PbTe, see Ref.~\onlinecite{HsiehTCI} and SN~\ref{sec:DFTPbTe}] by double band inversions at the $R$ and $R'$ points  [${\bs k}_{R'} = {\bs b}_{2}/2$] in the tetragonal supercell between two pairs of Kramers pairs of states with opposite parity eigenvalues [four valence bands and four conduction bands become inverted at both $R$ and $R'$].  The four band inversions drive the bulk into a fourfold ``rotation-anomaly'' TCI phase~\cite{HOTIBernevig,ChenRotation,ChenTCI} with a nontrivial weak (partial) fragile index vector $\bs{M}^{\mathrm{F}}_{\nu} = ({\bs b}_{1} + {\bs b}_{2})/2$ [see SN~\ref{sec:weakFragile} and~\ref{sec:SnTeDFTSPECIFIC} and the text surrounding Eq.~(\ref{eq:HEND})].}
\label{fig:SnTePristineMain} 
\end{figure}

\textbf{Topological 0D Defect States in 3D Insulators} -- Having analytically established the existence of a new weak index for 2D fragile (and OAL) topology in 3D crystals with $\mathcal{I}$ symmetry -- $\bs{M}_{\nu}^{\mathrm{F}}$ -- we will now numerically confirm the presence of anomalous HEND dislocation states in 3D insulators with $\bs{B}$ and $\bs{M}_{\nu}^{\mathrm{F}}$ vectors that satisfy Eq.~(\ref{eq:HEND}).  We begin by considering a magnetic 3D insulator with only orthorhombic lattice translations $T_{x,y,z}$ and $\mathcal{I}$ symmetry, such that the system respects the symmetries of magnetic space group (SG) 2.4 $P\bar{1}$~\cite{MTQC} (Fig.~\ref{fig:AXI}, numerical details provided in SN~\ref{sec:3Dscrew}).  We take the pristine crystal to initially be furnished with two occupied, uncoupled, spinful $s$ orbitals and two unoccupied, uncoupled, spinful $p$ orbitals -- all at the origin of each unit cell.  This implies that initially, the electronic structure at each TRIM point consists of two occupied states with positive parity eigenvalues and two unoccupied states with negative parity eigenvalues~\cite{QuantumChemistry,MTQC}.

Next, by tuning model parameters to drive double band inversions at different TRIM points~\cite{YoungkukMonopole,HOTIBismuth,TMDHOTI}, we may realize several different 3D insulating phases, including chiral HOTIs (AXIs) and weak stacks of 2D FTIs.  Specifically, if there is an odd total number of double band inversions (recalling that single band inversions give rise to Weyl semimetal phases~\cite{MTQC}), and if the bulk is gapped and all weak Chern numbers vanish, then the system is an $\mathcal{I}$-symmetry-indicated AXI~\cite{HOTIBernevig,WladTheory,FanHOTI,TMDHOTI,MTQC,VDBHOTI,WiederAxion,BarryBenCDW}.   In an AXI phase, the bulk topology can generically be expressed as a pumping cycle of a 2D FTI or OAL with $\pm e/2$-charged 0D corner modes, where the 3D spectral flow of each 0D corner mode manifests as a 1D chiral hinge state~\cite{WiederAxion,TMDHOTI}.  Hence, in an AXI, the weak fragile index $\bs{M}_{\nu}^{\mathrm{F}}$ indicates whether the 2D BZ planes in which the Hamiltonians characterize 2D FTIs and OALs with anomalous corner modes lie in the BZ boundary.

We next insert two $\bs{B}=\hat{z}$ screw dislocations of opposite chiralities (SN~\ref{sec:kpScrew}) at $\mathcal{I}$-related positions into the four-band model taken with hollow-doughnut boundary conditions (HDBC, see Fig.~\ref{fig:AXI}~$\bs{h,i}$) for each of the occupied parity eigenvalue configurations in Fig.~\ref{fig:AXI}~$\bs{c}$-$\bs{e}$.  The HDBC geometry is closely related to the ``Corbino doughnut'' employed in Ref.~\onlinecite{FuKaneInversion} to characterize 3D TIs; however, in this work, we will introduce screw dislocations (and later flux tubes) in a different arrangement than in Ref.~\onlinecite{FuKaneInversion}.  In Fig.~\ref{fig:AXI}~$\bs{f,g}$, we plot the HDBC spectra of the three insulators with the parity eigenvalues listed in Fig.~\ref{fig:AXI}~$\bs{c}$-$\bs{e}$, which respectively are an AXI with $\bs{M}_{\nu}^{\mathrm{F}}=\bs{0}$, an AXI with $\bs{M}_{\nu}^{\mathrm{F}}=\pi\hat{z}$, and a weak $z$-directed stack~\cite{ZhidaFragileAffine} of an $\mathcal{I}$-symmetric 2D FTI, where the weak FTI stack also exhibits $\bs{M}_{\nu}^{\mathrm{F}}=\pi\hat{z}$.  To draw connection with previous works, we note that the $\mathcal{I}$-symmetric weak FTI in Fig.~\ref{fig:AXI}~$\bs{e}$, when cut into a rod geometry, exhibits the same flat-band-like floating hinge states (per spin) as a spinless (spin-doubled) $\mathcal{I}\times\mathcal{T}$-symmetric 3D Stiefel-Whitney insulator~\cite{YoungkukMonopole}.  In Fig.~\ref{fig:AXI}~$\bs{d,e}$, but not $\bs{c}$, alternating ends of the screw dislocations bind filling-anomalous, $\pm e/2$-charged 0D HEND states (Fig.~\ref{fig:AXI}~$\bs{i}$).

This result can be understood by focusing on the weak FTI stack whose occupied parity eigenvalues are shown in Fig.~\ref{fig:AXI}~$\bs{e}$.  In the limit in which the weak FTI is adiabatically deformed into decoupled layers of 2D FTIs and the screw dislocations replaced with edge dislocations (see SN~\ref{sec:kpEdge},~\ref{sec:defecttopology}, and~\ref{subsec:weakFTIstack}), the plane between the dislocations represents a ``leftover'' FTI that may be adiabatically decoupled from the bulk crystal (Fig.~\ref{fig:AXI}~$\bs{h,i}$), analogous to the previous ``leftover'' SSH chain in Fig.~\ref{fig:2D}~$\bs{c}$.  Hence, the HEND states in Fig.~\ref{fig:AXI} are equivalent to the corner charges of the 2D FTI that comprises each layer of the weak stack.  Furthermore, because the gapped 1D edges of 2D $\mathcal{I}$-symmetric FTIs carry anomalous halves of the $e/2$ polarization of an isolated SSH chain when global $\mathcal{I}$ symmetry is enforced~\cite{WiederAxion,TMDHOTI}, then each of the screw dislocations in Fig.~\ref{fig:AXI}~$\bs{i}$ carries only half of the fractionally charged end states of an isolated SSH chain.

To provide further support for the HEND-state response introduced in this work [Eq.~(\ref{eq:HEND})], we will next demonstrate the presence of anomalous HEND states on the corners of edge dislocations with the shortest possible integer Burgers vectors in 3D SnTe crystals.  Through first-principles calculations detailed in the Methods section and SN~\ref{sec:DFTSnTe}, we find in this work that 3D SnTe -- a well-established fourfold rotation-anomaly TCI with helical hinge states~\cite{HsiehTCI,HOTIBernevig,ChenRotation,ChenTCI} -- exhibits a nontrivial HEND-state response vector.  SnTe crystals respect the symmetries of the face-centered-cubic space group (SG) 225  $Fm\bar{3}m1'$.  We begin by, for geometric simplicity, artificially enlarging the unit cell of SnTe into a tetragonal supercell in SG 123 $P4/mmm1'$ [Fig.~\ref{fig:SnTePristineMain}~$\bs{a}$] with lattice vectors given by:
\begin{equation}
\bs{a}_1 = (1/2,-1/2,0),\  \bs{a}_2 = (1/2,1/2,0),\ \bs{a}_3 = (0,0,1),
\end{equation}
in units in which the lattice spacing $a=1$, and reciprocal lattice vectors given by:
\begin{equation}
\bs{b}_1 = 2\pi(1,-1,0),\ \bs{b}_2 = 2\pi(1,1,0),\ \bs{b}_3 = 2\pi(0,0,1).  
\label{eq:3DSnTeReciprocalVectorMain}
\end{equation}

\begin{figure}
\includegraphics[width=0.94\columnwidth]{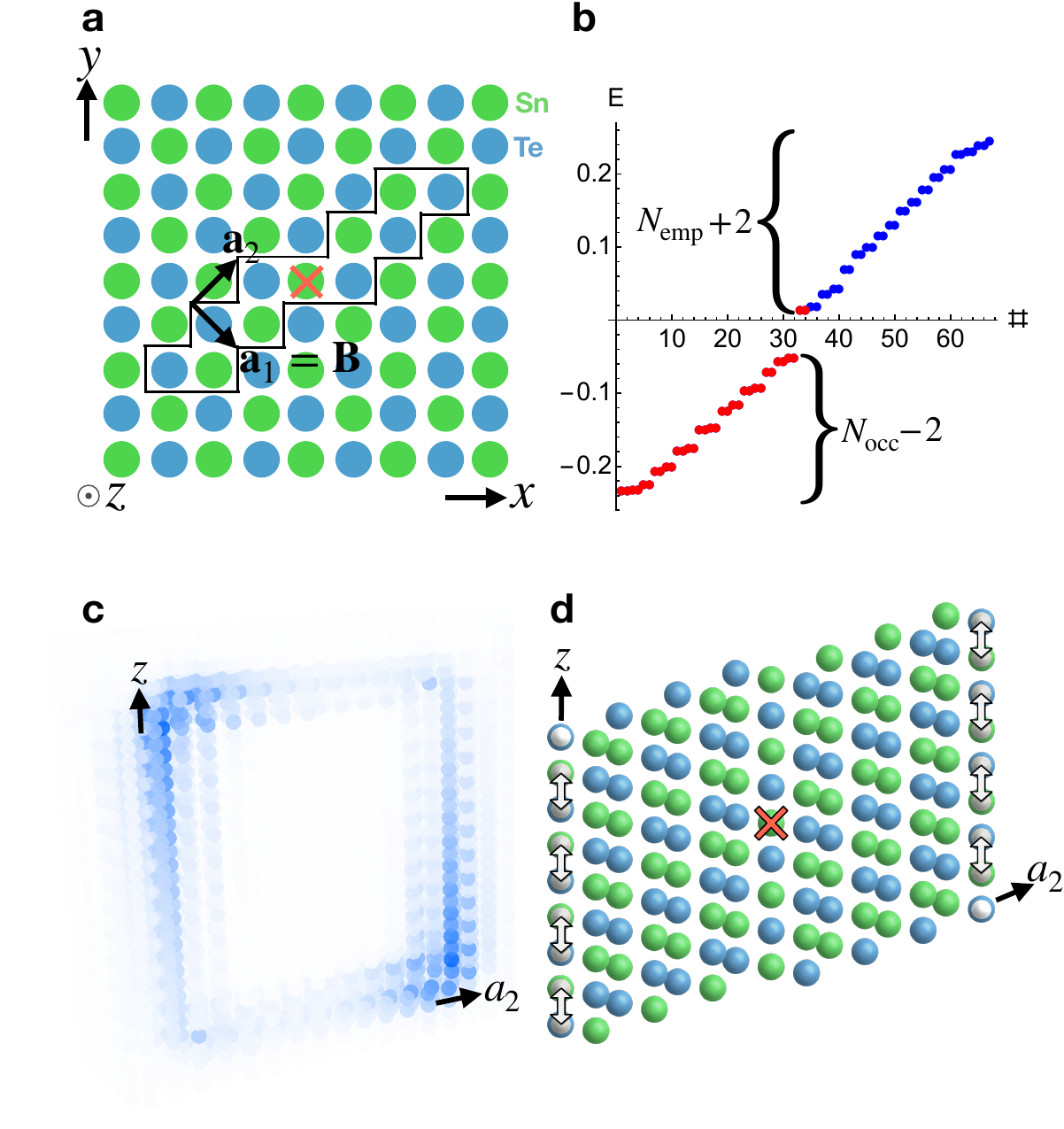}
\caption{{\bf 0D dislocation states in 3D SnTe crystals.} $\bs{a}$~Defect geometry for an $\mathcal{I}$-related pair of internal edge dislocations with ${\bs B}={\bs a}_{1}$ in 3D SnTe, where the $\mathcal{I}$ center is marked with a red $\times$ symbol.  In~$\bs{a}$, the sites enclosed within the black line have been removed in a finite number of layers in the tight-binding calculation to implement the pair of edge dislocations.  $\bs{b}$ The PBC dislocation spectrum of SnTe using the edge dislocation geometry in $\bs{a}$ exhibits four filling-anomalous states (two Kramers pairs), consistent with Eq.~(\ref{eq:HEND}) [see SN~\ref{sec:SnTeTB} for calculation details].  $\bs{c}$ The real-space profile of the four anomalous states in $\bs{b}$.  In $\bs{c}$, two total Kramers pairs of states are localized on $\mathcal{I}$-related dislocation corners (one Kramers pair of states is bound to every other corner).  When the HEND states in $\bs{c}$ are half-filled, each Kramers pair corresponds to a chargeless, spin-1/2 quasiparticle (\emph{i.e.} a spinon) that is equivalent to the corner state of an $\mathcal{I}$- and $\mathcal{T}$-symmetric 2D FTI (see SN~\ref{sec:3DscrewTRSsec} and Refs.~\onlinecite{TMDHOTI,WiederAxion}).  $\bs{d}$~The SnTe defect plane, for which a cross-sectional cut is enclosed by the black lines in~$\bs{a}$, schematically depicted as a stack of PbTe monolayer defect lines (Fig.~\ref{fig:monolayer}~$\bs{c,e}$).  In $\bs{d}$, each defect line has two 0D dislocations on its end, which each bind first-order 0D topological dislocation states.  We choose PbTe for the monolayers -- rather than SnTe -- because a decoupled stack of PbTe monolayers has the same $x,y$ components of the $\bs{M}_{\nu}^{\mathrm{F}}$ vector as a tetragonal supercell of 3D SnTe, whereas the interlayer coupling in realistic 3D PbTe drives additional band inversions [Eqs.~(\ref{eq:PbTe2DMain}) and~(\ref{eq:SnTe3DMain}), see Fig.~\ref{fig:monolayer} and SN~\ref{sec:DFT} for further details].  Hence, HEND dislocation states can be considered the result of stacking and symmetrically coupling (gray arrows in $\bs{d}$) an odd number of 2D monolayers that each contain first-order dislocation bound states.}
\label{fig:SnTeDefectMain} 
\end{figure}

In SN~\ref{sec:SnTeDFTSPECIFIC}, we show that 3D SnTe differs from an unobstructed atomic limit without corner or hinge states [\emph{i.e.} 3D PbTe, see Ref.~\onlinecite{HsiehTCI} and SN~\ref{sec:DFTPbTe}] by double band inversions at the $R$ point [${\bs k}_{R} = {\bs b}_{1}/2$] and at the symmetry-related point $R'$ [${\bs k}_{R'} = {\bs b}_{2}/2$] between two pairs of Kramers pairs of states with opposite parity eigenvalues [four valence bands and four conduction bands become inverted at $R$ and at $R'$, see Fig.~\ref{fig:SnTePristineMain}~$\bs{b,c}$].  The four Kramers pairs of band inversions drive SnTe into a fourfold rotation-anomaly TCI phase with a nontrivial weak (partial) fragile index vector  [see SN~\ref{sec:weakFragile} and~\ref{sec:SnTeDFTSPECIFIC} and the text surrounding Eq.~(\ref{eq:HEND})]:
\begin{equation}
\bs{M}^{\mathrm{F}}_{\nu} = ({\bs b}_{1} + {\bs b}_{2})/2,
\label{eq:SnTe3DMain}
\end{equation}
given in terms of the tetragonal supercell reciprocal lattice vectors in Eq.~(\ref{eq:3DSnTeReciprocalVectorMain}).

To probe the HEND-state dislocation response of SnTe, we begin with the tight-binding model introduced in Ref.~\onlinecite{HsiehTCI}, and then insert an $\mathcal{I}$-related pair of $\bs{B} = \bs{a}_1$ internal edge dislocations, as shown in Fig.~\ref{fig:SnTeDefectMain}~$\bs{a}$.  Notably, ${\bs a}_{1}$ is also a primitive lattice vector in the face-centered-cubic cell of 3D SnTe in SG 225 $Fm\bar{3}m1'$ (see Fig.~\ref{fig:SnTePristineMain}~$\bs{a}$).  Because the Frank energy criterion~\cite{FrankReadEnergy} for dislocation formation indicates that dislocations with larger values of $|{\bs B}|$ are energetically unfavorable, then dislocations with the smallest possible integer Burgers vectors -- such as the $\bs{B} = \bs{a}_{1}$ dislocations in our calculations -- may be energetically favorable and present in SnTe samples.  In the dislocation geometry with PBC, the energy spectrum is filling-anomalous (Fig.~\ref{fig:SnTeDefectMain}~$\bs{b}$), with alternating dislocation corners binding Kramers pairs of spin-charge-separated HEND states (Fig.~\ref{fig:SnTeDefectMain}~$\bs{c}$, see SN~\ref{sec:SnTeTB} for calculation details).  As discussed earlier and in SN~\ref{sec:kp}, the Kramers pairs of dislocation bound states in Fig.~\ref{fig:SnTeDefectMain}~$\bs{c}$ are equivalent to the corner states of an $\mathcal{I}$- and $\mathcal{T}$-symmetric 2D FTI~\cite{TMDHOTI,WiederAxion}, which are themselves equivalent to the end states of an $\mathcal{I}$- and $\mathcal{T}$-symmetric spinful SSH chain~\cite{HeegerReview}.  The appearance of filling-anomalous dislocation bound states in an $\mathcal{I}$-symmetric defect geometry in 3D SnTe provides further evidence for a HEND-state dislocation response in 3D insulators whose pristine electronic structure and dislocation Burgers vectors satisfy Eq.~(\ref{eq:HEND}).

Lastly, as shown in Fig.~\ref{fig:SnTeDefectMain}~$\bs{d}$, the HEND states in SnTe can be understood as the result of stacking and pairwise coupling monolayers of 2D PbTe (Fig.~\ref{fig:monolayer}), where each layer is shifted by $(\bs{a}_1+\bs{a}_2)/2$ with respect to the layer underneath and contains 0D dislocations with first-order dislocation bound states at the same in-plane position.  In an $\mathcal{I}$-symmetric stack, the 0D dislocations evolve into 1D dislocations, and neighboring 0D states pairwise annihilate in an $\mathcal{I}$-symmetric fashion, leaving two filling-anomalous HEND states.  We choose 2D PbTe for the monolayers -- rather than SnTe -- because the interlayer coupling in realistic 3D PbTe drives additional band inversions, whereas a tetragonal supercell of 3D SnTe has the same $x,y$ components of the $\bs{M}_{\nu}^{\mathrm{F}}$ vector as a decoupled stack of PbTe monolayers [Eqs.~(\ref{eq:PbTe2DMain}) and~(\ref{eq:SnTe3DMain}), see SN~\ref{sec:DFT} for further details].  Hence, in the same sense that a helical HOTI is equivalent to an $\mathcal{I}$-symmetric stack of 2D TIs (with an odd total number of layers)~\cite{ChenTCI,HOTIBernevig,HOTIBismuth,TMDHOTI,MTQC}, HEND dislocation states can be considered the result of stacking and symmetrically coupling an odd number of 2D monolayers that each contain first-order dislocation bound states.  Furthermore, if an additional layer were added to the top of Fig.~\ref{fig:SnTeDefectMain}~$\bs{d}$, global $\mathcal{I}$ symmetry would be relaxed, but each surface would still carry only one HEND state.  Hence in more realistic material geometries without global $\mathcal{I}$ symmetry, we more generally expect a 3D insulator with $\bs{M}^{\mathrm{F}}_{\nu}\neq\bs{0}$ to exhibit a random configuration of HEND states in which, on the average, every other end or corner of a dislocation satisfying Eq.~(\ref{eq:HEND}) carries a spin-charge-separated HEND state.  This is analogous to the helical hinge modes in the HOTI bismuth, which appear in STM probes on every other surface step edge, despite the absence of perfect global point group symmetries~\cite{HOTIBismuth}.

\textbf{0D Flux States in 3D Insulators} -- We now shift focus to the closely related problem of static $\pi$-flux bound states in crystals with nontrivial band topology.  As shown in several previous works~\cite{QHZ,QiFlux,AshvinFlux,FranzWormhole,Vlad2D}, $\pi$-flux cores can bind anomalous 0D solitons with the same fractional charge or spin-charge separation as the 0D HEND dislocation states introduced earlier in this work.  Specifically, $\pi$-fluxes in Chern insulators (2D TIs) bind solitons with $\pm e/2$ charge (spin-charge separation).  We have numerically confirmed the static $\pi$-flux responses of 2D Chern insulators and TIs in SN~\ref{sec:2DVORTEXtoysubsec} and~\ref{sec:2DVORTEXtoysubsecTRS}, respectively.

As previously for dislocation bound states, in this work, we recognize that the anomalous 0D $\pi$-flux bound states in Chern insulators and 2D TIs represent specific cases of a more general phenomenon.  Rather than probing the BZ-boundary topology, as is done by dislocations (see the text above, as well as SN~\ref{sec:kpEdge},~\ref{sec:kpScrew},~\ref{sec:defecttopology}, and~\ref{sec:FrankScrew}), we find that fluxes in 2D [3D] insulators bind anomalous states deriving from the summed topologies of all BZ lines [planes].  The topological boundary states of the summed topological phase correspondingly appear at the boundary of the real-space line [plane] connecting two flux tubes.  More succinctly, whereas crystal defects are sensitive to weak indices, we find that the $\pi$-flux response of an insulator is only sensitive to strong topological indices, in agreement with the results of previous works~\cite{TeoKaneDefect,QiDefect2,Vlad2D,QiFlux,AshvinFlux,Radzihovsky2019}.  Crucially, although the location of the position-space line [plane] between the flux cores [tubes] is sensitive to the gauge of the electromagnetic vector potential, the locations of the anomalous states on its boundaries -- the flux cores [tubes] -- are gauge-independent, as the flux cores [tubes] contribute a measurable Aharonov–Bohm phase shift.  Our recognition that magnetic fluxes probe bulk stable topology is supported by extensive numerical calculations (SN~\ref{sec:numerics2}), as well as rigorous analytic proofs, which are summarized in the Methods section, and provided in complete detail in SN~\ref{sec:kpFlux} and~\ref{sec:fluxfluxtubetopologymapping}.

Our analytic calculations suggest that in 3D AXIs and HOTIs, which can respectively be represented as pumping cycles of $\mathcal{T}$-broken and $\mathcal{T}$-symmetric 2D FTIs with anomalous 0D corner states~\cite{WiederAxion,TMDHOTI}, $\pi$-flux tubes will bind anomalous 0D HEND states.  To confirm this result, we have respectively in SN~\ref{sec:3Dvortex} and~\ref{sec:3DvortexTRS} numerically computed the $\pi$-flux-tube responses of $\mathcal{I}$-symmetric AXIs and $\mathcal{I}$- and $\mathcal{T}$-symmetric helical HOTIs.

\begin{figure*}[t]
\includegraphics[width=0.7\textwidth]{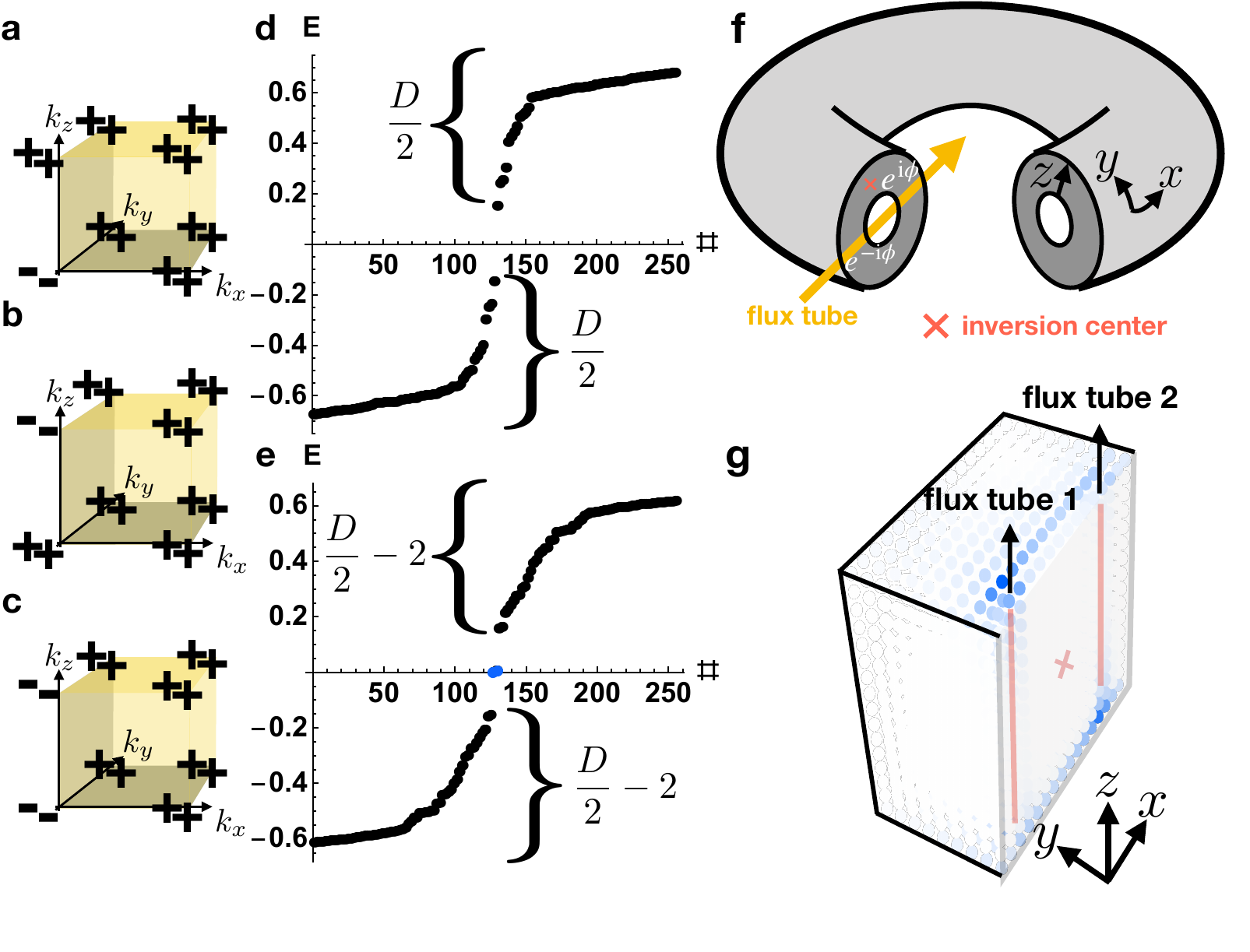}
\caption{{\bf $\pi$-flux signatures of helical higher-order topological insulators.} $\bs{a}$-$\bs{c}$~Bulk parity eigenvalues per Kramers pair of $\mathcal{I}$- and $\mathcal{T}$-symmetric 3D insulators with four occupied bands.  $\bs{a,b}$~The occupied parity eigenvalues of helical HOTIs formed from double band inversion~\cite{YoungkukMonopole,TMDHOTI} about $\Gamma$ and $Z$, respectively.  $\bs{c}$~A weak stack of the $\mathcal{I}$- and $\mathcal{T}$-symmetric 2D FTI from Ref.~\onlinecite{TMDHOTI}, which is equivalent to two superposed, $\mathcal{T}$-reversed copies of the magnetic 2D FTI introduced in Refs.~\onlinecite{TMDHOTI,WiederAxion}.  $\bs{f}$~We place the insulators in $\bs{a}$-$\bs{c}$ in the hingeless HDBC geometry detailed in Fig.~\ref{fig:AXI} and the surrounding text, and then pierce the doughnut with magnetic flux $\phi$, creating $\bs{g}$ a pair of flux tubes related by global $\mathcal{I}$ symmetry (red $\times$ symbol in~$\bs{f,g}$).  Plotting the HDBC spectra of the insulators in $\bs{a}$-$\bs{c}$ for $\phi=\pi$ flux tubes, we observe a filling anomaly~\cite{TMDHOTI,WiederAxion,WladCorners,HingeSM} $\bs{e}$ for the helical HOTIs in cases $\bs{a}$ and $\bs{b}$ and a trivial spectrum $\bs{d}$ for the 3D weak FTI in case $\bs{c}$.  For the helical HOTIs in $\bs{a,b}$ with threaded $\pi$-flux tubes,~$\bs{g}$ only one end of each $\pi$-flux tube binds a Kramers pair of spin-charge separated HEND states, such that each surface carries only a single Kramers pair.  Per surface, this represents half of the $\pi$-flux response of an isolated 2D TI (see Refs.~\onlinecite{TeoKaneDefect,QiDefect2,Vlad2D,QiFlux,AshvinFlux,Radzihovsky2019} and SN~\ref{sec:kpFlux}), implying that gapped helical HOTI surfaces carry anomalous ``half'' quantum spin Hall states.  Furthermore, because each flux tube is equivalent to the gapped 1D edge of an $\mathcal{I}$- and $\mathcal{T}$-symmetric 2D FTI~\cite{TMDHOTI}, then the flux tubes each carry an anomalous half of the nontrivial partial (time-reversal) polarization of a spinful SSH chain~\cite{TRPolarization} in the case in which global $\mathcal{I}$ symmetry is enforced, as it is in our numerics for the purpose of detecting filling anomalies.  This suggests that helical HOTIs carry a novel bulk response that represents the 3D generalization of 1D time-reversal polarization.}
\label{fig:flux}
\end{figure*}

In the case of an AXI, our numerical calculations reproduce the established result that a pair of parallel $\pi$-flux tubes in an AXI carries a total bulk $e/2$ polarization density along the direction of the tubes~\cite{FranzWormhole}.  This represents a signature that the bulk is a TCI with a nontrivial axion angle (magnetoelectric polarizability) $\theta=\pi$, where $\theta$ is the coefficient of the magnetoelectric response ${\bs E}_{e}\cdot{\bs B}_{e}$.  Specifically, the nontrivial axion angle $\theta=\pi$ indicates that as a flux quantum $\phi$ is adiabatically threaded from $\phi=0$ to $2\pi$ into an AXI cut into a cylindrical geometry (where the flux tube is aligned with the cylinder axis and open boundary conditions are taken in all directions), a charge $|e|$ is pumped from the flux tube ($r = 0$ in cylinder coordinates) to the boundary ($r = R$) of the top and bottom surfaces in a manifestation of the bulk topological magnetoelectric effect~\cite{QHZ}.  This observation is consistent with the appearance in our analytic and numerical calculations of an $|e|/2$-charged, anomalous midgap state bound to the end of the flux tube at the midpoint of the pumping cycle $\phi=\pi$~\cite{QHZ,VDBAxion}.  Specifically, on both the top and bottom surfaces of the cylinder (which are related by $\mathcal{I}$ symmetry), a charge $|e|/2$ is pumped from the flux tube to the boundary, consistent with the anomalous $\sigma_{xy} = e^2/(2h)$ Hall conductivity of gapped AXI surfaces.

Returning to the HDBC geometry employed in this work, in which there are (untwisted) PBC in the directions perpendicular to the threaded magnetic flux (Fig.~\ref{fig:flux}~$\bs{f}$), we note that a lattice model cannot be constructed with a $\phi$-flux tube unless a second tube with a flux $-\phi$ is inserted elsewhere into the system.  Hence, in the case numerically investigated in this work of an AXI with two threaded flux tubes and HDBC, a charge $|e|$ is instead pumped from one flux tube to the other as $\phi$ is advanced from $0$ to $2\pi$.  Lastly, we note that because there are two flux tubes with opposite fluxes $\pm \phi$, then, even if the locations of the flux tubes are related by a global $\mathcal{I}$ center, neither flux tube lies exactly on the global $\mathcal{I}$ center, as this would require the flux tubes to lie at the same position.  Hence, the HDBC flux-tube geometry itself generically violates $\mathcal{I}$ symmetry, except at the $\mathcal{I}$- and $\mathcal{T}$-invariant flux values $\phi=0,\pi$.

Unlike AXIs~\cite{WilczekAxion,VDBAxion,VDBHOTI,QHZ,AndreiInversion,AshvinAxion2,WuAxionExp}, $\mathcal{I}$- and $\mathcal{T}$-symmetric helical HOTIs exhibit trivial axion angles $\theta\text{ mod }2\pi=0$, and it is currently unknown -- and of great theoretical and experimental interest -- whether there exist 3D bulk or 2D surface quantized response effects that distinguish trivial insulators from $\mathcal{T}$-symmetric HOTIs.  In this work, we discover for the first time that $\pi$-flux tubes in $\mathcal{I}$- and $\mathcal{T}$-symmetric 3D HOTIs bind Kramers pairs of spin-charge-separated HEND states on only one end (Fig.~\ref{fig:flux}~$\bs{g}$).  Specifically, on a lattice terminated in the hingeless, $\mathcal{I}$-symmetric HDBC geometry in Fig.~\ref{fig:flux}~$\bs{f}$, both helical HOTIs and trivial insulators (FTIs) exhibit fully gapped spectra.  However, when we pierce a hollow doughnut of the topologically distinct insulators with $\pi$-flux tubes that preserves an $\mathcal{I}$ center (red $\times$ symbol in Fig.~\ref{fig:flux}~$\bs{f,g}$), the HOTI exhibits a filling-anomalous~\cite{TMDHOTI,WiederAxion,WladCorners,HingeSM} HDBC spectrum (Fig.~\ref{fig:flux}~$\bs{e}$), whereas the trivial insulator (FTI) does not (Fig.~\ref{fig:flux}~$\bs{d}$).  Crucially, because two $\pi$-flux cores threaded into an isolated 2D TI each bind a Kramers pair of states corresponding to a spin-charge-separated soliton~\cite{TeoKaneDefect,QiDefect2,Vlad2D,QiFlux,AshvinFlux,Radzihovsky2019}, then relaxing global $\mathcal{I}$ symmetry by ``gluing'' additional 2D TIs onto the surface does not change the number of free-angle surface spinons modulo $2$ (in the case in which the system remains half-filled).  Hence on each 2D surface, pairs of $\pi$-flux tubes bind only a single spin-charge-separated soliton between them, indicating that each gapped surface carries an anomalous ``half'' of the static $\pi$-flux response of a 2D TI.  This implies that even without global $\mathcal{I}$ symmetry, each surface of an $\mathcal{I}$- and $\mathcal{T}$-symmetric 3D HOTI is topologically equivalent to ``half'' of a quantum spin Hall insulator -- \emph{i.e.} two $\mathcal{T}$-reversed copies of the anomalous half-integer quantum Hall state of a gapped AXI surface~\cite{FuKaneInversion,AndreiInversion,QHZ,FanHOTI,VDBAxion,DiracInsulator,VDBHOTI,WiederAxion}.

To understand this result, we first recognize that the surfaces of HOTIs derive from unpaired fourfold Dirac fermions~\cite{TMDHOTI}, which cannot be stabilized in isolated $\mathcal{T}$-symmetric 2D semimetals, as discussed in SN~\ref{sec:kpFlux} and Ref.~\onlinecite{DiracInsulator}.  Because each fourfold Dirac fermion in 2D, when gapped without breaking $\mathcal{T}$ symmetry, provides half of the contribution towards the bulk being a 2D TI or trivial insulator (\emph{i.e.} a half unit of spin Hall conductivity in the limit of $s^{z}$-spin symmetry)~\cite{KaneMeleZ2,AndreiTI,LevinSternSpinHallPRL}, then the gapped 2D surface states of $\mathcal{I}$- and $\mathcal{T}$-symmetric HOTIs cannot be either 2D TIs or trivial insulators, and must instead be anomalous ``halves'' of a quantum spin Hall insulator.  We refer to the anomalous 2D surface phase as a half-integer quantum spin Hall insulator, as opposed to half of a 2D TI (which is a more precise designation, because $s^{z}$ spin is not generically a conserved quantity in solid-state materials with spin-orbit coupling [SOC]~\cite{AndreiTI,KaneMeleZ2,TRPolarization}), to draw connection with the more familiar half-integer quantum Hall insulators present on gapped AXI surfaces~\cite{FuKaneInversion,AndreiInversion,QHZ,FanHOTI,VDBAxion,VDBHOTI,WiederAxion}, as well as with earlier works~\cite{HalfQSH}.  Specifically, the half-integer quantum spin Hall state was previously predicted to appear on the top surfaces of weak TIs~\cite{HalfQSH}; however in this work, we recognize the anomalous half-integer quantum spin Hall state to more generally manifest on all gapped surfaces of $\mathcal{I}$- and $\mathcal{T}$-symmetric HOTIs.

Unlike the surfaces of AXIs -- which are physically distinguishable by their anomalous Hall conductivities~\cite{VDBAxion,VDBHOTI} $\pm e^{2}/2h$ -- it is currently unknown whether HOTI surfaces with anomalous halves of a quantum spin Hall state can similarly be distinguished in a gauge-invariant manner in the absence of $s^{z}$-spin-conservation symmetry, both from each other, and from 2D trivial insulators.  However in the artificial limit of $s^{z}$-spin conservation symmetry, half-integer quantum spin Hall phases may straightforwardly be differentiated by the signs of their spin Hall conductivities~\cite{JuvenSpinChargeAxion,LevinSternSpinHallPRL}.  Additionally, because the surface states of weak TIs and non-axionic TCIs with $2+4n$ twofold surface Dirac cones are equivalent to $1+2n$ (massive or massless) anomalous fourfold Dirac fermions upon BZ folding~\cite{ChenTCI,AshvinTCI,DiracInsulator,ChenRotation,TMDHOTI}, then our observation of a surface half quantum spin Hall state suggests that previous studies of Anderson localization and topological order on interacting weak TI and TCI surfaces~\cite{JensSurfaceWTI,Radzihovsky2019} should be revisited in the context of higher-order topology and crystal-symmetry-enhanced fermion doubling.  Specifically, our observation of an anomalous $\pi$-flux response on helical HOTI surfaces implies that when the surface Dirac fermions of a TCI phase are gapped by breaking a crystal symmetry while preserving $\mathcal{T}$, the resulting gapped surface, despite its vanishing Hall conductivity, is not necessarily featureless, as assumed in some of the earlier literature.  Lastly, because previous constructions of strongly-interacting topological phases have exploited the half-quantized surface quantum Hall effect of 3D TIs~\cite{AdyReview}, then our identification of a half-quantized surface quantum spin Hall effect in HOTIs may also provide further insight towards the theoretical construction of $\mathcal{T}$-symmetric fractional TIs and other phases with anomalous topological order~\cite{ZouSurfaceSpinLiquid}.

The presence of HEND states bound to $\pi$-flux tubes in a helical HOTI -- but not in a trivial insulator (see Fig.~\ref{fig:flux} and SN~\ref{sec:3DvortexTRS}) -- additionally provides the first example of a bulk response effect that distinguishes helical HOTIs from trivial insulators.  Specifically, because each flux tube in Fig.~\ref{fig:flux}~$\bs{f,g}$ is equivalent to the gapped 1D edge of an $\mathcal{I}$- and $\mathcal{T}$-symmetric 2D FTI~\cite{TMDHOTI}, then, in the presence of global $\mathcal{I}$ symmetry, the flux tubes each carry an anomalous half of the time-reversal polarization of an isolated spinful SSH chain (SN~\ref{sec:nested} and Ref.~\onlinecite{TRPolarization}), in that each flux tube binds a spin-charge-separated Kramers pair on only one end.  This implies that the bulk exhibits a novel form of quantized nontrivial MSP -- a spin-charge-separated generalization of the magnetoelectric polarizability of AXIs~\cite{WilczekAxion,VDBAxion,VDBHOTI,QHZ,AndreiInversion,AshvinAxion2,WuAxionExp}.

We may further understand the MSP by recognizing that an $\mathcal{I}$-symmetric, finite-sized sample of an $\mathcal{I}$- and $\mathcal{T}$-symmetric helical HOTI is equivalent to a stack (layer construction) of 2D TIs in which the edge states have been pairwise gapped, leaving behind sample-encircling helical hinge modes~\cite{AshvinIndicators,AshvinTCI,ChenTCI,MTQC}.  In the limit of $s^{z}$-spin conservation, it has previously been established that 2D TIs in a Corbino disc geometry with adiabatically threaded magnetic flux pass a quantized spin current from the inner region to the outer region in a manifestation of the quantum spin Hall effect~\cite{AndreiTI,JuvenSpinChargeAxion,LevinSternSpinHallPRL}.  Hence, we can conclude that in the $s^{z}$-conserving limit, adiabatically threading a single magnetic flux from $\phi=0$ to $2\pi$ through an $\mathcal{I}$- and $\mathcal{T}$-symmetric helical HOTI in a finite cylindrical geometry can transport a quantized amount of spin from the flux tube ($r = 0$ in cylinder coordinates) to the boundary ($r = R$) of the top and bottom surfaces, representing a higher-order generalization of the quantum spin Hall effect.  This is consistent with the appearance in our numerical calculations of spin-charge-separated HEND states bound to one end of each flux tube at the midpoint of the pumping cycle $\phi=\pi$ (see Fig.~\ref{fig:flux}~$\bs{f,g}$).  It is important to note that in the absence of $s^{z}$-spin conservation symmetry, there is no guarantee that the MSP implies a magnetic-field-dependent quantized spin accumulation.  We leave the exciting questions of a Berry-connection formulation of the MSP, the $\theta$-like topological field theory for the MSP, and whether the MSP can be computed \emph{ab initio} for future works.

\begin{figure*}[t]
\includegraphics[width=\textwidth]{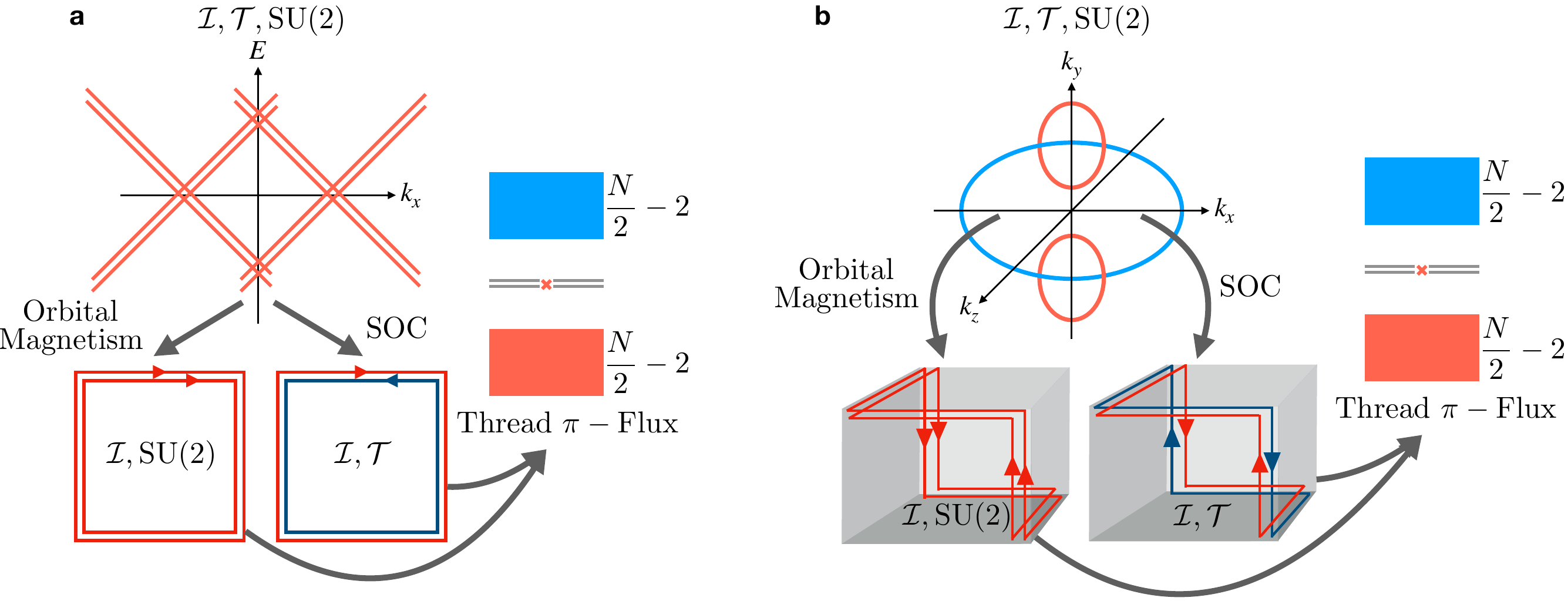}
\caption{{\bf Patterns of identical static $\pi$-flux response in topologically distinct insulators.} $\bs{a}$ (Top) A 2D graphene-like topological semimetal with two fourfold Dirac cones protected by $\mathcal{I}$, $\mathcal{T}$, and SU(2)-spin-rotation symmetries~\cite{YoungkukLineNode} can gap into two topologically distinct insulators.  (Bottom, left) Applying $\mathcal{I}$-symmetric orbital (Haldane) magnetism gaps the Dirac semimetal in $\bs{a}$ into a $|C|=2$ spin-degenerate Chern insulator with $\mathcal{I}$ and SU(2) symmetries~\cite{HaldaneModel}.  (Bottom, right) Conversely, $\mathcal{I}$-symmetric spin-orbit coupling (SOC) gaps the Dirac semimetal in $\bs{a}$ into a 2D TI with $\mathcal{I}$ and $\mathcal{T}$ symmetries~\cite{AndreiTI,KaneMeleZ2}.  (Right) However, $|C|=2$ spin-degenerate Chern insulators and 2D TIs exhibit the same $\pi$-flux response.  In both 2D insulators, $\pi$-flux cores each bind a twofold-degenerate, spin-charge-separated 0D soliton, where the twofold flux-state degeneracy in the Chern insulator [2D TI] is protected by SU(2) [$\mathcal{T}$] symmetry (see SN~\ref{sec:2DVORTEXtoysubsec} and~\ref{sec:2DVORTEXtoysubsecTRS}).  Nevertheless, $|C|=2$ spin-degenerate Chern insulators and 2D TIs are still physically distinguishable by their $\mathbb{Z}$-valued Hall conductivities, where the Hall conductivity of the Chern insulator [2D TI] is given by $\sigma^{H}=2e^{2}/h$ [$\sigma^{H}=0$]~\cite{AndreiTI,KaneMeleZ2,QHZ}.  $\bs{b}$ (Top) A 3D monopole-charged nodal-line semimetal (MNLSM) with a $\mathcal{T}$-reversed pair of nodal lines (red ellipses) that are locally protected by $\mathcal{I}$, $\mathcal{T}$, and SU(2) symmetries~\cite{YoungkukLineNode} and carry nontrivial $\mathbb{Z}_{2}$ monopole charges~\cite{TMDHOTI,YoungkukMonopole}.  (Bottom, left) Applying $\mathcal{I}$-symmetric orbital magnetism gaps the MNLSM in $\bs{b}$ into a $\theta\text{ mod }2\pi=0$ spin-doubled AXI with $\mathcal{I}$ and SU(2) symmetries~\cite{TMDHOTI,YoungkukMonopole}.  (Bottom, right) Conversely, $\mathcal{I}$-symmetric SOC gaps the MNLSM in $\bs{b}$ into a $\theta\text{ mod }2\pi=0$ helical HOTI with $\mathcal{I}$ and $\mathcal{T}$ symmetries~\cite{TMDHOTI}.  (Right)  However, like the $|C|=2$ spin-doubled Chern insulator and 2D TI in $\bs{a}$, the spin-doubled AXI and helical HOTI in $\bs{b}$ exhibit the same $\pi$-flux response.  In both 3D non-axionic HOTIs, $\pi$-flux tubes each bind a twofold-degenerate, spin-charge-separated 0D soliton on only one end, where the twofold surface flux-state degeneracy in the spin-doubled AXI [helical HOTI] is protected by SU(2) [$\mathcal{T}$] symmetry (see SN~\ref{sec:3Dvortex} and~\ref{sec:3DvortexTRS}).}
\label{fig:itsu2response}
\end{figure*}

\textbf{Identical $\pi$-Flux States in Topologically Distinct Insulators} -- Lastly, we will briefly discuss the limitations of static $\pi$-flux insertion as a complete diagnostic of bulk topology, suggesting interesting directions for future study.  We begin by considering a 2D graphene-like topological semimetal with two fourfold Dirac cones protected locally by $\mathcal{I}$, $\mathcal{T}$, and SU(2) spin-rotation symmetry~\cite{YoungkukLineNode} (Fig.~\ref{fig:itsu2response}~$\bs{a}$, top).  The bulk may either be gapped by $\mathcal{I}$-symmetric orbital (Haldane) magnetism into a $|C|=2$ spin-degenerate Chern insulator with $\mathcal{I}$ and SU(2) symmetries~\cite{HaldaneModel} (Fig.~\ref{fig:itsu2response}~$\bs{a}$, bottom left), or by $\mathcal{I}$-symmetric SOC into a 2D TI with $\mathcal{I}$ and $\mathcal{T}$ symmetries~\cite{AndreiTI,KaneMeleZ2} (Fig.~\ref{fig:itsu2response}~$\bs{a}$, bottom right).  However, from our earlier discussions and the numerical calculations performed in SN~\ref{sec:2DVORTEXtoysubsec} and~\ref{sec:2DVORTEXtoysubsecTRS}, we deduce that $|C|=2$ spin-degenerate Chern insulators and 2D TIs exhibit the same $\pi$-flux response, despite being topologically distinct phases of matter.  Specifically, when $\pi$-flux is threaded into $|C|=2$ spin-doubled Chern insulators and 2D TIs, each flux core binds a twofold-degenerate, spin-charge-separated 0D soliton, where the twofold flux-state degeneracy in the Chern insulator [2D TI] is protected by SU(2) [$\mathcal{T}$] symmetry (Fig.~\ref{fig:itsu2response}~$\bs{a}$, center right).  Nevertheless, $|C|=2$ spin-degenerate Chern insulators and 2D TIs are still physically distinguishable by their $\mathbb{Z}$-valued Hall conductivities, where the Hall conductivity of the Chern insulator [2D TI] is given by $\sigma^{H}=2e^{2}/h$ [$\sigma^{H}=0$]~\cite{AndreiTI,KaneMeleZ2,QHZ}.

In this work, we discover a similar pattern of identical static $\pi$-flux responses in two topologically distinct non-axionic 3D HOTIs that originate from the same semimetallic quantum critical point.  We begin our analysis of 3D HOTIs by considering a 3D topological semimetal with a time-reversed pair of nodal lines at the Fermi level, where each nodal line is locally protected by $\mathcal{I}$, $\mathcal{T}$, and SU(2) symmetries~\cite{YoungkukLineNode}, and carries a nontrivial $\mathbb{Z}_{2}$ monopole charge~\cite{TMDHOTI,YoungkukMonopole} (Fig.~\ref{fig:itsu2response}~$\bs{b}$, top).  Monopole nodal-line semimetals (MNLSMs) represent the 3D, higher-order-topological~\cite{HingeSM} generalizations of graphene, and MNLSM phases have been demonstrated to occur in 3D graphdiyne~\cite{YoungkukMonopole,GraphdiyneMonopole1} and $\beta$-MoTe$_2$~\cite{TMDHOTI} when the effects of SOC are neglected. Like graphene, 3D MNLSMs represent the quantum critical points between topologically distinct insulating phases~\cite{TMDHOTI}.  A 3D MNLSM may be gapped by $\mathcal{I}$-symmetric orbital magnetism into an $\mathcal{I}$- and SU(2)-symmetric spin-doubled (spinless) AXI with two co-propagating chiral hinge modes and gapped 2D surfaces with anomalous SU(2)-symmetric $|C|=1$ Chern insulating phases [where each spin sector contributes an anomalous half-integer surface Hall conductivity of $\sigma^{H}=e^{2}/(2h)$]~\cite{TMDHOTI,YoungkukMonopole,MTQC} (Fig.~\ref{fig:itsu2response}~$\bs{b}$, bottom left).  Alternatively, a 3D MNLSM may be gapped by $\mathcal{I}$-symmetric SOC into an $\mathcal{I}$- and $\mathcal{T}$-symmetric helical HOTI~\cite{TMDHOTI} with helical hinge modes and gapped 2D surfaces with anomalous $\mathcal{T}$-invariant halves of 2D TI phases (Fig.~\ref{fig:itsu2response}~$\bs{b}$, bottom left), as demonstrated in this work (see Fig.~\ref{fig:flux}~$\bs{e}$,$\bs{g}$).  However, from our discussions above and the numerical calculations performed in SN~\ref{sec:3Dvortex} and~\ref{sec:3DvortexTRS}, we deduce that like the $|C|=2$ spin-doubled Chern insulator and 2D TI in Fig.~\ref{fig:itsu2response}~$\bs{a}$, $\mathcal{I}$- and SU(2)-symmetric spin-doubled AXIs and $\mathcal{I}$- and $\mathcal{T}$-symmetric helical HOTIs exhibit the same $\pi$-flux response, despite being topologically distinct phases of matter.  Specifically, when $\pi$-flux tubes are threaded into spin-doubled AXIs and helical HOTIs, each flux tube binds a twofold-degenerate, spin-charge-separated 0D soliton on only one end, where the twofold surface flux-state degeneracy in the spin-doubled AXI [helical HOTI] is protected by SU(2) [$\mathcal{T}$] symmetry (Fig.~\ref{fig:itsu2response}~$\bs{b}$, center right).  Distinctly unlike the $|C|=2$ spin-doubled Chern insulator and 2D TI in Fig.~\ref{fig:itsu2response}~$\bs{a}$, spin-doubled AXIs and helical HOTIs both exhibit trivial $\mathbb{Z}_{2}$-valued axion angles $\theta\text{ mod }2\pi=0$, and are therefore non-axionic.

It remains an open and urgent theoretical question whether there exists a quantized response effect beyond the axionic magnetoelectric effect and static $\pi$-flux insertion that can distinguish between spin-doubled AXIs and helical HOTIs.  While it is clear that adiabatically threading a flux quantum can pump a charge $|2e|$ [quantized spin] from the bulk of a flux tube to the boundary of a spin-doubled AXI [helical HOTI in the $s^{z}$-conserving limit], neither effect is characterized by a well-established quantized response theory in noninteracting spinful topological (crystalline) insulators, such as the magnetoelectric effect.  Specifically, the $\mathbb{Z}_{2}$-valued, axionic magnetoelectric response can only distinguish between pumping cycles that pass even and odd numbers of electron charges $|e|$ per threaded flux quantum~\cite{QHZ,VDBAxion}, and therefore cannot distinguish between spin-doubled AXIs, helical HOTIs, and trivial insulators.

 {
\vspace{0.08in}
\centerline{\bf Discussion}
\vspace{0.08in}
}

The HEND states proposed in this work may be observable through STM probes of the corners of edge dislocations and the surface terminations of screw dislocations and flux tubes (solenoids) in 3D insulators that, respectively, satisfy Eq.~(\ref{eq:HEND}) or exhibit stable higher-order topology.  For the case of flux-induced HEND states, it is important to note that for most solid-state topological materials~\cite{AndreiMaterials2}, an unrealistically strong magnetic field would be required to generate one $\pi$-flux per unit cell.  However, because a 3D helical HOTI phase can be constructed by layering 2D TI states~\cite{ChenTCI}, then by layering and twisting 2D TI layers to generate a Moir\'{e} potential, one could construct a HOTI with a much larger unit cell in which a proportionately smaller magnetic field is required to produce a $\pi$-flux.  Twisted transition metal dichalcogenide few-layers have been theoretically predicted to host quantum spin Hall states~\cite{TrithepTMDTwistedPNAS}, and may hence represent a promising platform for probing the flux-induced HEND states and MSP HOTI response identified in this work.  Additionally, in AXIs, the bulk magnetoelectric and anomalous surface Hall responses can be probed in optical experiments performed under applied magnetic fields significantly weaker than one $\pi$-flux per unit cell~\cite{QHZ,WuAxionExp}.  There may also exist analogous optical signatures of the anomalous surface half quantum spin Hall states in helical HOTIs predicted in this work, which we leave as an exciting direction for future investigations.

The recent theoretical and experimental identification of HOTI phases in materials including bismuth~\cite{HOTIBismuth}, the transition metal dichalcogenides MoTe$_2$ and WTe$_2$~\cite{TMDHOTI,DavidMoTe2Exp,PhuanOngMoTe2Hinge,MazWTe2Exp}, BiBr~\cite{SYBiBr,BiBrFanHOTI,BiBrHOTIExp,FanZahidRoomTempBiBrExp}, the Ba$_3$Cd$_2$As$_4$ family~\cite{ZhijunRotationHOTI}, the Sr$_3$PbO family of perovskites~\cite{JenHOTImaterial}, as well as in recently established vast databases of topological materials~\cite{AndreiMaterials2,ChenMaterials,AshvinMaterials1} indicates particular promise for future experimental investigations of flux and defect HEND states.  Spin-charge-separated HEND dislocation states may also be observable in weak FTI phases, for which several material candidates~\cite{AndreiMaterials2} were recently discovered through the symmetry-based indicators of fragile topology introduced in Refs.~\onlinecite{ZhidaFragileAffine,FragileKoreanInversion}.  3D OAL phases have recently been identified in electrides~\cite{MurakamiElectride} and other stoichiometric insulators~\cite{AndreiMainOALDatabase2}, and may also exhibit nontrivial HEND-state dislocation responses.  For HEND states that carry chargeless spin, the spinon excitations may be detectable through nonlinear spectroscopy~\cite{PeterSpinonDetect,YongBaekSpinonDetect}.  Additionally, recent investigations have revealed that $\mathcal{T}$-symmetric topological semimetals gapped with charge-density waves exhibit the same low-energy theory as helical HOTIs~\cite{BarryBenCDW,CDWWeyl}, suggesting an intriguing future venue for investigating the spin-charge-separated defect and flux response effects introduced in this work.  Furthermore, though we have focused on solid-state materials, metamaterials can also exhibit nontrivial defect and flux responses~\cite{DisclinationNatureMeta1,FluxCycleSonicCrystal}, and may therefore provide an additional platform for realizing HEND states.  Lastly, it was recently demonstrated that dislocations in $d$-D crystals can also map interacting $(d-1)$-D topological phases to real space~\cite{XCubeFoliatedScrew}, suggesting that the interplay of crystal defects and topological order is a promising direction for future study.

{
\vspace{0.03in}
\centerline{\bf Methods}
\vspace{0.05in}
}

We will here summarize our analytic proofs of the criteria for generating 0D dislocation and flux HEND states in 3D insulators (see Table~\ref{tb:ResultsMain}).  Our proofs are supported by extensive numerical calculations of 0D dislocation and flux bound states, which we respectively detail in SN~\ref{sec:numerics} and~\ref{sec:numerics2}.  We will then detail our first-principles and tight-binding calculations demonstrating a nontrivial first-order dislocation response in 2D PbTe monolayers and a nontrivial HEND-state dislocation response in 3D SnTe.

\textbf{Summary of Analytic HEND Dislocation State Proofs} -- In this work we have formulated two alternative and equivalent sets of proofs demonstrating that integer dislocations map lower-dimensional topology from momentum space to position space.  We have crucially demonstrated that dislocations can map not just stable topological phases with 1D edge modes, but also FTIs and OALs with anomalous 0D corner states.  Our proofs further reproduce the results of all previous studies of crystal dislocation bound states with integer $\bs{B}$~\cite{AshvinScrewTI,TeoKaneDefect,QiDefect2,Vlad2D,TanakaDefect,VladScrewTI}.  First, building upon the ``cutting'' and ``gluing'' construction of topological defect states developed in Ref.~\onlinecite{AshvinScrewTI} to predict helical dislocation modes in weak TIs~\cite{FuKaneInversion}, we have employed $k\cdot p$ theory to predict 0D HEND states in 3D crystals (see SN~\ref{sec:kpEdge} and~\ref{sec:kpScrew}).  Next, in SN~\ref{sec:defecttopology} and~\ref{sec:FrankScrew}, we use more general arguments based on second-quantized expressions for noninteracting (topological) ground states to demonstrate that $(d-2)$-D dislocations in $d$-D crystals can map $(d-1)$-D BZ surfaces to $(d-1)$-D real-space surfaces, leading in 3D crystals to the presence of 1D and 0D topological defect states.  Below, we will outline the $k\cdot p$-level proof, leaving the more general case for SN~\ref{sec:defecttopology} and~\ref{sec:FrankScrew}.

For simplicity and without loss of generality, we will focus here on $\mathcal{I}$-symmetric, $\mathcal{T}$-broken insulators with edge dislocations.  Because an $\mathcal{I}$- and $\mathcal{T}$- symmetric HOTI can be formed by superposing a time-reversed pair of $\mathcal{I}$-symmetric AXIs, the results derived here for magnetic AXIs (and $\mathcal{I}$-symmetric, $\mathcal{T}$-broken FTIs) can also be straightforwardly extended to helical HOTIs (and $\mathcal{I}$- and $\mathcal{T}$-symmetric FTIs), as shown in SN~\ref{sec:kpEdge} and~\ref{sec:kpScrew}.  To begin the summary of our $k\cdot p$ derivation of anomalous HEND-state dislocation response, the low-energy $k\cdot p$ Bloch Hamiltonian of an $\mathcal{I}$-symmetric insulator can be expressed as:
\begin{equation}
\mathcal{H}(\bs{q}) = \bigoplus_{a}\mathcal{H}_{a}(\bs{q}),
\label{eq:Htotal_main}
\end{equation}
where $a$ runs over the TRIM points $\bs{k}_{D,a}$ whose bands are inverted relative to those of the atomic insulator formed from the occupied atomic orbitals when all hoppings are taken to vanish~\cite{QuantumChemistry}, and where $\bs{q} = \bs{k}-\bs{k}_{D,a}$.  We next take the simplifying assumption that the $k\cdot p$ Hamiltonian at each TRIM point $\bs{k}_{D,a}$ has the form of the low-energy theory of the Bernevig-Hughes-Zhang model of a 3D TI~\cite{AndreiTI,QHZ,AndreiInversion}:
\begin{equation}
\mathcal{H}_{a}(\bs{q}) = m_{a}\tau^{z} + \sum_{i=x,y,z}v_{i}q_{i}\tau^{x}\sigma^{i},
\label{eq:BHZtrim_main}
\end{equation}  
where $\tau^{x,y,z}$ and $\sigma^{x,y,z}$ are Pauli matrices, and where we have employed the shorthand notation $\tau^i \otimes \sigma^j \equiv \tau^i \sigma^j$. We emphasize that in a four-band model with singly degenerate bands (such as a model with only $\mathcal{I}$ symmetry), we must invert two bands in order to ensure a band gap throughout the BZ: a single band inversion about a TRIM point instead gives rise to a Weyl semimetal phase~\cite{MTQC}.  As we are in this work focusing on gapped topological phases, the minimal realization of Eq.~\eqref{eq:BHZtrim_main} relevant to the dislocation responses analyzed in this work hence involves a $4 \times 4$ $k\cdot p$ Hamiltonian.

Next, we construct a long-wavelength description of a pair of edge dislocations whose Burgers vectors lie along a crystallographic axis.  As prescribed in Ref.~\onlinecite{AshvinScrewTI}, we model an internal loop of edge dislocations by cutting the insulator described by $\mathcal{H}(\bs{q})$ [Eq.~(\ref{eq:Htotal_main})] into two pieces with $\pm\hat{z}$-normal (top and bottom) surfaces, and then ``gluing'' the two pieces back together with $|\bs{B}|/c$ extra rows of unit cells in the region between the edge dislocations, where $c$ is the lattice spacing in the $z$ direction.  We initially implement the gluing with $\mathcal{I}$- and $\mathcal{T}$-symmetric coupling between the top and bottom surfaces, and then later relax $\mathcal{T}$ symmetry.  At each TRIM point in the bulk at which bands are inverted, the top and bottom surfaces each contribute a twofold Dirac-cone surface state to the interface.  This implies that the combined top and bottom surfaces carry one effective fourfold Dirac fermion in 2D for each band inversion in the bulk, where each fourfold Dirac fermion admits a single, $\mathcal{T}$-symmetric mass term~\cite{DiracInsulator}.  To account for the presence or absence of nonzero $\bs{B}$, we derive in SN~\ref{sec:kpEdge} a consistent, intrinsic phase for the coupling mass at each band-inverted TRIM point, finding in particular that the relative sign across the dislocation of the mass of the fourfold Dirac cone induced from the TRIM $a$ is proportional to $\cos(\bs{k}_{D,a}\cdot\bs{B})$.  Hence, the edge dislocation loop effectively realizes an interface between two gapped fourfold Dirac cones, where the relative sign of the gap is given by $\cos(\bs{k}_{D,a}\cdot\bs{B})$.  If the relative sign is negative, then the Dirac mass switches sign, and the resulting domain wall binds a helical pair of defect-localized states~\cite{JackiwRebbi,HingeSM}.  This implies that each band-inverted bulk TRIM point $\bs{k}_{D,a}$ will only contribute helical modes to an edge dislocation if $\bs{k}_{D,a}\cdot\bs{B}$ is an odd multiple of $\pi$.  Lastly, we relax $\mathcal{T}$ symmetry while preserving $\mathcal{I}$ symmetry.  From the analysis of $\mathcal{I}$-symmetric 2D insulators with anomalous corner modes in Refs.~\onlinecite{TMDHOTI,WiederAxion}, we can immediately deduce that each of the bulk TRIM points that previously contributed a pair of helical modes at the edge dislocation will necessarily now contribute an anomalous number of 0D $\pm e/2$-charged (anti)solitons under the introduction of $\mathcal{I}$-symmetric magnetism.  As discussed in Refs.~\onlinecite{TMDHOTI,WiederAxion,WladCorners,HingeSM}, this conclusion is crucially not reliant on particle-hole symmetry, which is not present in real materials~\cite{AndreiMaterials2}.

\textbf{Summary of Analytic HEND Flux State Proofs} -- In this work we have also formulated two alternative and equivalent sets of proofs demonstrating that $\pi$-flux tubes in 3D insulators bind anomalous 1D and 0D states, including HEND states,  if and only if the bulk is a stable TI or TCI.  Our proofs reproduce the results of Refs.~\onlinecite{QHZ,QiFlux,AshvinFlux,FranzWormhole,Vlad2D}, as well as suggest the presence of a novel quantized $\pi$-flux response in $\mathcal{I}$- and $\mathcal{T}$-symmetric helical (non-axionic) HOTIs.  As previously for integer dislocation bound states, our flux-state proofs were performed both within the $k\cdot p$ approximation for 3D insulators (SN~\ref{sec:kpFlux}), and using more general arguments based on second-quantized expressions for the noninteracting (topological) ground states of $d$-D insulating crystals (SN~\ref{sec:fluxfluxtubetopologymapping}).  Below, we will detail the $k\cdot p$-level proof, leaving the more general case for SN~\ref{sec:fluxfluxtubetopologymapping}.

We will again here focus on the response of $\mathcal{I}$-symmetric, $\mathcal{T}$-broken 3D insulators.  Because an $\mathcal{I}$- and $\mathcal{T}$-symmetric helical HOTI can be formed by superposing a time-reversed pair of $\mathcal{I}$-symmetric AXIs~\cite{TMDHOTI}, then the $\pi$-flux-tube response derived here for magnetic AXIs can straightforwardly be extended to helical HOTIs, as detailed and performed in SN~\ref{sec:kpFlux}.  We begin the summary of our $k\cdot p$ derivation of anomalous HEND-state $\pi$-flux response by again considering a 3D insulator with (initially) $\mathcal{I}$ and $\mathcal{T}$ symmetries.  We take the 3D insulator to differ from a trivial atomic insulator by a series of band inversions at a set of TRIM points $\{\bs{k}_{D,a}\}$ between Kramers pairs of states with opposite parity eigenvalues.  The low-energy Hamiltonian of the insulator in the absence of threaded magnetic flux is hence again given by Eqs.~(\ref{eq:Htotal_main}) and~(\ref{eq:BHZtrim_main}).

Next, we construct a long-wavelength description of magnetic flux threaded into the 3D insulator through two parallel 1D tubes with opposite field strengths $\pm \phi$ located at $\mathcal{I}$-related positions.  To implement the pair of flux tubes, we cut the insulator described by $\mathcal{H}(\bs{q})$  [Eq.~(\ref{eq:Htotal_main})] into two pieces with $\pm \hat{x}_{\perp}$-normal surfaces, and again glue the pieces back together.  In the region between the flux tubes, we multiply all couplings between the top and bottom surface states by $e^{\mathrm{i \phi}}$.  We emphasize that the effective $\pm \phi/2$ phase rotation per surface only represents a local gauge transformation on each surface in the limit in which the surfaces are considered separately -- however when the two surfaces are coupled, the $\phi$ phase difference between the surfaces corresponds to the gauge-invariant insertion of $\pm\phi$-fluxes along the boundaries of the region between the flux tubes.

As previously for integer dislocations, the interface between the top and bottom surfaces contains an effective fourfold Dirac cone from the two twofold surface Dirac cones contributed by each bulk band inversion at $\bs{k}_{D,a}$ (one twofold Dirac cone from each of the top and bottom surfaces).  However, unlike for integer dislocations, the $\mathcal{T}$-symmetric mass of the fourfold Dirac cone carries a relative phase of $e^{\mathrm{i} \phi}$ between the regions inside and outside of the pair of flux tubes.  Hence crucially, and unlike in the previous case of edge and screw dislocations, the relative sign of the fourfold Dirac mass for $\pi$-flux tubes is independent of $\bs{k}_{D,a}$.  This implies that when $\phi=\pi$, the two flux tubes bind an odd (anomalous) number of helical pairs of modes if the bulk contains an odd total number of band inversions between Kramers pairs of states at TRIM points such that -- through the Fu-Kane parity criterion -- the bulk is a 3D TI~\cite{FuKaneInversion}.  Alternatively, this result may be summarized through the statement that $\pi$-flux tubes bind anomalous helical modes in an $\mathcal{I}$- and $\mathcal{T}$-symmetric 3D insulator if the 2D momentum-space Hamiltonian in only one of the $k_{x_{\perp}}=0,\pi$ BZ planes is equivalent to a 2D TI, because a 3D TI  can be expressed as a helical pump of a 2D TI~\cite{FuKaneInversion,QHZ}.  As shown in SN~\ref{sec:kpFlux} and~\ref{sec:fluxfluxtubetopologymapping}, we find more generally that a parallel pair of $x_{\parallel 2}$-directed $\pi$-flux tubes separated by a distance along $x_{\parallel 1}$ sums the 2D topology of all of the momentum-space Hamiltonians in the $k_{x_{\perp}}$-indexed BZ planes of the pristine insulating crystal [see Eq.~(\ref{eq:Htotal_main}), and note that $x_{\parallel 1,2}$ span the plane perpendicular to $x_{\perp}$].  The summed 2D momentum-space topology is then projected onto the real-space surface spanning the flux tubes.

From this result, it is straightforward to derive the $\pi$-flux response of $\mathcal{I}$-symmetric AXIs.  Numerous previous works~\cite{FuKaneInversion,AndreiInversion,QHZ,FanHOTI,VDBAxion,VDBHOTI,WiederAxion} have shown that an $\mathcal{I}$-symmetric 3D strong TI gaps into an AXI under the introduction of $\mathcal{I}$-symmetric magnetism.  Furthermore, it was shown in recent works~\cite{TMDHOTI,WiederAxion} that, because an $\mathcal{I}$-symmetric 2D TI gaps into a 2D FTI with anomalous $\pm e/2$-charged corner modes, then an AXI is equivalent to an odd, chiral pumping cycle of an $\mathcal{I}$-symmetric 2D FTI.  Hence, when $\mathcal{T}$ symmetry is relaxed in an $\mathcal{I}$-symmetric 3D TI with two $\pi$-flux tubes, the Dirac-cone surface states and helical flux states become gapped, but there remain an anomalous number of $\pm e/2$-charged 0D states bound to the loop formed from the two flux tubes and the crystal surfaces.  Hence, $\pi$-flux tubes in an AXI necessarily bind anomalous $\pm e/2$-charged 0D HEND states, which appear in our numerical calculations on $\mathcal{I}$-related flux tube ends (see SN~\ref{sec:3Dvortex} and~\ref{sec:3DvortexTRS}).

Because an $\mathcal{I}$- and $\mathcal{T}$-symmetric helical HOTI is equivalent to the superposition of a time-reversed pair of $\mathcal{I}$-symmetric AXIs~\cite{TMDHOTI}, then the previous derivation of flux-tube HEND states in AXIs also implies the $\pi$-flux response of helical HOTIs.  Specifically, as detailed in SN~\ref{sec:kpFlux}, we discover in this work that $\pi$-flux tubes threaded into helical HOTIs bind Kramers pairs of spin-charge-separated 0D HEND states, rather than $\pm e/2$ end charges.

\textbf{First-Principles and Tight-Binding Calculation Details for PbTe Monolayers} -- We will here detail our first-principles and tight-binding calculations for 2D PbTe monolayers (see SN~\ref{sec:DFTPbTe} for complete calculation details).  To obtain the crystal structure of a single, pristine monolayer of PbTe, we start with a 3D crystal of rock-salt-structure PbTe [SG 225 $Fm\bar{3}m1'$, Inorganic Crystal Structure Database (ICSD)~\cite{ICSD} No. 194220, further details available at~\url{https://topologicalquantumchemistry.com/#/detail/194220}~\cite{QuantumChemistry,AndreiMaterials,AndreiMaterials2,BCS1,BCS2}], increase the lattice spacing in the $z$ ($c$-axis) direction to isolate a single plane of Pb and Te atoms, and then restrict the system symmetry to layer group (LG)~\cite{BigBook,MagneticBook,subperiodicTables,WiederLayers,SteveMagnet,DiracInsulator,HingeSM} $p4/mmm1'$.  We next perform fully relativistic DFT calculations of the electronic structure using the Vienna Ab initio Simulation Package (VASP)~\cite{VASP1,VASP2} employing the projector-augmented wave (PAW) method~\cite{PAW1,PAW2} and the Perdew, Burke, and Ernzerhof generalized-gradient approximation (GGA-PBE)~\cite{PBE-1996} for the exchange-correlation functional.  In our first-principles calculations, we have used the primitive unit cell shown in Fig.~\ref{fig:monolayer}~$\bs{a}$, which contains one Pb atom at $(x,y) = (0,0)$ and one Te atom at $(1/2,0)$.  The lattice vectors of the primitive cell (see Fig.~\ref{fig:monolayer}~$\bs{a}$) are given by:
\begin{equation}
\bs{a}_1 = (1/2,-1/2),\ \bs{a}_2 = (1/2,1/2),
\end{equation} 
and the reciprocal lattice vectors are given by:
\begin{equation}
\bs{b}_1 = 2\pi(1,-1),\ \bs{b}_2 = 2\pi(1,1). 
\label{eq:monolayerGvecsMethods}
\end{equation}
Lastly, we have allowed the in-plane lattice spacing $a_{1}=a_{2}=a$ to relax from its experimental value to an equilibrium length of $a=4.483$ \AA.

To determine the topological indices of the PbTe monolayer, we use the~\href{https://github.com/stepan-tsirkin/irrep}{IrRep} program~\cite{StepanIrRepCode} to first deduce the small corepresentations (coreps) of the six highest valence and the two lowest conduction bands, which are shown in Fig.~\ref{fig:monolayer}~$\bs{c,d}$ and labeled employing the convention of the~\href{http://www.cryst.ehu.es/cgi-bin/cryst/programs/representations.pl?tipogrupo=dbg}{REPRESENTATIONS DSG} tool on the BCS~\cite{QuantumChemistry,Bandrep1} for the $k_{z}=0$ plane of SG 123 $P4/mmm1'$, the index-2 supergroup of LG $p4/mmm1'$ generated by adding lattice translations in the $z$ direction.

Next, to determine the dislocation response of PbTe monolayers, we calculate the weak (partial) SSH invariant vector $\bs{M}_{\nu}^{\mathrm{SSH}}$, which is defined in the text surrounding Eq.~(\ref{eq:SSH}).  $\bs{M}_{\nu}^{\mathrm{SSH}}$ can be obtained by counting the number of parity-eigenvalue-exchanging band inversions by which a set of bands differs from an unobstructed (trivial) atomic limit with a trivial dislocation response.  As shown in Fig.~\ref{fig:monolayer}~$\bs{c,d}$, PbTe monolayers differ from an unobstructed atomic limit through band inversion at the $X$ point [${\bs k}_{X}={\bs b}_{1}/2 = (\pi,-\pi)$] between bands labeled by the small coreps $\bar{X}_{5,6}$ of the little group at $X$.  The small coreps $\bar{X}_{5,6}$ correspond to doubly-degenerate pairs of states with the same parity ($\mathcal{I}$) eigenvalues within each pair, such that:
\begin{equation}
\chi_{\bar{X}_{5}}(\mathcal{I})=2,\ \chi_{\bar{X}_{6}}(\mathcal{I})=-2,
\label{eq:parityCharacter}
\end{equation}
where $\chi_{\rho}(h)$ is the character of the unitary symmetry $h$ in the corep $\rho$, and is equal to the sum of the eigenvalues of $h$ in $\rho$.  Because the $X$ and symmetry-equivalent $X'$ [${\bs k}_{X'}=C_{4z}{\bs k}_{X}\text{ mod }{\bs b}_{1}\text{ mod }{\bs b}_{2}={\bs b}_{2}/2 = (\pi,\pi)$] points lie along the BZ-edge $XM$ and $X'M$ lines, then we conclude that PbTe monolayers exhibit a nontrivial weak partial (time-reversal) SSH invariant vector:
\begin{equation}
\bs{M}_{\nu}^{\mathrm{SSH}} = \frac{1}{2}(\bs{b}_1 + \bs{b}_2) = (2\pi,0).
\label{eq:PbTeWeakIndex}
\end{equation}
We emphasize that, despite $\nu^{\mathrm{SSH}}_{x}\text{ mod }2\pi = \nu^{\mathrm{SSH}}_{y}\text{ mod }2\pi = 0$ in Eq.~(\ref{eq:PbTeWeakIndex}), $\bs{M}_{\nu}^{\mathrm{SSH}}$ is still nontrivial, because $(2\pi,0)$ and $(0,2\pi)$ are not reciprocal lattice vectors [Eq.~(\ref{eq:monolayerGvecsMethods})] in the rotated coordinates employed in our calculations.

To confirm the nontrivial dislocation response of a PbTe monolayer, we next insert a pair of 0D dislocations into an eight-band tight-binding model obtained from maximally-localized, symmetric Wannier functions through~\textsc{Wannier90}~\cite{Mostofi2008,Mostofi2014}.  In practice, when mapping a DFT calculation to a tight-binding model, one must choose a cutoff distance for hopping interactions.  Surprisingly, even though the band inversion in PbTe monolayers is relatively strong (the negative band gap at the $X$ and $X'$ points is roughly $\sim 260$ meV)~\cite{PbTeMonolayer1,PbTeMonolayer2,LiangTCIMonolayer}, we find that the strong and weak partial-polarization topology of a PbTe monolayer is only reproduced in a tight-binding model that is truncated to a minimum range of sixth-nearest-neighbor hopping.  As detailed in SN~\ref{sec:DFTPbTe} and shown in Fig.~\ref{fig:monolayer}~$\bs{d,e}$, computing the PBC spectrum of our Wannier-based tight-binding model with a pair of $\bs{B} = \bs{a}_1$ dislocations, we observe four filling-anomalous dislocation bound states, confirming the nontrivial first-order dislocation response of PbTe monolayers.

\textbf{First-Principles and Tight-Binding Calculation Details for 3D SnTe} -- We will next detail our first-principles and tight-binding calculations demonstrating a nontrivial HEND-state dislocation response in 3D SnTe crystals (see SN~\ref{sec:DFTSnTe} for complete calculation details).  To draw a comparison with SnTe, we have also performed analogous calculations on the isostructural compound PbTe, which we find to exhibit a trivial dislocation response.  We begin by performing fully-relativistic DFT calculations of the electronic structure of 3D SnTe and PbTe using VASP~\cite{VASP1,VASP2} employing the PAW method~\cite{PAW1,PAW2} and GGA-PBE~\cite{PBE-1996} for the exchange-correlation functional.  The lattice parameters of the rock-salt structure [SG 225 $Fm\bar{3}m1'$] were fixed to their experimental values~\cite{latt-SnTePbTe} $a=6.32$ {\AA} for SnTe and $a=6.46$ {\AA} for PbTe.

Below, we will specifically compute the dislocation response for the shortest possible dislocation Burgers vectors -- \emph{i.e.} dislocations for which the Burgers vector ${\bs B}$ is equal to one of the primitive, face-centered-cubic lattice vectors of SnTe or PbTe.  For geometric simplicity and because 3D SnTe and PbTe are cubic, we without loss of generality form a tetragonal supercell in which the ${\bs a}_{1}$ and ${\bs a}_{2}$ primitive lattice vectors are also lattice vectors in the face-centered cubic cell, but in which ${\bs a}_{3}$ is $\sqrt{2}$ times the length of a face-centered-cubic primitive lattice vector [see Fig.~\ref{fig:SnTePristineMain}~$\bs{a}$].  The tetragonal cell specifically contains two Sn/Pb atoms at $(x,y,z) = (0,0,0)$ and $(1/2,1/2,1/2)$ and two Te atoms at $(0,0,1/2)$ and $(1/2,1/2,0)$, and respects the symmetries of SG 123 $P4/mmm1'$.  The lattice and reciprocal lattice vectors of the tetragonal supercell are shown in Fig.~\ref{fig:SnTePristineMain}~$\bs{a}$ and detailed in Eq.~(\ref{eq:3DSnTeReciprocalVectorMain}) and the surrounding text.  In our first-principles calculations, we only incorporate valence-shell states -- hence, our calculations only include the $5p$ orbitals of Te and $5p$ ($6p$) orbitals of Sn (Pb), as well as twelve total empty conduction bands from higher-shell (empty) valence orbitals.    Therefore, at each TRIM point in Fig.~\ref{fig:SnTePristineMain}~$\bs{c}$, the lower twelve (upper twelve) bands are occupied (unoccupied) [the bands in Fig.~\ref{fig:SnTePristineMain}~$\bs{c}$ are fourfold degenerate due to the combined effects of spinful $\mathcal{I}\times\mathcal{T}$ symmetry and supercell BZ folding].

$\bs{M}^{\mathrm{F}}_{\nu}$ can be obtained by counting the number of parity-eigenvalue-exchanging band inversions by which a set of bands differs from an unobstructed atomic limit with a trivial dislocation response.  We first establish, in agreement with previous works~\cite{HsiehTCI}, that 3D PbTe realizes an unobstructed atomic limit in which three Kramers pairs of states are located on each of the four Te atoms in the tetragonal supercell.  Our calculations indicate that 3D SnTe differs from 3D PbTe by double band inversions at the $R$ point [${\bs k}_{R} = {\bs b}_{1}$] and at the symmetry-related point $R'$ [${\bs k}_{R'} = C_{4z}{\bs k}_{R}\text{ mod }{\bs b}_{1}\text{ mod }{\bs b}_{2} = {\bs b}_{2}/2$] between two pairs of Kramers pairs of states with opposite parity eigenvalues [four valence states become inverted with four conduction states at $R$ and at $R'$].

To determine the dislocation response of SnTe, we first establish that $\bs{M}^{\mathrm{F}}_{\nu} = {\bs 0}$ in PbTe, because PbTe is an unobstructed atomic limit.  Hence, because SnTe differs from PbTe by double band inversions at the $R$ and $R'$ points in the tetragonal supercell (see Fig.~\ref{fig:SnTePristineMain}), the HEND-state response of SnTe is nontrivial:
\begin{equation}
\bs{M}^{\mathrm{F}}_{\nu} = (\bs{b}_1 + \bs{b}_2)/2 = (2 \pi,0,0).
\label{eq:SnTe3DHENDresponseMethods}
\end{equation}
We emphasize that, despite $\nu^{\mathrm{F}}_{x}\text{ mod }2\pi = \nu^{\mathrm{F}}_{y}\text{ mod }2\pi = 0$ in Eq.~(\ref{eq:SnTe3DHENDresponseMethods}), $\bs{M}_{\nu}^{\mathrm{F}}$ is still nontrivial, because $(2\pi,0,0)$ and $(0,2\pi,0)$ are not reciprocal lattice vectors in the tetragonal supercell of SnTe [Eq.~(\ref{eq:3DSnTeReciprocalVectorMain})].

We next explicitly confirm the nontrivial defect response of 3D SnTe.  To model an edge dislocation in SnTe, we use the tight-binding model from Ref.~\onlinecite{HsiehTCI}, with the parameters listed in Ref.~\onlinecite{FulgaCoupledLayers}.  We first enlarge the model unit cell to obtain the tetragonal supercell shown in Fig.~\ref{fig:SnTePristineMain}~$\bs{a}$.  We then determine the locations of the $\mathcal{I}$ centers in the supercell from the mirror symmetry representations given in Ref.~\onlinecite{FulgaCoupledLayers} -- in real space, the Sn and Te atoms in the model in Ref.~\onlinecite{HsiehTCI} occupy inversion centers that coincide with lines of $C_{4z}$ (fourfold rotation) symmetry in the tetragonal supercell (Fig.~\ref{fig:SnTePristineMain}~$\bs{a}$).
Next, we implement an internal edge dislocation with $\bs{B} = \bs{a}_1$, as shown in Fig.~\ref{fig:SnTeDefectMain}~$\bs{a}$ and detailed in SN~\ref{sec:SnTeTB}.  Importantly, in order to use filling anomalies to diagnose the nontrivial HEND-state dislocation response, we must implement the defect plane in an $\mathcal{I}$-symmetric manner, which we accomplish with the alternating pattern of site removal depicted in Fig.~\ref{fig:SnTeDefectMain}~$\bs{a}$.

To provide a reference for our numerical analysis of the defect response in 3D SnTe, we have also implemented a $\bs{B} = \bs{a}_1$ pair of edge dislocations in a tight-binding model of 3D PbTe.  To construct the tight-binding model, we have increased the on-site energy difference between the two inequivalent atoms in the primitive unit cell [specifically, in the notation of Ref.~\onlinecite{FulgaCoupledLayers}, we have changed the parameter $m$ from $1.65$ to $3$ in Eq.~(16) in Ref.~\onlinecite{FulgaCoupledLayers}].  Increasing the on-site energies reverses the pair of double band inversions at $R$ and $R'$, and reproduces the first-principles-derived parity eigenvalues and electronic structure of PbTe.  The on-site potential can also be understood as a chemical potential that localizes all of the electrons on the Te atoms of PbTe.  Because PbTe is isostructural to SnTe, then the real-space defect geometry for our tight-binding model of PbTe is identical to the defect geometry previously employed in SnTe (depicted in Fig.~\ref{fig:SnTeDefectMain}~$\bs{a}$).

In Fig.~\ref{fig:SnTeDefectMain}~$\bs{b}$, we plot the PBC defect spectrum for SnTe, and in SN~\ref{sec:SnTeTB}, we plot the analogous defect spectrum for PbTe.  The dislocation spectrum of PbTe exhibits a large gap and is trivial, whereas the defect spectrum of SnTe is conversely filling-anomalous, specifically exhibiting four midgap states (two Kramers pairs corresponding to the circled states in Fig.~\ref{fig:SnTeDefectMain}~$\bs{c}$).  This result validates our first-principles bulk identification of a nontrivial HEND-state dislocation response vector in 3D SnTe, and a trivial HEND-state response vector in 3D PbTe.

{
\vspace{0.08in}
\centerline{\bf Data Availability}
\vspace{0.08in}
}

The data supporting the findings of this study are available from the corresponding authors upon reasonable request.

{
\vspace{0.08in}
\centerline{\bf Acknowledgments}
\vspace{0.08in}
}

$^\ddag$Primary address for B. A. B.: Department of Physics,
Princeton University,
Princeton, NJ 08544, USA.  Intuition for the interpretation of our flux-threading data was derived from conversations with Barry Bradlyn, Giandomenico Palumbo, Kuan-Sen Lin, Gregory A. Fiete, and Charles L. Kane.  We further acknowledge helpful discussions with Jennifer Cano, Raquel Queiroz, Senthil Todadri, Zhijun Wang, and Binghai Yan.  B. A. B. and B. J. W. were supported by NSF-MRSEC Grant No. DMR-2011750, Simons Investigator Grant No. 404513, ONR Grant No. N00014-20-1-2303, the Schmidt Fund for Innovative Research, the BSF Israel US Foundation Grant No. 2018226, and the Gordon and Betty Moore Foundation through Grant No. GBMF8685 towards the Princeton theory program and Grant No. GBMF11070 towards the EPiQS Initiative.  B. A. B. acknowledges additional support through the European Research Council (ERC) under the European Union's Horizon 2020 research and innovation program (Grant Agreement No. 101020833) and the Princeton Global Network Fund.  S. S. T. and T. N. acknowledge support from the European Union’s Horizon 2020 research and innovation program (ERC-StG-Neupert-757867-PARATOP).  S. S. T. also acknowledges support from the Swiss National Science Foundation (grant number: PP00P2\_176877).  F. S. was supported by a fellowship at the Princeton Center for Theoretical Science.  F. S. also wishes to thank the Kavli Institute for Theoretical Physics, which is supported by the National Science Foundation under Grant No. NSF PHY-1748958, for hosting during some stages of this work.  Concurrent with the preparation of this work and consistent with our findings, a bulk spin-magnetoelectric response and anomalous surface half quantum spin Hall states were numerically identified in helical HOTIs in Ref.~\onlinecite{PartialAxionHOTINumerics}.  Further discussions of related investigations performed during the long preparation of this work are provided in SN~\ref{sec:recentRefs}.

{
\vspace{0.08in}
\centerline{\bf Author Contributions}
\vspace{0.08in}
}

$\\$
All authors contributed equally to the intellectual content of this work.  The extensive tight-binding calculations of defect and flux states were performed by F. S.  The possibility of a defect response in HOTIs was recognized by B. A. B. and T. N. in consultation with F. S. and B. J. W.  The possibility of realizing 0D defect and flux states in 3D insulators, and the connection to lower-dimensional insulators with corner states, was recognized by B. J. W.  The presence of anomalous half 2D TI states on helical HOTI surfaces was recognized by B. J. W., and their detection through magnetic flux insertion was proposed by B. A. B. in consultation with B. J. W.  Bulk index theorems for defect and flux HEND states were derived by B. J. W. using nested Jackiw-Rebbi calculations and nested Wilson loops, and by F. S. and T. N. through a ground-state mapping between momentum space and real space.  Material candidates were identified by B. J. W., F. S., and S. S. T.  Material analysis and first-principles calculations were performed by S. S. T.  The manuscript was written by B. J. W. and F. S. with input from all of the authors.  B. J. W. was responsible for the overall research direction.

{
\vspace{0.08in}
\centerline{\bf Competing Interests}
\vspace{0.08in}
}

The authors declare no competing interests.

\onecolumngrid

\bibliographystyle{naturemag}
\bibliography{../defectBibs}

\end{document}


\title{Supplementary Information for ``Topological Zero-Dimensional Defect and Flux States\\ in Three-Dimensional Insulators''}

\author{Frank Schindler}
\thanks{Corresponding author: \url{schindler@princeton.edu}}
\affiliation{Princeton Center for Theoretical Science, Princeton University, Princeton, NJ 08544, USA}

\author{Stepan S. Tsirkin}
\affiliation{Department of Physics, University of Zurich, Winterthurerstrasse 190, 8057 Zurich, Switzerland}
            
\author{Titus Neupert}
\affiliation{Department of Physics, University of Zurich, Winterthurerstrasse 190, 8057 Zurich, Switzerland}
            
\author{B. Andrei Bernevig$^\ddag$}
\affiliation{Department of Physics,
Princeton University,
Princeton, NJ 08544, USA
	}
\affiliation{Donostia International Physics Center, P. Manuel de Lardizabal 4, 20018 Donostia-San Sebastian, Spain}
\affiliation{IKERBASQUE, Basque Foundation for Science, Bilbao, Spain}

\author{Benjamin J. Wieder}
\thanks{Corresponding author: \url{bwieder@mit.edu}}
\affiliation{Department of Physics, Massachusetts Institute of Technology, Cambridge, MA 02139, USA}
\affiliation{Department of Physics, Northeastern University, Boston, MA 02115, USA}
\affiliation{Department of Physics,
Princeton University,
Princeton, NJ 08544, USA
	}

\date{\today}
\maketitle

\onecolumngrid

\clearpage
\tableofcontents
\clearpage

\section*{\bf Supplementary Notes}

\section{Review of Recent Related Literature}
\label{sec:recentRefs}

There has been tremendous progress in recent years linking the band topology of pristine crystals~\cite{CharlieZahidReview,WiederReview,AndreiTI,AndreiSpinHallPRL,CharlieTI,KaneMeleZ2,FuKaneMele,FuKaneInversion,MooreBalentsWeak,AndreiInversion,
CavaHasanTI,HgTeExp,QHZ,LiangTCI,TeoFuKaneTCI,HermeleSymmetry,ZhidaHermele,HsiehTCI,FangFuNSandC2,HarukiRotation,SlagerNatPhysDisorder,SlagerSymmetry,HourglassInsulator,
Cohomological,DiracInsulator,ChenRotation,SYBiBr,BiBrFanHOTI,ZhijunRotationHOTI,WladTheory,HOTIBernevig,HOTIChen,WiederAxion,S4Guys,HigherOrderTIPiet,HigherOrderTIPiet2,
TynerGoswamiSU2Winding,SaavanthTaylorSSHDisorder,WeakQI,TaylorHingeSC,TitusInteractHOTI,BarryBenCDW,JiabinBenCDW,BarryKuansenCDW,CDWWeyl,AxionCDWExperiment,
DisclinationNatureMeta1,DisclinationNatureMeta2,
GrapheneReview,meleDirac,GuidoOrtixInversionOALEdge,WeylReview,BinghaiReview,NaDirac,SchnyderDirac,NagaosaDirac,ZJDirac,SyDiracSurface,CavaDirac1,SteveDirac,AshvinWeyl1,AndreiWeyl,YoungkukLineNode,ChenWithWithout,DDP,KramersWeyl,NewFermions,RhSi,CoSi,HingeSM} to the anomalous electronic states that can be bound by crystal defects~\cite{AshvinScrewTI,TeoKaneDefect,TitusSnTeDefect,QiDefect1,QiDefect2,Vlad2D,TanakaDefect,WrayDefect,JenDefect,VladScrewTI,WeylDefectTaylor,WeylDefectJP1,WeylDefectJP2,DiracDefectRussia,
MadeleineDefect1,MadeleineDefect2,AffleckBLG,WiederBLG,KoshelevSovietDefect,ZhidaAXIDisorder,HingeMetaVortexExp,2DpointDefectDatabase,TaylorEmbeddedTIs,TaylorEmbeddedSMs}.  For the topological crystalline insulating phases~\cite{LiangTCI,TeoFuKaneTCI,HsiehTCI,FangFuNSandC2,HarukiRotation,HourglassInsulator,Cohomological,DiracInsulator,ChenRotation,SYBiBr,BiBrFanHOTI,ZhijunRotationHOTI} that have become known as higher-order TIs (HOTIs)~\cite{multipole,WladTheory,HOTIBernevig,HOTIChen,HigherOrderTIPiet,HigherOrderTIPiet2,FanHOTI,EzawaMagneticHOTI,ZeroBerry,FulgaAnon,TMDHOTI,HarukiLayers,HOTIBismuth,ChenTCI,AshvinIndicators,AshvinTCI,S4Guys,TitusInteractHOTI,VDBHOTI,WiederAxion,KoreanAxion,NicoDavidAXI2,IvoAXI1,IvoAXI2,TaylorHingeSC}, these efforts have had only recent, incipient success, motivating the present study.  Furthermore, though HOTI material candidates are readily accessible, experimental studies of candidate HOTIs have yielded results that have attracted an array of --  at times contradictory -- explanations~\cite{HOTIBismuth,AliEarlyBismuth,BismuthSawtooth,BismuthSawtooth2,ZeroHallExp,DavidMoTe2Exp,PhuanOngMoTe2Hinge,MazWTe2Exp,WTe2HingeStep,OtherWTe2Hinge,MazNatPhysHOTIReview,BiBrHOTIExp,BiBrHingeExp1,BiBrHingeExp2,FanZahidRoomTempBiBrExp,BismuthScrewProposal,JenDefect,BismuthSurfaceBallisticExp}.  In this work, we have performed extensive theoretical and numerical calculations demonstrating that dislocations with integer Burgers vectors~\cite{MerminReview} in 3D insulators can bind 0D higher-order end (HEND) states with anomalous charge and spin as a consequence of the bulk topology [see Supplementary Notes (SN)~\ref{sec:kpEdge},~\ref{sec:kpScrew},~\ref{sec:defecttopology},~\ref{sec:FrankScrew},~\ref{sec:indices}, and~\ref{sec:numerics}].  Using first-principles calculations, we have shown that PbTe monolayers~\cite{PbTeMonolayer1,PbTeMonolayer2,LiangTCIMonolayer} have a nontrivial first-order defect-state response (SN~\ref{sec:DFTPbTe}), and that the 3D TCI and HOTI SnTe (but not PbTe)~\cite{ChenRotation,HOTIChen,HOTIBernevig,AndreiMaterials,AndreiMaterials2,BarryPbTe} carries a nontrivial HEND-state dislocation response (SN~\ref{sec:DFTSnTe}).  HEND dislocation states may also be observable in ``fragile'' topological insulators (TIs)~\cite{AshvinFragile,JenFragile1,HingeSM,AdrianFragile,BarryFragile,ZhidaBLG,AshvinBLG1,AshvinBLG2,AshvinFragile2,ZhidaFragileTwist1,ZhidaFragileTwist2,YoungkukMonopole,TMDHOTI,HarukiFragile,OrtixTRealSpace,BouhonMagneticFragile1,BouhonMagneticFragile2,KoreanFragile,WiederAxion,KoreanAxion,NicoDavidAXI2,IvoAXI1,IvoAXI2,
FragileKoreanInversion,ZhidaFragileAffine} and obstructed atomic limits~\cite{QuantumChemistry,Bandrep2,Bandrep3,MTQC,JenOAL}.  Numerous candidate fragile TIs have recently been predicted through high-throughput material searches~\cite{ZhidaFragileAffine,AndreiMaterials2}, and 3D OAL phases have recently been identified in electrides~\cite{MurakamiElectride,AndreiElectride,ZhijunTQCElectride} and other stoichiometric insulators~\cite{AndreiFEOALDatabase,ZhijunOALDatabase,AndreiMainOALDatabase1,AndreiMainOALDatabase2}.

In this work, we have additionally analytically and numerically compared the electronic states bound to crystal defects to the related problem of magnetic flux insertion (see SN~\ref{sec:kpFlux},~\ref{sec:fluxfluxtubetopologymapping}, and~\ref{sec:numerics2}), which has previously been shown to provide a probe of bulk stable topology~\cite{QiFlux,AshvinFlux,AdyFlux,FranzWormhole,WormholeNumerics,MirlinFlux,CorrelatedFlux,Vlad2D,FluxCycleSonicCrystal}.  Building from the previous recognition that the gapped 2D surfaces of 3D magnetic axion insulators (chiral HOTIs)~\cite{HOTIBernevig,HOTIChen,WladTheory,AxionZhida,EslamInversion,FanHOTI,TMDHOTI,HarukiLayers,MTQC,MTQCmaterials,AshvinMagnetic,EzawaMagneticHOTI,VDBHOTI,WiederAxion,KoreanAxion,NicoDavidAXI2,IvoAXI1,IvoAXI2} carry anomalous half-integer quantum Hall states that can be detected by magnetic flux insertion~\cite{FranzWormhole,WormholeNumerics}, we applied the techniques of magnetic flux insertion and filling anomalies~\cite{TMDHOTI,WiederAxion,WladCorners,HingeSM} to $\mathcal{I}$- and $\mathcal{T}$-symmetric helical HOTIs, for which the bulk and surface theories are unknown.  Our numerical calculations reveal in particular that the gapped surfaces of helical HOTIs are neither trivial nor integer topological states, but rather carry anomalous halves of $\mathcal{T}$-invariant quantum \emph{spin} Hall (QSH) states (SN~\ref{sec:kpFlux} and~\ref{sec:3DvortexTRS}).  Similar half-integer QSH states were previously predicted to appear on the top surfaces of weak TIs~\cite{HalfQSH}, and in this work, we recognize the anomalous half-integer QSH state to more generally manifest on \emph{all} gapped surfaces of $\mathcal{I}$- and $\mathcal{T}$-symmetric HOTIs.  This finding raises several interesting questions for future works, including whether halves of QSH states can be distinguished in the absence of $s^{z}$-spin symmetry~\cite{KaneMeleZ2,AndreiTI,LevinSternSpinHallPRL,TitusFTI,CharlieTI,XGWenZooReview}, and whether studies of Anderson localization on weak TI and TCI surfaces~\cite{FuKaneFugacity,JensSurfaceWTI,Radzihovsky2019,MirlinAndresonGraphene,GianAdolphoGaugeTCI}
 should be revisited in light of the existence of anomalous gapped surface states with trivial (charge) Hall conductivities (\emph{i.e.} half QSH states).   Additionally, previous constructions of strongly-interacting topological phases have exploited the half-quantized surface quantum Hall effect of 3D TIs~\cite{MetlitskiKaneSurface,AdyReview,SCZFTI,AliceaSurfaceOrder}, and our identification of a half-quantized surface quantum spin Hall effect in HOTIs may hence also provide further insight towards the theoretical construction of $\mathcal{T}$-symmetric fractional TIs and other phases with anomalous topological order~\cite{SenthilSurfaceSpinLiquid,SenthilSurfaceSpinLiquid2,ZouSurfaceSpinLiquid}.  Lastly, we also explored the limitations of static $\pi$-flux insertion as a complete diagnostic of bulk topology by theoretically demonstrating the presence of identical $\pi$-flux-tube spectra in two physically distinct phases of matter: an $\mathcal{I}$- and $\mathcal{T}$-protected helical HOTI~\cite{HOTIBernevig,TMDHOTI,HOTIBismuth,ChenTCI,AshvinTCI,EslamInversion} with anomalous surface half quantum spin Hall states, and an $\mathcal{I}$- and SU(2)-symmetric spin-doubled (spinless) axion insulator with anomalous surface integer quantum Hall states~\cite{TMDHOTI,YoungkukMonopole,AshvinBosonicTCI,MaxCharlieBosonicTCI,MTQC}.  Though we have largely focused in this work on helical HOTIs, for which numerous material candidates are known~\cite{HOTIBismuth,TMDHOTI,DavidMoTe2Exp,PhuanOngMoTe2Hinge,MazWTe2Exp,ZhijunRotationHOTI,JenHOTImaterial,AndreiMaterials,AndreiMaterials2,ChenMaterials,AshvinMaterials1,AshvinMaterials3}, an $\mathcal{I}$- and SU(2)-symmetric spin-doubled (spinless) axion insulator phase can be realized by inducing $\mathcal{T}$-symmetric orbital (Haldane) magnetism in a monopole-charged nodal-line semimetal state~\cite{TMDHOTI,YoungkukMonopole,ChenWithWithout} like that predicted in 3D graphdiyne~\cite{GraphdiyneMonopole1,GraphdiyneMonopole2}.

During the long preparation of this work, several other works also studied defect and flux responses in topological systems and closely related theoretical methods.  First, concurrent with the preparation of this work and consistent with our findings, a bulk spin-magnetoelectric response and anomalous surface half QSH states were numerically identified in helical HOTIs in Supplementary Reference (SRef.)~\onlinecite{PartialAxionHOTINumerics}.  Next, an analysis of helical modes bound to partial defects with fractional Burgers vectors in $\mathcal{T}$-symmetric HOTIs was performed in SRef.~\onlinecite{JenDefect}; the analysis in SRef.~\onlinecite{JenDefect} is complementary to, but does not overlap with, the results of the present work.  Disclination states in TCIs and higher-order topological superconductors were recently explored in SRef.~\onlinecite{FulgaTCIDefect}, and the presence of gapped helical modes in screw dislocations in $\mathcal{T}$-invariant stable HOTIs was also recently demonstrated in SRef.~\onlinecite{VladHOTIDefect}, but was not attributed to the (fragile or OAL) topological indices introduced in this work.  Spinon excitations in corner-mode phases were also recently demonstrated in SRef.~\onlinecite{YizhiQuadSpinon}.  Additionally, the possibility of topological 0D defect states in 3D insulators was briefly suggested in SRef.~\onlinecite{BJYangVortex} as a means of completing the search for experimentally accessible defect states as classified by Teo and Kane~\cite{TeoKaneDefect}; the HEND states introduced in this work are consistent with this suggestion.  A Wilson-loop formulation of partial nested Berry phase was also introduced during the preparation of this work in SRef.~\onlinecite{KooiPartialNestedBerry}; the authors of SRef.~\onlinecite{KooiPartialNestedBerry} use $C_{2}$ and $\mathcal{T}$ symmetries to obtain a formulation of partial nested Berry phase that is equivalent to the definition that we introduce in SN~\ref{sec:nested}, but do not relate the new bulk topological quantity to the corner (defect) spectrum, as we do in this work.  Closely related concentric Wilson loop invariants for $\mathcal{T}$- and rotation-symmetric insulators were also recently introduced in SRef.~\onlinecite{FragilePartialConcentricWilson}.  During the preparation of this work, the authors of SRef.~\onlinecite{HOTIFractonYizhi19} analyzed the flux response of interacting $C_{4}\times\mathcal{T}$-symmetric AXIs, which are distinct from the $\mathcal{I}$-symmetric AXIs and helical HOTIs considered in this work.  Threaded magnetic flux was also recently theoretically proposed as a means of probing the sample-geometry dependence of the anomalous surface Dirac fermions of the TCI and HOTI SnTe~\cite{RoniSnTeHOTIFlux}.  Lastly, after the submission of this work, a semiclassical treatment of a spinor-axion response was explored in relation
to HOTIs in SRef.~\onlinecite{ZilberbergSpinAxion}.

\section{Mapping Topology from Momentum Space to Position Space with Defects and $\pi$-Flux}
\label{sec:proofs}

In this section, we will provide proofs demonstrating the relationship between the topology of pristine crystals and the electronic states bound to line-like defects and $\pi$-flux tubes.  We will specifically show that dislocations with integer Burgers vectors~\cite{MerminReview} and threaded magnetic $\pi$-flux can map the momentum-space topology of lower-dimensional surfaces of the BZs of pristine crystals to surfaces in position space.  We will analytically demonstrate that, under the preservation of specific symmetries, this mapping reveals the presence of anomalous 0D states bound to the ends of line-like dislocations and flux tubes in 3D insulators.

To reconcile our results with previous works, we will provide two alternative and equivalent sets of proofs demonstrating the aforementioned dislocation and $\pi$-flux topological mapping from momentum space to position space.  First, in SN~\ref{sec:kp}, building upon the ``cutting'' and ``gluing'' construction of topological defect states developed in SRef.~\onlinecite{AshvinScrewTI} to predict dislocation helical modes in weak topological insulators~\cite{FuKaneMele,FuKaneInversion,AdyWeak,MooreBalentsWeak}, we will employ $k\cdot p$ theory to predict HEND states in 3D crystals.  Then, in SN~\ref{sec:frankProofs}, we will use more general arguments to demonstrate that edge dislocations and flux tubes in $d$-dimensional crystals can map $(d-1)$-dimensional [$(d-1)$-D] BZ surfaces to $(d-1)$-D real-space surfaces, leading in 3D crystals to the presence of 1D and 0D topological defect states.  Specifically, in SN~\ref{sec:defecttopology}, we will prove that generalized edge dislocations in $d$-D, $\mathcal{I}$-symmetric crystals map the $(d-1)$-D BZ-surface topology of pristine crystals to the topology of $(d-1)$-D real-space surfaces between crystal defects, and in SN~\ref{sec:fluxfluxtubetopologymapping}, we will employ similar arguments to analyze the topology of $\mathcal{I}$-symmetric arrangements of flux tubes in $d$-D crystals.  Finally, in SN~\ref{sec:FrankScrew}, we will extend the arguments in SN~\ref{sec:defecttopology} and~\ref{sec:fluxfluxtubetopologymapping} to the experimentally relevant case of (1D) screw dislocations in 3D crystals.

\subsection{Defect and Flux Bound States in 3D $\mathcal{I}$-Symmetric Insulators from $k\cdot p$ Theory}
\label{sec:kp}

\begin{figure}[h]
\includegraphics[width=.5\textwidth]{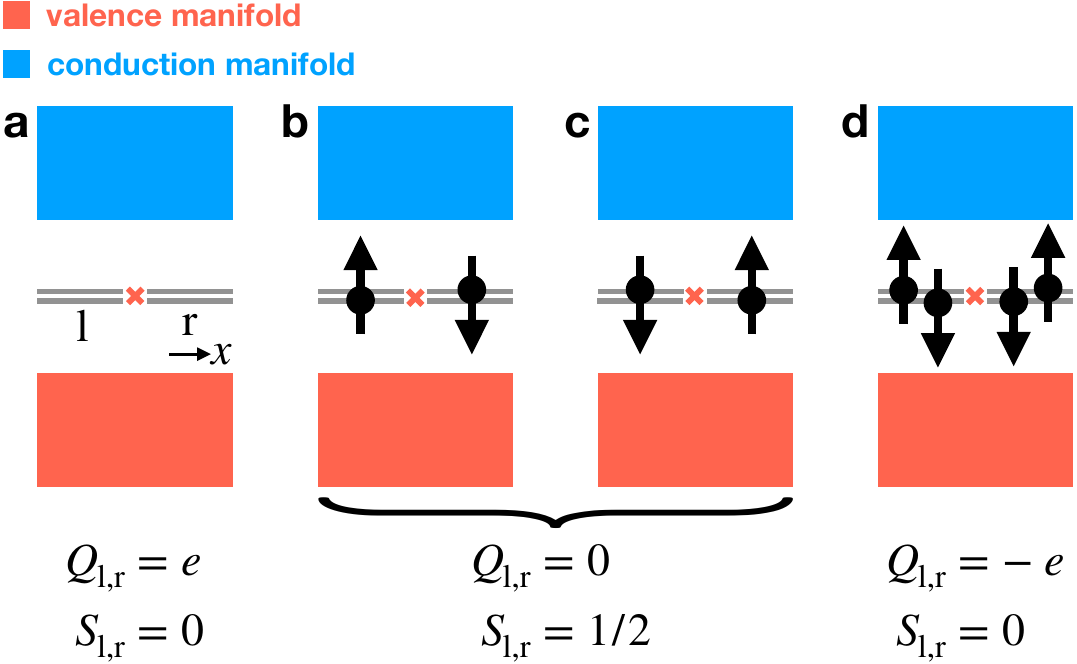}
\caption{{\bf Spin-charge-separated Kramers pairs of defect or flux states.} $\bs{a-d}$~An inversion- ($\mathcal{I}$-) related pair of Kramers pairs of defect or flux states in a spinful, time-reversal- ($\mathcal{T}$-) symmetric insulator (where the $\mathcal{I}$ center is represented with a red $\times$ symbol in~$\bs{a-d}$).  $\bs{b,c}$~When the entire system lies at charge neutrality, each Kramers pair is filled by only a single electron, and therefore carries an excess chargeless spin-1/2 moment.  Hence, at half filling and taking the spins of the electrons occupying each pair of states to point in opposite directions, $\mathcal{I}$ (which relates the positions of the Kramers pairs) and $\mathcal{T}$ symmetries are ``softly'' broken~\cite{WilczekAxion,GoldstoneWilczek,NiemiSemenoff,WiederAxion,HingeSM}, and each half-filled Kramers pair of states forms an effective spinon quasiparticle with a free-angle spin-1/2 moment (depicted in~$\bs{b,c}$ in configurations that preserve $\mathcal{I}\times\mathcal{T}$ symmetry).  By~$\bs{a}$ removing or~$\bs{d}$ adding two electrons to the system (one electron per Kramers pair), we may realize a system configuration in which each Kramers pair respectively carries a net charge of $\pm e$ (taking electrons to carry a charge $-e$), but carries a net-zero spin.  Hence, each Kramers pair of defect states either carries chargeless spin or spinless charge, and therefore exhibits the same reversed spin-charge relations as the solitons in polyacetylene~\cite{RiceMele,SSH,SSHspinon,HeegerReview}.}
\label{fig:spinCharge}
\end{figure}

In this section, we will employ $k\cdot p$ theory to demonstrate the presence of HEND bound states on the corners and ends of crystal defects and flux tubes in two classes of HOTIs: inversion- ($\mathcal{I}$-) symmetric axion insulators (AXIs)~\cite{HOTIBernevig,HOTIChen,WladTheory,AxionZhida,EslamInversion,FanHOTI,TMDHOTI,HarukiLayers,MTQC,MTQCmaterials,AshvinMagnetic,EzawaMagneticHOTI,VDBHOTI,WiederAxion,KoreanAxion,NicoDavidAXI2,IvoAXI1,IvoAXI2}, which are 3D magnetic insulators with anomalous, gapped 2D surfaces and gapless hinges with 1D chiral modes~\cite{FuKaneMele,FuKaneInversion,AndreiInversion,QHZ,FanHOTI,VDBAxion,DiracInsulator,MulliganAnomaly,DrewPotterPeel,TitusHOTISurfaceAnomaly,VDBHOTI,WiederAxion}, and time-reversal- ($\mathcal{T}$-) and $\mathcal{I}$-symmetric HOTIs~\cite{WladTheory,AshvinIndicators,AshvinTCI,ChenTCI,TMDHOTI} with anomalous, gapped ($\mathcal{T}$-symmetric) surfaces~\cite{DiracInsulator} and gapless hinges with 1D \emph{helical} modes (see SN~\ref{sec:kpFlux} for more details).  We will also show that crystal defects, but crucially \emph{not} $\pi$-flux tubes, bind anomalous HEND states in weak stacks of $\mathcal{I}$-symmetric \emph{fragile} topological insulators (FTIs)~\cite{AshvinFragile,JenFragile1,HingeSM,AdrianFragile,BarryFragile,ZhidaBLG,AshvinBLG1,AshvinBLG2,AshvinFragile2,ZhidaFragileTwist1,ZhidaFragileTwist2,YoungkukMonopole,TMDHOTI,HarukiFragile,OrtixTRealSpace,BouhonMagneticFragile1,BouhonMagneticFragile2,KoreanFragile,WiederAxion,KoreanAxion,NicoDavidAXI2,IvoAXI1,IvoAXI2,
FragileKoreanInversion,ZhidaFragileAffine}  and obstructed atomic limits (OALs)~\cite{QuantumChemistry,Bandrep2,Bandrep3,MTQC}.

Because AXIs and $\mathcal{I}$- and $\mathcal{T}$-symmetric HOTIs can be diagnosed by their bulk parity ($\mathcal{I}$) eigenvalues (through $\mathbb{Z}_{2}$-~\cite{FuKaneMele,FuKaneInversion,AndreiInversion,QHZ,VDBAxion,EslamInversion,MTQC,MTQCmaterials,AshvinMagnetic,WiederAxion,KoreanAxion,NicoDavidAXI2,IvoAXI1,IvoAXI2,MTQC,MTQCmaterials} and $\mathbb{Z}_{4}$-valued~\cite{AshvinIndicators,AshvinTCI,ChenTCI,TMDHOTI} indices, respectively), then we can capture the bulk topology of AXIs and $\mathcal{I}$- and $\mathcal{T}$-symmetric HOTIs that differ from trivial insulators by (double) band-inversion at a time-reversal-invariant crystal momentum (TRIM point) $k_{\alpha,\beta,\gamma}=0,\pi$ through $k\cdot p$ Hamiltonians.  For simplicity, in this section, we will only consider crystals with orthorhombic lattice vectors~\cite{BigBook}, for which $\alpha,\beta,\gamma=x,y,z$, though our results can be extended to body- and face-centered geometries through generalizations of the BZ folding arguments in SRef.~\onlinecite{HingeSM}.  First, in SN~\ref{sec:kpEdge}, we will extend the $k\cdot p$ construction from SRef.~\onlinecite{AshvinScrewTI} to predict $\pm e/2$-charged HEND states bound to the corners of edge dislocations in AXIs and 3D weak FTIs.  In SN~\ref{sec:kpScrew}, we will then adapt the arguments from SN~\ref{sec:kpEdge} to also predict HEND states bound to the ends of screw dislocations in AXIs and 3D weak FTIs.  Finally, in SN~\ref{sec:kpFlux}, we will further utilize the $k\cdot p$ construction in SN~\ref{sec:kpEdge} and~\ref{sec:kpScrew} to analyze flux threading in 3D insulators.  In particular, in SN~\ref{sec:kpFlux}, we will show that the ``wormhole'' helical modes of strong TIs~\cite{QiFlux,AshvinFlux,AdyFlux,FranzWormhole,WormholeNumerics,MirlinFlux,CorrelatedFlux} with threaded $\pi$-flux evolve into anomalous HEND states in AXIs under the introduction of $\mathcal{I}$-symmetric magnetism.

An $\mathcal{I}$- and $\mathcal{T}$- symmetric HOTI can be formed by superposing a time-reversed pair of AXIs.  Hence, the results derived in SN~\ref{sec:kpEdge},~\ref{sec:kpScrew}, and~\ref{sec:kpFlux} for AXIs (and $\mathcal{I}$-symmetric, $\mathcal{T}$-broken FTIs) can also be straightforwardly extended to helical HOTIs (and $\mathcal{I}$- and $\mathcal{T}$-symmetric FTIs).  Specifically, we will show that $\mathcal{I}$- and $\mathcal{T}$-symmetric insulators with crystal defects or threaded $\pi$-flux bind \emph{Kramers pairs} of HEND states that are equivalent to higher-order generalizations~\cite{multipole,TMDHOTI,HingeSM,WiederAxion,KoreanFragile} of the spin-charge separated solitons (\emph{i.e.} spinons or \emph{fluxons}) discussed in SRefs.~\onlinecite{RiceMele,SSHspinon,HeegerReview,QiFlux,AshvinFlux,AdyFlux,MirlinFlux,CorrelatedFlux}.

To understand the spin-charge separation, we consider two $\mathcal{I}$-related pairs of topological defects or flux tubes in a 2D or 3D insulator that each binds a Kramers pair of 0D states (four degenerate single-particle states in total), taking each Kramers pair of states to be half-filled at charge neutrality [Supplementary Figure (SFig.)~\ref{fig:spinCharge}~$\bs{b,c}$], so that the ground state is degenerate.  Enforcing $\mathcal{I}$ symmetry (where we have denoted a global $\mathcal{I}$ center with a red $\times$ symbol in SFig.~\ref{fig:spinCharge}), there is one filled state per Kramers pair.  Hence, each Kramers pair carries a balanced (net-zero) charge, but necessarily ``softly'' breaks $\mathcal{T}$ symmetry, because each pair of states is filled with an unpaired spin-1/2 degree of freedom (without spin conservation symmetry such as $s^{z}$, however, the unpaired electron is not required to exhibit a quantized spin projection along a particular high-symmetry axis).  When the system is doped away from charge neutrality by adding two more electrons in an $\mathcal{I}$-symmetric fashion, $\mathcal{I}$ and $\mathcal{T}$ symmetries can be satisfied (SFig.~\ref{fig:spinCharge}~$\bs{d}$).  In the system configuration with two extra electrons, each fully-filled Kramers pair of states carries a charge $-e$ (taking electrons to have charge $-e$ measured from charge neutrality).  Unlike in the previous system configuration with chargeless spin-1/2 0D states at zero doping in SFig.~\ref{fig:spinCharge}~$\bs{b,c}$, at a system doping of $-2e$, each Kramers pair of states is charged, but exhibits a net-zero spin, because $\mathcal{T}$ symmetry pairs electrons with reversed spins.  In this $\mathcal{I}$- and $\mathcal{T}$-symmetric system configuration, the ground state is unique and has no degeneracy.  Similarly, if we remove one electron from each Kramers pair of states in SFig.~\ref{fig:spinCharge}~$\bs{b,c}$, then we realize a system configuration in which there is a total charge of $+2e$, implying that each fully empty pair of states carries a charge $+e$ and does not carry an electron spin (SFig.~\ref{fig:spinCharge}~$\bs{a})$.  Hence, the 0D Kramers pairs of states exhibit the same well-established spin-charge separation and reversed spin-charge relations as the solitons in polyacetylene~\cite{RiceMele,SSH,SSHspinon,HeegerReview}.

\subsubsection{$k\cdot p$ Derivation of HEND States on Edge Dislocation Corners in AXIs and Weak Fragile TIs}
\label{sec:kpEdge}

To derive the edge dislocation response of $\mathcal{I}$-symmetric 3D magnetic insulators, we will begin by formulating the $k\cdot p$ Hamiltonian of an $\mathcal{I}$- and $\mathcal{T}$-symmetric 3D insulator.  After we complete this derivation, we will then relax $\mathcal{T}$ symmetry and observe the emergence of HEND states on edge dislocation corners.  We consider the topology of the 3D $\mathcal{I}$- and $\mathcal{T}$-symmetric insulator to be fully captured by a set of band inversions at TRIM points between the two highest valence bands and the two lowest conduction bands~\cite{FuKaneMele,FuKaneInversion,AndreiInversion}, and therefore, for now, we exclude HOTIs and weak FTIs formed from \emph{double} band inversion at the same TRIM point~\cite{YoungkukMonopole,WladTheory,HOTIBernevig,AshvinIndicators,HOTIBismuth,TMDHOTI,WiederAxion}.  Later, in SN~\ref{sec:kpFlux}, we will reintroduce more general symmetry and counting arguments that capture HOTIs formed through double band inversion.

\begin{figure}[t]
\includegraphics[width=.89\textwidth]{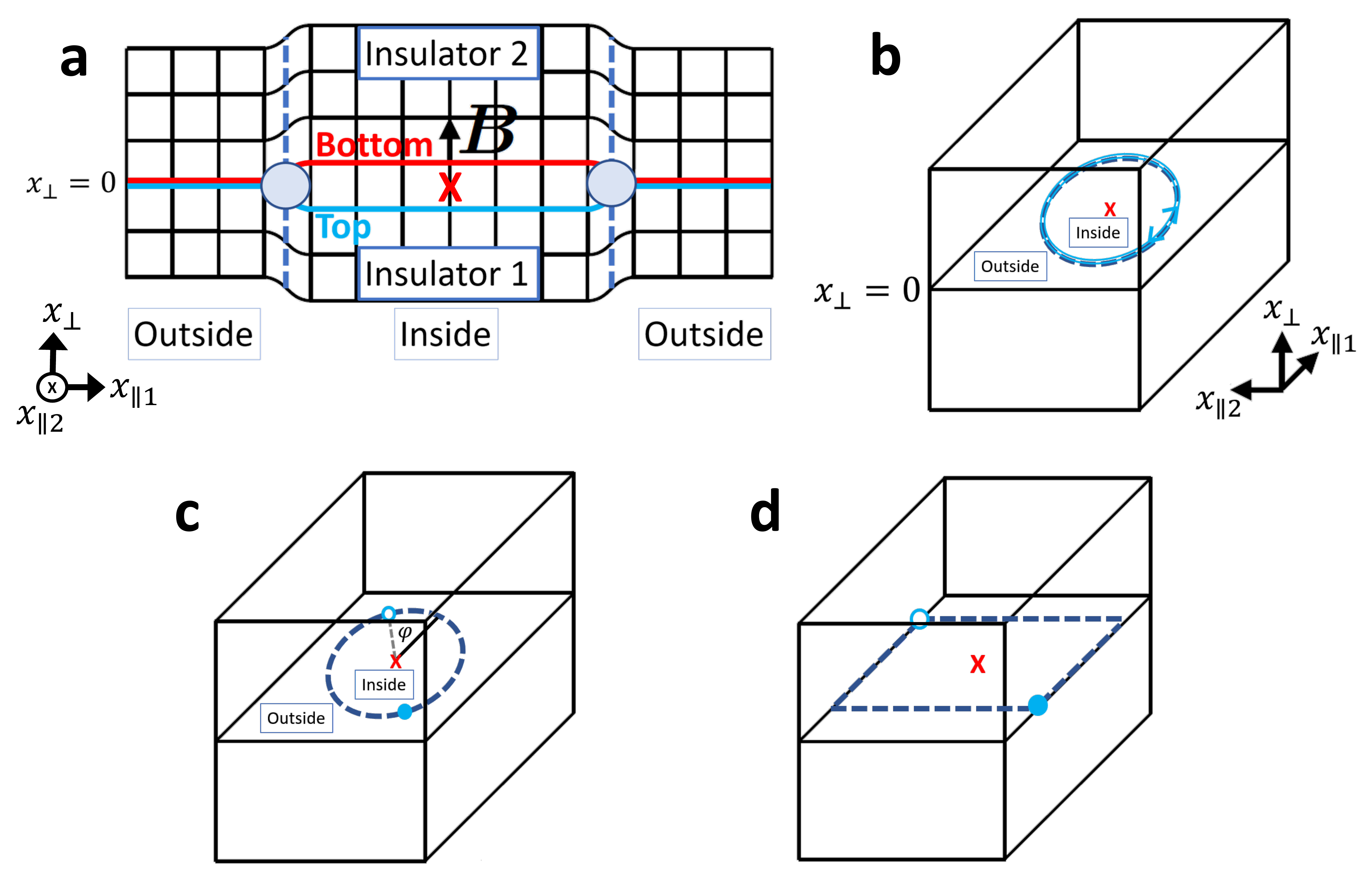}
\caption{{\bf 0D edge dislocation states in a 3D crystal.} $\bs{a}$ The Burgers vector $\bs{B}$ of the two dislocations (blue circles) has a length $|\bs{B}|=a_{\perp}$, the lattice spacing in the $x_{\perp}$ direction.  Following the procedure in SRef.~\onlinecite{AshvinScrewTI}, we divide the system by cutting along the red, blue, and dashed lines.  The red and blue lines separate two insulators with the same bulk topology, whose top and bottom surfaces are glued together with (inside) and without (outside) an additional row of unit cells.  $\bs{b}$  A 3D crystal with an internal edge dislocation formed by taking the two dislocations in $\bs{a}$ and closing them into a circle in the $x_{\perp} = 0$ plane.  As shown in SRefs.~\onlinecite{AshvinScrewTI,TeoKaneDefect,QiDefect2,Vlad2D,TanakaDefect,VladScrewTI} and rederived in SEq.~(\ref{eq:BurgersAppendix}), if insulators 1,2 are inversion- ($\mathcal{I}$-) and time-reversal- ($\mathcal{T}$-) symmetric and feature nontrivial weak indices $\bs{M}_{\nu}$, then, if $\bs{B}\cdot \bs{M}_{\nu}\text{ mod }2\pi = \pi$, a helical pair of modes will be bound to the circular edge dislocation (dashed line) in $\bs{b}$.  However, we can further extend the set of possible dislocation responses by requiring that an inversion center (red $\times$) is preserved in the plane surrounded by the dislocation.  $\bs{c}$  In this case, under the relaxation of $\mathcal{T}$ symmetry in an $\mathcal{I}$-symmetric manner, the helical modes gap into $2+4n$ (filling) anomalous~\cite{TMDHOTI,WiederAxion,WladCorners,HingeSM} 0D states with charges $\pm e/2$ [SEq.~(\ref{eq:FTImass})].  Specifically, the 0D states are equivalent to the corner modes of the 2D FTIs detailed in SRefs.~\onlinecite{TMDHOTI,WiederAxion}.  $\bs{d}$  We next smoothly grow the internal dislocation until it meets with the crystal boundary, at which point the edge dislocations coincide with exterior crystal surfaces, realizing an effective ``stacking'' defect within the region bounded by the edge dislocations (dashed lines).  In $\bs{d}$, two of the four corners between edge dislocations bind $\pm e/2$-charged higher-order 0D (\emph{i.e.} HEND) states.  Because the process of breaking $\mathcal{T}$ symmetry while preserving $\mathcal{I}$ converts $\bs{M}_{\nu}$ to the weak fragile (and OAL) index $\bs{M}^{\mathrm{F}}_{\nu}$ [defined in SEq.~(\ref{eq:NandNu}) and the surrounding text in SN~\ref{sec:weakFragile}], then we conclude that anomalous HEND states are present in $\mathcal{T}$-broken, $\mathcal{I}$-symmetric 3D insulators with edge dislocations for which $\bs{B}\cdot \bs{M}^{\mathrm{F}}_{\nu}\text{ mod }2\pi = \pi$ [SEq.~(\ref{eq:BurgersFragileAppendix})].}
\label{fig:kpDefect}
\end{figure}

The low-energy Hamiltonian of an $\mathcal{I}$-symmetric insulator can be expressed as:
\begin{equation}
\mathcal{H}(\bs{q}) = \bigoplus_{a}\mathcal{H}_{a}(\bs{q}),
\label{eq:Htotal}
\end{equation}
where $a$ runs over the TRIM points $\bs{k}_{D,a}$ whose bands are inverted relative to those of the atomic insulator formed from the occupied atomic orbitals when all hoppings are taken to vanish~\cite{QuantumChemistry}, and where $\bs{q} = \bs{k}-\bs{k}_{D,a}$.  In anticipation of later breaking translation symmetry through the insertion of a crystal defect, we note that in Supplementary Equation (SEq.)~(\ref{eq:Htotal}), the summation is equivalent to folding the Hamiltonians of the band-inverted TRIM points onto a single $k$ point and expanding in a $k\cdot p$ momentum range ${\bs q}$ around that point.  We take the $k\cdot p$ Hamiltonian of each TRIM point $\bs{k}_{D,a}$ to have the form of the low-energy theory of the Bernevig-Hughes-Zhang (BHZ) model~\cite{AndreiTI,QHZ,AndreiInversion} of a 3D TI (where we have neglected a $\bs{q}^2$ contribution to the bulk mass, because it does not affect the topology):
\begin{equation}
\mathcal{H}_{a}(\bs{q}) = m_{a}\tau^{z} + \sum_{i=x,y,z}v_{i}q_{i}\tau^{x}\sigma^{i}.
\label{eq:BHZtrim}
\end{equation}
SEq.~(\ref{eq:BHZtrim}) is invariant under $\mathcal{I}$ and $\mathcal{T}$ symmetries, whose action we represent as:
\begin{equation}
\mathcal{I}:\ \mathcal{H}_{a}(\bs{q})\rightarrow\tau^{z}\mathcal{H}_{a}(-\bs{q})\tau^{z},\ \mathcal{T}:\ \mathcal{H}_{a}(\bs{q})\rightarrow\sigma^{y}\mathcal{H}^{*}_{a}(-\bs{q})\sigma^{y}.
\label{eq:bulkIT}
\end{equation}

Next, we construct a long-wavelength description of a loop of edge dislocations whose Burgers vectors $\bs{B}=(B_{x},B_{y},B_{z})$ lie along one of the lattice vectors (SFig.~\ref{fig:kpDefect}~$\bs{a}$), where $B_{x,y,z}$ are integer multiples of the lattice spacing.  For the purposes of the following proof, we take $\bs{B}\parallel \hat{z}$, such that:
\begin{equation}
x_{\perp} = z,\ x_{\parallel 1} = x,\ x_{\parallel 2} = y,
\end{equation}
in the notation of SFig.~\ref{fig:kpDefect}.  As prescribed in SRef.~\onlinecite{AshvinScrewTI}, we model an internal loop of edge dislocations by cutting the insulator described by $\mathcal{H}(\bs{q})$ [SEq.~(\ref{eq:Htotal})] into two pieces with $\pm\hat{z}$-normal surfaces, and then ``gluing'' the two pieces back together with and without extra rows of unit cells in the region (labeled ``inside'' in SFig.~\ref{fig:kpDefect}) between the edge dislocations.

Using the labeling in SFig.~\ref{fig:kpDefect}~$\bs{a}$, we begin by solving for the Hamiltonian of the top surface of insulator $1$.  We first form a Jackiw-Rebbi domain wall~\cite{JackiwRebbi,HingeSM} by taking the mass at each band-inverted bulk TRIM point to spatially vary $m_{a}\rightarrow m^T(z)$ [SEq.~(\ref{eq:BHZtrim})]. Specifically, we take $m^T(z)$ to be large and negative for $z<z^{T}_{\mu}$ and large and positive for $z>z^{T}_{\mu}$, where $z^{T}_{\mu}$ is the $z$-coordinate of the top surface of insulator $1$ in either the ``inside'' or ``outside'' region in SFig.~\ref{fig:kpDefect}, indexed by $\mu$.  This distribution of $m^T(z)$ can be summarized as:
\begin{equation}
\sgn[m^T(z)] = \sgn[z-z_{\mu}^{T}].
\label{eq:mzTop}
\end{equation}
Next, taking $q_{x}=q_{y}=0$, we Fourier transform:
\begin{equation}
q_{z}\rightarrow -(\bs{k}_{D,a}\cdot\hat{z}) -i\partial_{z},
\label{eq:bulkFourier}
\end{equation}
and search for zero-energy, normalizable bound states $|\psi^{T}_{a,\mu}(z)\rangle$ that satisfy:
\begin{equation}
[m^T(z)\tau^{z}-v_{z}\tau^{x}\sigma^{z}(\bs{k}_{D,a}\cdot\hat{z} +i\partial_{z})]|\psi^{T}_{a,\mu}(z)\rangle=0.
\label{eq:topJackiw}
\end{equation}
SEq.~(\ref{eq:topJackiw}) can be solved by left-multiplying by $\tau^{x}\sigma^{z}$ and integrating:
\begin{equation}
|\psi^{T}_{a,\mu}(z)\rangle_{i} = \frac{1}{\sqrt{N}}e^{i(\bs{k}_{D,a}\cdot\hat{z})(z-z^{T}_{\mu})}e^{-\frac{1}{v_{z}}\int_{z^{T}_{\mu}}^{z}m^T(z')dz'}|+\rangle_{i} \equiv e^{i(\bs{k}_{D,a}\cdot\hat{z})(z-z^{T}_{\mu})}\mathcal{Z}^T(z-z^{T}_{\mu})|+\rangle_{i},
\label{eq:topBound}
\end{equation}
where $N$ is a normalization constant and $|+\rangle_{i}$ is an eigenstate of $\tau^{y}\sigma^{z}$ with eigenvalue $+1$(SFig.~\ref{fig:kpDefect}~$\bs{a}$):
\begin{equation}
|+\rangle_{1}=\frac{1}{\sqrt{2}}\left(\begin{array}{c}
0 \\
i \\
0 \\
1\end{array}\right),\ |+\rangle_{2}=\frac{1}{\sqrt{2}}\left(\begin{array}{c}
-i \\
0 \\
1 \\
0\end{array}\right).
\label{eq:topEvals}
\end{equation}
To restore dispersion in the $q_{x,y}$ directions, we follow the procedure employed in SRefs.~\onlinecite{HOTIBismuth,HingeSM} for higher-order TIs and semimetals and project the linear terms in SEq.~(\ref{eq:BHZtrim}) into the basis of $|\psi^{T}_{a,\mu}(z)\rangle_{i}$:
\begin{eqnarray}
[\mathcal{H}^{T}_{a,\mu}(q_{x},q_{y})]_{ij} &=& {}_{i}\langle\psi^{T}_{a,\mu}(z)| v_{x}q_{x}\tau^{x}\sigma^{x} + v_{y}q_{y}\tau^{x}\sigma^{y}|\psi^{T}_{a,\mu}(z)\rangle_{j} = \int_{-\infty}^{\infty}dz |\mathcal{Z}^T(z-z^{T}_{\mu})|^{2} {}_{i}\langle +|\mathcal{H}_{a}(q_{x},q_{y},0)|+\rangle_{j}, \nonumber \\
\mathcal{H}^{T}_{a,\mu}(q_{x},q_{y}) &=& v_{x}q_{x}s^{y}+v_{y}q_{y}s^{x},
\label{eq:Htop}
\end{eqnarray}
where $s$ is a matrix in the $2\times 2$ basis of $|+\rangle_{1,2}$, and where we have exploited that $|\mathcal{Z}^T(z-z^{T}_{\mu})|^{2}\rightarrow \delta(z-z^{T}_{\mu})$ if $m(z)$ rapidly saturates at a large value relative to~\cite{JackiwRebbi,HingeSM} $v_{z}$. Though SEq.~(\ref{eq:Htop}) is actually exact, because $\mathcal{Z}^T(z-z^{T}_{\mu})$ is already normalized in SEq.~(\ref{eq:topBound}), we will find the approximation $\mathcal{Z}^T(z-z^{T}_{\mu})\sim\sqrt{\delta(z-z^{T}_{\mu})}$ useful for subsequent calculations. $\mathcal{H}^{T}_{a,\mu}(q_{x},q_{y})$ has the expected form of the $k\cdot p$ Hamiltonian of the linear (twofold Dirac) surface fermion of a 3D TI~\cite{FanTISurface}, and correspondingly, respects $\mathcal{T}$ symmetry, here represented by the action:
\begin{equation}
\mathcal{T}:\ \mathcal{H}^{T}_{a,\mu}(q_{x},q_{y})\rightarrow s^{y}[\mathcal{H}^{T}_{a,\mu}(-q_{x},-q_{y})]^{*}s^{y}.
\label{eq:Ttop}
\end{equation}

Next, we solve for the bottom surface Hamiltonian of the insulator described by SEq.~(\ref{eq:Htotal}). The derivation is equivalent to the previous derivation for the top surface up to the replacement $m^T(z) \rightarrow m^B(z) = m^T(-z)$ (where we have chosen the origin of the $z$ coordinate such that $z^{B}_{\mu} = - z^{T}_{\mu}$).  Then,
\begin{equation}
\sgn[m^B(z)] = \sgn[-z+z_{\mu}^{B}].
\label{eq:mzBottom}
\end{equation}
Due to the change in sign of the $z$ coordinate, the normalizable solutions for the bottom surface are instead proportional to eigenstates of $\tau^{y}\sigma^{z}$ with eigenvalue $-1$, as opposed to the eigenstates with eigenvalue $+1$ previously obtained for the top surface Hamiltonian [SEq.~(\ref{eq:topEvals})].  For the bottom surface states, we obtain:
\begin{equation}
|\psi^{B}_{a,\mu}(z)\rangle_{i} = \frac{1}{\sqrt{N}}e^{i(\bs{k}_{D,a}\cdot\hat{z})(z-z^{B}_{\mu})}e^{+\frac{1}{v_{z}}\int_{z^{B}_{\mu}}^{z}m^B(z')dz'}|-\rangle_{i}=e^{i(\bs{k}_{D,a}\cdot\hat{z})(z-z^{B}_{\mu})}\mathcal{Z}^B(z-z^{B}_{\mu})|-\rangle_{i},
\label{eq:bottomBound}
\end{equation}
where $N$ is a normalization constant and $|-\rangle_{i}$ is an eigenstate of $\tau^{y}\sigma^{z}$ with eigenvalue $-1$ (SFig.~\ref{fig:kpDefect}~$\bs{a}$):
\begin{equation}
|-\rangle_{1}=\frac{1}{\sqrt{2}}\left(\begin{array}{c}
0 \\
i \\
0 \\
-1\end{array}\right),\ |-\rangle_{2}=\frac{1}{\sqrt{2}}\left(\begin{array}{c}
-i \\
0 \\
-1 \\
0\end{array}\right).
\end{equation}
To restore dispersion in the $q_{x,y}$ directions, we follow the procedure in SEq.~(\ref{eq:Htop}) and project the linear terms in SEq.~(\ref{eq:BHZtrim}) into the basis of the bottom surfaces states $|\psi^{B}_{a,\mu}(z)\rangle_{i}$:
\begin{eqnarray}
\mathcal{H}^{B}_{a,\mu}(q_{x},q_{y}) = -v_{x}q_{x}s^{y}-v_{y}q_{y}s^{x}=-\mathcal{H}^{T}_{a,\mu}(q_{x},q_{y}),
\label{eq:Hbottom}
\end{eqnarray}
where $s$ is a matrix in the $2\times 2$ basis of $|-\rangle_{1,2}$. Like $\mathcal{H}^{T}_{a,\mu}(q_{x},q_{y})$, $\mathcal{H}^{B}_{a,\mu}(q_{x},q_{y})$ respects $\mathcal{T}$ symmetry, which is represented by the same action as in SEq.~(\ref{eq:Ttop}).

Now that we have established the forms of the top- and bottom-surface Hamiltonians [SEqs.~(\ref{eq:Htop}) and~(\ref{eq:Hbottom})], we next explicitly form an $\mathcal{I}$-symmetric interface between the top and bottom of two insulators that each exhibit the same bulk band ordering and topology as SEq.~(\ref{eq:Htotal}) (SFig.~\ref{fig:kpDefect}~$\bs{a}$).  This is equivalent to cutting a single copy of this insulator into two halves~\cite{AshvinScrewTI} related by $\mathcal{I}$ symmetry (red $\times$ in SFig.~\ref{fig:kpDefect}).  Redefining the $z$ coordinates, we take the inversion center of the cut region to lie at $z=0$, and consider two regions: an ``outside'' ($\mu=out$) region in which the top and bottom surfaces are flush with each other, and an ``inside'' ($\mu=in$) region in which there are $|\bs{B}|$ extra unit cells in the $z$ direction between the top and bottom surfaces of the two insulators (SFig.~\ref{fig:kpDefect}~$\bs{a}$).  The 1D interface between the ``inside,'' ``outside,'' ``top,'' and ``bottom'' regions in SFig.~\ref{fig:kpDefect}~$\bs{a}$ is therefore spanned by edge dislocations with Burgers vector $\bs{B}$.  Before allowing coupling between the top and bottom surfaces, we combine the two Kramers pairs of linear surface fermions in SEqs.~(\ref{eq:Htop}) and~(\ref{eq:Hbottom}) in both the ``inside'' and ``outside'' regions to form the Hamiltonian of a fourfold-degenerate, 2D Dirac fermion~\cite{Steve2D,WiederLayers,SteveMagnet,DiracInsulator}:
\begin{equation}
\mathcal{H}^{D}_{a,\mu}(q_{x},q_{y}) = v_{x}q_{x}\xi^{z}s^{y}+v_{y}q_{y}\xi^{z}s^{x},
\label{eq:defectHam}
\end{equation}
where $\xi$ is a matrix that acts on the ``top'' and ``bottom'' degrees of freedom and $s$ continues to act on the indices $i=1,2$ in $|\pm\rangle_{i}$ within each of the two surfaces of the insulator described by SEq.~(\ref{eq:Htotal}).  In this geometry (SFig.~\ref{fig:kpDefect}~$\bs{a}$), $\mathcal{I}$ is represented through the action:
\begin{equation}
\mathcal{I}:\ \mathcal{H}^{D}_{a,\mu}(q_{x},q_{y})\rightarrow \xi^{x}\mathcal{H}^{D}_{a,\mu}(-q_{x},-q_{y})\xi^{x}.
\label{eq:defectInversion}
\end{equation}
SEq.~(\ref{eq:defectHam}) also remains invariant under $\mathcal{T}$ symmetry, which is represented in the $4\times 4$ basis of $\xi\otimes s$ by:
\begin{equation}
\mathcal{T}:\ \mathcal{H}^{D}_{a,\mu}(q_{x},q_{y})\rightarrow s^{y}[\mathcal{H}^{D}_{a,\mu}(-q_{x},-q_{y})]^{*}s^{y}.
\label{eq:Tdefect}
\end{equation}

Finally, we allow for $\mathcal{I}$- and $\mathcal{T}$-symmetric coupling between the top and bottom surfaces.  Because $\mathcal{H}^{D}_{a,\mu}(q_{x},q_{y})$ in SEq.~(\ref{eq:defectHam}) is equivalent to the Hamiltonian of a fourfold Dirac fermion in 2D~\cite{Steve2D,WiederLayers,SteveMagnet,DiracInsulator}, then $\mathcal{H}^{D}_{a,\mu}(q_{x},q_{y})$ admits a single, $\mathcal{T}$-symmetric, anticommuting mass term, which here is proportional to $\xi^{x}$.  To account for the presence or absence of nonzero $\bs{B}$ in the regions indexed by $\mu$, we derive a consistent phase for the coupling mass by fixing the defect displacements to be centered about the interface at $z=0$ (SFig.~\ref{fig:kpDefect}~$\bs{a}$).  We then define a spinless interface coupling potential:
\begin{align}
[V^{\mathrm{C}}_{a,\mu}]^{TB}_{ij} &= m_{\mathrm{C}}\delta_{ij}\int_{-\infty}^{\infty}dz\ \left[|\psi^{T}_{a,\mu}(z)\rangle_{i}\langle\psi^{B}_{a,\mu}(z)|_{j} + \mathrm{ h.c.}\right], \nonumber \\
[V^{\mathrm{C}}_{a,\mu}]^{TT}_{ij} &= [V^{\mathrm{C}}_{a,\mu}]^{BB}_{ij} = 0,
\label{eq:couplingM}
\end{align}
where $m_{\mathrm{C}}$ is the strength of the coupling; $i,j=1,2$; and $\delta_{ij}$ enforces $\mathcal{T}$ symmetry [SEq.~(\ref{eq:Tdefect})] by excluding magnetic coupling interactions (for now).  Substituting SEqs.~(\ref{eq:topBound}) and~(\ref{eq:bottomBound}) into SEq.~(\ref{eq:couplingM}), we arrive at the more specific expression:
\begin{equation}
[V^{\mathrm{C}}_{a,\mu}]^{TB}_{ij} = m_{\mathrm{C}}\delta_{ij} e^{i(\bs{k}_{D,a}\cdot\hat{z})(z^{B}_{\mu}-z^{T}_{\mu})} \left[\int_{-\infty}^{\infty}dz\ \mathcal{Z}^T(z-z^{T}_{\mu})\mathcal{Z}^B(z-z^{B}_{\mu}) \right] |+\rangle_{i}{}_{j}\langle -| +\mathrm{ h.c.}.
\label{eq:couplingM2}
\end{equation}

In the $\mu=out$ region, for which:
\begin{equation}
z^{T}_{out}=z^{B}_{out}=0,
\end{equation}
SEq.~(\ref{eq:couplingM2}) greatly simplifies:
\begin{eqnarray}
[V^{\mathrm{C}}_{a,out}]^{TB}_{ij}&=&m_{\mathrm{C}}\delta_{ij}\int_{-\infty}^{\infty}dz\ \delta(z) = m_{\mathrm{C}}\delta_{ij}, \nonumber \\
V^{\mathrm{C}}_{a,out} &=& m_{\mathrm{C}}\xi^{x},
\label{eq:outsideCoupling}
\end{eqnarray}
for which we have exploited that both $\mathcal{Z}^{T,B}(z)$ can be taken to go to zero sufficiently rapidly away from $z=0$  [see the text after SEq.~(\ref{eq:Htop})] to approximate that $[\mathcal{Z}^{T,B}(z)]^2 \rightarrow \delta(z)$.  For the $\mu=in$ region, however, for which:
\begin{equation}
z^{T}_{in}=-|\bs{B}|/2,\   z^{B}_{in}=|\bs{B}|/2,
\label{eq:edgeburgersin}
\end{equation}
SEq.~(\ref{eq:couplingM2}) acquires an additional phase:
\begin{equation}
[V^{\mathrm{C}}_{a,in}]^{TB}_{ij}=m_{\mathrm{C}}\delta_{ij} e^{i(\bs{k}_{D,a}\cdot\hat{z})|\bs{B}|} \int_{-\infty}^{\infty}dz\ C\delta(z),
\label{eq:insideCouplingPre}
\end{equation}
where we have approximated $\mathcal{Z}^T(z+|\bs{B}|/2)\mathcal{Z}^{B}(z-|\bs{B}|/2)$ as a single delta function localized at $z=0$ weighted by the real constant $C\geq 0$, which captures the overlap between the two offset wavefunctions $\mathcal{Z}^{T,B}(z)$.  This approximation is valid if $|\bs{B}|$ is small compared to the decay lengths of $\mathcal{Z}^{T,B}(z)$.  While the limit of SEq.~(\ref{eq:insideCouplingPre}) necessitates that $C <1$ (and possibly $C\ll 1$), because $|\mathcal{Z}^{T,B}(z)|^{2}$ was earlier approximated as a delta function $\delta(z)$, we can still take $m_{\mathrm{C}}C$ to be sufficiently large compared to the dispersions $v_{x,y}$ in SEq.~(\ref{eq:defectHam}) to ensure that higher-order, nested Jackiw-Rebbi domain-wall constructions~\cite{TMDHOTI,HingeSM} remain valid predictors of 1D (and 0D) bound states.  Next, as we are specializing to integer Burgers vectors, $\mathcal{I}$ and $\mathcal{T}$ symmetry as represented in SEqs.~(\ref{eq:defectInversion}) and~(\ref{eq:Tdefect}), respectively, lead to the restriction that SEq.~(\ref{eq:insideCouplingPre}) only admits real phases:
\begin{equation}
V^{\mathrm{C}}_{a,in} = m_{\mathrm{C}}C\cos\left(\bs{k}_{D,a}\cdot\bs{B}\right)\xi^{x},
\label{eq:insideCoupling}
\end{equation}
where we have used the previous simplification that $\bs{B}=|\bs{B}|\hat{z}$ to reexpress the argument of the cosine in its more general form.  During the long preparation of this manuscript, a discussion of topological response effects in HOTIs with partial defects characterized by fractional Burgers vectors appeared in SRef.~\onlinecite{JenDefect}; we will not further address the case of fractional $|\bs{B}|$ in the present work.

We will now briefly discuss the bulk origin of $[V^{\mathrm{C}}_{a,\mu}]^{TB}_{ij}$.  First, the top and bottom surface states are given by SEqs.~\eqref{eq:topBound} and~\eqref{eq:bottomBound}, respectively.  Projecting the bulk Hamiltonian into the interface basis spanned by $|\psi_{a,\mu}(z)\rangle_{\alpha} = \{|\psi^{T}_{a,\mu}(z)\rangle_{1},|\psi^{T}_{a,\mu}(z)\rangle_{2},|\psi^{B}_{a,\mu}(z)\rangle_{1},|\psi^{B}_{a,\mu}(z)\rangle_{2}\}_\alpha$, $\alpha = 1 \dots 4$, we obtain:
\begin{align}
\left[\mathcal{H}^{\mathrm{C}}_{a,\mu}(q_{x},q_{y})\right]_{\alpha\beta} &= \int_{-\infty}^{\infty} {}_{\alpha} \langle \psi_{a,\mu}(z) |\mathcal{H}_{a}(q_{x},q_{y},0)| \psi_{a,\mu}(z)\rangle_{\beta} \, dz \nonumber \\
&= \left[ \xi^z \left(v_{x}q_{x}s^{y}+v_{y}q_{y}s^{x} \right) + m_a \xi^{x}\left( \delta_{\mu,out}  + C\delta_{\mu,in} e^{i(\bs{k}_{D,a}\cdot\hat{z})(z^{T}_{\mu}-z^{B}_{\mu})}\right) \right]_{\alpha \beta},
\label{eq:FrankBulkEdge}
\end{align}
where we have taken the $4 \times 4$ matrices $\xi^i s^j$, $i,j=0,x,y,z$, to act on the basis indexed by $\alpha = 1 \dots 4$.  This recovers our earlier result: $\mathcal{H}^{\mathrm{C}}_{a,\mu}(q_{x},q_{y}) = \mathcal{H}^{\mathrm{D}}_{a,\mu}(q_{x},q_{y}) + V^{\mathrm{C}}_{a,\mu}$.  From SEq.~(\ref{eq:FrankBulkEdge}), we identify the bulk origin of the coupling mass as $m_{\mathrm{C}} = m_a$ (\emph{i.e.}, the BHZ mass at ${\bs k}_{D,a}$).

With SEqs.~(\ref{eq:outsideCoupling}) and~(\ref{eq:insideCoupling}) established, we will now determine the conditions under which defect bound states are present.  We consider an internal edge dislocation at $z=0$ with a circular geometry and radius $R$ (SFig.~\ref{fig:kpDefect}~$\bs{b}$).  The geometry of the circular dislocation is more naturally addressed in cylindrical coordinates, and so we first Fourier transform $\mathcal{H}^{D}_{a}(q_{x},q_{y})$ [SEq.~(\ref{eq:defectHam})] by taking $q_{x,y}\rightarrow -i\partial_{x,y}$, where we have suppressed (for the in-plane coordinates $x,y$) the factors of $\bs{k}_{D,a}$ that previously appeared in SEq.~(\ref{eq:bulkFourier}), because they will ultimately only contribute to gauge-dependent phases that cancel when physical observables (\emph{e.g.} bound state locations and charges) are calculated.  We then convert to cylindrical coordinates:
\begin{eqnarray}
\partial_{x} &=& \cos(\varphi)\partial_{r} - \frac{1}{r}\sin(\varphi)\partial_{\varphi}, \nonumber \\
\partial_{y} &=& \sin(\varphi)\partial_{r} + \frac{1}{r}\cos(\varphi)\partial_{\varphi},
\end{eqnarray}
such that SEq.~(\ref{eq:defectHam}), taking the isotropic limit that $v_{x}=v_{y}=v$, now takes the form:
\begin{equation}
\mathcal{H}^{D}_{a}(r,\varphi) = -iv\xi^{z}\left(s^{1}(\varphi)\partial_{r} + \frac{1}{r}s^{2}(\varphi)\partial_{\varphi}\right),
\label{eq:polarDispersion}
\end{equation}
where:
\begin{eqnarray}
s^{1}(\varphi) &=& \sin(\varphi)s^{x} + \cos(\varphi)s^{y}=\left(\begin{array}{cc}
0 & -ie^{i\varphi} \\
ie^{-i\varphi} & 0 \end{array}\right), \nonumber \\
s^{2}(\varphi) &=& \cos(\varphi)s^{x} - \sin(\varphi)s^{y}=\left(\begin{array}{cc}
0 & e^{i\varphi} \\
e^{-i\varphi} & 0 \end{array}\right), \nonumber \\
\{s^{1}(\varphi),s^{2}(\varphi)\} &=& 0,\ s^{1}(\varphi)s^{2}(\varphi)=-is^{z}.
\end{eqnarray}
The ``inside'' and ``outside'' index $\mu$ has been suppressed in SEq.~(\ref{eq:polarDispersion}); we instead capture $\mu$ through the introduction of an $r$-dependent coupling mass between the top and bottom surfaces inside and outside of the defect:
\begin{equation}
V^{\mathrm{C}}_{a}(r,\varphi) = m_{\mathrm{C}}(r)\xi^{x},
\label{eq:polarCoupling}
\end{equation}
where $m_{\mathrm{C}}(r)$ is large and positive for $r>R$ and large with a sign given by $\cos\left(\bs{k}_{D,a}\cdot\bs{B}\right)$ for $r<R$ compared to the other parameters in the Hamiltonian [SEq.~(\ref{eq:insideCoupling})].  We then form the combined Hamiltonian:
\begin{equation}
\mathcal{H}^{C}_{a}(r,\varphi) = \mathcal{H}^{D}_{a}(r,\varphi) + V^{\mathrm{C}}_{a}(r,\varphi),
\label{eq:polarTI}
\end{equation}
which is invariant under the cylindrical-coordinate representations of $\mathcal{I}$ and $\mathcal{T}$ symmetry:
\begin{equation}
\mathcal{I}:\ \mathcal{H}^{C}_{a}(r,\varphi)\rightarrow \xi^{x}\mathcal{H}^{C}_{a}(r,\varphi+\pi)\xi^{x},\ \mathcal{T}:\ \mathcal{H}^{C}_{a}(r,\varphi)\rightarrow s^{y}(\mathcal{H}^{C}_{a}(r,\varphi))^{*}s^{y}.
\label{eq:polarITsym}
\end{equation}

Because SEq.~(\ref{eq:polarTI}) is equivalent to the low-energy theory of a 2D TI~\cite{AndreiTI,CharlieTI,KaneMeleZ2} (or a gapped anomalous fourfold Dirac surface fermion~\cite{DiracInsulator}), domain walls between regions with opposite signs of the $\mathcal{T}$-symmetric mass $\xi^{x}$ will bind 1D helical modes.  In the geometry in SFig.~\ref{fig:kpDefect}~$\bs{b}$, such a domain wall lies at $r=R$ if $\cos\left(\bs{k}_{D,a}\cdot\bs{B}\right)=-1$.  This implies that each band-inverted bulk TRIM point $\bs{k}_{D,a}$ will only contribute helical modes to $R$ if $\bs{k}_{D,a}\cdot\bs{B}$ is an odd multiple of $\pi$.  For example, for the Burgers vector in this section $\bs{B}=|\bs{B}|\hat{z}$, only TRIM points in the $k_{z}=\pi$ plane can possibly contribute helical modes.  Recalling that SEq.~(\ref{eq:Htotal}), the full position-space Hamiltonian at $z=0$ is given by:
\begin{equation}
\mathcal{H}^{C}(r,\varphi) = \bigoplus_{a}\mathcal{H}^{C}_{a}(r,\varphi), 
\label{eq:HelicalTotal}
\end{equation}
we determine that the total number of pairs of helical modes bound at $r=R$ is given by the total number of band-inverted bulk TRIM points for which:
\begin{equation}
\bs{k}_{D,a}\cdot\bs{B} = \pi\text{ mod }2\pi.
\label{eq:TRIMKBT}
\end{equation}
SEq.~(\ref{eq:TRIMKBT}) implies that only edge dislocations with Burgers vectors equal to odd-integer lattice translations and band inversions at TRIM points away from $\Gamma$ ($k_{x}=k_{y}=k_{z}=0$) can contribute to dislocation helical modes.  Furthermore, because an even number of pairs of 1D helical modes on the edge of a 2D insulator can be pairwise pushed into the valence and conduction manifolds, absent additional crystal symmetries (\emph{i.e.} mirror reflections)~\cite{KaneMeleZ2,FuKaneInversion,TeoFuKaneTCI}, then we further require that an odd number of TRIM points satisfy SEq.~(\ref{eq:TRIMKBT}) in order for anomalous helical modes to be bound at $r=R$.  Taken together, this implies that if $\bs{B}$ contains an odd number of primitive lattice vectors in the $k_{i}$ direction, then the Hamiltonian of the $k_{i}=\pi$ plane must differ from that of an atomic insulator by an odd number of band inversions between Kramers pairs of bands with opposite parity eigenvalues, implying that the weak index $M_{i}$ is nontrivial~\cite{FuKaneMele,FuKaneInversion} [where $M_{i}$ counts the number of band inversions per Kramers pair in the BZ plane satisfying SEq.~\eqref{eq:TRIMKBT}].  From this, we recover the main result of SRefs.~\onlinecite{AshvinScrewTI,TeoKaneDefect,QiDefect2,Vlad2D,TanakaDefect,VladScrewTI}: an $\mathcal{I}$- and $\mathcal{T}$-symmetric insulator with an edge dislocation with an integer Burgers vector $\bs{B}$ will bind anomalous helical modes along the dislocation if:
\begin{equation}
\bs{B}\cdot \bs{M}_{\nu}\text{ mod }2\pi = \pi.
\label{eq:BurgersAppendix}
\end{equation}
We note that $\mathcal{I}$ symmetry is not required for this result to hold; $\mathcal{T}$ symmetry alone is sufficient.  However, exploiting recent advances in higher-order (polarization)~\cite{multipole,WladTheory,HOTIChen,ZeroBerry,FulgaAnon,EmilCorner,HingeSM,OrtixCorners,CornerWarning,TMDHOTI,KoreanFragile,AshvinFragile2,ZhidaFragileTwist1,ZhidaFragileTwist2,HarukiFragile,OrtixTRealSpace,BouhonMagneticFragile1,BouhonMagneticFragile2,WiederAxion,WladCorners,FrankCorners,YoungkukBLG,CaseWesternCorners,JenOAL} and fragile~\cite{AshvinFragile,JenFragile1,HingeSM,AdrianFragile,BarryFragile,ZhidaBLG,AshvinBLG1,AshvinBLG2,AshvinFragile2,ZhidaFragileTwist1,ZhidaFragileTwist2,YoungkukMonopole,TMDHOTI,HarukiFragile,OrtixTRealSpace,BouhonMagneticFragile1,BouhonMagneticFragile2,KoreanFragile,WiederAxion,KoreanAxion,NicoDavidAXI2,IvoAXI1,IvoAXI2,
FragileKoreanInversion,ZhidaFragileAffine} topology, we will show below that \emph{filling anomalies} (SRefs.~\onlinecite{TMDHOTI,WiederAxion,WladCorners,HingeSM} and SN~\ref{sec:numerics}) can be exploited to further extend SEq.~(\ref{eq:BurgersAppendix}).

Beginning with $\mathcal{H}^{C}_{a}(r,\varphi)$ [SEq.~(\ref{eq:polarTI})], we now relax $\mathcal{T}$ symmetry while preserving $\mathcal{I}$ symmetry [SEq.~(\ref{eq:polarITsym})].  Without $\mathcal{T}$ symmetry, SEq.~(\ref{eq:polarTI}) admits a set of $\mathcal{I}$-symmetric, $r$-independent masses that gap all edge states~\cite{HingeSM}:
\begin{equation}
V^{A}_{a}(r,\varphi) = \sum _{n,F} \sum_{\bar{\gamma}_n} m_{n,F,\bar{\gamma}_n}\Gamma_{\mathrm{F}}\cos(L_{z,n}^{\mathrm{FTI}}\varphi + \bar{\gamma}_n),
\label{eq:FTImass}
\end{equation}
where:
\begin{equation}
L_{z,n}^{\mathrm{FTI}} = 1 + 2n \text{ where }n\in\{\mathbb{Z}^{+},0\},\ \Gamma_{\mathrm{F}}=\xi^{y},\ \xi^{z}s^{z},
\label{eq:fragileAnomalyCounting}
\end{equation}
$\bar{\gamma}_n$ is a free angle, and the sum in SEq.~(\ref{eq:FTImass}) runs over all possible-symmetry allowed values of angular momenta ($L_{z,n}^{\mathrm{FTI}}$), angles ($\bar{\gamma}_n$) per angular momentum, and Dirac matrices ($\Gamma_{\mathrm{F}}$).  As rigorously shown in SRefs.~\onlinecite{TMDHOTI,WiederAxion,HingeSM}, a Hamiltonian of the form:
\begin{equation}
\mathcal{H}^{A}_{a}(r,\varphi) = \mathcal{H}^{C}_{a}(r,\varphi) + V^{A}_{a}(r,\varphi),
\label{eq:2DaxionFTI}
\end{equation}
binds singly-degenerate, 0D (anti)solitons at the zeroes of $\cos(L_{z,n}^{\mathrm{FTI}}\varphi + \bar{\gamma}_n)$, which respectively acquire charges $\pm e/2$ under the ``soft'' relaxation of $\mathcal{I}$ symmetry~\cite{WilczekAxion,GoldstoneWilczek,NiemiSemenoff,WiederAxion,HingeSM} (SFig.~\ref{fig:kpDefect}~$\bs{c}$).  Because $\cos(L_{z,n}^{\mathrm{FTI}}\varphi + \bar{\gamma}_n)$ has $2L_{z,n}^{\mathrm{FTI}}$ zeroes on a circle, and $2L_{z,n}^{\mathrm{FTI}} = 4n+2$ (anti)solitons are necessarily filling-anomalous under the presence of $\mathcal{I}$ symmetry (SRefs.~\onlinecite{TMDHOTI,WiederAxion,WladCorners,HingeSM} and SN~\ref{sec:numerics}), then SEq.~(\ref{eq:2DaxionFTI}) necessarily implies that each of the bulk TRIM points that previously contributed a pair of helical modes at $r=R$ through SEq.~(\ref{eq:TRIMKBT}) will necessarily \emph{now} contribute an anomalous number of 0D (anti)solitons at $r=R$ under the introduction of $\mathcal{I}$-symmetric magnetism.  As shown in SRefs.~\onlinecite{TMDHOTI,WiederAxion,WladCorners,HingeSM}, this conclusion is crucially \emph{not} reliant on particle-hole symmetry, which is not present in real materials~\cite{AndreiMaterials,AndreiMaterials2}.  The presence of $\bar{\gamma}_n$ in SEq.~(\ref{eq:2DaxionFTI}) reflects that, until sharp ``corners'' are externally imposed, the anomalous 0D boundary modes are free to lie at any $\mathcal{I}$-related pair of angles $\varphi$ and $\varphi+\pi$ [SEq.~(\ref{eq:polarITsym})] along $r=R$, because the 2D (layer~\cite{WiederLayers,SteveMagnet,DiracInsulator}) point group with only $\mathcal{I}$ symmetry has no fixed points (angles) on the boundary of a circle~\cite{BilbaoPoint,PointGroupTables,MagneticBook,BCS1,BCS2}.  The effects of curvature on 0D boundary solitons in a cylindrical geometry are explored in detail in SRef.~\onlinecite{HingeSM}.  As previously with SEqs.~(\ref{eq:Htotal}) and~(\ref{eq:HelicalTotal}), the full, $\mathcal{T}$-broken position-space Hamiltonian at $z=0$ is:
\begin{equation}
\mathcal{H}^{A}(r,\varphi) = \bigoplus_{a}\mathcal{H}^{A}_{a}(r,\varphi), 
\label{eq:HENDtotal}
\end{equation}  
implying that the total number of 0D boundary states is given by the summed contributions of each of the TRIM points $\bs{k}_{D,a}$ that previously contributed helical modes at $r=R$ before $\mathcal{T}$ was relaxed [SEq.~(\ref{eq:TRIMKBT})].  This result can be understood by recognizing that SEq.~(\ref{eq:polarTI}) is equivalent to the $k\cdot p$ Hamiltonian of an $\mathcal{I}$- and $\mathcal{T}$-symmetric 2D TI~\cite{AndreiTI,CharlieTI,KaneMeleZ2}, and that $\mathcal{I}$- and $\mathcal{T}$-symmetric 2D TIs transition into 2D magnetic FTIs with anomalous corner modes under the application of $\mathcal{I}$-symmetric magnetism~\cite{TMDHOTI,WiederAxion}.

Next, absent additional rotation~\cite{multipole,WladTheory,WladCorners,HingeSM} or magnetic antiunitary~\cite{TMDHOTI,WiederAxion,KoreanFragile,KoreanAxion,NicoDavidAXI2,IvoAXI1,IvoAXI2} symmetries, two superposed copies of the $k\cdot p$ Hamiltonian in SEq.~(\ref{eq:2DaxionFTI}) combine to contribute a non-anomalous number of 0D boundary states~\cite{TMDHOTI,WiederAxion,WladCorners}.  Hence, we conclude that an $\mathcal{I}$-symmetric, $\mathcal{T}$-broken 3D insulator with a closed set of $\mathcal{I}$-symmetric edge dislocations with Burgers vector $\bs{B}$ will bind anomalous 0D states at $\mathcal{I}$-related locations along the set of dislocations if:
\begin{equation}
\bs{B}\cdot \bs{M}_{\nu}^{\mathrm{F}}\text{ mod }2\pi = \pi,
\label{eq:BurgersFragileAppendix}
\end{equation}
where $\bs{M}_{\nu}^{\mathrm{F}}$ is a new weak index vector characterizing which of the BZ-boundary planes host Hamiltonians that are topologically equivalent to the $\mathcal{I}$-symmetric FTI characterized in SRefs.~\onlinecite{TMDHOTI,WiederAxion}, or the obstructed atomic limit (OAL) that results from adding trivial bands without anomalous corner charges.  Specifically, in the case in which the FTI is trivialized to an OAL with corner charges, the valence manifold on the BZ boundary is instead composed of fully trivial bands (unobstructed atomic limits) and fragile bands that as a whole combine to form a (Wannierizable) 2D OAL~\cite{AshvinFragile,JenFragile1,HingeSM,AdrianFragile,BarryFragile,AshvinFragile2,ZhidaFragileTwist1,ZhidaFragileTwist2,TMDHOTI,HarukiFragile,OrtixTRealSpace,BouhonMagneticFragile1,BouhonMagneticFragile2,KoreanFragile,QuantumChemistry,
FragileKoreanInversion,ZhidaFragileAffine,schindler2021noncompact}; nevertheless, in this case, $\bs{M}_{\nu}^{\mathrm{F}}$ remains a valid indicator of the higher-order dislocation (HEND state) response derived in this section (see SN~\ref{sec:weakFragile} for an explicit numerical demonstration).

In SN~\ref{sec:indices}, we formally define $\bs{M}_{\nu}^{\mathrm{F}}$ in terms of elementary band representations (EBRs)~\cite{ZakBandrep1,ZakBandrep2,QuantumChemistry,Bandrep1,Bandrep2,Bandrep3,MTQC,JenFragile1,BarryFragile} and nested Wilson loops~\cite{AndreiXiZ2,Fidkowski2011,ArisInversion,Cohomological,HourglassInsulator,DiracInsulator,BarryFragile,multipole,WladTheory,TMDHOTI,HingeSM,KoreanFragile,WiederAxion}.  In SN~\ref{sec:FTIwithT}, we show that, because the anomalous solitons of SEq.~(\ref{eq:2DaxionFTI}) can remain anomalous if doubled under $\mathcal{T}$ symmetry, then there also exists a $\mathcal{T}$-symmetric generalization of SEq.~(\ref{eq:BurgersFragileAppendix}) that predicts whether or not edge dislocations in $\mathcal{I}$- and $\mathcal{T}$-symmetric crystals bind Kramers pairs of 0D states.  We emphasize that $\bs{M}_{\nu}^{\mathrm{F}}$ does not distinguish between 3D insulators for which both the $k_{i}=0,\pi$ planes have topologically equivalent 2D Hamiltonians and insulators for which $k_{i}$ indexes a pumping cycle between a 2D trivial insulator at $k_i = 0$ and a 2D FTI at $k_i = \pi$~\cite{WiederAxion,KoreanAxion,NicoDavidAXI2,IvoAXI1,IvoAXI2}.  Furthermore, FTIs can be trivialized by the introduction of appropriately chosen trivial bands~\cite{AshvinFragile,JenFragile1,HingeSM,AdrianFragile,BarryFragile,ZhidaBLG,AshvinBLG1,AshvinBLG2,AshvinFragile2,ZhidaFragileTwist1,ZhidaFragileTwist2,YoungkukMonopole,TMDHOTI,HarukiFragile,OrtixTRealSpace,BouhonMagneticFragile1,BouhonMagneticFragile2,KoreanFragile,WiederAxion,KoreanAxion,NicoDavidAXI2,IvoAXI1,IvoAXI2,
FragileKoreanInversion,ZhidaFragileAffine}. Therefore, we conclude that the dislocation response of 3D insulators does not distinguish between strong and weak (fragile) topology.  This implies that 3D insulators may still exhibit an anomalous crystal defect response even if they are Wannierizable (\emph{i.e.}, a 3D OAL)~\cite{QuantumChemistry,Bandrep1,Bandrep2,Bandrep3,MTQC,JenFragile1,BarryFragile,TKNN,ThoulessWannier,ThoulessPump,AndreiXiZ2,AlexeyVDBWannier,AlexeyVDBTI}.  This is in stark contrast to the flux response, which we will show in SN~\ref{sec:kpFlux} and~\ref{sec:fluxfluxtubetopologymapping} to only be anomalous in strong topological (crystalline) insulators.  Finally, to connect the circular internal edge dislocation used throughout this section (SFig.~\ref{fig:kpDefect}~$\bs{c}$) to surface-terminating edge dislocations in real materials, we imagine smoothly growing the internal dislocation until it meets the sample boundary, and only then creating sharp corners between the dislocations and the surface (SFig.~\ref{fig:kpDefect}~$\bs{d}$).  In this geometry, the ends (corners) of the surface-facing edge dislocations (which now appear as surface point dislocations, see SFig.~\ref{fig:kpDefect}~$\bs{d}$), will carry higher-order 0D (HEND) states, consistent with our calculations of the dislocation bound states in tight-binding models (SN~\ref{sec:numerics}) and in 3D SnTe crystals (SN~\ref{sec:DFTSnTe}).

\subsubsection{$k\cdot p$ Derivation of HEND States on Screw Dislocation Ends in AXIs and Weak Fragile TIs}
\label{sec:kpScrew}

\begin{figure}[b]
\includegraphics[width=0.83\textwidth]{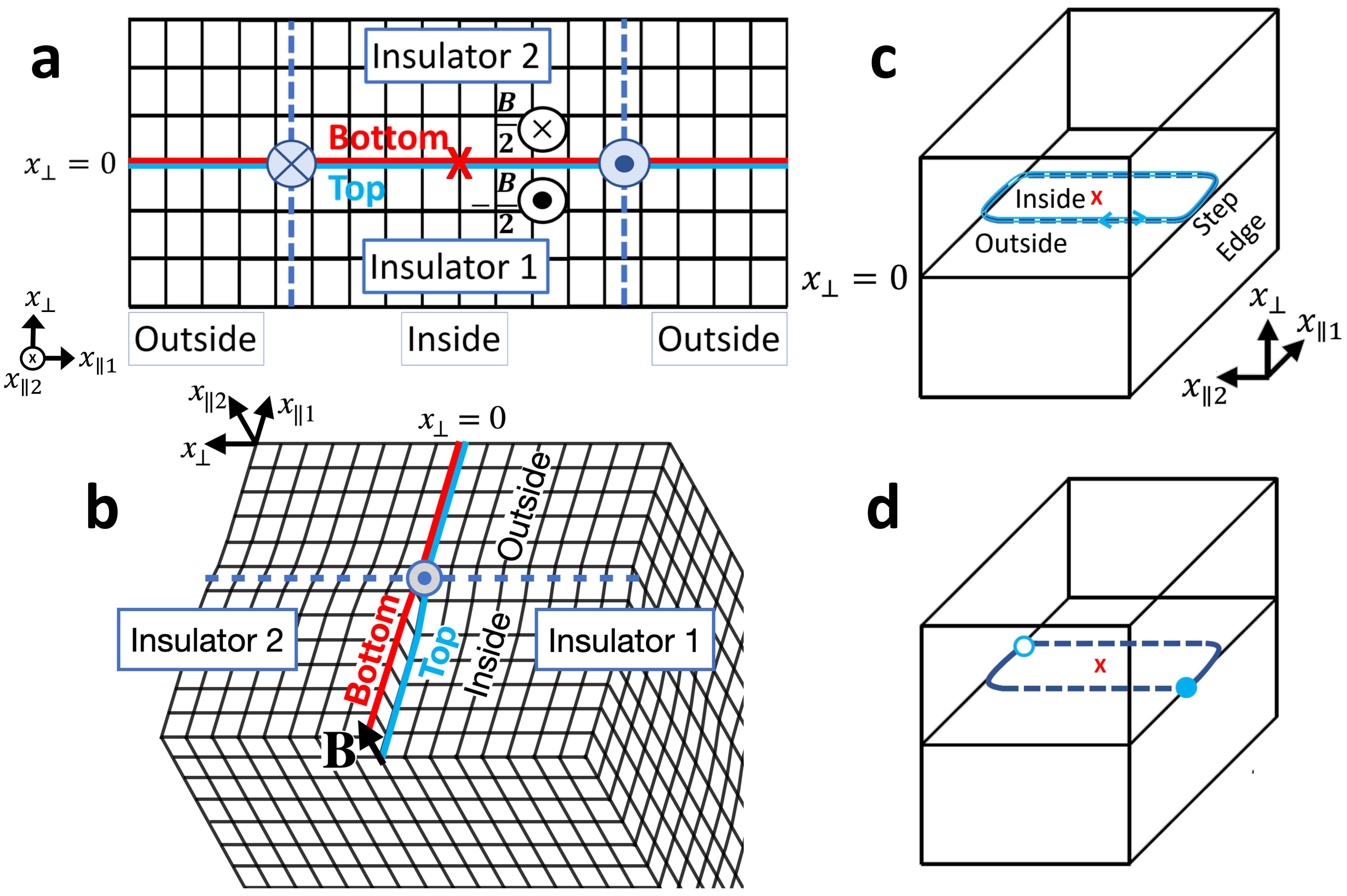}
\caption{{\bf 0D screw dislocation states in a 3D crystal.} $\bs{a}$ A cross-sectional cut of an $\mathcal{I}$- and $\mathcal{T}$-symmetric 3D crystal with a pair of screw dislocations with opposite sense vectors $\bs{s}$~\cite{MerminReview}, indicated by the blue vectors into and out of the page on the left and right, respectively.   The Burgers vectors $\bs{B}$ of both dislocations point into the page, which we take to be the positive $x_{\parallel 2}$ direction.  Therefore, as indicated by $\sgn[\bs{B}\cdot \bs{s}]$ [SEq.~(\ref{eq:chiralityScrew})], the screw on the left (right) in $\bs{a}$ is right- (left-) handed.   Following the procedure in SRef.~\onlinecite{AshvinScrewTI} and SN~\ref{sec:kpEdge}, we divide the system in $\bs{a}$ by cutting along the red, blue, and dashed lines.  As previously with the edge dislocation in SFig.~\ref{fig:kpDefect}~$\bs{a}$, the red and blue lines separate two insulators with the same bulk topology.  However, unlike for an edge dislocation, we instead glue the top and bottom surfaces of the two insulators together with (inside) and without (outside) an extra lattice displacement in the $x_{\parallel 2}$ direction.  To preserve an inversion center (red $\times$), we split the displacement evenly over insulators $1$ and $2$, such that insulator $1$ ($2$) is displaced by $|\bs{B}/2|$ in the $-x_{\parallel 2}$ ($+x_{\parallel 2}$) direction.  $\bs{b}$ A 3D view of the $\hat{x}_{\parallel 2}$-normal surface of the crystal with screw dislocations in panel~$\bs{a}$.  $\bs{c}$ A 3D crystal with the pair of screw dislocations from $\bs{a}$ (dashed lines).  Unlike previously in SFig.~\ref{fig:kpDefect}~$\bs{b,c}$, we are unable to consider a fully internal screw dislocation (\emph{i.e.}, a screw dislocation that does not terminate on a surface), because, as the Burgers vectors of screw dislocations are parallel to the dislocation line (sense vector)~\cite{MerminReview}, such a dislocation would have a position-dependent Burgers vector.  Because we expect the electronic structure of screw dislocations with spatially varying Burgers vectors to be considerably more complicated than that of screw dislocations with spatially constant Burgers vectors, we do not further examine screw dislocations with spatially-varying Burgers vectors in this work.  Instead, we simply consider two parallel screw dislocations with the same Burgers vector $\bs{B}$, and avoid the curvature effects of sharp corners~\cite{HingeSM} by lightly ``sanding'' away some of the crystal near the screw dislocations so that the dislocations curve (relatively) smoothly into surface step edges (solid blue lines).  In this geometry (panel $\bs{c}$), the screw dislocations and step edges form a closed loop that is topologically equivalent to the rounded internal edge dislocation in SFig.~\ref{fig:kpDefect}~$\bs{b}$.  If the Burgers vectors of the screw dislocations and the bulk electronic structure satisfy SEq.~(\ref{eq:BurgersAppendixScrew}), then the screw and step-edge loop in $\bs{c}$ will bind an odd (anomalous) number of helical modes, as previously shown in SRefs.~\onlinecite{AshvinScrewTI,TeoKaneDefect,QiDefect2,Vlad2D,TanakaDefect,VladScrewTI}.  $\bs{d}$ If an inversion center is also preserved throughout the formation of the screw dislocations, then, upon the introduction of $\mathcal{I}$-symmetric magnetism, $\bs{M}_{\nu}$ will again be converted to the weak fragile (and OAL) index $\bs{M}_{\nu}^{\mathrm{F}}$ [see SEq.~(\ref{eq:BurgersFragileAppendixScrew}) and SN~\ref{sec:weakFragile}], and the helical modes will gap into an anomalous number of HEND states.  In $\bs{d}$, we depict the two filling-anomalous HEND states appearing on alternating, $\mathcal{I}$-related ends of the screw dislocations in $\bs{a}$, as they do in our numerical calculations in SN~\ref{sec:numerics}.  The four ends of the two screw dislocations in $\bs{d}$ are thus equivalent to the corners of the $\mathcal{I}$-symmetric 2D FTI characterized in SRefs.~\onlinecite{TMDHOTI,WiederAxion}.}
\label{fig:kpScrew}
\end{figure}

In this section, we will adapt the arguments derived in the previous section for HEND states bound to edge dislocations to analyze the electronic structure of screw dislocations in AXIs and 3D weak FTIs.  To begin, we again consider, as previously in SN~\ref{sec:kpEdge}, a 3D insulator with $\mathcal{I}$ and $\mathcal{T}$ symmetries that differs from an atomic insulator by a series of band inversions at the Fermi energy between only the two highest valence bands and the two lowest conduction bands.  We take these band inversions to occur at a set of TRIM points $\bs{k}_{D,a}$ between Kramers pairs of states with opposite parity eigenvalues.  The Hamiltonian of the bulk insulator, when pristine, is given by SEqs.~(\ref{eq:Htotal}) and~(\ref{eq:BHZtrim}).  Next, we insert a pair of screw dislocations with Burgers vector $\bs{B}$ (SFig.~\ref{fig:kpScrew}~$\bs{a}$), where $|\bs{B}|$ is an integer multiple of the lattice spacing in the $\hat{B}$ direction.  For this section, unlike previously for the edge dislocations in SN~\ref{sec:kpEdge}, we take $\bs{B}\parallel\hat{y}$, such that:
\begin{equation}
x_{\perp}=z,\ x_{\parallel 1}=x,\ x_{\parallel 2}=y,
\end{equation}
in the notation of SFig.~\ref{fig:kpScrew}.

We will find it useful to briefly introduce the notion of screw chirality.  For a screw dislocation, the direction of net displacement is indicated by a sense vector~\cite{MerminReview} $\bs{s}$ that is parallel to $\bs{B}$.  The chirality $\mathsf{C}$ of a particular screw dislocation is then indicated by:
\begin{equation}
\mathsf{C} = \sgn[\bs{B}\cdot\bs{s}].
\label{eq:chiralityScrew}
\end{equation}
We take the screw dislocation on the left (right) in SFig.~\ref{fig:kpScrew}~$\bs{a}$ to have a sense vector $\bs{s}$ that points in the $\hat{y}$ ($-\hat{y}$) direction, such that the screw on the left (right) is right- (left-) handed.  If the insertion of the screw dislocations preserves an inversion center (red $\times$ in SFig.~\ref{fig:kpScrew}), then the action of $\mathcal{I}$ exchanges the locations and chiralities of the two screws, in agreement with well-established notion that $\bs{s}$ transforms as a vector ($\mathcal{I}$-odd) and $\bs{B}$ transforms as a pseudovector ($\mathcal{I}$-even), such that $\mathsf{C}$ transforms as a pseudoscalar~\cite{MerminReview}.

As was previously done in SN~\ref{sec:kpEdge}, we model the pair of screw dislocations by cutting the insulator described by $\mathcal{H}(\bs{q})$  [SEq.~(\ref{eq:Htotal})] into two pieces with $\pm \hat{z}$-normal surfaces, and then ``gluing'' the pieces back together.  However, instead of inserting an extra row of unit cells in the $\hat{z}$ direction to create edge dislocations, as we did previously in SFig.~\ref{fig:kpDefect}~$\bs{a}$, we implement a pair of screw dislocations by gluing the upper portion of the cut insulator (labeled insulator $2$ in SFig.~\ref{fig:kpScrew}~$\bs{a}$) to the lower portion (labeled insulator $1$) with an extra relative displacement of $\bs{B}$ along the $y$ direction in the region between the two insulators (labeled ``inside'' in SFig.~\ref{fig:kpScrew}).  To preserve $\mathcal{I}$ symmetry, we implement this displacement evenly between the two insulators, such that insulator $1$ ($2$) is displaced by $|\bs{B}/2|$ in the $-y$ ($+y$) direction (a 3D rendering of this geometry, viewed from the exterior of the crystal with screw dislocations, is provided in SFig.~\ref{fig:kpScrew}~$\bs{b}$).

We next solve for the top surface states of insulator $1$.  Here, because the crystal with the two screw dislocations still preserves $y$ direction lattice translation at this stage of the calculation (SFig.~\ref{fig:kpScrew}~$\bs{a}$), then the analysis from the previous section [SEqs.~(\ref{eq:mzTop}) through (\ref{eq:Htop})] can still be used with only minor modifications.  To determine the effect of the screw dislocations on the electronic structure, we begin with the original bound states in SEq.~(\ref{eq:topBound}), and then, setting $z^{T}_{\mu}=0$, act with $\hat{T}_{y^{T}_{\mu}}$, the $y$ direction lattice translation operator.  The operation of translation $\hat{T}_{y^{T}_{\mu}}$ results in a phase shift of the original states in SEq.~(\ref{eq:topBound}):
\begin{equation}
|\psi^{T}_{a,\mu}(z)\rangle_{i} =e^{i(\bs{k}_{D,a}\cdot\hat{y})y^{T}_{\mu}}e^{i(\bs{k}_{D,a}\cdot\hat{z})z}\mathcal{Z}^T(z)|+\rangle_{i},
\label{eq:topBoundScrew}
\end{equation}
where $y^{T}_{\mu}$ is the displacement of insulator $1$ in the region $\mu$, and $\mathcal{Z}^T(z)$ and $|+\rangle_{i}$ are defined in SEq.~(\ref{eq:topBound}) and the surrounding text.  We next derive the dispersion (velocity) terms for the top-surface Hamiltonian by projecting the $q_{x,y}$ dispersion terms from SEq.~(\ref{eq:BHZtrim}) into the basis of $|\psi^{T}_{a,\mu}(z)\rangle_{i}$, as was previously done in SEq.~(\ref{eq:Htop}):
\begin{equation}
\mathcal{H}^{T}_{a,\mu}(q_{x},q_{y}) = v_{x}q_{x}s^{y}+v_{y}q_{y}s^{x},
\label{eq:HtopScrew}
\end{equation}
where $s^{i}$ is a $2\times 2$ matrix that indexes the states $|+\rangle_{i}$.  SEq.~(\ref{eq:HtopScrew}) is invariant under $\mathcal{T}$ symmetry, which remains represented by SEq.~(\ref{eq:Ttop}).  Next, following SEq.~(\ref{eq:bottomBound}), we obtain the bottom-surface bound states of insulator $2$:
\begin{equation}
|\psi^{B}_{a,\mu}(z)\rangle_{i} = e^{i(\bs{k}_{D,a}\cdot\hat{y})y^{B}_{\mu}}e^{i(\bs{k}_{D,a}\cdot\hat{z})z}\mathcal{Z}^B(z)|-\rangle_{i}.
\label{eq:bottomBoundScrew}
\end{equation} 
We then project the bulk $q_{x,y}$ dispersion terms of SEq.~(\ref{eq:BHZtrim}) into the basis of $|\psi^{B}_{a,\mu}(z)\rangle_{i}$ to realize the bottom-surface Hamiltonian:
\begin{equation}
\mathcal{H}^{B}_{a,\mu}(q_{x},q_{y}) = -v_{x}q_{x}s^{y}-v_{y}q_{y}s^{x} = - \mathcal{H}^{T}_{a,\mu}(q_{x},q_{y}),
\label{eq:HbottomScrew}
\end{equation}
where $s^{x,y,z}$ are Pauli matrices that act on the states $|-\rangle_{i}$, $i=1,2$. SEq.~(\ref{eq:HbottomScrew}) is invariant under $\mathcal{T}$ symmetry, which is also represented by SEq.~(\ref{eq:Ttop}).

With SEqs.~(\ref{eq:HtopScrew}) and (\ref{eq:HbottomScrew}) established, we can again form the Hamiltonian in the plane between the defects ($x_{\perp}=0$ in SFig.~\ref{fig:kpScrew}).  We begin by forming the uncoupled Hamiltonian of both the top and bottom surfaces, which is, as previously in the case of edge dislocations [see the text surrounding SEq.~(\ref{eq:defectHam})], the Hamiltonian of a fourfold-degenerate, 2D Dirac fermion~\cite{Steve2D,WiederLayers,SteveMagnet,DiracInsulator}:
\begin{equation}
\mathcal{H}^{D}_{a,\mu}(q_{x},q_{y}) = v_{x}q_{x}\xi^{z}s^{y}+v_{y}q_{y}\xi^{z}s^{x}.
\label{eq:defectHamScrew}
\end{equation}
In SEq.~(\ref{eq:defectHamScrew}), $\xi^{i}$ is a Pauli matrix that indexes the ``top'' and ``bottom'' degrees of freedom and $s$ continues to index $i=1,2$ in $|\pm\rangle_{i}$ within each of the two surfaces of the insulator described by SEq.~(\ref{eq:Htotal}).  In this geometry, $\mathcal{I}$ and $\mathcal{T}$ are respectively represented through the actions in SEqs.~(\ref{eq:defectInversion}) and~(\ref{eq:Tdefect}).

Finally, we allow for $\mathcal{I}$- and $\mathcal{T}$-symmetric coupling between the top and bottom surfaces of the two insulators.  Once again, SEqs.~(\ref{eq:defectInversion}),~(\ref{eq:Tdefect}), and~(\ref{eq:defectHamScrew}) imply that there will only be a single coupling mass, and that it will be proportional to $\xi^{x}$.  We account for the presence of the screw dislocations by defining a consistent coupling phase about $y=z=0$ (red $\times$ in SFig.~\ref{fig:kpScrew}~$\bs{a}$) in the spinless coupling interaction:
\begin{align}
[V^{\mathrm{C}}_{a,\mu}]_{ij}^{TB} &= m_{\mathrm{C}}\delta_{ij}\int_{-\infty}^{\infty}dz\ \left[|\psi^{T}_{a,\mu}(z)\rangle_{i}\langle\psi^{B}_{a,\mu}(z)|_{j} + \mathrm{ h.c.}\right], \nonumber \\
[V^{\mathrm{C}}_{a,\mu}]_{ij}^{TT} &= [V^{\mathrm{C}}_{a,\mu}]_{ij}^{BB} = 0,
\label{eq:couplingMScrew}
\end{align}
where $m_{\mathrm{C}}$ is the strength of the coupling; $i,j=1,2$; and $\delta_{ij}$ enforces $\mathcal{T}$ symmetry [SEq.~(\ref{eq:Tdefect})] by again excluding spinful (magnetic) interactions.  Substituting SEqs.~(\ref{eq:topBoundScrew}) and~(\ref{eq:bottomBoundScrew}) into SEq.~(\ref{eq:couplingMScrew}), we arrive at the expression:
\begin{eqnarray}
[V^{\mathrm{C}}_{a,\mu}]^{TB}_{ij} &=& m_{\mathrm{C}}\delta_{ij}\int_{-\infty}^{\infty}dz\ \left[e^{i(\bs{k}_{D,a}\cdot\hat{y})(y^{T}_{\mu}-y^{B}_{\mu})}\mathcal{Z}^T(z)\mathcal{Z}^{B}(z)|+\rangle_{i}{}_{j}\langle -| +\mathrm{ h.c.}\right], \nonumber \\
&=& m_{\mathrm{C}} \delta_{ij} e^{i(\bs{k}_{D,a}\cdot\hat{y})(y^{T}_{\mu}-y^{B}_{\mu})}|+\rangle_{i}{}_{j}\langle -| +\mathrm{ h.c.},
\label{eq:couplingM2Screw}
\end{eqnarray}
where, in the second line, we have again employed the approximation that $\mathcal{Z}^T(z)\mathcal{Z}^{B}(z)\rightarrow\delta(z)$ [see the text preceding SEq.~(\ref{eq:edgeburgersin})].

In the $\mu=out$ region:
\begin{equation}
y^{T}_{out}=y^{B}_{out}=0,
\end{equation}
such that:
\begin{equation}
V^{\mathrm{C}}_{a,out} = m_{\mathrm{C}}\xi^{x}.
\label{eq:outsideCouplingScrew}
\end{equation}
For the $\mu=in$ region, however, for which:
\begin{equation}
y^{T}_{in}= -|\bs{B}|/2,\   y^{B}_{in}=|\bs{B}|/2,
\end{equation}
an additional nonzero phase factor persists from SEq.~(\ref{eq:couplingM2Screw}):
\begin{equation}
[V^{\mathrm{C}}_{a,in}]^{TB}_{ij} = m_{\mathrm{C}} \delta_{ij} e^{-i(\bs{k}_{D,a}\cdot\hat{y})|\bs{B}|}|+\rangle_{i}{}_{j}\langle -| +\mathrm{ h.c.},
\label{eq:insideCouplingPreScrew}
\end{equation}
just as it previously did for the edge dislocation in SEq.~\eqref{eq:insideCouplingPre}.  As discussed in the text surrounding SEq.~(\ref{eq:insideCoupling}), we next restrict consideration to integer Burgers vectors, and require $\mathcal{I}$ and $\mathcal{T}$ symmetry, which allows us to discard the imaginary phases in SEq.~(\ref{eq:insideCouplingPreScrew}).  Imposing this restriction on SEq.~(\ref{eq:insideCouplingPreScrew}) leads to the simplified expression:
\begin{equation}
V^{\mathrm{C}}_{a,in} = m_{\mathrm{C}}\cos\left(\bs{k}_{D,a}\cdot\bs{B}\right)\xi^{x},
\label{eq:insideCouplingScrew}
\end{equation}
where we have used the previous simplification that $\bs{B}=|\bs{B}|\hat{y}$ to reexpress the argument of the cosine in a more general form that is independent of our choice of the screw-dislocation Burgers vector direction.  For each band-inverted bulk TRIM point $\bs{k}_{D,a}$, the screw dislocations will therefore form domain walls in the sign of $V^{\mathrm{C}}_{a}$ if and only if $\cos\left(\bs{k}_{D,a}\cdot\bs{B}\right)=-1$, which is only satisfied by the previous conditions in SEq.~(\ref{eq:TRIMKBT}).  Because SEq.~(\ref{eq:defectHamScrew}) is equivalent to the Hamiltonian of a 2D, $\mathcal{T}$-symmetric, fourfold Dirac fermion~\cite{Steve2D,WiederLayers,SteveMagnet,DiracInsulator}, the domain walls necessarily bind an anomalous pair of helical modes (SFig.~\ref{fig:kpScrew}~$\bs{c}$), which are equivalent to the edge states of a 2D TI~\cite{AndreiTI,CharlieTI,KaneMeleZ2} located in the plane between the two dislocations (the ``inside'' region in SFig.~\ref{fig:kpScrew}).  The total number of helical modes is thus determined by summing over all of the bulk TRIM points [SEq.~(\ref{eq:HelicalTotal})], again resulting in the statement that a closed loop formed from two screw dislocations and two step edges in an $\mathcal{I}$- and $\mathcal{T}$-symmetric crystal will bind anomalous helical modes if the dislocation Burgers vector $\bs{B}$ and the weak index vector $\bs{M}_\nu$ of the pristine crystal satisfy:
\begin{equation}
\bs{B}\cdot\bs{M}_{\nu}\text{ mod }2\pi =\pi,
\label{eq:BurgersAppendixScrew}
\end{equation}
reproducing the central result of SRefs.~\onlinecite{AshvinScrewTI,TeoKaneDefect,QiDefect2,Vlad2D,TanakaDefect,VladScrewTI}.

As previously in SN~\ref{sec:kpEdge}, we can immediately extend SEq.~(\ref{eq:BurgersAppendixScrew}) to the $\mathcal{T}$-broken regime through the recent recognition that the edge states of an $\mathcal{I}$-symmetric 2D TI necessarily gap into the anomalous corner modes of a 2D FTI under the introduction of $\mathcal{I}$-symmetric magnetism~\cite{TMDHOTI,WiederAxion}.  Unlike previously for edge dislocations, we are unable to form a fully internal loop of screw dislocations (\emph{i.e.}, screw dislocations that do not terminate on a surface), because the Burgers vector of a pair of screw dislocations is parallel to the dislocation, and we wish to avoid the confounding complications of position-dependent Burgers vectors.  However, we can consider the closed loop of two screw dislocations and two step edges in SFig.~\ref{fig:kpScrew} to be sufficiently smooth for the previous analysis in SEqs.~(\ref{eq:FTImass}) through~(\ref{eq:HENDtotal}) to remain applicable if the ends of the screw dislocations are taken to continuously evolve into the surface step edges in SFig.~\ref{fig:kpScrew}~$\bs{b,c,d}$ in an $\mathcal{I}$-symmetric manner with nonsingular local curvature~\cite{HingeSM}.  In this case, the closed loop of the dislocations and step edges becomes equivalent to the smooth boundary of an $\mathcal{I}$-symmetric FTI~\cite{TMDHOTI,WiederAxion}, and binds an anomalous number of 0D HEND states if:
\begin{equation}
\bs{B}\cdot \bs{M}_{\nu}^{\mathrm{F}}\text{ mod }2\pi = \pi,
\label{eq:BurgersFragileAppendixScrew}
\end{equation}
where $\bs{M}_{\nu}^{\mathrm{F}}$ is the new weak index previously introduced in SEq.~(\ref{eq:BurgersFragileAppendix}) to characterize which of the BZ-boundary planes of the pristine bulk insulator host Hamiltonians equivalent to (possibly trivialized copies of) 2D $\mathcal{I}$-symmetric FTIs.  In SN~\ref{sec:FTInoT} and~\ref{sec:FTIwithT}, we will rigorously define $\bs{M}_{\nu}^{\mathrm{F}}$ and its $\mathcal{T}$-symmetric generalization using EBRs and nested Wilson loops.

\subsubsection{$k\cdot p$ Derivation of HEND States on Flux Tube Ends in AXIs and HOTIs}
\label{sec:kpFlux}

\begin{figure}[h]
\includegraphics[width=\textwidth]{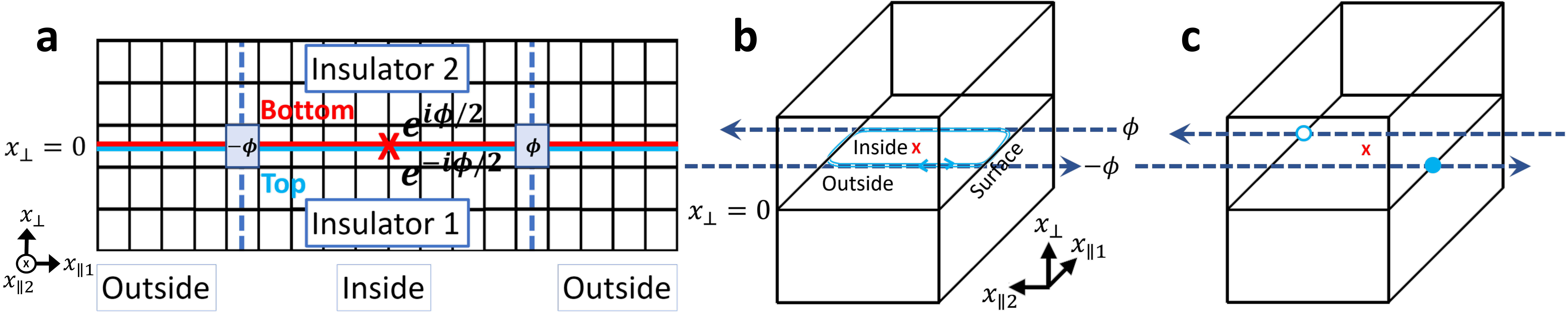}
\caption{{\bf 0D $\pi$-flux states in a 3D crystal.} $\bs{a}$ A cross-sectional cut of an $\mathcal{I}$- and $\mathcal{T}$-symmetric 3D crystal with a pair of threaded flux tubes with opposite magnetic flux $\pm \phi$ directed in the $x_{\parallel 2}$ direction, indicated by the blue plaquettes.  Following the procedure used for screw dislocations in SN~\ref{sec:kpScrew}, we divide the system in $\bs{a}$ by cutting the lattice along the red, blue, and dashed lines.  As previously with the screw dislocations in SFig.~\ref{fig:kpScrew}~$\bs{a}$, the red and blue lines separate two insulators with the same bulk topology.  To implement the effects of the $\pm \phi$-flux tubes, we multiply all couplings between the top surface states of insulator 1 and the bottom surface states of insulator 2 by  $e^{i\phi}$ in the ``inside'' region between the two flux tubes.  The tubes are inserted at locations related by a bulk inversion center (red $\times$), and importantly, $\mathcal{I}$ is only a system symmetry when $\phi\text{ mod }2\pi=0,\pi$.  $\bs{b}$ An $\mathcal{I}$-symmetric strong TI with the pair of flux tubes from $\bs{a}$ (dashed lines).  When $\phi=\pi$ per unit cell (depicted as plaquettes in panel~$\bs{a}$), each tube binds an anomalous pair of helical modes, forming a ``wormhole'' between the $\pm \hat{x}_{\parallel 2}$-normal surfaces~\cite{QiFlux,AshvinFlux,AdyFlux,FranzWormhole,WormholeNumerics,MirlinFlux,CorrelatedFlux}.  Though the wormhole states merge with the surface states, we can consider the closed loop of helical modes across the flux tubes and parallel surfaces to be equivalent to the anomalous edge of a 2D TI.  $\bs{c}$ Relaxing $\mathcal{T}$ symmetry while preserving $\mathcal{I}$, the bulk of both insulators 1 and 2 necessarily transitions into an $\mathcal{I}$-symmetric axion insulator (AXI)~\cite{FuKaneMele,FuKaneInversion,AndreiInversion,QHZ,FanHOTI,VDBAxion,DiracInsulator,MulliganAnomaly,DrewPotterPeel,TitusHOTISurfaceAnomaly,VDBHOTI,WiederAxion}.  Relying on the previous recognition from SRefs.~\onlinecite{TMDHOTI,WiederAxion} that the edge states of an $\mathcal{I}$-symmetric 2D TI gap into the anomalous corner modes of a 2D FTI in the presence of $\mathcal{I}$-symmetric magnetism, we conclude that $\pi$-flux tubes threaded through an AXI bind 0D HEND states that, like the HEND states bound to the edge and screw dislocations previously discussed in SN~\ref{sec:kpEdge} and~\ref{sec:kpScrew} respectively, are equivalent to the $\pm e/2$-charged corner modes of an $\mathcal{I}$-symmetric 2D FTI.  In $\bs{c}$, we depict two filling-anomalous HEND flux states appearing on alternating, $\mathcal{I}$-related ends of the flux tubes in $\bs{a}$, as they do in our numerical calculations in SN~\ref{sec:numerics2}.}
\label{fig:kpFlux}
\end{figure}

In this section, we will exploit the previous derivation in SN~\ref{sec:kpScrew} of HEND states bound to screw dislocations to determine the flux response of $\mathcal{I}$-symmetric axion insulators (AXIs).  As in the previous two sections (SN~\ref{sec:kpEdge} and~\ref{sec:kpScrew}), we begin by considering a 3D insulator with $\mathcal{I}$ and $\mathcal{T}$ symmetries that differs from an atomic insulator by a series of band inversions at the Fermi energy between only the two highest valence bands and the two lowest conduction bands.  We take these band inversions to occur at a set of TRIM points $\bs{k}_{D,a}$ between Kramers pairs of states with opposite parity eigenvalues.  The Hamiltonian of the insulator in the absence of an external magnetic field is given by SEqs.~(\ref{eq:Htotal}) and~(\ref{eq:BHZtrim}).  Next, we thread magnetic flux into the insulator through two small parallel tubes with opposite field strengths $\pm \phi$ located at $\mathcal{I}$-related positions (SFig.~\ref{fig:kpFlux}~$\bs{a}$).  We take the magnetic field in the right (left) tube to be directed in the $\hat{y}$ ($-\hat{y}$) direction, such that:
\begin{equation}
x_{\perp} = z,\ x_{\parallel 1} = x,\ x_{\parallel 2} = y,
\end{equation}
in the notation of SFig.~\ref{fig:kpFlux}.  The tubes are therefore localized in the $z=0$ plane at $\mathcal{I}$-related $x$ coordinates.  We note, however, that this flux configuration is not invariant under $\mathcal{I}$ and $\mathcal{T}$ symmetries for generic values of $\phi$, because $\mathcal{T}\phi = -\phi$, and because the action of $\mathcal{I}$ exchanges the locations of the flux tubes, which generically carry opposite magnetic fluxes $\pm \phi$.  Later in this section, we will specialize to the cases of $\phi\text{ mod }2\pi=0,\pi$, at which $\mathcal{I}$ and $\mathcal{T}$ symmetries are restored.

To implement the pair of flux tubes, we cut the insulator described by $\mathcal{H}(\bs{q})$  [SEq.~(\ref{eq:Htotal})] into two pieces with $\pm \hat{z}$-normal surfaces, and again glue the pieces back together. Instead of gluing together the region between the flux tubes (labeled ``inside'' in SFig.~\ref{fig:kpFlux}) with a relative translation between the top and bottom surfaces of the two insulators, we instead rotate the phase of the top (bottom) surface states of the $\mu=in$ region of insulator $1$ ($2$) by $-\phi/2$ ($\phi/2$).  Formally, this is accomplished through the substitution:
\begin{equation}
(\bs{k}_{D,a}\cdot\hat{y})y_{\mu}^{T,B}\rightarrow\phi^{T,B}_{\mu}, \quad \phi^{T,B}_{\mu} = \mp \frac{\phi}{2},
\end{equation}
in SEqs.~(\ref{eq:topBoundScrew}) and~(\ref{eq:bottomBoundScrew}), respectively, resulting in:
\begin{eqnarray}
|\psi^{T}_{a,\mu}(z)\rangle_{i} &=& e^{i\phi^{T}_{\mu}}e^{i(\bs{k}_{D,a}\cdot\hat{z})z}\mathcal{Z}^T(z)|+\rangle_{i}, \nonumber \\
|\psi^{B}_{a,\mu}(z)\rangle_{i} &=& e^{i\phi^{B}_{\mu}}e^{i(\bs{k}_{D,a}\cdot\hat{z})z}\mathcal{Z}^B(z)|-\rangle_{i}.
\label{eq:topbottomBoundFlux}
\end{eqnarray}
Projecting the $q_{x,y}$ dispersion terms in SEq.~(\ref{eq:BHZtrim}) into the basis of the top and bottom surface states in SEq.~(\ref{eq:topbottomBoundFlux}), and, for now, ignoring coupling between the top and bottom surfaces, we again realize the Hamiltonian of a 2D fourfold Dirac fermion~\cite{Steve2D,WiederLayers,SteveMagnet,DiracInsulator}: 
\begin{equation}
\mathcal{H}^{D}_{a,\mu}(q_{x},q_{y}) = v_{x}q_{x}\xi^{z}s^{y}+v_{y}q_{y}\xi^{z}s^{x},
\label{eq:fluxHam}
\end{equation}
as we did previously in the cases of edge and screw dislocations [SEqs.~(\ref{eq:defectHam}) and~(\ref{eq:defectHamScrew}), respectively].  In SEq.~(\ref{eq:fluxHam}), $\xi$ is a matrix that acts on the ``top'' and ``bottom'' degrees of freedom, and $s$ continues to act on the indices $i=1,2$ in $|\pm\rangle_{i}$ within each of the two surfaces of the insulator described by SEq.~(\ref{eq:Htotal}).  Because SEq.~(\ref{eq:fluxHam}) is independent of $\phi$, then SEq.~(\ref{eq:fluxHam}) remains invariant under $\mathcal{I}$ and $\mathcal{T}$ for all values of $\phi$.

However and importantly, because magnetic flux is odd under $\mathcal{T}$, and the locations of the flux tubes in SFig.~\ref{fig:kpFlux} are exchanged by the action of $\mathcal{I}$ (red $\times$), then the full Hamiltonian of the ``glued'' interface at $z=0$ ($x_{\perp}=0$), which includes symmetry-allowed couplings, will \emph{not} be invariant under $\mathcal{I}$ and $\mathcal{T}$ symmetries for generic values of $\phi$, but will rather only be invariant under the product $\mathcal{I}\times\mathcal{T}$.  We capture $\mathcal{I}$- and $\mathcal{T}$-symmetry-breaking at generic flux strengths by introducing a coupling interaction $V^{\mathrm{C}}_{a,\mu}(\phi)$ between the top and bottom surfaces of the two insulators in SFig.~\ref{fig:kpFlux}:
\begin{equation}
\mathcal{H}^{C}_{a,\mu}(q_{x},q_{y},\phi) = \mathcal{H}^{D}_{a,\mu}(q_{x},q_{y}) + V^{\mathrm{C}}_{a,\mu}(\phi),
\label{eq:coupledFluxHamHam}
\end{equation}
where $\mathcal{H}^{C}_{a,\mu}(q_{x},q_{y},\phi)$ is now a three-parameter \emph{family} of Hamiltonians indexed by $q_{x,y}$ and $\phi$.  In the specific flux-tube geometry in SFig.~\ref{fig:kpFlux}:
\begin{equation}
\mathcal{I}\phi = -\phi,\ \mathcal{T}\phi = -\phi,
\label{eq:symmetryOnFlux}
\end{equation}
such that the action of $\mathcal{I}$ symmetry from SEq.~(\ref{eq:defectInversion}) becomes:
\begin{equation}
\mathcal{I}:\ \mathcal{H}^{C}_{a,\mu}(q_{x},q_{y},\phi) \rightarrow \xi^{x}\mathcal{H}^{C}_{a,\mu}(-q_{x},-q_{y},-\phi)\xi^{x},
\label{eq:newfluxInversion}
\end{equation}
and the action of $\mathcal{T}$ symmetry from SEq.~(\ref{eq:Tdefect}) becomes:
\begin{equation}
\mathcal{T}:\ \mathcal{H}^{C}_{a,\mu}(q_{x},q_{y},\phi) \rightarrow s^{y}[\mathcal{H}^{C}_{a,\mu}(-q_{x},-q_{y},-\phi)]^{*}s^{y}.
\label{eq:newTflux}
\end{equation} 
SEqs.~(\ref{eq:symmetryOnFlux}),~(\ref{eq:newfluxInversion}), and~(\ref{eq:newTflux}), along with the recognition that $\phi$ is only implemented as the argument of the exponential $e^{\mathrm{i} \phi}$ [see the text surrounding SEq.~(\ref{eq:topbottomBoundFlux})], imply that the system can only be $\mathcal{I}$ and $\mathcal{T}$ invariant at $\phi\text{ mod }2\pi = 0,\pi$.

From SEq.~(\ref{eq:fluxHam}), we recognize that $V^{\mathrm{C}}_{a,\mu}(\phi)$ can only be guaranteed to open a gap if it contains terms proportional to $\xi^{x}$, $\xi^{y}$, $\xi^{z}s^{z}$, or $s^{z}$.  Furthermore, of these possibilities, only $\xi^{x}$ is invariant under $\mathcal{I}$ and $\mathcal{T}$ symmetries as represented in SEqs.~(\ref{eq:newfluxInversion}) and~(\ref{eq:newTflux}), respectively.  Hence, as in this work we are considering the restrictions on flux bound states from $\mathcal{I}$ and $\mathcal{T}$ symmetries, then we may focus without loss of generality on spinless coupling terms containing $\xi^{x}$, which are captured by:
\begin{align}
[V^{\mathrm{C}}_{a,\mu}(\phi)]^{TB}_{ij} &= m_{\mathrm{C}}\delta_{ij}\int_{-\infty}^{\infty}dz\ \left[|\psi^{T}_{a,\mu}(z)\rangle_{i}\langle\psi^{B}_{a,\mu}(z)|_{j} + \mathrm{ h.c.}\right], \nonumber \\
[V^{\mathrm{C}}_{a,\mu}(\phi)]^{TT}_{ij} &= [V^{\mathrm{C}}_{a,\mu}(\phi)]^{BB}_{ij}=0,
\label{eq:couplingMFlux}
\end{align}
where $m_{\mathrm{C}}$ is the strength of the coupling.  Substituting SEq.~(\ref{eq:topbottomBoundFlux}) into SEq.~(\ref{eq:couplingMFlux}), we arrive at the expression:
\begin{eqnarray}
[V^{\mathrm{C}}_{a,\mu}(\phi)]^{TB}_{ij} &=& m_{\mathrm{C}}\delta_{ij}\int_{-\infty}^{\infty}dz\ \left[e^{i(\phi^{T}_{\mu}-\phi^{B}_{\mu})}\mathcal{Z}^T(z)\mathcal{Z}^{B}(z)|+\rangle_{i}{}_{j}\langle -| +\mathrm{ h.c.}\right], \nonumber \\
&=& m_{\mathrm{C}} \delta_{ij} e^{i(\phi^{T}_{\mu}-\phi^{B}_{\mu})}|+\rangle_{i}{}_{j}\langle -| +\mathrm{ h.c.}.
\label{eq:couplingM2Flux}
\end{eqnarray}
where, in the second line, we have again employed the approximation that $\mathcal{Z}^T(z)\mathcal{Z}^{B}(z)\rightarrow\delta(z)$ described in the text before SEq.~(\ref{eq:edgeburgersin})

In the $\mu=out$ region:
\begin{equation}
\phi^{T}_{out}=\phi^{B}_{out}=0,
\end{equation}
such that:
\begin{equation}
V^{\mathrm{C}}_{a,out}(\phi) = m_{\mathrm{C}}\xi^{x}.
\label{eq:outsideCouplingFlux}
\end{equation}
For the $\mu=in$ region, however, for which:
\begin{equation}
\phi^{T}_{in}= -\phi/2,\ \phi^{B}_{in}=\phi/2,
\end{equation}
an additional nonzero phase factor persists from SEq.~(\ref{eq:couplingM2Flux}):
\begin{eqnarray}
[V^{\mathrm{C}}_{a,in}(\phi)]^{TB}_{ij} &=& m_{\mathrm{C}}\delta_{ij} e^{-i\phi}|+\rangle_{i}{}_{j}\langle -| +\mathrm{ h.c.} \nonumber \\
V^{\mathrm{C}}_{a,in}(\phi) &=& m_{\mathrm{C}}\left[\cos(\phi)\xi^{x} + \sin(\phi)\xi^{y}\right].
\label{eq:insideCouplingPreFlux}
\end{eqnarray} 
While the family of coupling terms characterized by SEq.~(\ref{eq:insideCouplingPreFlux}) is as a set invariant under $\mathcal{I}$ and $\mathcal{T}$ symmetries as represented by SEqs.~\eqref{eq:newfluxInversion} and~\eqref{eq:newTflux}, $\phi$ as a parameter is itself only invariant under $\mathcal{I}$ and $\mathcal{T}$ symmetries at $\phi\text{ mod }2\pi=0,\pi$ [see the text surrounding SEq.~(\ref{eq:symmetryOnFlux})].  Hence, at generic values of $\phi$ away from $0$ and $\pi$, SEq.~(\ref{eq:insideCouplingPreFlux}) is only invariant under the magnetic symmetry $\mathcal{I}\times\mathcal{T}$, as expected for the flux tube geometry shown in SFig.~\ref{fig:kpFlux}~$\bs{a}$.

If we restrict consideration to flux tubes with $\phi=\pi$, then SEq.~(\ref{eq:insideCouplingPreFlux}) simplifies to an $\mathcal{I}$- and $\mathcal{T}$-symmetric form:
\begin{equation}
V^{\mathrm{C}}_{a,out}(\pi) = -m_{\mathrm{C}}\xi^{x}.
\label{eq:insideCouplingFlux}
\end{equation}
Thus, the $\pi$-flux tubes represent domain walls in the mass of a $\mathcal{T}$-symmetric 2D fourfold Dirac fermion~\cite{Steve2D,WiederLayers,SteveMagnet,DiracInsulator} for \emph{every} bulk band-inverted TRIM point $\bs{k}_{D,a}$, \emph{distinctly unlike} the previous $k$-dependent masses for the edge and screw dislocations in SN~\ref{sec:kpEdge} and~\ref{sec:kpScrew}, respectively [SEqs.~(\ref{eq:insideCoupling}) and~(\ref{eq:insideCouplingScrew}), respectively].  Hence, each $\bs{k}_{D,a}$ contributes an anomalous pair of helical modes to each flux tube (SFig.~\ref{fig:kpFlux}~$\bs{b}$).  As the full coupled Hamiltonian is given by:
\begin{equation}
\mathcal{H}^{C}_{\mu}(q_{x},q_{y},\phi) = \bigoplus_{a}\mathcal{H}^{C}_{a,\mu}(q_{x},q_{y},\phi),
\label{eq:fluxSum}
\end{equation}
then the total number of pairs of helical modes is simply equal to the number of bulk band-inverted TRIM points, and specifically, is only anomalous if this number is odd.  Therefore, $\pi$-flux tubes will only bind anomalous helical modes (\emph{i.e.} ``wormholes'' between surfaces~\cite{FranzWormhole,WormholeNumerics}) in a 3D $\mathcal{I}$- and $\mathcal{T}$-symmetric insulator that differs from an atomic insulator by an odd number of band inversions between bands with opposite parity eigenvalues.  Through the Fu-Kane parity criterion~\cite{FuKaneMele,FuKaneInversion}, we recognize that such an insulator is necessarily a strong 3D TI.  We have thus reproduced the result of SRefs.~\onlinecite{QiFlux,AshvinFlux,AdyFlux,FranzWormhole,WormholeNumerics,MirlinFlux,CorrelatedFlux} that $\pi$-flux tubes in 3D strong TIs bind anomalous helical modes.

More generally, we can rearrange the sum in SEq.~(\ref{eq:fluxSum}) to gain further insight: 
\begin{equation}
\mathcal{H}^{C}_{\mu}(q_{x},q_{y},\phi) = \bigoplus_{k_{z}}\bigoplus_{\bs{k}_{D,a}\cdot\hat{z}=k_{z}}\mathcal{H}^{C}_{a,\mu}(q_{x},q_{y},\phi) \equiv \bigoplus_{k_{z} = 0, \pi}\mathcal{H}_{k_{z},\mu}^{C}(q_{x},q_{y},\phi),
\label{eq:fluxSum2}
\end{equation} 
where we have separated the low-energy contributions of the TRIM points by $k_{z}$.  The sum in SEq.~(\ref{eq:fluxSum2}) implies that $\pi$-flux tubes bind anomalous helical modes in a 3D $\mathcal{I}$- and $\mathcal{T}$-symmetric insulator that contains an odd number of $k_{z}$-indexed BZ planes with Hamiltonians that differ from 2D atomic insulators by an odd number of band inversions that exchange parity eigenvalues (\emph{i.e.} are topologically equivalent to 2D TIs~\cite{AndreiTI,CharlieTI,KaneMeleZ2}).  Therefore, we can consider the $\pi$-flux tubes to be summing the 2D topologies of \emph{all} of the Hamiltonians of the $k_{z}$-indexed planes of the pristine crystal [SEq.~(\ref{eq:Htotal})], and projecting them to the real-space surface spanning the flux tubes (labeled ``inside'' in SFig.~\ref{fig:kpFlux}~$\bs{a,b}$).  As previously with the screw dislocations in SFig.~\ref{fig:kpScrew}~$\bs{c,d}$, we can imagine lightly bending the flux tubes as they approach the crystal surface to avoid any curvature effects associated with sharp corners~\cite{HingeSM}.  In this construction, and for now ignoring the complications of merging surface and flux-tube states, an $\mathcal{I}$- and $\mathcal{T}$-symmetric crystal with two flux tubes will therefore host a real-space surface between the tubes that is equivalent to a 2D TI if and only if the bulk contains an odd number of (here $k_{z}$-indexed) 2D surfaces with the topology of a 2D TI.  Specifically, because the superposition of two 2D TIs does not exhibit anomalous helical modes (though it does exhibit anomalous corner states if $\mathcal{I}$ symmetry is additionally present~\cite{TMDHOTI,HingeSM}), then helical modes only appear on the tubes if an odd number of 2D TIs contribute to the summation in SEq.~(\ref{eq:fluxSum2}).

Furthermore, because a 3D TI can be reexpressed as a pumping cycle of a 2D TI with odd, helical winding~\cite{FuKaneMele,FuKaneInversion,QHZ,AndreiXiZ2}, the edges of the real-space surface between the $\pi$-flux tubes (\emph{i.e.}, the tubes themselves, as well as two lines on the crystal surface [SFig.~\ref{fig:kpFlux}~$\bs{b,c}$] bind anomalous helical modes if and only if the bulk pristine crystal is a 3D strong TI.  Crucially, while the location of the 2D real-space surface is electromagnetic gauge-dependent (here, we take it to simply lie at $z=0$), the locations of its edges, the $\pi$-flux tubes, are gauge-independent because we have fixed their positions in real space, and because they give rise to a gauge-invariant Aharonov-Bohm phase shift.  This pumping interpretation of $\pi$-flux response is explored for more general Hamiltonians in SN~\ref{sec:fluxfluxtubetopologymapping}.

From this result, it becomes straightforward to derive the $\pi$-flux response of $\mathcal{I}$-symmetric AXIs.  Numerous previous works~\cite{FuKaneMele,FuKaneInversion,AndreiInversion,QHZ,FanHOTI,VDBAxion,DiracInsulator,MulliganAnomaly,DrewPotterPeel,TitusHOTISurfaceAnomaly,VDBHOTI,WiederAxion} have shown that an $\mathcal{I}$-symmetric 3D strong TI gaps into an AXI under the introduction of $\mathcal{I}$-symmetric magnetism.  Furthermore, it was shown in recent works~\cite{TMDHOTI,WiederAxion,KoreanAxion,NicoDavidAXI2,IvoAXI1,IvoAXI2} that, because an $\mathcal{I}$-symmetric 2D TI gaps into an FTI with anomalous corner modes under breaking $\mathcal{T}$ symmetry with $\mathcal{I}$-symmetric magnetism, then an AXI is equivalent to an odd, chiral pumping cycle of an $\mathcal{I}$-symmetric FTI.  For example, in the AXI models in SRef.~\onlinecite{WiederAxion}, the Hamiltonian in the $k_{z}=0$ BZ plane is equivalent to a $\mathcal{I}$-symmetric 2D FTI with (filling-) anomalous corner modes, whereas the Hamiltonian in the $k_{z}=\pi$ plane is equivalent to 2D trivial atomic insulator.  Thus, when $\mathcal{T}$ symmetry is relaxed in an $\mathcal{I}$-symmetric 3D TI with two $\pi$-flux tubes, the helical surface and flux states gap, resulting in an anomalous number of fractionally charged 0D states bound to the loop formed from the two flux tubes and the crystal surfaces (SFig.~\ref{fig:kpFlux}~$\bs{c}$).  Therefore, $\pi$-flux tubes in an AXI necessarily bind anomalous HEND states, which appear in our numerical calculations on $\mathcal{I}$-related flux tube ends (SN~\ref{sec:numerics2}).

This can also be understood through the pumping interpretation of the flux response described in the text surrounding SEq.~(\ref{eq:fluxSum2}).  Because $y$-directed $\pi$-flux tubes separated by an $x$-directed displacement project the superposed topologies of the momentum-space Hamiltonians of $k_{z}$-indexed planes of a pristine AXI, and, because there are necessarily an odd number of $k_{z}$-indexed BZ planes with corner modes in an AXI pumping cycle~\cite{WiederAxion}, then the $\pi$-flux tubes necessarily bind HEND states.  Furthermore, because the flux-tube and surface loop (SFig.~\ref{fig:kpFlux}~$\bs{c}$) exhibits the largest curvature~\cite{HingeSM} where the tubes meet the surfaces, then at least two HEND states will necessarily be localized on $\mathcal{I}$-related tube ends on opposing surfaces.  This is in agreement with the results of previous works~\cite{QiDefect2,QHZ,FranzWormhole,WormholeNumerics,FuKaneInversion} that showed that the surfaces of an AXI accumulate fractional charge under threaded $\pi$ flux in a manifestation of the nontrivial bulk magnetoelectric polarizability~\cite{WilczekAxion,VDBAxion,VDBHOTI,QHZ,AndreiInversion,AshvinAxion1,AshvinAxion2,WuAxionExp,WiederAxion}.  Furthermore, because $\pi$ fluxes in $|C|=1$ Chern insulators bind 0D states with $\pm e/2$ fractional charge (SRef.~\onlinecite{QiDefect2} and SN~\ref{sec:numerics2}), then the presence of isolated HEND states on $\mathcal{I}$-related surfaces represents a signature that AXI surfaces are topologically equivalent to anomalous ``halves'' of the integer quantum Hall effect~\cite{FuKaneMele,FuKaneInversion,AndreiInversion,QHZ,FanHOTI,VDBAxion,DiracInsulator,MulliganAnomaly,DrewPotterPeel,TitusHOTISurfaceAnomaly,VDBHOTI,WiederAxion}.  Specifically, each surface of an AXI exhibits only \emph{half} of the $\pi$-flux response of a $|C|=1$ Chern insulator, because only one of the two flux cores per surface binds a HEND state (with the $\mathcal{I}$-related HEND state being localized on the opposite surface).

Exploiting the recent recognition that $\mathcal{I}$- and $\mathcal{T}$-symmetric higher-order TIs (HOTIs) are equivalent to two superposed, $\mathcal{T}$-reversed copies of AXIs~\cite{WiederAxion,TMDHOTI}, we can use the above conclusions to determine the $\pi$-flux response of HOTIs.  Because an $\mathcal{I}$- and $\mathcal{T}$-symmetric FTI exhibits corner modes that are equivalent to the spin-charge separated solitons of the spinful Su-Schrieffer-Heeger (SSH) chain~\cite{TMDHOTI,WiederAxion,RiceMele,SSHspinon,HeegerReview}, and are themselves formed from Kramers pairs of the corner (anti)solitons of $\mathcal{T}$-broken FTIs~\cite{TMDHOTI,WiederAxion}, then we immediately conclude that $\pi$-flux tubes in HOTIs bind Kramers pairs of anomalous spin-charge separated HEND states (\emph{i.e.} fluxons~\cite{QiFlux,AshvinFlux,AdyFlux,MirlinFlux,CorrelatedFlux}).  Like the $\pi$-flux HEND states in AXIs, the 0D $\pi$-flux states in HOTIs appear on alternating, $\mathcal{I}$-related ends of pairs of tubes (SFig.~\ref{fig:kpFlux}~$\bs{c}$).  Crucially, because $\pi$ fluxes in 2D TIs bind 0D fluxons (SRefs.~\onlinecite{QiFlux,AshvinFlux,AdyFlux,MirlinFlux,CorrelatedFlux,Vlad2D} and SN~\ref{sec:numerics2}), then the presence of unpaired HEND states on $\mathcal{I}$-related surfaces represents a signature that gapped HOTI surfaces are not topologically equivalent to 2D trivial or topological insulators, but are rather equivalent to anomalous ``halves'' of the quantum \emph{spin} Hall effect, a relatively unexplored phase of matter.  Specifically, each surface of a HOTI only exhibits half the $\pi$-flux response of a 2D TI, because the two flux cores on each surface only bind one total fluxon (which manifests in our numerical calculations as a Kramers pair of HEND states on the end of only one of the two flux tubes).  While the half quantum spin Hall (QSH) effect was previously predicted to occur on the top surfaces of weak TIs~\cite{HalfQSH}, in this work, we recognize the half QSH effect to more generally manifest on \emph{all} (gapped) surfaces of $\mathcal{I}$- and $\mathcal{T}$-symmetric HOTIs.

To understand why the surfaces of $\mathcal{I}$- and $\mathcal{T}$-symmetric HOTIs host half-integer QSH phases, instead of full-integer 2D trivial or topological insulators, we draw connection to recent discussions of symmetry-enhanced fermion doubling in TCIs.  Specifically, the 2D surfaces of 3D HOTIs are generically formed from fourfold, $\mathcal{T}$-symmetric Dirac cones~\cite{TMDHOTI} that become gapped by symmetry-allowed perturbations in the absence of specific surface symmetries, such as perpendicular glide reflections~\cite{DiracInsulator}.  When located at a surface TRIM point, each fourfold Dirac cone allows a single $\mathcal{T}$-symmetric anticommuting mass, whose sign has been shown in previous works~\cite{HOTIBernevig,HOTIChen,AshvinIndicators,AshvinTCI,ChenTCI,WiederAxion} to be an intrinsic property of the bulk topology.  As we will detail below, if one were to try to describe the topology of any individual gapped 2D HOTI surface, one would be unable to regularize the continuum theory derived from the single gapped fourfold Dirac cone, due to an obstruction from symmetry-enhanced fermion doubling~\cite{DiracInsulator}; the theory could only be regularized if another (gapped) surface fourfold Dirac cone were also present~\cite{Steve2D}.  This is analogous to the gapped -- yet not regularizable -- surface theory of the massive twofold Dirac cone appearing on AXI surfaces, which exhibits an anomalous $\sigma_{xy} = e^2/(2h)$ Hall conductivity.  Indeed, $\mathcal{I}$- and $\mathcal{T}$-symmetric HOTIs have been shown to admit an $s^{z}$-preserving limit~\cite{TMDHOTI}, and in this limit, it is clear that the obstruction to lattice-regularizing the surface theory originates from the parity anomaly (anomalous half-integer surface Hall conductivity) per spin sector.

More generally, in the presence of spin-orbit coupling (SOC) that removes the $s^{z}$ sector labels, the obstruction to regularizing the gapped surface theory of $\mathcal{I}$- and $\mathcal{T}$-symmetric HOTIs can be understood by considering the role of crystal symmetries.  First, it has been shown that an $\mathcal{I}$- and $\mathcal{T}$-symmetric HOTI can be expressed as the real-space superposition (formally a ``stack'' or Kronecker sum) of $2+4n$ ($n\in\{\mathbb{Z}^{+},0\}$) $\mathcal{I}$-symmetric 3D TIs~\cite{EslamInversion,AshvinTCI,ChenTCI}.  We emphasize that the term superposition in real space does \emph{not} imply that the HOTI ground state wavefunction is a superposition of the ground state wavefunctions of $2+4n$ 3D TIs -- rather, in the decoupled limit, the HOTI ground state wavefunction is a product of the individual ground state wavefunctions of the $2+4n$ superposed (summed) 3D TIs.  Next, considering the case of two superposed, initially decoupled 3D TIs with $\mathcal{I}$ symmetry that host band inversions at the same bulk TRIM point, each surface carries an unpaired, gapless fourfold Dirac cone that is stabilized by $\mathcal{T}$ symmetry and an artificial (``sublattice'') symmetry that indexes the two decoupled, superposed TIs~\cite{HOTIBernevig,TMDHOTI,HOTIBismuth,ChenTCI,AshvinTCI,DiracInsulator}.  When $\mathcal{I}$- and $\mathcal{T}$-symmetric coupling between the two superposed TIs is introduced to break the artificial symmetry, the fourfold surface Dirac cones develop a gap.  Crucially, SRefs.~\onlinecite{EslamInversion,AshvinTCI,ChenTCI} demonstrated that the sign of the $\mathcal{T}$-invariant Dirac mass is opposite on surfaces with $\mathcal{I}$-related Miller indices.  Hence, in $\mathcal{I}$- and $\mathcal{T}$-symmetric HOTIs, there exists a crystal symmetry ($\mathcal{I}$ in the simplest HOTIs~\cite{HOTIBernevig,TMDHOTI,HOTIBismuth,ChenTCI,AshvinTCI}) that relates the gapped surface phases resulting from either sign of the $\mathcal{T}$-symmetric Dirac mass~\cite{DiracInsulator}.  Additionally, as discussed in SRefs.~\onlinecite{AndreiTI,Steve2D,DiracInsulator}, two $\mathcal{T}$-symmetric 2D insulators that differ only by the sign of the mass of a fourfold Dirac fermion at a TRIM point must differ topologically by a relative Kane-Mele $\mathbb{Z}_{2}$ index.  However, it has previously been established that 2D TIs and trivial insulators are \emph{not} related by crystal symmetry operations (\emph{c.f.} SM 1B of SRef.~\onlinecite{DiracInsulator}).  Hence, absent a second fourfold surface Dirac cone that is symmetry-stabilized in the decoupled limit and pinned to a TRIM point, the continuum theory of an unpaired, gapped, surface, fourfold Dirac cone that was previously symmetry-stabilized cannot describe \emph{any} isolated $\mathcal{T}$-invariant 2D insulator (\emph{i.e.} a 2D insulator that is not connected to a 3D bulk)~\cite{DiracInsulator}, because the 2D insulator and its $\mathcal{I}$-symmetry-related partner on the opposing surface must \emph{both} differ by a relative $\mathbb{Z}_{2}$ index \emph{and} be related by a crystal symmetry.  Therefore, $\mathcal{I}$-related HOTI surfaces, whose topologies are determined by the sign of only a \emph{single} (fourfold) Dirac mass, are neither completely trivial nor topological 2D insulators, but must instead represent anomalous ``halves'' of a 2D TI.  Similar to the half-integer surface quantum Hall effect in AXIs~\cite{FuKaneMele,FuKaneInversion,AndreiInversion,QHZ,FanHOTI,VDBAxion,DiracInsulator,MulliganAnomaly,DrewPotterPeel,TitusHOTISurfaceAnomaly,VDBHOTI,WiederAxion}, domain walls between halves of the QSH effect still bind topological 1D modes~\cite{DiracInsulator,HalfQSH} (though the modes are helical, instead of chiral).  The 1D helical modes are, in fact, the characteristic hinge states of HOTIs~\cite{HOTIBismuth,TMDHOTI,AshvinTCI,ChenTCI}.  From this perspective, the anomalous halves of the QSH effect can be considered superposed, $\mathcal{T}$-reversed copies of the half-integer quantum Hall effect that remain anomalous under the preservation of global $\mathcal{T}$ and bulk $\mathcal{I}$ symmetries.

The presence of spin-charge-separated Kramers pairs of states (fluxons)~\cite{QiFlux,AshvinFlux,AdyFlux,MirlinFlux,CorrelatedFlux} on HOTI surfaces with threaded $\pi$-flux (specifically spinons at half filling~\cite{RiceMele}) implies that, even though the magnetoelectric polarizability $\theta$ of HOTIs is trivial ($\theta\text{ mod }2\pi=0$)~\cite{HOTIBernevig,HOTIChen,AshvinIndicators,AshvinTCI,ChenTCI,WiederAxion}, there is still a quantized bulk response that may be observable in experiment.  We leave further analysis of this response for future works~\cite{PartialAxionHOTINumerics}.

\subsection{Ground State Mapping Proofs for Defect and $\pi$-Flux States}
\label{sec:frankProofs}

In this section, we will use the many-body (Slater-determinant) ground states of pristine crystals to determine defect and flux responses beyond the $k\cdot p$ approximation previously employed in SN~\ref{sec:kp}.  For edge dislocations (SN~\ref{sec:defecttopology}) and $\pi$-flux tubes (SN~\ref{sec:fluxfluxtubetopologymapping}) in particular, our arguments can be straightforwardly generalized to crystals with arbitrarily large dimensionality $d$.

We will here model each dislocation or flux tube as an interface between two thermodynamically large, $d$-D bulk systems, whose occupied electronic states in momentum space differ by the Brillouin zones of $(d-1)$-D phases.  We will then subsequently deduce the presence of $(d-2)$-D or $(d-3)$-D bound states via the bulk-boundary correspondences of the $(d-1)$-D phases.  Throughout this section, we will assume that any symmetries that enforce the topology of the $(d-1)$-D momentum-space subsystems are also preserved by the position-space dislocation or flux-tube geometry itself.  Because we are assuming that the crystal without defects is insulating, any bound states are necessarily localized on the defects and flux tubes (or on their corners and ends), and decay exponentially away from them.

\subsubsection{Edge Dislocations in $d$-D Crystals}
\label{sec:defecttopology}

Here, we derive the topological relations governing the appearance of bound states arising from the insertion of a set of $\mathcal{I}$-symmetric edge dislocations into an insulating system with arbitrary dimensionality ($d$-D).  We first focus on an edge dislocation with Burgers vector $\bs{B} = \hat{y}$ for clarity [see SFig.~\ref{fig:2D}~$\bs{b}$ in the main text], before generalizing to arbitrary Burgers vectors.

Let $S$ be a $d$-dimensional insulating system defined on a lattice $\Lambda$ of size $V_S = L_y V_S^\perp$, where $L_y$ is the linear extent in $y$ direction (we take the lattice spacing to be $a=1$ along each of the lattice vectors) and $V_S^\perp$ denotes the $(d-1)$-dimensional volume spanning the remaining directions.  We take $S$ to be governed by a gapped single-particle Hamiltonian $\mathcal{H}$, and first consider the pristine system without dislocations.  Making translation symmetry in the $y$ direction explicit, and leaving translation in the remaining perpendicular directions implicit, we Fourier transform $\mathcal{H}$ to realize a Bloch Hamiltonian $\mathcal{H}(k_y)$ that acts on the degrees of freedom of $(d-1)$-dimensional slices (layers or rows) with size $V_{S}^{\perp}$ of the lattice.  With periodic boundary conditions, and taking $L_y$ to be an even integer without loss of generality, $k_y$ is an element of the discrete Brillouin zone (BZ):
\begin{equation}
\label{eq:evenfinitesizequantization}
\mathrm{BZ}_S = \frac{2\pi}{L_y} \left\{-\left(\frac{L_y}{2} -1\right), \dots, 0, \dots, \left(\frac{L_y}{2} -1\right), \frac{L_y}{2}\right\}.
\end{equation}
In the case of $N$ occupied bands, the eigenstates of $\mathcal{H}(k_{y})$ are given by:
\begin{equation}
\mathcal{H}(k_y) \ket{u_{\alpha}(k_y)} = E_\alpha(k_y) \ket{u_{\alpha}(k_y)}, 
\end{equation}
where $\alpha = 1 \dots N$.  Re-expressing the eigenstates of $\mathcal{H}(k_y)$ using second-quantization:
\begin{equation}
\ket{u_\alpha(k_y)} = c_\alpha^\dagger(k_y) \ket{0}, 
\end{equation}
where the orthonormality of the occupied eigenstates $\ket{u_{\alpha}(k_y)}$ implies that:
\begin{equation}
 \{c_\alpha(k_y),c^\dagger_\beta(k_y)\} = \delta_{\alpha \beta}. 
 \end{equation}
In this construction, the ground state of $S$ takes the form: 
\begin{equation}
\ket{\mathrm{GS}_S} = \prod_{k_y}^{\mathrm{BZ}_S} \prod_{\alpha =1}^N c_\alpha^\dagger(k_y) \ket{0}.
\label{eq:TempGSDislocation}
\end{equation}
We note that in SEq.~(\ref{eq:TempGSDislocation}), the degrees of freedom along the perpendicular directions in $V_S^\perp$ are covered by the band index $\alpha$, so that $N$ may be very large.

\begin{figure}[t]
\begin{center}
\includegraphics[width=0.5 \textwidth]{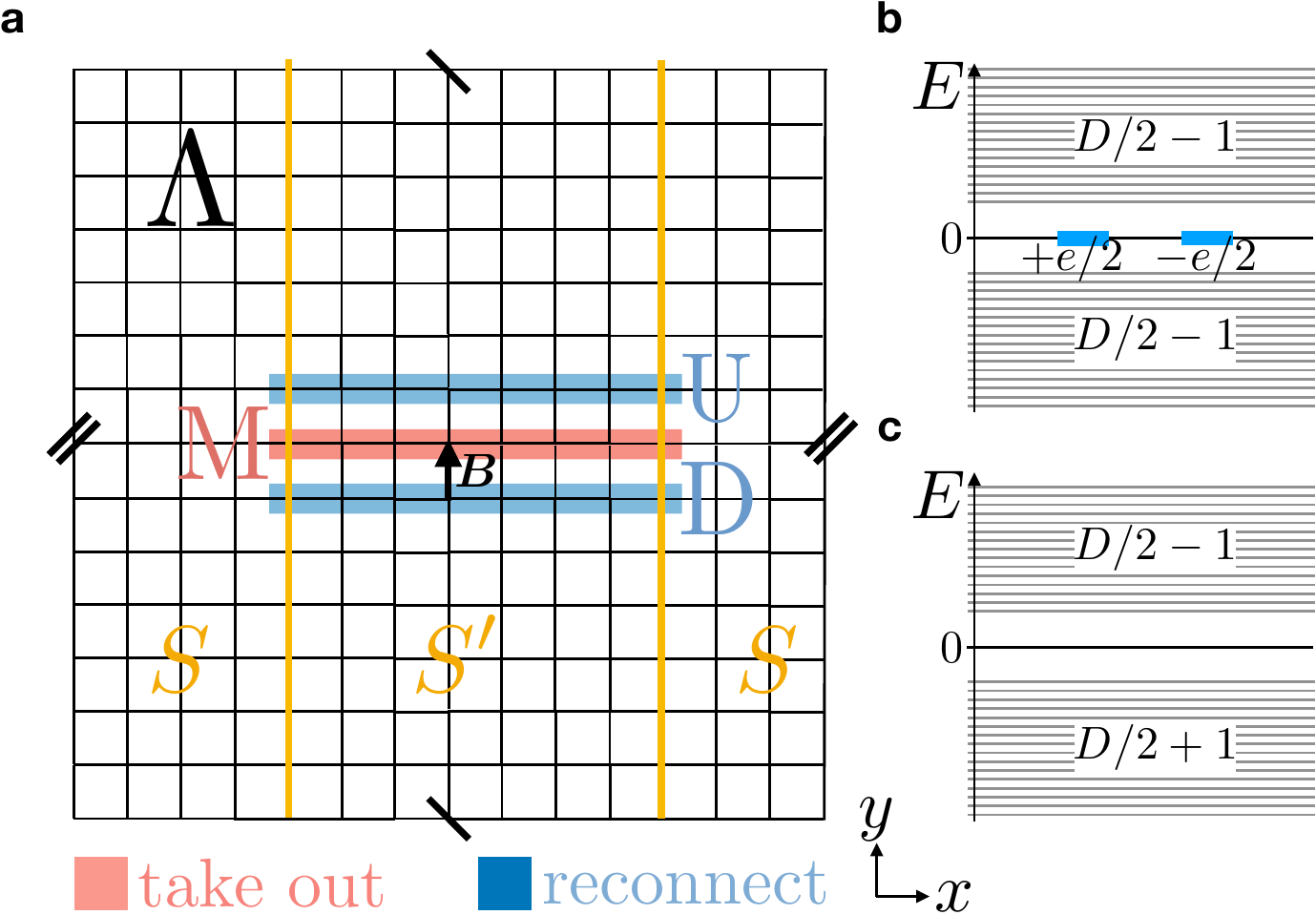}
\caption{{\bf Schematic edge dislocation implementation.} $\bs{a}$ To create a pair of edge dislocations with Burgers vectors $\bs{B}= \hat{y}$ in a 2D crystal defined on a lattice $\Lambda$, we remove the sites belonging to an $x$-directed line $M$, and reconnect the sites above $M$ -- belonging to the line $U$ -- with those below $M$ -- belonging to the line $D$, as detailed in the text surrounding SEq.~\eqref{eq:dislocalgo}.  $\bs{b}$ In the spectrum of a 2D $\mathcal{I}$-symmetric system calculated with open boundary conditions, a pair of dislocations can carry nontrivial topology that is indicated via dislocation bound states pinned to zero energy by a chiral (unitary particle-hole) symmetry $\Pi$, which is absent in real materials.  In the topologically nontrivial phase, the conductance and valence bands are both missing one state with respect to the same spectrum calculated with periodic boundary conditions.  $\bs{c}$. When $\Pi$ is relaxed, however, the midgap states can be shifted out of the bulk gap.  Nevertheless, the nontrivial bulk topology is still present in the system with the pair of dislocations, and is instead more generally indicated via a filling anomaly~\cite{WiederAxion,HingeSM,WladCorners,AshvinFragile2,ZhidaFragileTwist1,ZhidaFragileTwist2}.  Specifically, at half filling, there is a ground state degeneracy, arising from the choice of which midgap state should be occupied. Then, when one electron is added or removed, the system becomes gapped, but there is a mismatch in the number of states above and below the gap (depicted in~$\bs{b,c}$ as centered at $E=0$).}
\label{fig:2Dimplementation}
\end{center}
\end{figure}

An edge dislocation with $\bs{B} = \hat{y}$, located at a given $y$ coordinate, realizes an interface -- oriented parallel to the $y$ direction -- of $S$ with a system $S'$ that has size $V_S' = (L_y-1) V_S^\perp$, but which otherwise has the same symmetry and topology as $S$.  Here, we assume lattice relaxation away from the edge dislocation, so that both $S$ and $S'$ have translation symmetry in $y$ direction, and also in all remaining directions away from the dislocation. While dislocations in real materials generically do not fully satisfy this requirement, the topological dislocation bound states that we uncover and analyze in this work are insensitive to the scale of translation-restoring lattice relaxation, provided that the bulk and interface gaps are large, such that the 0D defect states are well localized.  Under the preservation of $y$-direction translation symmetry, $S'$ is described by the Bloch Hamiltonian $\mathcal{H}(k'_y)$, where $k'_y$ is instead an element of the discrete BZ: 
\begin{equation}
\mathrm{BZ}_{S'} = \frac{2\pi}{L_y-1} \left\{-\left(\frac{L_y}{2} -1\right), \dots, 0, \dots, \left(\frac{L_y}{2} -1\right) \right\}.
\label{eq:oddfinitesizequantization}
\end{equation}
The ground state of $S'$ ($\ket{\mathrm{GS}_{S'}}$) is formed from the occupied subspaces of $H({k}'_y)$ with ${k}'_y$ being an element of SEq.~(\ref{eq:oddfinitesizequantization}), such that $\ket{\mathrm{GS}_{S'}}$ is given by:
\begin{equation}
\label{eq:primedGS}
\ket{\mathrm{GS}_{S'}} = \prod_{k'_y}^{\mathrm{BZ}_{S'}} \prod_{\alpha =1}^N c_\alpha^\dagger(k'_y) \ket{0}.
\end{equation}
Importantly, in SEqs.~(\ref{eq:oddfinitesizequantization}) and~(\ref{eq:primedGS}), the $k'_y = \pi$ subspace of the continuum BZ is never sampled, and thus does not contribute, even though $\mathrm{BZ}_{S'}$ approaches $\mathrm{BZ}_{S}$ as $L_{y}$ goes to infinity.  To restate, as long as $L_{y}$ is discretely valued, only $\mathrm{BZ}_{S}$ contains $k_{y}=\pi$.  Therefore, a $\bs{B} = \hat{y}$ edge dislocation forms an interface between two systems $S$ and $S'$, of which only one system has a ground state in which $k_y = \pi$ is occupied.

Comparing SEqs.~\eqref{eq:evenfinitesizequantization} and~\eqref{eq:oddfinitesizequantization} establishes that there is a mismatch between $\ket{\mathrm{GS}_{S}}$ and $\ket{\mathrm{GS}_{S'}}$ that is given by the $(d-1)$-D (possibly topological) contribution of $\mathcal{H}(\pi)$ to the ground state of $S$:
\begin{equation}
\ket{\mathrm{GS}_S} \approx \prod_{\alpha =1}^N c_\alpha^\dagger(\pi) \ket{\mathrm{GS}_{S'}},
\label{eq:piplanedofs}
\end{equation}
where by $\approx$ we mean that the states on either side of the equation are adiabatically related to each other.  Namely, the $\approx$ symbol indicates that the states on the two sides of SEq.~(\ref{eq:piplanedofs}) only differ by the precise BZ location of the low-symmetry $k_y$ momenta (neither $k_y=0$ nor $k_y = \pi$) contained in the ground states of $S$ and $S'$.  BZ manifolds at such choices of $k_y$ break $\mathcal{I}$ symmetry, and therefore do not contribute any $\mathcal{I}$-protected topological indices to the summed topology.  The $(d-1)$-D contribution of $\mathcal{H}(\pi)$ can be trivial or (crystalline or polarization) topological; in the latter case, the edge dislocation acts as a boundary that can host topologically protected states by the bulk-boundary (domain-wall) correspondence between $S$ and $S'$.

To be precise, the ground state of $S$ is adiabatically related to a state given by the (appropriately antisymmetrized) tensor product of the ground state of $\mathcal{H}(k_y=\pi)$ with that of $S'$. We relate this to the bulk-boundary (domain-wall) correspondence between $S$ and $S'$ by noting that, under the tensor product, which is physically implemented by stacking (\emph{i.e.} superposition), free-fermion topological phases form an additive group~\cite{TurnerAshvinReview}, which allows us to ``subtract'' ($\ominus$) $\ket{\mathrm{GS}_{S'}}$ from both sides of the dislocation in SFig.~\ref{fig:2Dimplementation}~$\bs{a}$.  We then conclude that the ground state of the subsystem of the dislocation is, via this subtraction, adiabatically related to the ground state of $\mathcal{H}(k_y=\pi)$, which is given by:
\begin{equation}
\ket{\mathrm{GS}_{S}} \ominus \ket{\mathrm{GS}_{S'}} \approx \prod_{\alpha =1}^N c_\alpha^\dagger(\pi) \ket{0},
\label{eq:ominusGS}
\end{equation} 
with respect to the vacuum $\ket{0}$.

The condition for anomalous bound states to be present is that all (unitary and antiunitary) symmetries that protect the topology of $\mathcal{H}(\pi)$ remain enforced by the defect geometry. Importantly, we do not require individual edge dislocations to preserve these symmetries on their own, but we do require the set of dislocations to be arranged in a manner in which the $(d-1)$-D surface connecting them does preserve the global symmetries enforcing the topology of $\mathcal{H}(k_{y})$.

We also emphasize that there exist several distinct notions of topological nontriviality.  In the language of Topological Quantum Chemistry (TQC)~\cite{QuantumChemistry,Bandrep1,Bandrep2,Bandrep3,MTQC}, the occupied bands of topological (crystalline) insulators (TIs and TCIs) cannot be represented in terms of maximally localized, symmetric Wannier functions~\cite{MarzariReview}.  However, the results in this section also apply if the ground state of $\mathcal{H}(k_{y}=\pi)$ [SEq.~(\ref{eq:ominusGS})] carries the topology of an OAL, whose occupied bands \emph{can} be represented in terms of maximally localized Wannier functions, but not on the positions of the underlying atoms~\cite{QuantumChemistry}.  Specifically, if the ground state of $\mathcal{H}(k_{y})$ in SEq.~(\ref{eq:piplanedofs}) is an OAL, then a dislocation still acts as a boundary for it, and can still host protected (0D) states due to excess charge (or spin)~\cite{SSH,SSHspinon,HeegerReview,RiceMele,multipole,WladTheory,HingeSM,WladCorners}.  As shown in recent works~\cite{HingeSM,WiederAxion,WladCorners,TMDHOTI}, fragile TIs (FTIs) also carry the same boundary spin and charge as OALs, and thus our analysis applies to FTIs as well. For simplicity, in this section, we group strong TIs, TCIs, OALs, and FTIs together under the general label ``topological,'' unless stated otherwise.

In order to generalize to arbitrary Burgers vectors, we phrase the above in coordinate-independent terms. Making the full translation symmetry of $\mathcal{H}$ explicit, we consider its Fourier transform $\mathcal{H}(\bs{k})$, where $\bs{k}$ is an element of the $d$-dimensional Brillouin zone.  Hence, an edge dislocation with Burgers vector $\bs{B}$ forms an interface between two systems $S$ and $S'$ that differ by the ground state of a $(d-1)$-dimensional momentum-space subsystem with the Bloch Hamiltonian: 
\begin{equation}
\label{eq:dislocationtheory}
\mathcal{H}(\bs{k}_\perp), \quad \bs{k}_\perp \in \{\bs{k} \, | \, \bs{B} \cdot \bs{k} = \pi \}.
\end{equation}

\subsubsection{$\pi$-Flux States in $d$-D Crystals}
\label{sec:fluxfluxtubetopologymapping}

In this section, we will derive the topological relations governing the appearance of bound states arising from the insertion of a set of $\pi$-flux tubes into an insulating system with arbitrary dimensionality ($d$-D).  We first focus on the physically relevant cases of two and three dimensions before briefly generalizing to insulators with arbitrarily large dimensionality $d$.

\begin{figure}[h]
\begin{center}
\includegraphics[width=0.75 \textwidth]{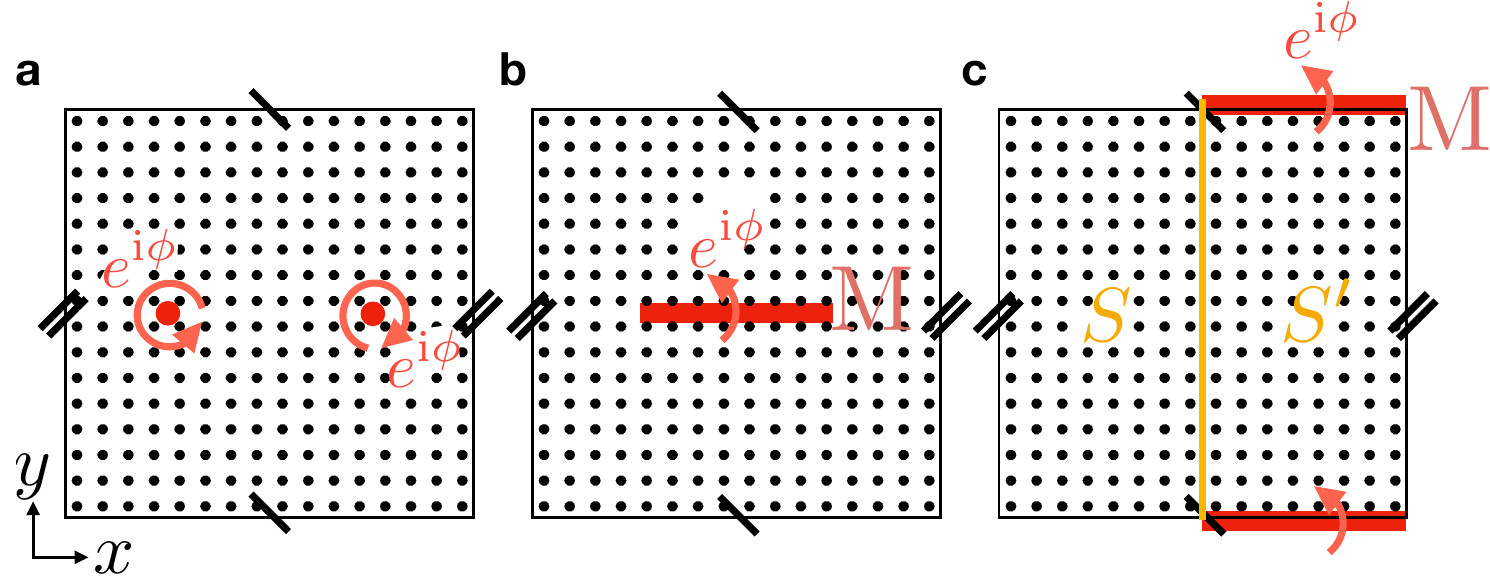}
\caption{{\bf Schematic magnetic flux implementation.} $\bs{a}$-$\bs{c}$ A pair of magnetic $ \pm\phi$-fluxes (flux tubes in 3D) in the $(x,y)$-plane, shown in $\bs{a}$, can be implemented by multiplying all hoppings across a line (plane) with length $L_{M}$ in the $x$ direction connecting the two flux cores (tubes) by $e^{\mathrm{i} \phi}$, as shown in $\bs{b}$.  This is known as the Peierls substitution~\cite{PeierlsSubstitution}, and corresponds to a gauge choice for the electromagnetic vector potential.  In a system with otherwise periodic boundary conditions, the Peierls substitution in $\bs{b}$ is equivalent to implementing twisted boundary conditions in a part of the system, as shown in $\bs{c}$. The flux cores (tubes) therefore form an interface between the system with standard periodic boundary conditions and the system with twisted boundary conditions (shown in the $y$ direction in $\bs{c}$).}
\label{fig:2DimplementationFlux}
\end{center}
\end{figure}

A pair of magnetic fluxes (flux tubes) with strength $\pm\phi$ can be introduced in a 2D (3D) system $S$, containing $L_x L_y$ $(L_x L_y L_z)$ lattice sites, by multiplying all of the hoppings across a line (plane) $M$ of sites in the crystal lattice by a factor of $e^{\mathrm{i} \phi}$ in one direction and $e^{-\mathrm{i} \phi}$ in the other.  This is known as the Peierls substitution~\cite{PeierlsSubstitution}, and here specifically represents a convenient gauge choice: the only physical (gauge-invariant) requirement is that the hoppings encircling a flux core (tube) in real space accumulate a phase of $e^{\pm\mathrm{i}\phi}$.  We orient the line (plane) $M$ containing the $e^{\mathrm{i} \phi}$ phase shift (red line in SFig.~\ref{fig:2DimplementationFlux}~$\bs{b}$) along the $x$ direction in 2D (along the $x$ and $z$ directions in 3D), and choose its linear extent to be half of the linear size (here $L_{M}$) of our full system (see also SFig.~\ref{fig:2DimplementationFlux}~$\bs{a}$-$\bs{c}$).  While the proofs in this section and in SN~\ref{sec:kpFlux} employ this particular straight-line (plane) electromagnetic gauge, ultimately, the physical observables obtained in both sections (bound charge, spin, or chiral or helical modes) appear at gauge-invariant locations with topologically quantized values (or multiplicities).

As shown in SFig.~\ref{fig:2DimplementationFlux}~$\bs{c}$, in the thermodynamic limit, the flux core (tube) realizes a $(d-2)$-D interface between the original, $d$-D unmodified system $S$ (with periodic boundary conditions in the $y$ direction), and a system $S'$ with twisted boundary conditions in the $y$ direction (in the straight-line gauge in SFig.~\ref{fig:2DimplementationFlux}).  This follows from noting that, in $S'$ realized with the gauge in SFig.~\ref{fig:2DimplementationFlux}, there are a total of $L_M$ bonds in the positive $y$ direction that cross $M$ and carry a rotated hopping phase of $e^{\mathrm{i} \phi}$, and that there are $L_M$ bonds in the negative $y$ direction with a hopping phase rotated by $e^{-\mathrm{i} \phi}$.

\begin{figure}[h]
\begin{center}
\includegraphics[width=0.5 \textwidth]{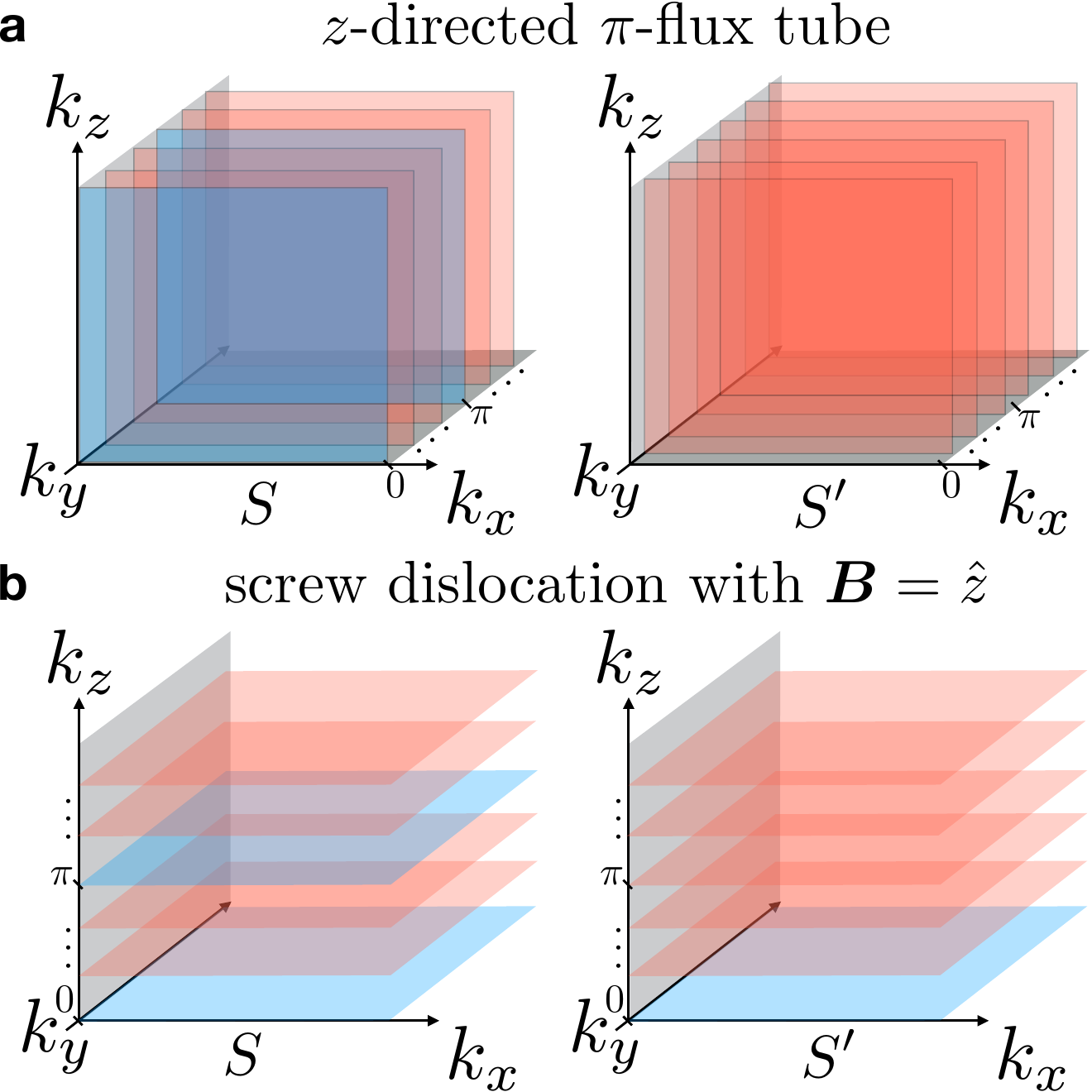}
\caption{{\bf Flux and dislocation systems in discretized momentum space.} 3D Brillouin zones for the systems $S$ and $S'$ as respectively defined in SN~\ref{sec:fluxfluxtubetopologymapping} and~\ref{sec:FrankScrew} for the cases of $\bs{a}$ a $z$-directed $\pi$-flux tube and $\bs{b}$ a screw dislocation with Burgers vector $\bs{B}=\hat{z}$.  $\bs{a}$ A $z$-directed flux tube lies at the boundary between a pristine system $S$ and, in the gauge in SFig.~\ref{fig:2Dimplementation}, a system $S'$ in which all momenta $k_y$ are shifted by $\pi/L_y$ [SEq.~(\ref{eq:fluxSPrime})].  Assuming $L_y$ to be even and $S$ and $S'$ to be $\mathcal{I}$-symmetric, $S$ contains the (possibly topologically nontrivial) BZ planes at $k_y = 0,\pi$ (indicated in blue), in addition to planes at intermediate momenta (indicated in red), which are generically equivalent to 2D insulators without edge or corner states [discounting for now the case of 3D weak Chern (quantum anomalous Hall) insulators].  In contrast, the discrete BZ of $S'$ [SEq.~(\ref{eq:fluxFiniteBZ})] only contains the intermediate values $k_y\neq0,\pi$.  Therefore, an interface between $S$ and $S'$ whose boundary is a $\pi$-flux tube will exhibit (given the preservation of bulk symmetries) the (gauge-invariant) boundary modes of the combined (superposed~\cite{TurnerAshvinReview}) topologies of the Hamiltonians of the $k_{y}=0,\pi$ planes of $S$ [SEq.~(\ref{eq:ominusFlux})].  $\bs{b}$ A screw dislocation with Burgers vector $\hat{z}$ forms an interface between a pristine system $S$ and a system $S'$ in which the momenta $k_y$ are shifted by $k_{z}$-dependent amounts $k_{z}/L_{y}$ [SEq.~(\ref{eq:screwFluxshift})].  Assuming $L_{y,z}$ to be even, $S$ contains the (possibly topologically nontrivial) BZ planes at $k_z = 0,\pi$ (indicated in blue), in addition to planes at intermediate momenta (indicated in red), which are, like those in $\bs{a}$, generically equivalent to 2D insulators without edge or corner states.  In contrast to that of $S$, the discrete BZ of $S'$ only contains the $\mathcal{I}$-symmetric plane $k_{z}=0$, and \emph{does not contain} $k_{z}=\pi$ [SEq.~(\ref{eq:newKz})].  Therefore, an interface between $S$ and $S'$ whose boundary is a screw dislocation with an odd-integer Burgers vector will exhibit (given the preservation of bulk symmetries) the (gauge-invariant) boundary modes of the $k_{z}=\pi$ plane of the pristine insulator $S$ [SEq.~(\ref{eq:ominusScrew})].}
\label{fig:FluxScrewMapping}
\end{center}
\end{figure}

We will now individually analyze $S$ and $S'$ in order to deduce the (gauge-invariant) electronic structure of the boundary between $S$ and $S'$ by using the bulk-boundary (domain-wall) correspondence between $S$ and $S'$.  To begin, let $S$ be described by a gapped Bloch Hamiltonian $\mathcal{H}(k_y)$, where we suppress other momentum labels for notational simplicity. Let $L_y$ be even without loss of generality.  As in our previous analysis in SN~\ref{sec:defecttopology}, with periodic boundary conditions, $k_{y}$ is here an element of the discrete BZ:
\begin{equation}
\label{eq:FLUXESevenfinitesizequantization}
\mathrm{BZ}_S = \frac{2\pi}{L_y} \left\{-\left(\frac{L_y}{2} -1\right), \dots, 0, \dots, \left(\frac{L_y}{2} -1\right), \frac{L_y}{2}\right\},
\end{equation}
where for simplicity, we have chosen units in which the lattice spacings $a_{x,y,z}=1$.  In the case of $N$ occupied bands, the occupied eigenstates of $\mathcal{H}(k_{y})$ are given by:
\begin{equation}
\mathcal{H}(k_y) \ket{u_{\alpha}(k_y)} = E_\alpha(k_y) \ket{u_{\alpha}(k_y)}, 
\end{equation}
where $\alpha = 1 \dots N$.  Reexpressing the eigenstates of $\mathcal{H}(k_y)$ using second-quantization:
\begin{equation}
\ket{u_\alpha(k_y)} = c_\alpha^\dagger(k_y) \ket{0}, 
\end{equation}
where the orthonormality of the occupied eigenstates $\ket{u_{\alpha}(k_y)}$ implies that:
\begin{equation}
 \{c_\alpha(k_y),c^\dagger_\beta(k_y)\} = \delta_{\alpha \beta}. 
 \end{equation}
In this construction, the ground state of $S$ again takes the form: 
\begin{equation}
\ket{\mathrm{GS}_S} = \prod_{k_y}^{\mathrm{BZ}_S} \prod_{\alpha =1}^N c_\alpha^\dagger(k_y) \ket{0}.
\label{eq:fluxgroundstateofS}
\end{equation}

Conversely, in $S'$, the twisted boundary conditions that arise from the flux insertion shift all momenta $k_y$, such that:
\begin{equation}
\mathrm{BZ}_{S'} = \frac{\phi}{L_y} + \mathrm{BZ}_S.
\label{eq:fluxSPrime}
\end{equation}
Specializing to the ($\mathcal{T}$-invariant) case of $\phi=\pi$ highlighted in this work, SEqs.~(\ref{eq:FLUXESevenfinitesizequantization}) and~(\ref{eq:fluxSPrime}) combine to realize a finite BZ in which all $k_{y}$ have become shifted by $\pi/{L_y}$:
\begin{equation}
\mathrm{BZ}_{S'} = \frac{2\pi}{L_y} \left\{-\left(\frac{L_y-1}{2}\right), \dots,  -\frac{1}{2}, \frac{1}{2}, \dots, \left(\frac{L_y-1}{2}\right)\right\}.
\label{eq:fluxFiniteBZ}
\end{equation}
Crucially, the finite set $\mathrm{BZ}_{S'}$ includes neither $k_{y}=0$ nor $k_{y}=\pi$ (SFig.~\ref{fig:FluxScrewMapping}~$\bs{a}$).  Using SEq.~(\ref{eq:fluxFiniteBZ}), the corresponding ground state of $S'$ is then given by:
\begin{equation}
\label{eq:fluxgroundstateofSprime}
\ket{\mathrm{GS}_{S'}} = \prod_{k_y}^{\mathrm{BZ}_{S'}} \prod_{\alpha =1}^N c_\alpha^\dagger(k_y) \ket{0}.
\end{equation}
As previously in SN~\ref{sec:defecttopology}, comparing the ground states of $S$ and $S'$, we find that a flux core (or tube) can host topologically protected boundary states if $\ket{\mathrm{GS}_{S}}$ and $\ket{\mathrm{GS}_{S'}}$ ``differ'' ($\ominus$) by a topologically nontrivial phase [as defined in the text surrounding SEq.~(\ref{eq:ominusGS})].

In this work, we are focused on insulators with spatial inversion symmetry ($\mathcal{I}$).  The operation of $\mathcal{I}$ takes $k_{y}\rightarrow -k_{y}$, such that in both 2D and 3D insulators, there are only two $\mathcal{I}$-invariant values $k_{y}=0,\pi$.  We may also re-express 2D (3D) insulators as (possibly trivial) $k_{y}$-indexed pumping cycles of 1D (2D) insulators in which only the values of $k_{y}=0,\pi$ correspond to 1D (2D) insulators with $\mathcal{I}$ (and optionally $\mathcal{T}$) symmetry~\cite{ThoulessPump,AlexeyVDBWannier,AlexeyVDBTI,TRPolarization,FuKaneMele,FuKaneInversion,QHZ,WladTheory,HOTIChen,WiederAxion,HingeSM}.  Hence, if the weak Chern numbers of a 3D $\mathcal{I}$-symmetric insulator are zero [\emph{i.e.} the bulk is not a quantum anomalous Hall insulator (QAH) insulator~\cite{QiDefect2,BarryBenCDW,MTQC,MTQCmaterials}], then the only possible topologically nontrivial contributions to the difference $\ket{\mathrm{GS}_{S}}\ominus \ket{\mathrm{GS}_{S'}}$ can come from $\mathcal{H}(k_{y}=0,\pi)$  [SEqs.~(\ref{eq:fluxgroundstateofS}) and~(\ref{eq:fluxgroundstateofSprime})].  Furthermore, because there are no 1D topological phases in the absence of crystal symmetry~\cite{AZClass,KitaevClass,SSH,SSHspinon,HeegerReview,RiceMele,TRPolarization}, then, in an $\mathcal{I}$- (and optionally $\mathcal{T}$-) symmetric 2D insulator, $\ket{\mathrm{GS}_{S}}\ominus \ket{\mathrm{GS}_{S'}}$ also only contains contributions from $\mathcal{H}(k_{y}=0,\pi)$.  Specifically, in both 2D and 3D, the real-space interface between $S$ and $S'$ in SFig.~\ref{fig:2DimplementationFlux}~$\bs{c}$ carries the summed (superposed) $(d-1)$-D topologies of $\mathcal{H}(0)$ and $\mathcal{H}(\pi)$.  While the location of the interface is gauge-dependent, the location of its boundaries, the flux cores (and tubes), are gauge-independent.  Therefore, exploiting that topological phases form an additive group~\cite{TurnerAshvinReview}, we formally express the ground state of the interface between $S$ and $S'$ as:
\begin{equation}
\ket{\mathrm{GS}_{S}} \ominus \ket{\mathrm{GS}_{S'}} \approx \left[\prod_{\alpha =1}^N c_\alpha^\dagger(0) \otimes \prod_{\alpha =1}^N c_\alpha^\dagger(\pi)\right]\ket{0},
\label{eq:ominusFlux}
\end{equation} 
with respect to the vacuum $\ket{0}$.  To summarize, in $\mathcal{I}$- (and optionally $\mathcal{T}$-) symmetric 2D and 3D insulators [in which the 3D insulators have vanishing weak Chern numbers], the flux tubes at the interface between $S$ and $S'$ will only exhibit (gauge-invariant) topological (anomalous) boundary modes if $\mathcal{H}(0)\otimes\mathcal{H}(\pi)$ is topologically equivalent to a phase with anomalous boundary modes (\emph{i.e}, a TI, TCI, or filling-anomalous OAL or FTI).

Unlike previously in SN~\ref{sec:defecttopology}, SEq.~(\ref{eq:ominusFlux}) implies the possibility that $\ket{\mathrm{GS}_{S}} \ominus \ket{\mathrm{GS}_{S'}}$ is topologically trivial even though $\mathcal{H}(0,\pi)$ are individually topologically nontrivial.  Specifically, if the topology of $\mathcal{H}(0,\pi)$ is diagnosed by a set of indices $\mathbb{Z}_{m_{1}}\otimes\mathbb{Z}_{m_{2}}\otimes\mathbb{Z}_{m_{3}}\otimes \ldots$ where $m_{i}\in\mathbb{Z}^{+}$, then, if both $\mathcal{H}(0)$ and $\mathcal{H}(\pi)$ have topological indices that sum to zero [\emph{e.g.} $\mathcal{H}(0)$ and $\mathcal{H}(\pi)$ both have nontrival $\mathbb{Z}_{2}$ 2D TI indices~\cite{AndreiTI,CharlieTI,KaneMeleZ2}], then $\mathcal{H}(0)\otimes\mathcal{H}(\pi)$ is necessarily topologically trivial.  Hence, SEq.~(\ref{eq:ominusFlux}) implies that $\ket{\mathrm{GS}_{S}} \ominus \ket{\mathrm{GS}_{S'}}$ is only topologically nontrivial if $\mathcal{H}(0)$ and $\mathcal{H}(\pi)$ do not have canceling topological indices, which, if the $\mathbb{Z}$-valued Chern numbers are zero, can only occur in a strong topological (insulating) phase in which $k_{y}$ indexes a nontrivial pumping cycle~\cite{ThoulessPump,AlexeyVDBTI,AlexeyVDBWannier,FuKaneMele,FuKaneInversion,QHZ,WladTheory,HOTIChen,WiederAxion}.

Finally, while in this section we have restricted to flux tubes in 2D and 3D insulators, it is straightforward to generalize our arguments to threaded $\pi$-flux tubes in insulators with arbitrarily large dimensionality $d$.  Specifically, to generalize the Peierls substitution employed in this section (SFig.~\ref{fig:2Dimplementation}) to higher dimensions, we replace the 1D line or 2D plane $M$ with modified hoppings in SFig.~\ref{fig:2DimplementationFlux}~$\bs{c}$ with a $(d-1)$-D \emph{hyperplane} with constant $y$-coordinate.  Even if the boundary between $S$ and $S'$ has a dimensionality larger than $2$, the analysis employed in this section to derive and analyze SEq.~(\ref{eq:ominusFlux}) still applies without further modification [though finer analysis is required if $\mathcal{H}(0,\pi)$ exhibit $\mathbb{Z}$- or $\mathbb{Z}_{n}$-valued ($n>2$) topological indices, analogous to the requirement in 3D that the QAH weak Chern numbers vanish, see the text surrounding SEq.~(\ref{eq:ominusFlux})].

\subsubsection{Screw Dislocations in 3D Crystals}
\label{sec:FrankScrew}

\begin{figure}[h]
\begin{center}
\includegraphics[width=0.75 \textwidth]{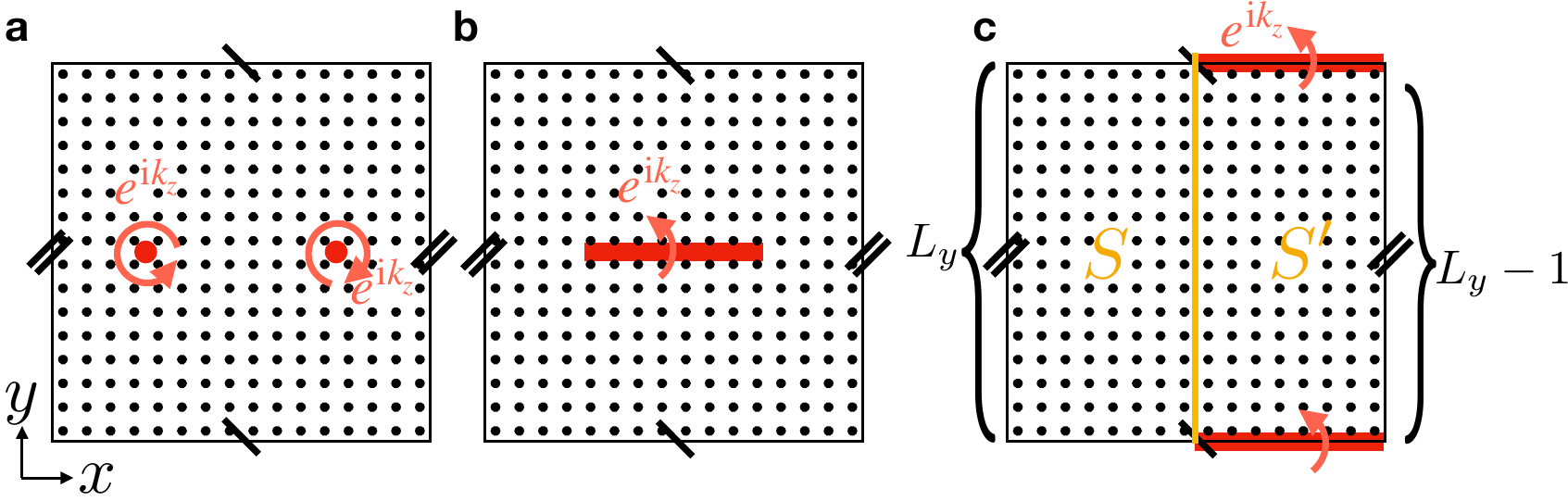}
\caption{{\bf Schematic screw dislocation implementation.} $\bs{a}$-$\bs{c}$ A pair of $\bs{B}=\hat{z}$ screw dislocations, shown in $\bs{a}$, can be implemented by multiplying all hoppings across a plane with length $L_{M}$ in the $x$ direction connecting the two screws by $e^{\mathrm{i} k_z}$, as shown in $\bs{b}$ (keeping periodic boundary conditions in $z$ direction, such that $k_z$ is a well-defined momentum quantum number).  This is equivalent to implementing $k_z$-dependent twisted boundary conditions in a part of the system, as shown in $\bs{c}$.  The screw dislocations therefore form an interface between the system with standard periodic boundary conditions and the system with $k_z$-dependent twisted boundary conditions (shown in the $y$ direction in $\bs{c}$, see also SFig.~\ref{fig:bzscrew}).}
\label{fig:2DimplementationDefect}
\end{center}
\end{figure}

In this section, we will derive the topological relations governing the appearance of bound states arising from the insertion of a set of screw dislocations into an insulating system.  Because the lattice displacements in screw dislocations are more complicated to visualize than those in edge dislocations~\cite{MerminReview}, we will restrict consideration in this section to the more familiar, experimentally relevant case of screw dislocations in 3D insulating crystals.  Unlike the edge dislocations analyzed in SN~\ref{sec:defecttopology}, a screw dislocation corresponds to a dislocation line of displacements that is parallel to its Burgers vector (see also SN~\ref{sec:kpScrew} and SRef.~\onlinecite{MerminReview}).  We consider a pair of $\bs{B} = \hat{z}$ screw dislocations, one left-handed and the other right-handed [see SEq.~(\ref{eq:chiralityScrew}) and the surrounding text].  To implement a right- (left-) handed screw dislocation with Burgers vector $\bs{B} = \hat{z}$, we multiply the lattice hoppings across a given line emanating from the dislocation in the $(x,y)$-plane by a phase of $e^{\mathrm{i}k_{z}}$ ($e^{-\mathrm{i}k_{z}}$), which originates from the net dislocation translation in the positive (negative) $z$ direction~\cite{MerminReview}.  This construction, shown in SFig.~\ref{fig:2DimplementationDefect}, is equivalent to the insertion of a pair of flux tubes (SFig.~\ref{fig:2DimplementationFlux}) under the replacement $\phi \rightarrow k_z$, and is hence reminiscent of the Peierls substitution for $\pi$-flux tubes~\cite{PeierlsSubstitution} previously employed in SN~\ref{sec:fluxfluxtubetopologymapping}.  Specifically, with periodic boundary conditions in the $z$ direction, a pair of screw dislocations is functionally equivalent to modifying the Bloch Hamiltonian $\mathcal{H}(\bs{k})$ with a momentum-dependent flux $\phi=k_{z}$.  Therefore, we can model a $\bs{B} = \hat{z}$ screw dislocation in a 3D insulator as inhabiting the (linear) boundary of a 2D interface between a system $S$ with periodic boundary conditions in all directions and a system $S'$ with twisted boundary conditions (TBC) in the $y$ direction for which the twist phase factor is momentum-dependent ($e^{\mathrm{i}k_{z}}$, see SFig.~\ref{fig:2DimplementationDefect}).

\begin{figure}[t]
\begin{center}
\includegraphics[width=0.6\textwidth]{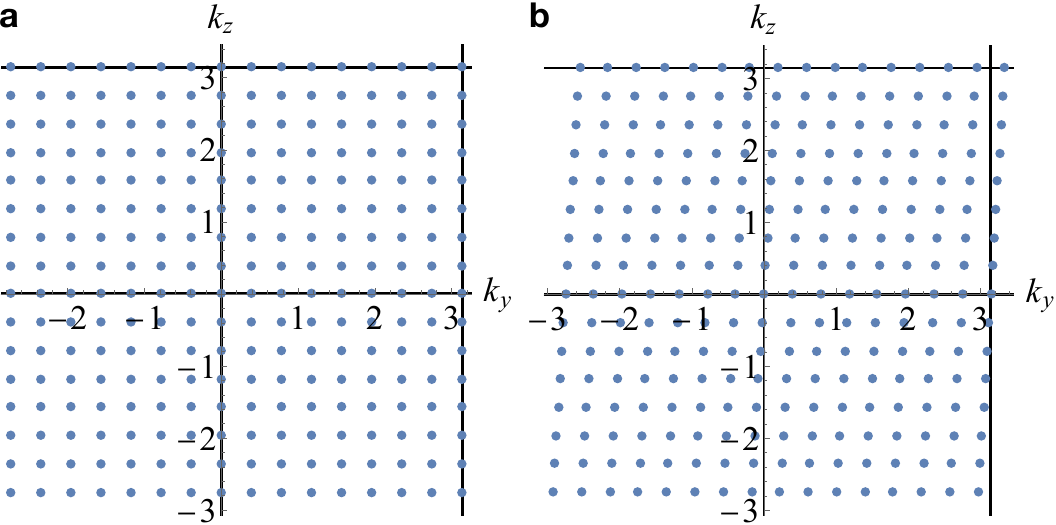}
\caption{{\bf Brillouin-zone mismatch due to $\bs{B} = \hat{z}$ screw insertion.} $\bs{a}$~The $(k_y,k_z)$ BZ of system $S$ in SN~\ref{sec:FrankScrew} for the choice $L_y = L_z = 16$ in SEqs.~\eqref{eq:regularSBZscrew} and~\eqref{eq:regularSBZscrewkz}.  The remaining momentum $k_x$ is measured perpendicular to the $(k_y,k_z)$-plane, and is not shown here, as it remains unaffected by our screw dislocation implementation.  Both the $k_z = 0$ and the $k_z = \pi$ planes contain four high-symmetry ($\mathcal{I}$-invariant) momenta (two each for the BZ lines containing the $\mathcal{I}$-symmetric momenta $k_x =0,\pi$).  $\bs{b}$~The $(k_y,k_z)$ BZ of system $S'$ in SN~\ref{sec:FrankScrew} for the choice $L_y = L_z = 16$ in SEqs.~\eqref{eq:regularSBZscrewkz} and~\eqref{eq:screwFluxshift}.  Crucially, only the $k_z = 0$ plane contains four high-symmetry ($\mathcal{I}$-invariant) momenta, while the $k_z = \pi$ plane does not contain any $\mathcal{I}$-invariant $k$ points.  As a consequence, the screw dislocation, which lies at the boundary between $S$ and $S'$, binds topological states governed by the domain-wall correspondence between $S$ and $S'$ \emph{in only the $k_z = \pi$ BZ plane}.}
\label{fig:bzscrew}
\end{center}
\end{figure}

As previously in SN~\ref{sec:fluxfluxtubetopologymapping}, we will now individually analyze $S$ and $S'$ in order to deduce the (gauge-invariant) electronic structure of the interface between $S$ and $S'$ by using the bulk-boundary (domain-wall) correspondence between $S$ and $S'$.  To begin, let $S$ be described by a gapped Bloch Hamiltonian $\mathcal{H}(k_y,k_z)$, where we suppress the momentum $k_x$ for notational simplicity. Let $L_y$ and $L_{z}$ be even without loss of generality.  With periodic boundary conditions, $k_{y}$ is an element of the discrete BZ:
\begin{equation}
\mathrm{BZ}_{S,k_{y}} = \frac{2\pi}{L_y} \left\{-\left(\frac{L_y}{2} -1\right), \dots, 0, \dots, \left(\frac{L_y}{2} -1\right), \frac{L_y}{2}\right\},
\label{eq:regularSBZscrew}
\end{equation}
where for simplicity, we have chosen units in which the lattice spacings $a_{x,y,z}=1$, and where the $k_{z}$ superscript in SEq.~(\ref{eq:regularSBZscrew}) implies a (here trivial) dependence of the discretization of $k_{y}$ on $k_{z}$.  Unlike previously in SN~\ref{sec:defecttopology} and~\ref{sec:fluxfluxtubetopologymapping}, $k_{z}$ is here also an element of a discrete BZ:
\begin{equation}
\mathrm{BZ}_{S,k_{z}} = \frac{2\pi}{L_z} \left\{-\left(\frac{L_z}{2} -1\right), \dots, 0, \dots, \left(\frac{L_z}{2} -1\right), \frac{L_z}{2}\right\}.
\label{eq:regularSBZscrewkz}
\end{equation}
In the case of $N$ occupied bands, the occupied eigenstates of $\mathcal{H}(k_{y},k_z)$ are given by:
\begin{equation}
\mathcal{H}(k_y,k_z) \ket{u_{\alpha}(k_y,k_z)} = E_\alpha(k_y,k_z) \ket{u_{\alpha}(k_y,k_z)}, 
\end{equation}
where $\alpha = 1 \dots N$.  Reexpressing the eigenstates of $\mathcal{H}(k_y,k_z)$ using second-quantization:
\begin{equation}
\ket{u_\alpha(k_y,k_z)} = c_\alpha^\dagger(k_y,k_z) \ket{0}, 
\end{equation}
where the orthonormality of the occupied eigenstates $\ket{u_{\alpha}(k_y,k_z)}$ implies that:
\begin{equation}
 \{c_\alpha(k_y,k_z),c^\dagger_\beta(k_y,k_z)\} = \delta_{\alpha \beta}. 
 \end{equation}
In this construction, the ground state of $S$ takes the form (see SFig.~\ref{fig:bzscrew}~$\bs{a}$): 
\begin{equation}
\ket{\mathrm{GS}_S} = \prod_{k_z}^{\mathrm{BZ}_{S,k_{z}}} \prod_{k_y}^{\mathrm{BZ}_{S,k_{y}}} \prod_{\alpha =1}^N c_\alpha^\dagger(k_y,k_z) \ket{0}.
\end{equation}
with respect to the vacuum $\ket{0}$, where all products are appropriately anti-symmetrized.

Conversely, in $S'$, the twisted boundary conditions that arise from the screw dislocations (as illustrated in SFig.~\ref{fig:2DimplementationDefect}) shift all momenta $k_y$ by a $k_z$-dependent amount, such that $k_{y}$ in $S'$ is an element of the discrete BZ:
\begin{equation}
\mathrm{BZ}_{S',k_{y}}^{k_z} = \frac{k_z}{L_y} + \mathrm{BZ}_{S,k_{y}}.
\label{eq:screwFluxshift}
\end{equation}
where $\mathrm{BZ}_{S,k_{y}}$ is given in SEq.~(\ref{eq:regularSBZscrew}).  SEq.~(\ref{eq:screwFluxshift}) implies that in system $S'$ at fixed $k_z$, $k_y$ is sampled from the $k_z$-dependent set $\mathrm{BZ}_{S',k_{y}}^{k_z}$, so that the tuple of momenta $(k_y,k_z)$ is drawn from 
\begin{equation}
\bigcup_{k_z}^{\mathrm{BZ}_{S',k_z}}  \bigcup_{k_y}^{\mathrm{BZ}_{S',k_{y}}^{k_z}} (k_y,k_z).
\end{equation}  
However, even though the screw dislocations in this section are $z$-directed (SFig.~\ref{fig:FluxScrewMapping}~$\bs{b}$), they \emph{do not} affect the discretization of $k_{z}$, such that:
\begin{equation}
\mathrm{BZ}_{S',k_{z}} = \mathrm{BZ}_{S,k_{z}},
\label{eq:BZkZScrew}
\end{equation}
where $\mathrm{BZ}_{S,k_{z}}$ is given in SEq.~(\ref{eq:regularSBZscrewkz}).  The corresponding ground state of $S'$ is then given by (see SFig.~\ref{fig:bzscrew}~$\bs{b}$):
\begin{equation}
\label{eq:ScrewgroundstateofSprime}
\ket{\mathrm{GS}_S'} = \prod_{k_z}^{\mathrm{BZ}_{S',k_z}} \prod_{k_y}^{\mathrm{BZ}_{S',k_{y}}^{k_z}} \prod_{\alpha =1}^N c_\alpha^\dagger(k_y,k_z) \ket{0}.
\end{equation}

Crucially, if we restrict to 3D insulators with only $\mathcal{I}$ (and optionally $\mathcal{T}$) symmetry, SEqs.~(\ref{eq:screwFluxshift}),~(\ref{eq:BZkZScrew}), and~(\ref{eq:ScrewgroundstateofSprime}) can be reformulated to realize a simple result.  Namely, first focusing on the $k_{z}=\pi$ plane in $S'$, SEq.~(\ref{eq:screwFluxshift}) implies that the lines $k_{y}=0,\pi$ are absent in the discrete BZ.  Therefore, the action of $\mathcal{I}$ (and $\mathcal{T}$) symmetry no longer takes the $k_{z}=\pi$ plane back to itself, such that in $S'$, the $k_{z}=\pi$ plane can no longer exhibit topology enforced by $\mathcal{I}$ or $\mathcal{T}$.  Hence, when restricted to $\mathcal{I}$-symmetric insulators, the ground state of $S'$ can be adiabatically deformed from SEq.~(\ref{eq:ScrewgroundstateofSprime}) to:
\begin{equation}
\label{eq:ScrewgroundstateofSprimeWithNoKz}
\ket{\mathrm{GS}_S'} = \prod_{\tilde{k}_z}^{\tilde{\mathrm{BZ}}_{S',\tilde{k_{z}}}} \prod_{\alpha =1}^N c_\alpha^\dagger(\tilde{k}_z) \ket{0},
\end{equation}
where $\tilde{k}_{z}$ is now an element of a discrete BZ:
\begin{equation}
\tilde{\mathrm{BZ}}_{S',\tilde{k}_{z}} = \frac{2\pi}{L_z} \left\{-\left(\frac{L_z}{2} -\frac{1}{2}\right), \dots, 0, \dots, \left(\frac{L_z}{2} -\frac{1}{2}\right) \right\}.
\label{eq:newKz}
\end{equation}
that \emph{does not} contain $\tilde{k}_{z}=\pi$.  Specifically, because the topology of a 3D $\mathcal{I}$- (and optionally $\mathcal{T}$-) symmetric insulator (that lacks additional rotation and reflection symmetries) is entirely determined by the topologies of the 2D Hamiltonians of the $k_{x,y,z}=0,\pi$ planes [excluding 3D quantum anomalous Hall insulators, see the discussion preceding SEq.~(\ref{eq:ominusFlux})], then the ground state of $S'$ can only contain topologically nontrivial contributions from the $\mathcal{I}$-symmetric Hamiltonian of the $k_{z}=0$ ($\tilde{k}_{z}=0$) plane.

As previously in SN~\ref{sec:defecttopology} and~\ref{sec:fluxfluxtubetopologymapping}, we next compare the topology of $S$ and $S'$.  Using the definition of ``subtraction'' ($\ominus$) established in the text surrounding SEq.~(\ref{eq:ominusGS}), we find that SEq.~(\ref{eq:ScrewgroundstateofSprime}) implies that a screw dislocation can host topologically protected boundary states if $\ket{\mathrm{GS}_{S}}$ and $\ket{\mathrm{GS}_{S'}}$ ``differ'' ($\ominus$) by a topologically nontrivial phase.  This can be summarized by stating that the gauge-invariant screw dislocations represent the boundary of a 2D insulator whose ground state is given by:
\begin{equation}
\ket{\mathrm{GS}_{S}} \ominus \ket{\mathrm{GS}_{S'}} = \prod_{\alpha =1}^N c_\alpha^\dagger(k_{z}=\pi)\ket{0}.
\label{eq:ominusScrew}
\end{equation} 
The screw dislocations therefore bind the 1D or 0D boundary modes of the Hamiltonian of the $k_{z}=\pi$ plane of S, provided that the dislocation geometry preserves $\mathcal{I}$ symmetry.  Generalizing to arbitrarily oriented screw dislocations, we recover SEq.~\eqref{eq:dislocationtheory}, in agreement with the results of SN~\ref{sec:kpScrew}.

\section{Weak Fragile and Obstructed-Atomic-Limit Indices for 2D Dislocation and 3D HEND States}
\label{sec:indices}

\subsection{The 2D Weak SSH Invariant $\textbf{\textit{M}}_{\nu}^{\mathrm{SSH}}$}
\label{sec:weakSSH}

\begin{figure}[h]
\begin{center}
\includegraphics[width=0.6\textwidth]{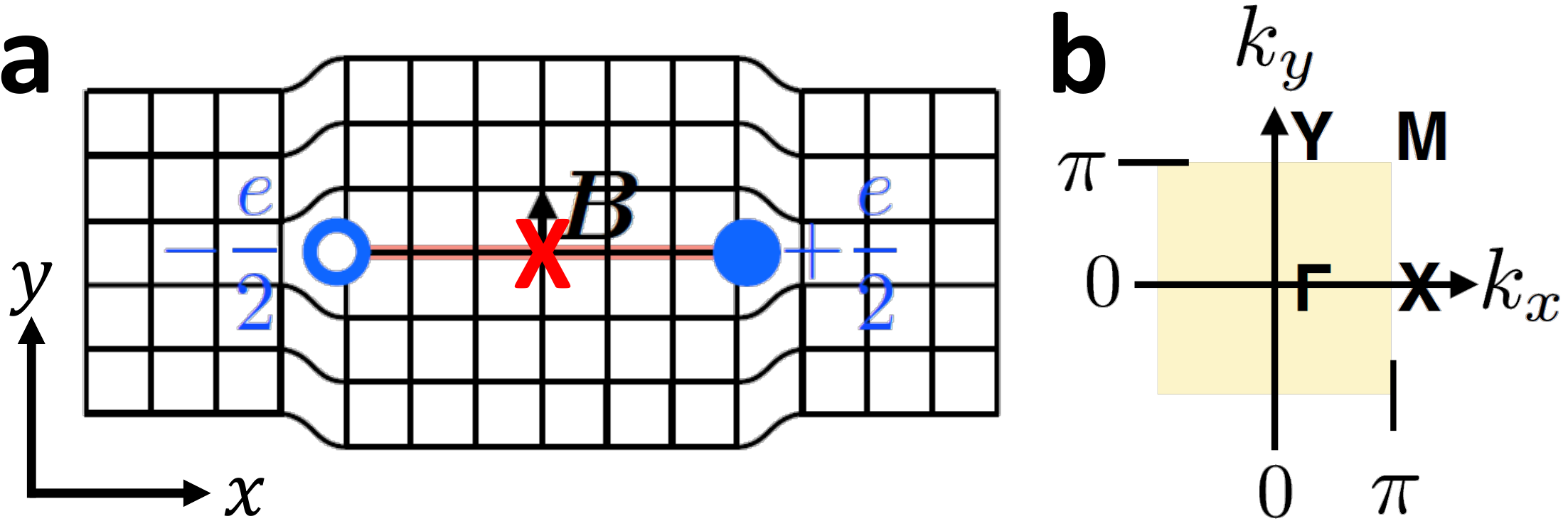}
\caption{{\bf 0D dislocation states in 2D crystals.} $\bs{a}$ A pair of dislocations with Burgers vectors $\bs{B}=\hat{y}$ in an $\mathcal{I}$-symmetric 2D insulator without spinful $\mathcal{T}$ symmetry.  $\bs{b}$ The 2D Brillouin zone (BZ) of the insulator in $\bs{a}$.  Under the preservation of $\mathcal{I}$ (red $\times$), the dislocations in $\bs{a}$ bind filling-anomalous (SRefs.~\onlinecite{TMDHOTI,WiederAxion,HingeSM,WladCorners} and SN~\ref{sec:defecttopology}) charges $\pm e/2$ if the $k_{y}=\pi$ BZ line exhibits a nontrivial polarization~\cite{SSH,SSHspinon,HeegerReview,RiceMele,ZakPhase,ArisInversion,TRPolarization,KaneMeleZ2} $n_{YM}$ when measured with respect to the $x$-coordinates of the valence atomic orbitals [SEq.~(\ref{eq:polarizationInvariant})].  More generally, an $\mathcal{I}$-symmetric dislocation pair with Burgers vector $\bs{B}$ binds anomalous charge if $\bs{B}\cdot\bs{M}_{\nu}^{\mathrm{SSH}}\text{ mod }2\pi=\pi$ [SEq.~(\ref{eq:SSHinvariant})].  If $\mathcal{T}$ symmetry is additionally present, then the dislocations will instead bind Kramers pairs of spin-charge-separated solitons, and specifically carry the same spin-charge relations as the solitons in polyacetylene (see SN~\ref{sec:kp} and SRefs.~\onlinecite{RiceMele,SSH,SSHspinon,HeegerReview}).  Using SEq.~(\ref{eq:polarizationInvariant}), we will specifically identify the presence of dislocation solitons in PbTe monolayers in SN~\ref{sec:DFTPbTe}.}
\label{fig:SSH}
\end{center}
\end{figure}

In this section, we will derive the 2D weak polarization invariant $\bs{M}_{\nu}^{\mathrm{SSH}}$.  As discussed in the main text and shown in SN~\ref{sec:defecttopology}, an $\mathcal{I}$-symmetric 2D insulator with rectangular lattice vectors can bind 0D electronic states on pairs of dislocations that are related by a bulk $\mathcal{I}$ center (red $\times$ in SFig.~\ref{fig:SSH}~$\bs{a}$).  Specifically, a pair of dislocations with defect Burgers vectors $\bs{B}=\hat{i}$ maps the momentum-space (polarization) topology of the $k_{i}=\pi$ line of the 2D BZ (SFig.~\ref{fig:SSH}~$\bs{b}$) to the real-space line spanning the dislocations (red line in SFig.~\ref{fig:SSH}~$\bs{a}$).  If the real-space crystal with two dislocations preserves the momentum-space symmetries of the $k_{i}=\pi$ line of the 2D BZ modulo reciprocal lattice translations (here $\mathcal{I}$), then the polarization topology of the Hamiltonian of $k_{i}=\pi$ implies the presence of end (dislocation) charges in real space.  That is, if the Hamiltonian of the $k_{i}=\pi$ line carries a nontrivial polarization (or \emph{time-reversal} polarization~\cite{TRPolarization,KaneMeleZ2}, if spinful $\mathcal{T}$ symmetry is also present) as measured relative to the unobstructed atomic limit of the valence atomic orbitals~\cite{QuantumChemistry}, then the dislocations will bind 0D states (or bound spin and charge in the more general case of relaxed particle-hole symmetry).   The momentum-space (weak) polarization topology can be summarized for an $\mathcal{I}$- (and optionally $\mathcal{T}$-) symmetric 2D system as:
\begin{equation}
\bs{M}_{\nu}^{\mathrm{SSH}}=\pi(n_{XM},n_{YM}),
\label{eq:SSHinvariant}
\end{equation}
where $n_{XM}/n_{YM}$ is the Su-Schrieffer-Heeger (SSH) polarization invariant~\cite{SSH,SSHspinon,HeegerReview,RiceMele,DiracInsulator} (or time-reversal polarization~\cite{TRPolarization,KaneMeleZ2}, if states are additionally doubled under $\mathcal{T}$ symmetry) of the BZ edge line $k_{x/y}=\pi$ (SFig.~\ref{fig:SSH}~$\bs{b}$).

We next define the polarization invariants $n_{XM}/n_{YM}$.  We will specifically focus on $n_{XM}$, with the understanding that $n_{YM}$ can analogously be derived by exchanging $x\leftrightarrow y$ in the following text.  It has been shown in numerous previous works~\cite{FuKaneInversion,AndreiInversion,ArisInversion} that the polarization (and time-reversal polarization) of a 1D Hamiltonian $\mathcal{H}(k_{i})$ is directly related to the Berry phase $\gamma$ (or \emph{partial} Berry phase if $\mathcal{T}$ is additionally present~\cite{TRPolarization,KaneMeleZ2}).  In 1D crystals, the presence of rod group $\mathcal{I}$ symmetry (\emph{i.e.} an inversion symmetry for which $\mathcal{I}^{2}=+1$ independent of the presence of spinful electrons)~\cite{subperiodicTables,ITCA,HingeSM} allows the immediate calculation of $\gamma$:
\begin{equation}
\gamma = \frac{\pi}{2}\left[1-\xi(0)\xi(\pi)\right],
\label{eq:BerryPhase}
\end{equation}
where $\xi(k_{D})$ is given by the product of the parity eigenvalues of the occupied bands at the 1D TRIM point $k_{D}=0,\pi$ (or the parity eigenvalues per Kramers pair if states are doubled by spinful $\mathcal{T}$~\cite{FuKaneInversion}).  Hence, for a pristine 2D insulator, the 1D Hamiltonian along $XM$ $\mathcal{H}_{XM}(k_{y})$ has a Berry phase:
\begin{equation}
\gamma_{XM}=\frac{\pi}{2}\left[1-\xi(X)\xi(M)\right].
\label{eq:LineBerry}
\end{equation}
Because all 1D Hamiltonians are Wannierizable~\cite{QuantumChemistry,ArisInversion,AlexeyVDBWannier,AlexeyVDBTI}, then we can immediately relate $\gamma$ to the locations of (hybrid) Wannier centers in the 2D insulator.  Specifically, in the hybrid Wannier~\cite{AlexeyVDBWannier,AlexeyVDBTI} basis of $y$ and $k_{x}$, $\gamma_{XM}\text{ mod }2\pi=\pi$ indicates that, for $k_{x}=\pi$, the number of hybrid Wannier centers (or Kramers pairs of Wannier centers under an additional $\mathcal{T}$ symmetry) at $y=1/2$ (\emph{i.e.} SSH Wyckoff position $1b$~\cite{QuantumChemistry}) and the number at $y=0$ (SSH Wyckoff position $1a$) differ by an \emph{odd} integer.

To complete our calculation of $n_{XM}$, we must also obtain the Berry phase $\gamma^{A}_{XM}$ of the valence atomic orbitals in the $y$ and $k_{x}=\pi$ hybrid Wannier basis, which we will shortly define below.  It is crucial to note that $\gamma^{A}_{XM}$ cannot be obtained from any momentum-space calculation, but must instead be determined \emph{a priori} through knowledge of the implementation of the tight-binding Hamiltonian.  Specifically, as described by Topological Quantum Chemistry (TQC)~\cite{QuantumChemistry,Bandrep1,Bandrep2,Bandrep3,MTQC,JenFragile1,BarryFragile}, position-space atomic orbitals induce EBRs, which are then subduced onto $k$-points to form momentum-space bands.  However, because the overlap between atomic orbitals can drive band inversions, the actual set of occupied bands may not necessarily exhibit the same symmetry eigenvalues and Berry phases as the ``uninverted'' atomic insulator with the same atomic orbitals.  Therefore, in order to determine $\gamma^{A}_{XM}$, one must have knowledge of which bands are inverted relative to the unobstructed atomic limit of the valence atomic orbitals.  Though this requirement may appear unrealistically stringent, in practice, most topological insulating materials with clean Fermi surfaces are narrow-gap semiconductors for which it is chemically straightforward to determine which bands from which atomic orbitals have become inverted~\cite{AndreiMaterials,AndreiMaterials2,ChenMaterials,AshvinMaterials1,AshvinMaterials2,AshvinMaterials3,TavazzaMaterials3D}.  For example, in SN~\ref{sec:DFTPbTe}, we will demonstrate that PbTe monolayers differ from unobstructed atomic limits by band inversion at just two (symmetry-equivalent) TRIM points.  Returning to the example in this section, we determine $\gamma^{A}_{XM}$ by projecting the positions of the valence atomic orbitals onto the $y$-axis, obtaining a set of $y$-coordinates $y_{a}$ for each valence orbital $a$.  From this, the Berry phase of the atomic orbitals is given by:
\begin{equation}
\gamma^{A}_{XM} = \left(2\pi\sum_{a}y_{a}\right)\text{ mod }2\pi,
\label{eq:atomicBerry}
\end{equation}
where the sum is restricted to atoms within a single unit cell.
Finally, we obtain $n_{XM}$ through SEqs.~(\ref{eq:LineBerry}) and~(\ref{eq:atomicBerry}):
\begin{equation}
\pi n_{XM} = \left(\gamma_{XM}-\gamma^{A}_{XM}\right)\text{ mod }2\pi.
\label{eq:polarizationInvariant}
\end{equation}
This definition follows in perfect analogy from the definition of 1D polarization in terms of electronic and atomic Berry phases~\cite{VDBBook}.  Substituting SEq.~(\ref{eq:polarizationInvariant}) into SEq.~(\ref{eq:SSHinvariant}) and determining $n_{YM}$ through the exchange $x\leftrightarrow y$ in SEqs.~(\ref{eq:LineBerry}) through (\ref{eq:polarizationInvariant}) and the surrounding text, we complete the definition of the weak polarization invariant $\bs{M}_{\nu}^{\mathrm{SSH}}$.

\subsection{The 3D Weak Fragile Invariant $\textbf{\textit{M}}_{\nu}^{\mathrm{F}}$}
\label{sec:weakFragile}

\begin{figure}[h]
\begin{center}
\includegraphics[width=0.6\textwidth]{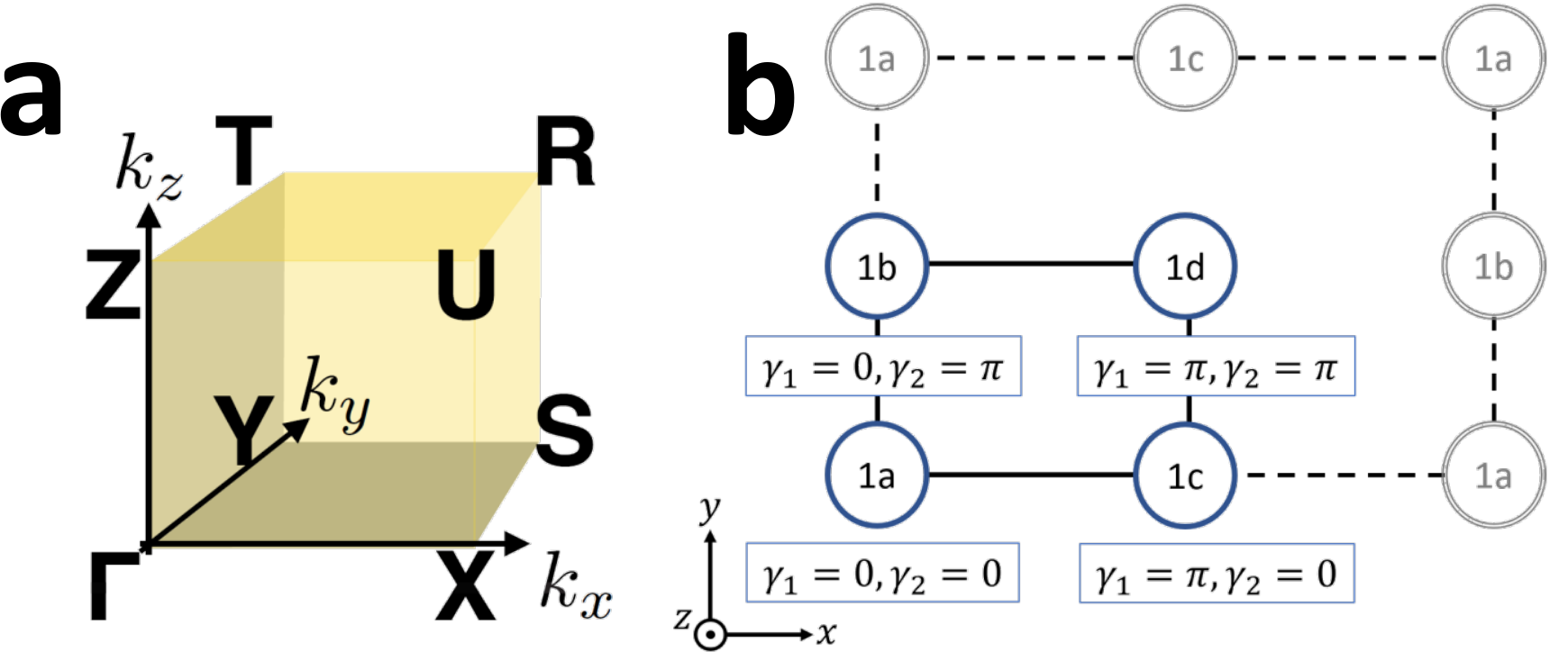}
\caption{{\bf Brillouin zone for orthorhombic 3D crystals and Berry phases in 2D.} $\bs{a}$ The 3D BZ of a crystal with orthorhombic lattice vectors~\cite{BigBook,BCS1,BCS2}.  $\bs{b}$ The 2D Wyckoff positions~\cite{BCS1,BCS2} of a layer group~\cite{BigBook,MagneticBook,subperiodicTables,WiederLayers,SteveMagnet,DiracInsulator,HingeSM} with only rectangular lattice translation $T_{x,y}$ and $\mathcal{I}$ symmetries (and optionally an additional spinful $\mathcal{T}$ symmetry).  As shown in SRef.~\onlinecite{WiederAxion}, a single Wannier orbital placed at one of the four $\mathcal{I}$ centers (maximal Wyckoff positions~\cite{QuantumChemistry}) exhibits the $x$-directed Berry phases $\gamma_{1}$ and $y$-directed nested Berry phases $\gamma_{2}$ shown in $\bs{b}$.  If $\mathcal{T}$ symmetry is additionally present, then all Wannier orbitals necessarily appear in Kramers pairs that exhibit a total $\gamma_{1,2}\text{ mod }2\pi =0$ at all of the maximal Wyckoff positions.  Nevertheless, as we will show in SN~\ref{sec:nested}, by expanding the concept of \emph{time-reversal polarization}~\cite{TRPolarization,KaneMeleZ2} to nested Wilson loops, we can define a partial \emph{nested} $\gamma_{2}$ that distinguishes the locations of Kramers pairs of Wannier orbitals.}
\label{fig:fragile}
\end{center}
\end{figure}

In this section, we will derive the new 3D weak invariant $\bs{M}_{\nu}^{\mathrm{F}}$ for $\mathcal{I}$- (and optionally spinful $\mathcal{T}$-) symmetric insulators to identify which of the BZ boundary planes host 2D Hamiltonians with anomalous corner charges.  We have previously shown in SN~\ref{sec:proofs} that these corner charges will manifest as dislocation HEND states if the respective BZ plane satisfies $\bs{k} \cdot \bs{B}\text{ mod }2\pi = \pi$, where $\bs{B}$ is the Burgers vector of the dislocation.  Because the simplest models of $\mathcal{I}$-symmetric 2D insulators with fractionally-charged corner states have fragile topological bands (specifically four-band models of $\mathcal{I}$-symmetric FTIs, see SRefs.~\onlinecite{TMDHOTI,WiederAxion}), then we will in this work associate $\bs{M}_{\nu}^{\mathrm{F}}$ with fragile topology.  However and crucially, because FTIs can be trivialized into OALs through the addition of appropriately chosen trivial bands~\cite{AshvinFragile,JenFragile1,HingeSM,AdrianFragile,BarryFragile,ZhidaBLG,AshvinBLG1,AshvinBLG2,AshvinFragile2,ZhidaFragileTwist1,ZhidaFragileTwist2,YoungkukMonopole,TMDHOTI,HarukiFragile,OrtixTRealSpace,BouhonMagneticFragile1,BouhonMagneticFragile2,KoreanFragile,WiederAxion,KoreanAxion,NicoDavidAXI2,IvoAXI1,IvoAXI2}, and because there also exist simple models of 2D $\mathcal{I}$-symmetric OALs with anomalous corner charges~\cite{WladCorners}, then we note for completeness that a nontrivial $\bs{M}_{\nu}^{\mathrm{F}}$ does not necessarily imply that the valence manifold taken as a whole exhibits fragile topology.  Like the established index $\bs{M}_{\nu}$ for weak TIs~\cite{FuKaneMele,FuKaneInversion,AdyWeak,MooreBalentsWeak}, $\bs{M}_{\nu}^{\mathrm{F}}$ has three components:
\begin{equation}
\bs{M}_{\nu}^{\mathrm{F}} = \pi\left[\nu_{x}^{\mathrm{F}}\hat{x} + \nu_{y}^{\mathrm{F}}\hat{y} + \nu_{z}^{\mathrm{F}}\hat{z}\right].
\label{eq:fragileMComponents}
\end{equation}
In this section, we will explicitly derive $\nu_{z}^{\mathrm{F}}$, with the understanding that $\nu_{x,y}^{\mathrm{F}}$ can similarly be obtained through the cyclic exchange $x\rightarrow y\rightarrow z \rightarrow x$.

To begin, we will first consider in SN~\ref{sec:FTInoT} a 3D insulator with orthorhombic lattice vectors and $\mathcal{I}$ symmetry.  Unlike previously in SN~\ref{sec:weakSSH}, we will start by explicitly requiring that spinful $\mathcal{T}$ symmetry is broken.  Later, in SN~\ref{sec:FTIwithT}, we will extend our arguments to $\mathcal{I}$- and $\mathcal{T}$-symmetric insulators.  Finally, in SN~\ref{sec:nested}, we will use nested Wilson loops~\cite{multipole,WladTheory,HOTIBernevig,HingeSM,WiederAxion,KoreanFragile,ZhidaBLG,TMDHOTI} to develop a new formulation of nested time-reversal polarization~\cite{TRPolarization,KaneMeleZ2} to diagnose 2D $\mathcal{T}$-symmetric OALs that display anomalous corner modes despite exhibiting trivial nested Berry phases (when taken modulo $2\pi$).

\subsubsection{$\mathcal{T}$-Broken EBR Formulation of $\textbf{\textit{M}}_{\nu}^{\mathrm{F}}$}
\label{sec:FTInoT}

In this section, we will first establish a definition of $\nu^{\mathrm{F}}_{z}$ for $\mathcal{I}$-symmetric, $\mathcal{T}$-broken 3D insulators, which, through SEq.~(\ref{eq:fragileMComponents}) and the cyclic exchange $x\rightarrow y\rightarrow z \rightarrow x$, is sufficient to also define the weak corner-mode invariant $\bs{M}_{\nu}^{\mathrm{F}}$.  As shown in SRefs.~\onlinecite{TMDHOTI,WiederAxion,WladCorners}, an $\mathcal{I}$-symmetric, $\mathcal{T}$-broken 2D insulator will exhibit (filling) anomalous corner charges if its valence bands $B$ are formed from an odd total number of FTIs and OALs with anomalous corner modes.  First, we will use the results of SRefs.~\onlinecite{WladCorners,HingeSM} to show that a single occupied band can be converted from an unobstructed atomic limit without corner modes to an OAL with $\mathcal{I}$-anomalous corner modes through band inversions at two distinct TRIM points.  Then, using the results of SRefs.~\onlinecite{TMDHOTI,WiederAxion}, we will show that there are eight possible FTIs with two occupied bands and $\mathcal{I}$-anomalous corner modes, which are each driven by double band inversion~\cite{YoungkukMonopole,TMDHOTI,WiederAxion} about the same TRIM point.  Taken together, these OALs and FTIs, along with (stable) Chern insulators, can be used to define $\nu^{\mathrm{F}}_{z}$ for a set of valence bands that differs from an unobstructed atomic limit through a known or controllable sequence of band inversions.  We note that the conclusions derived here no longer generically apply when reflection and high-fold rotation symmetries are also enforced, because higher symmetries can protect a more diverse set of stable and fragile topological phases than $\mathcal{I}$ by itself~\cite{LiangTCI,TeoFuKaneTCI,HsiehTCI,ChenBernevigTCI,multipole,WladTheory,HOTIChen,ZeroBerry,FulgaAnon,EmilCorner,HingeSM,OrtixCorners,CornerWarning,TMDHOTI,KoreanFragile,AshvinFragile2,ZhidaFragileTwist1,ZhidaFragileTwist2,HarukiFragile,OrtixTRealSpace,BouhonMagneticFragile1,BouhonMagneticFragile2,WiederAxion,WladCorners,
FragileKoreanInversion,ZhidaFragileAffine,FrankCorners,YoungkukBLG,CaseWesternCorners}.  However, the arguments presented in this section can be adapted to also predict electronic defect states with higher symmetry requirements, albeit with some effort to carefully treat the position-space action of more complicated (potentially magnetic~\cite{MTQC}) crystal symmetries.  We leave this generalization for future works.

To determine $\nu^{\mathrm{F}}_{z}$, we begin by restricting consideration to the $k_{z}=\pi$ plane of a 3D BZ (SFig.~\ref{fig:fragile}~$\bs{a}$).  The 2D Hamiltonian of the $k_{z}=\pi$ plane $\mathcal{H}_{k_{z}=\pi}(k_{x},k_{y})$ contains four TRIM points at which bands are generically singly degenerate and can be labeled according to their parity eigenvalues $\pm 1$.  First and foremost, in order for $\mathcal{H}_{k_{z}=\pi}(k_{x},k_{y})$ to exhibit only corner modes, its occupied bands $B$ must not carry a net Chern number $C_{B}$.  For 2D insulators with only $\mathcal{I}$ symmetry, $C_{B}\text{ mod }2$ can be determined through the parity eigenvalues of the occupied bands~\cite{ChenBernevigTCI}.  Specifically, for $\mathcal{H}_{k_{z}=\pi}(k_{x},k_{y})$:
\begin{equation}
C_{B}\text{ mod }2 = \frac{1}{2}\left[1-\xi(Z)\xi(U)\xi(T)\xi(R)\right],
\label{eq:ChernNumberZ}
\end{equation}
where $\xi(\bs{k}_{D})$ is the product of the parity eigenvalues of the occupied bands at the TRIM point $\bs{k}_{D}$.  

\begin{table}[h]
\begin{tabular}{|c|c|c|c|c|c|c|c|c|c|c|}
\hline
\multicolumn{11}{|c|}{EBRs of Magnetic Layer Group $p\bar{1}$} \\
\hline
\hline
 Orbitals & $\Gamma$ & $X$ & $Y$ & $M$ & \ \ \ \ \ \ \ \ \ \ \ &  Orbitals & $\Gamma$ & $X$ & $Y$ & $M$ \\
\hline
$(s)_{1a}$ & $+$ & $+$ & $+$ & $+$ & & $(p)_{1a}$ & $-$ & $-$ & $-$ & $-$ \\
\hline
$(s)_{1b}$ & $+$ & $+$ & $-$ & $-$ & & $(p)_{1b}$ & $-$ & $-$ & $+$ & $+$ \\
\hline
$(s)_{1c}$ & $+$ & $-$ & $+$ & $-$ & & $(p)_{1c}$ & $-$ & $+$ & $-$ & $+$ \\
\hline
$(s)_{1d}$ & $+$ & $-$ & $-$ & $+$ & & $(p)_{1d}$ & $-$ & $+$ & $+$ & $-$ \\
\hline
\end{tabular}
\caption{Elementary band representations in a rectangular 2D system with inversion symmetry.  Using~\href{https://www.cryst.ehu.es/cgi-bin/cryst/programs/bandrep.pl}{BANDREP}~\cite{QuantumChemistry,Bandrep1,Bandrep2,Bandrep3} on the Bilbao Crystallographic Server (BCS), we obtain the elementary band representations (EBRs)~\cite{ZakBandrep1,ZakBandrep2,QuantumChemistry,Bandrep1,Bandrep2,Bandrep3,MTQC,JenFragile1,BarryFragile} of magnetic layer group~\cite{MagneticBook,subperiodicTables,WiederLayers,SteveMagnet,DiracInsulator,HingeSM} $p\bar{1}$  (\emph{i.e.}, the 2D symmetry group generated by rectangular lattice translations $T_{x,y}$ and 3D inversion $\mathcal{I}$).  We employ a shorthand notation in which $(\rho)_{q}$ indicates the EBR induced from $\rho=s,p$ orbitals placed at the maximal Wyckoff position ($\mathcal{I}$ center) $q$ in SFig.~\ref{fig:fragile}~$\bs{b}$.  At each TRIM point in the 2D BZ of $p\bar{1}$ (SFig.~\ref{fig:SSH}~$\bs{b}$), a one-dimensional irreducible representation with a positive (negative) parity eigenvalue is subduced that is equivalent to the irreducible representation $\bar{A}_{g}$ ($\bar{A}_{u}$) of magnetic point group $\bar{1}$~\cite{BilbaoPoint,PointGroupTables,MagneticBook,BCS1,BCS2}, which we respectively denote with the shorthand $+$ ($-$).}
\label{tb:ebr}
\end{table}

\begin{table}[h]
\begin{tabular}{|c|c|c|c|c|}
\hline
Valence Bands & $\Gamma$ & $X$ & $Y$ & $M$ \\
\hline
$B$ & $+$ & $+$ & $+$ & $+$ \\
\hline
$B'$ & $-$ & $+$ & $+$ & $+$ \\
\hline
$B''$ & $-$ & $-$ & $+$ & $+$ \\
\hline
$B'''$ & $-$ & $-$ & $-$ & $+$ \\
\hline
$B''''$ & $-$ & $-$ & $-$ & $-$ \\
\hline
\end{tabular}
\caption{The parity ($\mathcal{I}$) eigenvalues of the occupied bands $B$--$B''''$ discussed in the text following SEq.~\eqref{eq:trimmappingto2D}.}
\label{tb:ebrsequence1}
\end{table}

If SEq.~(\ref{eq:ChernNumberZ}) indicates that $C_{B}\text{ mod }2=0$, then $B$ necessarily exhibits the same set of symmetry eigenvalues as a sum and difference of EBRs (\emph{i.e.}, bands induced from Wannier orbitals at the maximal Wyckoff positions~\cite{QuantumChemistry}).  In this case, we refer to $B$ as \emph{irreducible-representation-equivalent} ($\stackrel{I}{\equiv}$ in equations, irrep-equivalent in text)~\cite{JenFragile1,BarryFragile,HingeSM,AshvinFragile2,ZhidaFragileTwist1,ZhidaFragileTwist2} to a linear combination of EBRs.  In this section, the 2D Hamiltonian $\mathcal{H}_{k_{z}=\pi}(k_{x},k_{y})$ is invariant under magnetic layer group~\cite{MagneticBook,subperiodicTables,WiederLayers,SteveMagnet,DiracInsulator,HingeSM} $p\bar{1}$ (\emph{i.e.}, the 2D group generated by rectangular lattice translations $T_{x,y}$ and 3D inversion $\mathcal{I}$, for which $\mathcal{I}^{2}=+1$ whether or not SOC is relevant).  The complete set of EBRs of $p\bar{1}$ can be induced from $s$ and $p$ orbitals placed at the four maximal Wyckoff positions, \emph{i.e.}, the four $\mathcal{I}$ centers of the 2D unit cell (SFig.~\ref{fig:fragile}~$\bs{b}$).  We obtain the irreducible representations of the EBRs at each high-symmetry point of the 2D BZ of $p\bar{1}$ (SFig.~\ref{fig:SSH}~$\bs{b}$) by using the~\href{https://www.cryst.ehu.es/cgi-bin/cryst/programs/bandrep.pl}{BANDREP} tool on the Bilbao Crystallographic Server~\cite{QuantumChemistry,Bandrep1,Bandrep2,Bandrep3,MTQC} (BCS) for space group (SG) 2 $P\bar{1}1'$ taken without $\mathcal{T}$ symmetry, which is isomorphic to magnetic layer group $p\bar{1}$ modulo lattice translations~\cite{BigBook,MagneticBook}.  For convenience, the resulting EBRs are reproduced in Supplementary Table~\ref{tb:ebr}.  Because $\mathcal{H}_{k_{z}=\pi}(k_{x},k_{y})$ is (quasi-) 2D, then we will convert its TRIM points into the 2D notation in SFig.~\ref{fig:SSH}, both to facilitate the generalization of this calculation for the later determination of $\nu_{x,y}^{\mathrm{F}}$, and to draw connection with previous works~\cite{WiederLayers,WladCorners,TMDHOTI,HingeSM,WiederAxion}.  For $\nu_{z}^{\mathrm{F}}$, the conversion from 3D, zone-edge TRIM points (SFig.~\ref{fig:fragile}~$\bs{a}$) to 2D, layer group TRIM points (SFig.~\ref{fig:SSH}~$\bs{b}$ and Supplementary Table~\ref{tb:ebr}) is given by:
\begin{equation}
\label{eq:trimmappingto2D}
Z\rightarrow \Gamma,\ U\rightarrow X,\ T\rightarrow Y,\ R\rightarrow M.
\end{equation}

As a first step towards formulating $\nu^{\mathrm{F}}_{z}$, we will consider the simplest case where $B$ consists of a single occupied band.  Specifically, we begin with a band structure for which the valence and conduction bands are respectively induced from spinful $s$ and $p$ orbitals placed at the same Wyckoff position.  In this case, the band structure at each bulk TRIM point exhibits a gap between states with opposite parity eigenvalues, and $B$ characterizes an unobstructed atomic limit [the first row in Supplementary Table~\ref{tb:ebrsequence1} shows $B$ induced from the $1a$ position of magnetic layer group $p\bar{1}$ (Supplementary Table~\ref{tb:ebr})].  If we then invert bands at one of the TRIM points, a gap will necessarily open, because $\mathcal{I}$ symmetry cannot by itself protect crossings anywhere in a 2D BZ~\cite{WiederLayers,YoungkukLineNode,FangFuNSandC2} (though in a 3D BZ, Weyl points are allowed to form at generic crystal momenta~\cite{AshvinWeyl1}).  The band inversion results in a new insulator with an occupied band $B'$ (the second row in Supplementary Table~\ref{tb:ebrsequence1} shows $B'$ for a band inversion at $\Gamma$).  By the parity criterion in SEq.~(\ref{eq:ChernNumberZ}), $B'$ necessarily exhibits a Chern number $|C_{B'}|=1$.  Crucially, if we again invert bands, but this time at a different TRIM point than previously, we realize a valence manifold $B''$ (the third row in Supplementary Table~\ref{tb:ebrsequence1} shows $B''$, taking the second band inversion to occur at $X$) where symmetry-indicated Chern number of $B''$ is \emph{trivial}:
\begin{equation}
C_{B''}\text{ mod }2=0.
\end{equation}
Furthermore, comparing $B''$ to Supplementary Table~\ref{tb:ebr}, we observe that $B''$ is also irrep-equivalent to an EBR, but crucially \emph{not} the same EBR as the original valence band $B$.  As rigorously shown in SRefs.~\onlinecite{QuantumChemistry,HingeSM,WiederAxion,WladCorners}, if $C_{B''}=0$, which must be determined through an explicit Wilson-loop calculation, then, because $B''$ and the corresponding conduction band represent the shift of only two Wannier orbitals driven by one or more band inversions, $B''$ is necessarily an OAL.  Furthermore, as shown in SRefs.~\onlinecite{WiederAxion,WladCorners}, all OALs in $p\bar{1}$ with one occupied band necessarily exhibit filling-anomalous corner modes.

We continue by further inverting bands at a third TRIM point, yielding an occupied band $B'''$ (the fourth row in Supplementary Table~\ref{tb:ebrsequence1} shows $B'''$, taking the third band inversion to occur at $Y$).  Once, again, as with $B'$, SEq.~(\ref{eq:ChernNumberZ}) implies that $C_{B'''}\text{ mod 2}=1$.  Finally, if we invert bands at the last TRIM point, yielding an occupied band $B''''$ (the fifth row in Supplementary Table~\ref{tb:ebrsequence1} shows $B''''$, taking the final band inversion to occur at $M$), then again $C_{B''''}\text{ mod }2=0$ and $B''''$ is irrep-equivalent to an EBR.  However, all of the parity eigenvalues of $B''''$ are simply given by those of $B$ multiplied by $-1$.  By comparing $B''''$ to Supplementary Table~\ref{tb:ebr}, we discover that, while $B''''$ is irrep-equivalent to a different EBR than $B$, $B''''$ is still irrep-equivalent to an EBR induced from the same Wyckoff position as $B$.  In fact, this process of inverting bands at all four TRIM points, through our original construction, has merely exchanged the valence and conduction bands, which were indeed induced from $s$ and $p$ orbitals at the same Wyckoff position ($1a$ in Supplementary Table~\ref{tb:ebrsequence1}).  Therefore, $B''''$ does not exhibit anomalous corner modes, because it is not an OAL.  We have thus shown that if a generic valence manifold $B''$ consists of a single band and differs from an unobstructed atomic limit by two band inversions at different TRIM points between bands with opposite parity eigenvalues, then $B''$ will necessarily be irrep-equivalent to an OAL with anomalous corner modes.

\begin{table}[h]
\begin{tabular}{|c|c|c|c|c|}
\hline
Bands & $\Gamma$ & $X$ & $Y$ & $M$ \\
\hline
$\tilde{B}$ & $+,+$ & $+,+$ & $+,+$ & $+,+$ \\
\hline
$\tilde{B}'$ & $-,-$ & $+,+$ & $+,+$ & $+,+$ \\
\hline
$\tilde{\tilde{B}}'$ & $+,+$ & $-,-$ & $-,-$ & $-,-$ \\
\hline
$\tilde{B}''$ & $-,-$ & $-,-$ & $+,+$ & $+,+$ \\
\hline
\end{tabular}
\caption{The parity ($\mathcal{I}$) eigenvalues of the bands $\tilde{B}$--$\tilde{B}''$ discussed in the text surrounding SEq.~\eqref{eq:2Dfragilebands}.}
\label{tb:ebrsequence2}
\end{table}

Next, we consider the more complicated case in which the valence manifold $\tilde{B}$ consists of two occupied bands [the first row in Supplementary Table~\ref{tb:ebrsequence2} shows $\tilde{B}$ induced from two $s$ orbitals at the $1a$ position of magnetic layer group $p\bar{1}$ (see Supplementary Table~\ref{tb:ebr})].  For simplicity, we consider $\tilde{B}$ to be induced from two spinful $s$ orbitals at the same Wyckoff position (in the example shown in Supplementary Table~\ref{tb:ebrsequence2}, $\tilde{B}'$ is induced from the $1a$ position), and we consider the conduction bands to be induced from two spinful $p$ orbitals at the same Wyckoff position as the $s$ orbitals.  Hence, the two valence bands exhibit the same parity eigenvalues, and the two conduction bands both exhibit the same parity eigenvalues, opposite to those of the valence bands.  In this case, all band inversions at half filling necessarily occur between bands with opposite parity eigenvalues.  We next consider creating a set of occupied bands $\tilde{B}'$ through \emph{double band inversion}~\cite{YoungkukMonopole,TMDHOTI,HOTIBismuth}-- a process by which two valence bands are exchanged with two conduction bands at the same TRIM point (the second row in Supplementary Table~\ref{tb:ebrsequence2} shows $\tilde{B}'$, taking the double band inversion to occur at $\Gamma$).  Taking the valence (conduction) bands to be induced from pairs of spinful $s$ ($p$) orbitals at the $1a$ position and the double band inversion to occur about the $\Gamma$ point (SFig.~\ref{fig:SSH}~$\bs{b}$):
\begin{equation}
\tilde{B}' \stackrel{I}{\equiv} (p)_{1b} \oplus (p)_{1c} \oplus (p)_{1d} \ominus (p)_{1a}.
\label{eq:2Dfragilebands}
\end{equation}
The $\ominus$ in SEq.~(\ref{eq:2Dfragilebands}) indicates that $\tilde{B}'$ exhibits fragile topology~\cite{JenFragile1,BarryFragile,HingeSM,AshvinFragile2,ZhidaFragileTwist1,ZhidaFragileTwist2}.  As extensively shown in SRefs.~\onlinecite{TMDHOTI,WiederAxion,WladCorners}, $\tilde{B}'$ specifically characterizes an $\mathcal{I}$-symmetric FTI with filling-anomalous corner modes.

Next we will generalize SEq.~(\ref{eq:2Dfragilebands}) and generate all of the $\mathcal{I}$-symmetric FTIs with two occupied bands and anomalous corner modes.  First, we recognize that the previous double band inversion has created a set of two conduction bands $\tilde{\tilde{B}}'$ that are \emph{also} fragile~\cite{WiederAxion,TMDHOTI} (the third row in Supplementary Table~\ref{tb:ebrsequence2} shows $\tilde{\tilde{B}}'$, taking the double band inversion to occur at $\Gamma$).  Here, because all of the parity eigenvalues of $\tilde{\tilde{B}}'$ are the opposite of those of $\tilde{B}'$, we determine that $\tilde{\tilde{B}}'$ is also fragile and exhibits anomalous corner modes by setting $p \rightarrow s$ in SEq.~(\ref{eq:2Dfragilebands}) (Supplementary Table~\ref{tb:ebr}).  Next, because all of the maximal Wyckoff positions of $p\bar{1}$ have the same site-symmetry group ($\bar{1}$)~\cite{QuantumChemistry}, and because there are no other symmetries, then, in an infinite crystal, any two Wyckoff positions can be exchanged by a change of origin and coordinate definitions without loss of generality.  Specifically, this set of exchanges of origin and coordinate definitions in an infinite crystal should have no effect on the corner spectrum of the same crystal as long as the corner spectrum is calculated in a finite-sized geometry that still preserves $\mathcal{I}$~\cite{TMDHOTI,WiederAxion}.  Hence, the corner spectrum must be invariant with respect to the exchange of any two Wyckoff positions in SEq.~(\ref{eq:2Dfragilebands}).  In practice, exchanging $1b$, $1c$, or $1d$ with $1a$ in SEq.~(\ref{eq:2Dfragilebands}) simply generates the occupied bands of an FTI driven by double band inversion about a TRIM point other than $\Gamma$.  Therefore, to generate the eight possible $\mathcal{I}$-symmetric FTIs with two occupied bands and filling-anomalous corner modes, we can exchange $p\leftrightarrow s$ and $1a\leftrightarrow 1b$, $1c$, or $1d$ in SEq.~(\ref{eq:2Dfragilebands}).

Next, we will exclude insulators generated from double band inversion at the same TRIM point between pairs of bands with the same symmetry eigenvalues.  For example, if both the valence and conduction bands initially exhibit $\{+,-\}$ parity eigenvalues, then the valence and conduction bands will in this case \emph{also} exhibit $\{+,-\}$ parity eigenvalues after double band inversion.  Designating the initial (final) valence bands as $B$ ($B'$), double band inversion between sets of bands with $\{+,-\}$ eigenvalues, if bands do not simply avoid crossing, necessarily results in a situation for which $B\stackrel{I}{\equiv}B'$.  Therefore, double band inversion between bands with the same set of parity eigenvalues necessarily does not drive a symmetry-indicated phase transition.  We leave the possibility of non-symmetry-indicated band-inversion transitions for future works~\cite{JenOAL}.

Finally, we consider the process of further double band inverting from one of the four FTIs generated by the cyclic exchange of orbitals and Wyckoff positions in SEq.~(\ref{eq:2Dfragilebands}).  We take the second double band inversion to occur at a different TRIM point than the previous transition that drove $\tilde{B}$ to $\tilde{B}'$, and designate the occupied bands of the insulator with four total band inversions (occurring in pairs at two different TRIM points) $\tilde{B}''$ (the fourth row in Supplementary Table~\ref{tb:ebrsequence2} shows $\tilde{B}''$, taking the second double band inversion to occur at $X$).  Because the $k\cdot p$ Hamiltonian about the TRIM point of the second double band inversion (SRefs.~\onlinecite{TMDHOTI,HingeSM} and SN~\ref{sec:kpEdge}) is, by construction, unitarily related to the $k\cdot p$ Hamiltonian about the TRIM point of the first double band inversion, then $\tilde{B}''$ necessarily exhibits $2L_{z,n}^{\mathrm{FTI}}=2 + 4n$ (SEq.~\ref{eq:fragileAnomalyCounting}) \emph{more} corner modes than $\tilde{B}'$.  Therefore, $\tilde{B}''$ must exhibit a \emph{non-anomalous} number of corner modes $2L_{z}^{NI}$, where:
\begin{equation}
L_{z}^{NI} = 2n',\ n'\in\{\mathbb{Z}^{+},0\}.
\end{equation}
Hence, beginning with an unobstructed atomic limit, only odd numbers of double band inversions between sets of bands with opposite parity eigenvalues at the same TRIM point can result in insulators with $\mathcal{I}$-anomalous corner modes.

Combining the discussions in this section, we have shown for unobstructed atomic limits with one or two occupied bands and trivial corner spectra that $2L_{z,n}^{\mathrm{FTI}}=2+4n$ (SEq.~\ref{eq:fragileAnomalyCounting}) band inversions between bands with opposite parity eigenvalues necessarily results in an insulator with anomalous corner modes.  Therefore, for a 2D insulator in $p\bar{1}$, we can define a $\mathbb{Z}_{4}$-valued quantity $n_{4}$ that counts the number of parity-eigenvalue-exchanging band inversions required to transition between the occupied bands and a known unobstructed atomic limit without corner modes, corresponding to uninverted bands, modulo $4$.  When $n_{4}$ is odd, the occupied bands necessarily exhibit an odd Chern number [SEq.~(\ref{eq:ChernNumberZ})].  Finally, when $n_{4}=2$, the 2D system is either a non-symmetry-indicated Chern insulator~\cite{ChenBernevigTCI} with an even Chern number, or an insulator with gapped edges and anomalous corner states.

The $\mathbb{Z}_{4}$ nature of $n_{4}$ also provides some insight on the recent identification of $\mathbb{Z}_{4}$-valued symmetry-based indicators for centrosymmetric magnetic TIs and semimetals~\cite{MTQC,MTQCmaterials,AshvinMagnetic}.  To see this, consider a 3D insulator that is invariant under $\mathcal{I}$ and orthorhombic lattice translations $T_{x,y,z}$ (magnetic SG~\cite{BigBook,MagneticBook,BilbaoMagStructures,MTQC,MTQCmaterials,AshvinMagnetic} 2.4 $P\bar{1}$).  The Hamiltonians of both the $k_{z}=0,\pi$ planes are invariant under magnetic layer group $p\bar{1}$.  We fix $n_{4}=0$ for the Hamiltonian $\mathcal{H}_{k_{z}=\pi}(k_{x},k_{y})$ [$n_{4}(\pi)$] and allow $n_{4}$ of $\mathcal{H}_{k_{z}=0}(k_{x},k_{y})$ [$n_{4}(0)$] to vary.  When $n_{4}(0)\neq n_{4}(\pi)$, the 3D bulk can be reexpressed as a $k_{z}$-indexed periodic tuning cycle between two topologically distinct 2D insulators.  For odd $n_{4}(0)$, the Chern numbers of the occupied bands at $k_{z}=0$ and $k_{z}=\pi$ differ by an odd number [SEq.~(\ref{eq:ChernNumberZ})], and therefore the 3D system is necessarily a Weyl semimetal~\cite{NielsenNinomiya1}.  When $n_{4}(0)=2$, the bulk represents a gapless pumping cycle between an insulator with anomalous corner charges and one with no filling anomaly, and is thus a magnetic axion insulator (AXI) characterized by the axion (theta) angle $\theta = \pi$~\cite{WiederAxion,KoreanAxion,NicoDavidAXI2,IvoAXI1,IvoAXI2}.  Crucially, in the previous AXI, if we exchange $k_{z}=0,\pi$, then $n_{4}(\pi)=2$, which directly implies that $\nu_{z}^{\mathrm{F}}=1$.

Hence, $\nu^{\mathrm{F}}_{z}$ (as well as $\nu^{\mathrm{F}}_{x,y}$) can simply be calculated in a 3D $\mathcal{I}$-symmetric, $\mathcal{T}$-broken insulator by counting the number of parity-eigenvalue-exchanging band inversions modulo $4$ in the $k_{z}=\pi$ ($k_{x,y}=\pi$) plane, relative to an unobstructed atomic limit.  This is equivalent to calculating $n_{4}$ in the $k_{i}=\pi$ plane [$n_{4}(k_{i}=\pi)$].  If $n_{4}(k_{i}=\pi)$ is odd, then $\nu_{i}^{\mathrm{F}}$ is undefined, and if $n_{4}(k_{i}=\pi)$ is even, then:
\begin{equation}
\nu^{\mathrm{F}}_{i} = \frac{n_{4}(k_{i}=\pi)}{2}\text{ mod }2.
\label{eq:NandNu}
\end{equation}

We have thus completed our definition of $\bs{M}^{\mathrm{F}}_{\nu}$ for centrosymmetric, $\mathcal{T}$-broken 3D insulators [SEq.~(\ref{eq:fragileMComponents})].  While the discovery of magnetic topological materials is a fruitful field of ongoing study~\cite{BilbaoMagStructures,MTQC,MTQCmaterials,AshvinMagnetic,ChenMagnetic,BouhonMagneticFragile1,BouhonMagneticFragile2,ChenMagneticNodalLines,AndreiFlatBandTheory,SpinSGMagTIs,ChenSpinSpaceGroups}, we will only focus in this work on applying the $\mathcal{T}$-broken formulation of $\bs{M}^{\mathrm{F}}_{\nu}$ to simple tight-binding models -- which are provided in SN~\ref{sec:numerics},~\ref{sec:numerics2} -- as opposed to real-material magnetic topological (crystalline) insulators.  However, in SN~\ref{sec:FTIwithT}, we will shortly formulate a $\mathcal{T}$-symmetric extension of $\bs{M}^{\mathrm{F}}_{\nu}$, which can be used to predict spin-charge separated HEND states in nonmagnetic 3D insulators.

Lastly, because Wilson loop indices for filling anomalies in $\mathcal{I}$-symmetric, $\mathcal{T}$-broken 2D insulators were previously derived in SRef.~\onlinecite{WiederAxion}, we will not reintroduce them in this work.  We instead simply note that the nested Berry phase $\gamma_{2}$, as formulated in SRefs.~\onlinecite{TMDHOTI,WiederAxion}, along with the Chern number of the occupied bands, provides an alternative formulation of $n_{4}$.  In SN~\ref{sec:nested}, we will build upon the results of SRefs.~\onlinecite{TMDHOTI,WiederAxion} to develop analogous nested Wilson indicators for $\mathcal{I}$- and spinful $\mathcal{T}$-symmetric 2D insulators with anomalous corner states.

\subsubsection{$\mathcal{T}$-Symmetric EBR Formulation of $\textbf{\textit{M}}_{\nu}^{\mathrm{F}}$}
\label{sec:FTIwithT}

In this section, we will adapt the definition of $\nu_{z}^{\mathrm{F}}$ previously developed in SN~\ref{sec:FTInoT} to $\mathcal{I}$- and $\mathcal{T}$-symmetric 3D insulators, which, through SEq.~(\ref{eq:fragileMComponents}) and the cyclic exchange $x\rightarrow y \rightarrow z\rightarrow x$, is sufficient to also define the $\mathcal{T}$-symmetric weak corner-mode invariant $\bs{M}_{\nu}^{\mathrm{F}}$.  Because parity eigenvalues are real, then the addition of spinful $\mathcal{T}$ symmetry to the previous centrosymmetric magnetic insulators in SN~\ref{sec:FTInoT} simply doubles all of states at all crystal momenta through the combined symmetry~\cite{WiederLayers} $\mathcal{I}\times\mathcal{T}$ (if we maintain the previous simplifying assumption that high-fold rotation symmetries are absent).

To begin, rather than considering a 3D insulator in magnetic SG 2.4 $P\bar{1}$~\cite{BigBook,MagneticBook,BilbaoMagStructures,MTQC,MTQCmaterials,AshvinMagnetic}, we consider one in its $\mathcal{T}$-symmetric supergroup SG 2 $P\bar{1}1'$~\cite{BigBook,MagneticBook}.  Taking the lattice translations $T_{x,y,z}$ to be orthogonal without loss of generality, the Hamiltonians of the $k_{i}=0,\pi$ planes of the 3D BZ (SFig.~\ref{fig:fragile}~$\bs{a}$) are invariant under the $\mathcal{T}$-symmetric layer group $p\bar{1}1'$~\cite{WiederLayers,DiracInsulator}, whose generating elements are simply rectangular lattice translation, $\mathcal{I}$, and $\mathcal{T}$.  Layer group $p\bar{1}1'$, like its magnetic subgroup $p\bar{1}$, also has four maximal Wyckoff positions that coincide with the centers of $\mathcal{I}$ symmetry (SFig.~\ref{fig:fragile}~$\bs{b}$).  Therefore, each of the four maximal Wyckoff positions has the $\mathcal{T}$-symmetric site-symmetry group $\bar{1}1'$~\cite{BilbaoPoint,Bandrep1}, which has two, two-dimensional irreducible corepresentations:
\begin{equation}
\bar{\rho}^{+}\equiv\bar{A}_{g}\bar{A}_{g},\ \bar{\rho}^{-}\equiv\bar{A}_{u}\bar{A}_{u},
\end{equation}
that characterize Kramers pairs of states with positive and negative parity eigenvalues, respectively.

\begin{table}[h]
\begin{tabular}{|c|c|c|c|c|}
\hline
Valence Bands & $\Gamma$ & $X$ & $Y$ & $M$ \\
\hline
$B$ & $+$ & $+$ & $+$ & $+$ \\
\hline
$B'$ & $-$ & $+$ & $+$ & $+$ \\
\hline
$B''$ & $-$ & $-$ & $+$ & $+$ \\
\hline
\end{tabular}
\caption{The parity ($\mathcal{I}$) eigenvalues of the occupied bands $B$--$B''$ discussed in the text following SEq.~\eqref{eq:TIindexZ}.  Each $+$ and $-$ symbol represents a Kramers pair of states with the same parity eigenvalues.}
\label{tb:ebrsequence3}
\end{table}

Next, we will revisit all of the band-inversion arguments previously presented in SN~\ref{sec:FTInoT}.  Because all states in momentum space and localized states in real space are twofold degenerate due to spinful $\mathcal{I}\times\mathcal{T}$ symmetry, then we will in this section consider single band inversion to occur between \emph{Kramers pairs} of states, and double band inversion to occur between pairs of Kramers pairs (eight total) states, representing a doubling of the previous state counting arguments in SN~\ref{sec:FTInoT}.  We then recognize that, because $\mathcal{I}\times\mathcal{T}$ symmetry cannot protect semimetallic crossings in a 2D or 3D crystal for which $(\mathcal{I}\times\mathcal{T})^{2}=-1$~\cite{YoungkukLineNode,WiederLayers,FangFuNSandC2,Vafek14}, then all inversions between between bands with opposite parity eigenvalues still necessarily result in insulators, as they did for the 2D magnetic insulators in SN~\ref{sec:FTInoT}.  Beginning with an unobstructed atomic limit, we next recognize that an odd number of band inversions with $\mathcal{I}$ and spinful $\mathcal{T}$ symmetry necessarily yields the occupied bands of a 2D TI, rather than a symmetry-indicated Chern insulator with an odd Chern number [see the text surrounding SEq.~(\ref{eq:ChernNumberZ}) and SRef.~\onlinecite{ChenBernevigTCI}], as it did previously in SN~\ref{sec:FTInoT}.  Specifically, in the $\mathcal{I}$- and $\mathcal{T}$-symmetric 3D insulator discussed in this section, the 2D Hamiltonian of the $k_{z}=\pi$ plane $\mathcal{H}_{k_{z}=\pi}(k_{x},k_{y})$ exhibits a topology that can be partially diagnosed through the Fu-Kane index~\cite{FuKaneMele,FuKaneInversion}:
\begin{equation}
\nu_{x} = \frac{1}{2}\left[1-\xi(Z)\xi(U)\xi(T)\xi(R)\right],
\label{eq:TIindexZ}
\end{equation}
where $\xi(\bs{k}_{D})$ is the product of the parity eigenvalues per Kramers pair of the occupied bands at the TRIM point $\bs{k}_{D}$, and where $\nu_{x}=0$ ($\nu_{x}=1$) indicates a trivial insulator (2D TI). Therefore, if we begin with a 2D unobstructed atomic limit with the occupied bands $B$ (the first row in Supplementary Table~\ref{tb:ebrsequence3} shows $B$ induced from the $1a$ position of the $\mathcal{T}$-symmetric layer group $p\bar{1}1'$), a single band inversion between Kramers pairs of states with opposite parity eigenvalues results in an insulator whose occupied bands $B'$ necessarily characterize a 2D TI (the second row in Supplementary Table~\ref{tb:ebrsequence3} shows $B'$, taking the band inversion to occur at $\Gamma$).

From $B'$, we then consider a second band inversion, either at the same TRIM point or a different one, resulting in a set of occupied bands $B''$ (the third row in Supplementary Table~\ref{tb:ebrsequence3} shows $B''$, taking the second band inversion to occur at $X$).  Through SEq.~(\ref{eq:TIindexZ}), we recognize that $B''$, absent additional mirror or rotation symmetries~\cite{LiangTCI,TeoFuKaneTCI,HsiehTCI,ChenBernevigTCI,multipole,WladTheory,HOTIChen,ZeroBerry,FulgaAnon,EmilCorner,HingeSM,OrtixCorners,CornerWarning,TMDHOTI,KoreanFragile,AshvinFragile2,ZhidaFragileTwist1,ZhidaFragileTwist2,HarukiFragile,OrtixTRealSpace,BouhonMagneticFragile1,BouhonMagneticFragile2,WiederAxion,WladCorners,
FragileKoreanInversion,ZhidaFragileAffine,FrankCorners,YoungkukBLG,CaseWesternCorners}, is necessarily irrep-equivalent to a sum and difference of EBRs.  For $p\bar{1}1'$, we can use all of the previous EBRs of $p\bar{1}$ (Supplementary Table~\ref{tb:ebr}) by recognizing that both the initial spinful atomic orbitals and the subduced momentum-space degeneracies at the TRIM points have doubled without changing any of the parity eigenvalues~\cite{QuantumChemistry,Bandrep1,Bandrep2,Bandrep3,MTQC}.  We also recognize from the discussion in SRef.~\onlinecite{TMDHOTI} of weak-SOC (spinless), $\mathcal{I}$-symmetric FTIs with anomalous corner modes that the superposition of two time-reversed 2D insulators with $\mathcal{I}$-anomalous corner modes remains filling-anomalous under the preservation of $\mathcal{T}$.  Rather, the only qualitative differences between the $\mathcal{T}$-symmetric and $\mathcal{T}$-broken FTIs are a doubling of the state counting and a revised interpretation of the corner soliton spin and charge.  Specifically, in the $\mathcal{I}$- and $\mathcal{T}$-symmetric case, the corner modes manifest in Kramers pairs that exhibit the characteristic spin-charge separation of the solitons of the spinful SSH chain, rather than the fractional $e/2$ charges of the singly-degenerate solitons of the spinless (or $\mathcal{T}$-broken) SSH chain~\cite{SSH,SSHspinon,HeegerReview,RiceMele,WilczekAxion,GoldstoneWilczek,NiemiSemenoff,WiederAxion,HingeSM}.  Hence, $B''$, which resulted from two parity-eigenvalue-exchanging band inversions in an unobstructed atomic limit, necessarily characterizes a (possibly fragile) 2D insulator with anomalous corner modes.

Now that we have established that two band inversions necessarily drive an $\mathcal{I}$- and $\mathcal{T}$-symmetric unobstructed atomic limit into a 2D insulator with anomalous corner modes, all of the previous generalizations from SN~\ref{sec:FTInoT} to arbitrarily large numbers of band inversions also apply, without further modification.  Combining all of the analysis in this section and the previous discussion in SN~\ref{sec:FTInoT}, for a 2D insulator in $p\bar{1}1'$, we can again define a $\mathbb{Z}_{4}$-valued quantity $n_{4}$ that counts the number of parity-eigenvalue-exchanging band inversions required to transition between the occupied bands and a known unobstructed atomic limit, corresponding to uninverted bands, modulo $4$.  When $n_{4}$ is odd, the occupied bands necessarily characterize a 2D TI [SEq.~(\ref{eq:TIindexZ})].  Finally, when $n_{4}=2$, the 2D system is an insulator with gapped edges and anomalous \emph{Kramers pairs} of spin-charge-separated corner states.

The $\mathbb{Z}_{4}$ nature of $n_{4}$ in the $\mathcal{T}$-symmetric case also provides insight on the recent identification of a strong $\mathbb{Z}_{4}$ parity index for $\mathcal{I}$- and $\mathcal{T}$-symmetric 3D TIs and HOTIs~\cite{TMDHOTI,AshvinIndicators,ChenTCI,HOTIChen,AshvinTCI}.  To see this, consider a 3D insulator that is invariant under $\mathcal{I}$, $\mathcal{T}$, and orthorhombic lattice translations $T_{x,y,z}$ (SG 2 $P\bar{1}1'$).  The Hamiltonians of both the $k_{z}=0,\pi$ planes are invariant under the $\mathcal{I}$- and $\mathcal{T}$-symmetric layer group $p\bar{1}1'$.  We again fix $n_{4}=0$ for the Hamiltonian $\mathcal{H}_{k_{z}=\pi}(k_{x},k_{y})$ [$n_{4}(\pi)$] and allow $n_{4}$ of $\mathcal{H}_{k_{z}=0}(k_{x},k_{y})$ [$n_{4}(0)$] to vary.  When $n_{4}(0)\neq n_{4}(\pi)$, the 3D bulk can be reexpressed as a $k_{z}$-indexed periodic tuning cycle between two topologically distinct 2D insulators, analogous to the magnetic insulators previously discussed in the text surrounding SEq.~(\ref{eq:NandNu}).  For odd $n_{4}(0)$, the 2D TI indices of the occupied bands at $k_{z}=0$ and $k_{z}=\pi$ are distinct [SEq.~(\ref{eq:TIindexZ})].  Unlike previously in SN~\ref{sec:FTInoT}, this does not imply a Weyl semimetal, because the combined symmetry $\mathcal{I}\times\mathcal{T}$ prohibits the stabilization of conventional Weyl fermions (or any other form of chiral fermion) at generic crystal momenta~\cite{Vafek14}.  Instead, the 3D system represents a $\mathcal{T}$-symmetric pumping cycle between a 2D TI and a trivial insulator, which, in turn, characterizes a 3D TI~\cite{FuKaneMele,FuKaneInversion,QHZ}.  Finally, when $n_{4}(0)=2$, the bulk represents a gapped, helical pumping cycle between an insulator with a filling anomaly and a trivial insulator without corner modes, and is thus an $\mathcal{I}$- and $\mathcal{T}$-symmetric HOTI~\cite{TMDHOTI,WiederAxion}.  Crucially, for the aforementioned HOTI, if we exchange $k_{z}=0,\pi$, then $n_{4}(\pi)=2$, which directly implies that $\nu_{z}^{\mathrm{F}}=1$.

Therefore, $\nu^{\mathrm{F}}_{z}$ (as well as $\nu^{\mathrm{F}}_{x,y}$) can again simply be calculated in an $\mathcal{I}$- and $\mathcal{T}$-symmetric 3D insulator by counting the number of parity-eigenvalue-exchanging band inversions modulo $4$ in the $k_{z}=\pi$ ($k_{x,y}=\pi$) plane, relative to an unobstructed atomic limit.  This is equivalent to calculating $n_{4}$ in the $k_{i}=\pi$ plane [$n_{4}(k_{i}=\pi)$].  If $n_{4}(k_{i}=\pi)$ is odd, then $\nu_{i}^{\mathrm{F}}$ is undefined, but the conventional weak index $\nu_{i}$ is nontrivial~\cite{FuKaneMele,FuKaneInversion,AdyWeak,MooreBalentsWeak}.  And if $n_{4}(k_{i}=\pi)$ is even, then, if $n_{4}(k_{i}=\pi)=2L_{z,n}^{\mathrm{FTI}}\text{ mod } 4= 2$ [SEq.~(\ref{eq:fragileAnomalyCounting})], $\nu_{i}^{\mathrm{F}}$ is necessarily nontrivial.  This can be summarized as:
\begin{equation}
\nu_{i} = n_{4}(k_{i}=\pi)\text{ mod }2, 
\label{eq:weakTIIndex}
\end{equation}
and:
\begin{equation}
\nu_{i}^{\mathrm{F}} = \left\{
        \begin{array}{ll}
            \frac{n_{4}(k_{i}=\pi)}{2}\text{ mod }2, & \quad \nu_{i}=0 \\
            \text{undefined}, & \quad \nu_{i}=1.
        \end{array}
    \right.
\label{eq:NandNuTSymmetry}
\end{equation}

We have thus completed our definition of $\bs{M}^{\mathrm{F}}_{\nu}$ for $\mathcal{I}$- and $\mathcal{T}$-symmetric 3D insulators [SEq.~(\ref{eq:fragileMComponents})].  As noted previously in SN~\ref{sec:weakSSH}, because corner-mode phases in 2D reflect polarization (or fragile) topology~\cite{multipole,WladTheory,WladCorners,TMDHOTI,HingeSM,WiederAxion}, then, unlike strong topological phases, corner-mode phases cannot be diagnosed through their electronic structure alone, but instead must be identified by a comparison between obstructed and unobstructed atomic limits (or FTIs).  However, we note that most TIs, TCIs, and HOTIs with clean (bulk-insulating) Fermi surfaces are narrow-gap semiconductors for which it is chemically straightforward to identify which bands from which atomic orbitals have become inverted~\cite{AndreiMaterials,AndreiMaterials2,ChenMaterials,AshvinMaterials1,AshvinMaterials2,AshvinMaterials3,TavazzaMaterials3D}.  For example, in SN~\ref{sec:DFTSnTe}, we will demonstrate that 3D SnTe differs from an unobstructed atomic limit through double band inversions at two symmetry-related TRIM points, leading to a nontrivial $\bs{M}_{\nu}^{\mathrm{F}}$ vector, and, hence, a nontrivial HEND-state dislocation response.

\subsubsection{$\mathcal{T}$-Symmetric Nested Wilson Loop Formulation of Partial Nested Berry Phase}
\label{sec:nested}

In this section, we will use nested Wilson loops~\cite{multipole,WladTheory,HOTIBernevig,HingeSM,WiederAxion,KoreanFragile,ZhidaBLG,TMDHOTI} to rigorously relate the $\mathcal{T}$-symmetric corner-mode phases from SN~\ref{sec:FTIwithT} to a previously developed topological (polarization) index: the nested Berry phase $\gamma_{2}$.  We will show that, while $\gamma_{2}=0$ for both the filling-anomalous and trivial phases in SN~\ref{sec:FTIwithT}, a new \emph{partial} nested Berry phase $\tilde{\gamma}_{2}$ can be introduced to distinguish the two insulators.  During the preparation of this work, an equivalent Wilson-loop formulation of partial nested Berry phase was also introduced in SRef.~\onlinecite{KooiPartialNestedBerry}; however the authors of that work did not form a connection between the bulk invariant and corner modes (filling anomalies), which we will establish in this section.

In SRefs.~\onlinecite{TMDHOTI,WiederAxion}, the presence of anomalous corner modes in $\mathcal{I}$-symmetric, $\mathcal{T}$-broken 2D insulators was related to a nontrivial bulk (polarization) index: $\gamma_{2}$.  Specifically, in those works, beginning with atomic orbitals with trivial Berry phases $\gamma_{1}$ and nested Berry phases $\gamma_{2}$ [\emph{i.e.} orbitals placed at the $1a$ position in SFig.~\ref{fig:fragile}~$\bs{b}$ ($x,y=0$)], it was determined that an $\mathcal{I}$-symmetric 2D insulator with two or more occupied bands and either odd, fragile Wilson winding \emph{or} gapped Wilson loops and $\gamma_{2}=\pi$ necessarily exhibits filling-anomalous corner charges.

In SRef.~\onlinecite{TMDHOTI}, however, an $\mathcal{I}$- and $\mathcal{T}$-symmetric 2D FTI with anomalous \emph{Kramers pairs} of corner modes~\cite{FrankCorners} was also introduced.  Specifically, in Appendix~A of SRef.~\onlinecite{TMDHOTI}, the authors introduced a 2D FTI with $\mathcal{I}$, spinful $\mathcal{T}$, and SU(2) spin symmetry that represented the weak-SOC limit of the $\mathcal{I}$- and $\mathcal{T}$-symmetric FTI that is pumped in a generic $\mathcal{I}$- and $\mathcal{T}$-symmetric HOTI~\cite{WiederAxion} (equivalent to the $\mathcal{I}$- and spinful $\mathcal{T}$-symmetric FTIs previously discussed in SN~\ref{sec:FTIwithT}).  For the $\mathcal{I}$-, $\mathcal{T}$-, and SU(2)-symmetric 2D FTI in SRef.~\onlinecite{TMDHOTI}, the authors demonstrated that spin-degenerate pairs of corner states could be inferred, after the FTI was trivialized by adding trivial valence bands, by observing a nested Berry phase of $\gamma_{2}=\pi$ \emph{per spin}.  Though, taken over both spin sectors, $\gamma_{2}\text{ mod }2\pi$ was necessarily zero for both the filling-anomalous and trivial (unobstructed-atomic-limit) phases, the two phases could still be distinguished in the presence of additional SU(2) or $s^{z}$ symmetries.  Related corner-mode phases with net-zero $\gamma_{2}$ were also encountered in SRefs.~\onlinecite{FulgaAnon,GilManyBody}.

However, and crucially, breaking SU(2) symmetry while preserving spinful $\mathcal{T}$ symmetry \emph{does not} lift the filling anomaly in the model introduced in SRef.~\onlinecite{TMDHOTI}, though it does remove the ability to split the nested Wilson loop into spin sectors.  This raises the pertinent question of whether, in the absence of SU(2) or $s^{z}$ symmetries, there still exists a nested Wilson indicator of filling-anomalous corner states.  In the following text, we will definitively answer this question in the positive by formulating a new \emph{partial} nested Berry phase $\tilde{\gamma}_{2}$ that represents the higher-order generalization of the \emph{time-reversal polarization} introduced in SRefs.~\onlinecite{TRPolarization,KaneMeleZ2} to characterize 2D TIs as helical pumping cycles of the spinful SSH chain.

\begin{figure}[t]
\begin{center}
\includegraphics[width=\textwidth]{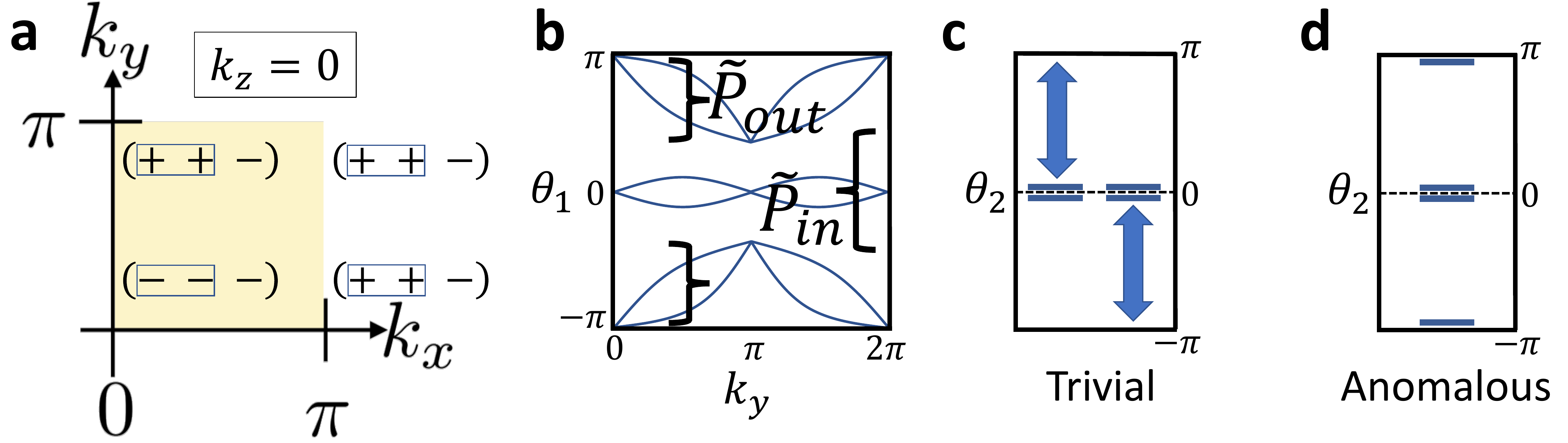}
\caption{{\bf Topological nested Wilson loops in 2D planes with time-reversal and inversion symmetry.} $\bs{a}$ The parity ($\mathcal{I}$) eigenvalues per Kramers pair of the six bulk bands in the $k_{z}=0$ plane of a 3D system.  The boxed parity eigenvalues in~$\bs{a}$ indicate that four of the bulk bands correspond in the $k_{z}=0$ plane to the $\mathcal{I}$- and $\mathcal{T}$-symmetric 2D FTI with anomalous corner states introduced in SRef.~\onlinecite{TMDHOTI}.  The remaining two bands, whose parity eigenvalues lie outside of the boxes in~$\bs{a}$, are induced from a Kramers pair of spinful $p$ orbitals at the $1a$ Wyckoff position ($x,y,z=0$).  $\bs{b}$ A schematic depiction of the $x$-directed Wilson loop spectrum $W_{1}(k_{y},k_{z})$ [SEq.~(\ref{eq:WilsonMain})] of an $\mathcal{I}$- and spinful $\mathcal{T}$-symmetric 3D system computed over six bulk bands with the parity eigenvalues in~$\bs{a}$ in the $k_{z}=0$ plane, drawn as a function of $k_{y}$ and evaluated at $k_{z}=0$.  $W_{1}(k_{y},k_{z})$ is Wilson particle-hole symmetric at each value of $k_{y,z}$ as a consequence of bulk $\mathcal{I}\times\mathcal{T}$ symmetry [SRef.~\onlinecite{TMDHOTI} and SEq.~(\ref{eq:ITWilson1})], and is also Wilson particle-hole symmetric at $\mathcal{I}$-related values of $k_{y,z}$ as a consequence of bulk $\mathcal{I}$ symmetry [SRef.~\onlinecite{WiederAxion} and SEq.~(\ref{eq:invWilson1})].  Bulk $\mathcal{T}$ symmetry acts as a spinful Wilson $\mathcal{T}$ symmetry that enforces Kramers' theorem at the $\mathcal{T}$-invariant values of $k_{y,z}$ [SEq.~(\ref{eq:TimeWilson1})].  $\bs{c,d}$~The two possible families of nested Wilson spectra computed symmetrically~\cite{TMDHOTI,WiederAxion} over four Wilson bands of a 2D insulator with $\mathcal{I}$ and spinful $\mathcal{T}$ symmetry [such as the $\tilde{P}_{out}$ Wilson bands at $k_{z}=0$ in $\bs{b}$].  The four nested Wilson bands in $\bs{c,d}$ exhibit nested Wilson particle-hole symmetry as a consequence of bulk $\mathcal{I}$ symmetry [SRef.~\onlinecite{WiederAxion} and SEq.~(\ref{eq:ITWilson2})], as well as Kramers degeneracies at all nested Wilson energies as a consequence of bulk spinful $\mathcal{T}$ symmetry [SEq.~(\ref{eq:TimeWilson1})].  As shown in SEq.~(\ref{eq:TruePartialParity}) and in the surrounding text, the nested Wilson spectrum in $\bs{c}$ can be deformed into the spectrum of a trivial insulator, because it has an even (here zero) number of Kramers pairs at $\theta_{2}=\pm \pi$.  Conversely, the nested Wilson spectrum in $\bs{d}$ \emph{cannot} be deformed into the trivial spectrum in $\bs{c}$ while preserving $\mathcal{I}$ and $\mathcal{T}$ symmetries.  In an $\mathcal{I}$- and spinful $\mathcal{T}$-symmetric 2D insulator whose bulk bands are induced from atoms at the $1a$ position, the nested Wilson spectrum in $\bs{d}$ also indicates the presence of a corner filling anomaly.  Crucially, our analysis of the nested Wilson spectra in $\bs{c}$ and $\bs{d}$ does not depend on $s^{z}$ spin symmetry labels, which are generically absent in spinful $\mathcal{T}$-symmetric insulators due to spin-orbit coupling~\cite{AndreiTI,CharlieTI,KaneMeleZ2,TRPolarization}.  As shown in this section, building upon the results of SRefs.~\onlinecite{TMDHOTI,WiederAxion,ArisInversion}, for an $\mathcal{I}$- and spinful $\mathcal{T}$-symmetric 2D insulator with six occupied bands that carry the parity eigenvalues in $\bs{a}$, the $x$-directed Wilson loop spectrum computed over the occupied bands will exhibit the Wilson band connectivity schematically depicted in $\bs{b}$, and the $y$-directed nested Wilson loop spectrum computed over the four $\tilde{P}_{out}$ Wilson bands in $\bs{b}$ will coincide with the anomalous (topologically nontrivial) nested Wilson loop spectrum depicted in $\bs{d}$.}
\label{fig:nested}
\end{center}
\end{figure}

To begin, consider a 2D insulator with $\mathcal{I}$-, spinful $\mathcal{T}$-, and weakly broken $s^{z}$ symmetry with four valence bands and four conduction bands.  We take the (valence) conduction bands to be induced from two Kramers pairs of spinful $s$ ($p$) orbitals at the $1a$ Wyckoff position of layer group $p\bar{1}1'$ (SFig.~\ref{fig:fragile}~$\bs{b}$)~\cite{BigBook,MagneticBook,subperiodicTables,WiederLayers,SteveMagnet,DiracInsulator,HingeSM}.  Using the symmetry labels in Supplementary Table~\ref{tb:ebr}, we determine that all four of the conduction (valence) bands have positive (negative) parity eigenvalues at all four TRIM points (SFig.~\ref{fig:SSH}~$\bs{b}$).  By construction, the occupied bands represent an unobstructed atomic limit~\cite{QuantumChemistry}.  Next, we double band invert at the $\Gamma$ point~\cite{YoungkukMonopole,TMDHOTI,HOTIBismuth}, after which all four of the occupied bands at $\Gamma$ now exhibit negative parity eigenvalues (boxed parity eigenvalues in SFig.~\ref{fig:nested}~$\bs{a}$).  As shown in SRefs.~\onlinecite{TMDHOTI,WiederAxion}, once the insulating gap is reopened, the occupied bands necessarily exhibit fragile topology; specifically, the occupied bands are irrep-equivalent to ($\mathcal{T}$-reversed pairs) of the linear combination (sum and difference) of EBRs in SEq.~(\ref{eq:2Dfragilebands}).  We then place an additional Kramers pair of occupied $p$ orbitals at the $1a$ position energetically far below the four fragile valence bands, resulting in a new valence manifold with six occupied bands that is irrep-equivalent to a linear combination of EBRs with only positive coefficients (SFig.~\ref{fig:nested}~$\bs{a}$).  As shown in SRefs.~\onlinecite{TMDHOTI,WiederAxion}, the six occupied bands no longer exhibit fragile topology, allowing the calculation of (nested) Wilson loops.

For the future generalization of our (nested) Wilson calculations to 3D systems, we next take the 2D trivialized FTI with six occupied bands in the previous paragraph to be described by the Hamiltonian at $k_{z}=0$ of an $\mathcal{I}$- and $\mathcal{T}$-symmetric 3D system with orthogonal lattice vectors.  Taking the six occupied bands of the 3D system, we will first calculate the discretized $x$-directed Wilson loop (holonomy) matrix~\cite{AndreiXiZ2,Fidkowski2011,ArisInversion,Cohomological,HourglassInsulator,DiracInsulator,BarryFragile}:
\begin{equation}
W_{1}(k_{y},k_{z}) =  V(2\pi\hat{x})\Pi(k_{x0},k_{y},k_{z}),
\label{eq:WilsonMain}
\end{equation}
where $V(2\pi\hat{x})$ is a sewing matrix that enforces the basepoint ($k_{x0}$) independence~\cite{HourglassInsulator} of SEq.~(\ref{eq:WilsonMain}) and $\Pi(k_{x0},k_{y},k_{z})$ is the product of projectors onto the occupied bands at each $\textbf{k}$ point along the line $(k_{x0} + 2\pi,k_{y},k_{z}) \rightarrow (k_{x0},k_{y},k_{z})$.  More formal definitions of $W_{1}(k_{y},k_{z})$ are provided in the appendices of SRefs.~\onlinecite{TMDHOTI,WiederAxion}.  The eigenvalues of $W_{1}(k_{y},k_{z})$ are gauge-independent phases $\exp[i\theta_{1}(k_{y},k_{z})]$~\cite{Fidkowski2011,ArisInversion,Cohomological,DiracInsulator}, allowing us to define a Hermitian \emph{Wilson Hamiltonian}:
\begin{equation}
H_{W_{1}}(k_{y},k_{z})= -i\ln[W_{1}(k_{y},k_{z})],
\label{eq:WilsonHam}
\end{equation}
whose eigenvalues take the form of real angles $\theta_{1}(k_{y},k_{z})$ and form smooth and continuous \emph{Wilson bands}~\cite{ArisInversion,Cohomological,DiracInsulator}.  We refer to the values of $\theta_{1}(k_{y},k_{z})$ as \emph{Wilson energies}.

The action of $\mathcal{I}\times\mathcal{T}$ and $\mathcal{I}$ on $H_{W_{1}}(k_{y},k_{z})$ were respectively determined in SRefs.~\onlinecite{TMDHOTI} and~\onlinecite{WiederAxion}, allowing us to also infer the action of spinful $\mathcal{T}$.  First, in SRef.~\onlinecite{TMDHOTI}, it was determined that spinful $\mathcal{I}\times\mathcal{T}$ acts on the Wilson Hamiltonian as an antiunitary particle-hole symmetry $\tilde{\Xi}$ that preserves the signs of $k_{y,z}$:
\begin{equation}
\tilde{\Xi} H_{W_{1}}(k_{y},k_{z})\tilde{\Xi}^{-1} = -\tilde{U}H^{*}_{W_{1}}(k_{y},k_{z})\tilde{U}^{\dag},
\label{eq:ITWilson1}
\end{equation}
where $\tilde{U}\tilde{U}^{*}=-\mathds{1}$.  SEq.~(\ref{eq:ITWilson1}) implies that, for every Wilson eigenstate at $k_{y,z}$ with eigenvalue $\theta_{1}(k_{y},k_{z})$, there is another eigenstate, also at $k_{y,z}$, with eigenvalue $-\theta_{1}(k_{y},k_{z})$.  Next, it was determined in SRef.~\onlinecite{WiederAxion} (in a reproduction of the analyses in SRefs.~\onlinecite{ArisInversion,DiracInsulator}) that $\mathcal{I}$ acts on the Wilson Hamiltonian as a unitary particle-hole symmetry $\tilde{\chi}$ that flips the signs of $k_{y,z}$:
\begin{equation}
\tilde{\chi} H_{W_{1}}(k_{y},k_{z})\tilde{\chi}^{-1} = -H_{W_{1}}(-k_{y},-k_{z}),
\label{eq:invWilson1}
\end{equation}
implying that, for every Wilson eigenstate at $k_{y,z}$ with eigenvalue $\theta_{1}(k_{y},k_{z})$, there is another eigenstate at $-k_{y,z}$ with eigenvalue $-\theta_{1}(k_{y},k_{z})$.  Taken together, SEqs.~(\ref{eq:ITWilson1}) and~(\ref{eq:invWilson1}) imply that spinful $\mathcal{T}$ symmetry acts on the Wilson Hamiltonian as an antiunitary time-reversal symmetry $\tilde{\Theta}$ that flips the signs of $k_{y,z}$:
\begin{equation}
\tilde{\Theta} H_{W_{1}}(k_{y},k_{z})\tilde{\Theta}^{-1} = \tilde{U}H^{*}_{W_{1}}(-k_{y},-k_{z})\tilde{U}^{\dag},
\label{eq:TimeWilson1}
\end{equation}
implying that, for every Wilson eigenstate at $k_{y,z}$ with eigenvalue $\theta_{1}(k_{y},k_{z})$, there is another eigenstate at $-k_{y,z}$ with the same eigenvalue $\theta_{1}(k_{y},k_{z})$, and that states are at least twofold degenerate~\cite{DiracInsulator,HourglassInsulator,Cohomological} by Kramers' theorem at the $\mathcal{T}$-invariant values of $k_{y,z}$.  In SFig.~\ref{fig:nested}~$\bs{b}$, we schematically depict the $x$-directed Wilson spectrum in the $k_{z}=0$ plane of an $\mathcal{I}$- and $\mathcal{T}$-symmetric 3D system.  Specifically, at $k_{z}=0$ in SFig.~\ref{fig:nested}~$\bs{b}$, the Wilson spectrum is the same as that of the six occupied bands of the 2D trivialized FTI discussed in this section, whose occupied parity eigenvalues (per Kramers pair) are shown in SFig.~\ref{fig:nested}~$\bs{a}$.  As drawn in SFig.~\ref{fig:nested}~$\bs{b}$, $W_{1}(k_{y},k_{z})$ satisfies the constraints imposed by $\mathcal{I}$ and spinful $\mathcal{T}$ symmetry [SEqs.~(\ref{eq:ITWilson1}) through~(\ref{eq:TimeWilson1})].

Next, we consider the discretized $y$-directed nested Wilson loop~\cite{multipole,WladTheory,WiederAxion,TMDHOTI} computed over spectrally isolated Kramers pairs of Wilson bands:
\begin{equation}
W_{2}(k_{z}) = \tilde{V}(2\pi\hat{y})\tilde{\Pi}(k_{x0},k_{y0},k_{z}),
\label{eq:nestedWilsonMain}
\end{equation}
where $\tilde{V}(2\pi\hat{y})$ is again a sewing matrix that enforces basepoint ($k_{y0}$) independence and $\tilde{\Pi}(k_{x0},k_{y0},k_{z})$ is the product of projectors onto a set of \emph{Wilson} bands along the line $(k_{x0},k_{y0} + 2\pi,k_{z}) \rightarrow (k_{x0},k_{y0},k_{z})$.  More formal expressions for SEqs.~(\ref{eq:WilsonMain}) and~(\ref{eq:nestedWilsonMain}) are also provided in the appendices of SRefs.~\onlinecite{TMDHOTI,WiederAxion}.  We will first determine the action of symmetries on $W_{2}(k_{z})$ before specifically analyzing the nested Wilson spectrum $W_{2}(0)$ of the $\mathcal{I}$- and $\mathcal{T}$-symmetric trivialized FTI highlighted in this section and previously in SN~\ref{sec:FTIwithT}.  The eigenvalues of $W_{2}(k_{y},k_{z})$ are gauge-independent phases $\exp[i\theta_{2}(k_{y},k_{z})]$, allowing us to define a Hermitian \emph{nested Wilson Hamiltonian}:
\begin{equation}
H_{W_{2}}(k_{z})= -i\ln[W_{2}(k_{z})],
\label{eq:nestedWilsonHam}
\end{equation}
whose eigenvalues take the form of real angles $\theta_{2}(k_{z})$ and form smooth and continuous \emph{nested Wilson bands}~\cite{WladTheory,HOTIBernevig,WiederAxion,TMDHOTI}.  We refer to the values of $\theta_{2}(k_{z})$ as \emph{nested Wilson energies}.   At each value of $k_{z}$, the sum of the nested Wilson energies modulo $2\pi$ is equal to the nested Berry phase:
\begin{equation}
\gamma_{2}(k_{z}) = \sum_{n=1}^{\tilde{n}_{\rm occ}}\theta^{n}_{2}(k_{z})\text{ mod }2\pi.
\label{eq:nestedSum}
\end{equation}

It was discovered in SRef.~\onlinecite{TMDHOTI} that, if $W_{2}(k_{z})$ is calculated using a $\tilde{\Xi}$-symmetric grouping of Wilson bands (\emph{i.e.}, using the nested Wilson projectors $\tilde{P}_{in,out}$ in SFig.~\ref{fig:nested}~$\bs{b}$), then $\mathcal{I}\times\mathcal{T}$ also imposes constraints on the \emph{nested} Wilson spectrum.  In SRef.~\onlinecite{WiederAxion}, the constraints imposed by $\mathcal{I}$ on $W_{2}(k_{z})$ (when calculated on a $\tilde{\chi}$-symmetric grouping of Wilson bands) were also determined.  Taken together, these constraints imply the action of $\tilde{\Theta}$, and thus spinful $\mathcal{T}$, on $W_{2}(k_{z})$.  Specifically, first, it was determined in SRef.~\onlinecite{TMDHOTI} that $\tilde{\Xi}$ (and therefore $\mathcal{I}\times\mathcal{T}$) acts on the nested Wilson Hamiltonian as an antiunitary particle-hole symmetry $\tilde{\tilde{\Xi}}$ that preserves the sign of $k_{z}$:
\begin{equation}
\tilde{\tilde{\Xi}} H_{W_{2}}(k_{z})\tilde{\tilde{\Xi}}^{-1} = -\tilde{\tilde{U}}H^{*}_{W_{2}}(k_{z})\tilde{\tilde{U}}^{\dag},
\label{eq:ITWilson2}
\end{equation}
where $\tilde{\tilde{U}}\tilde{\tilde{U}}^{*}=-\mathds{1}$.  SEq.~(\ref{eq:ITWilson2}) implies that, for every nested Wilson eigenstate at $k_{z}$ with eigenvalue $\theta_{2}(k_{z})$, there is another eigenstate, also at $k_{z}$, with eigenvalue $-\theta_{2}(k_{z})$.  Next, it was determined in SRef.~\onlinecite{WiederAxion} that $\tilde{\chi}$ (and therefore $\mathcal{I}$) acts on the nested Wilson Hamiltonian as a unitary particle-hole symmetry $\tilde{\tilde{\chi}}$ that flips the sign of $k_{z}$:
\begin{equation}
\tilde{\tilde{\chi}} H_{W_{2}}(k_{z})\tilde{\tilde{\chi}}^{-1} = -H_{W_{2}}(-k_{z}),
\label{eq:invWilson2}
\end{equation}
implying that, for every nested Wilson eigenstate at $k_{z}$ with eigenvalue $\theta_{2}(k_{z})$, there is another eigenstate at $-k_{z}$ with eigenvalue $-\theta_{2}(k_{z})$.  Taken together, SEqs.~(\ref{eq:ITWilson2}) and~(\ref{eq:invWilson2}) imply that $\tilde{\Theta}$ (and therefore spinful $\mathcal{T}$) acts on the nested Wilson Hamiltonian as an antiunitary time-reversal symmetry $\tilde{\tilde{\Theta}}$ that flips the sign of $k_{z}$:
\begin{equation}
\tilde{\tilde{\Theta}} H_{W_{2}}(k_{z})\tilde{\tilde{\Theta}}^{-1} = \tilde{\tilde{U}}H^{*}_{W_{2}}(-k_{z})\tilde{\tilde{U}}^{\dag},
\label{eq:TimeWilson2}
\end{equation}
implying that, for every nested Wilson eigenstate at $k_{z}$ with eigenvalue $\theta_{2}(k_{z})$, there is another eigenstate at $-k_{z}$ with the same eigenvalue $\theta_{2}(k_{z})$, and that states are at least twofold degenerate by Kramers' theorem at the $\mathcal{T}$-invariant values $k_{z}=0,\pi$.

In SFig.~\ref{fig:nested}~$\bs{c,d}$, we show the two possible families of $\mathcal{I}$- and spinful $\mathcal{T}$-symmetric nested Wilson spectra $W_{2}(0)$ symmetrically computed over four spinful Wilson bands (returning from $\tilde{\tilde{\chi}}$ and $\tilde{\tilde{\Theta}}$, respectively, to the symbols for bulk symmetries for notational simplicity, and focusing specifically on the $\mathcal{I}$- and $\mathcal{T}$-symmetric $k_{z}=0$ plane).  First, in SFig.~\ref{fig:nested}~$\bs{c}$, we show a nested Wilson spectrum with two Kramers pairs of states at $\theta_{2}=0$.  Satisfying nested Wilson time-reversal and particle-hole symmetries [SEqs.~(\ref{eq:ITWilson2}) through~(\ref{eq:TimeWilson2})], we can ``push'' the two nested Wilson eigenstates in SFig.~\ref{fig:nested}~$\bs{c}$ pairwise from $\theta_{2}=0$ to $\theta_{2}=\pi$.  In contrast, in SFig.~\ref{fig:nested}~$\bs{d}$, we show a nested Wilson spectrum with one Kramers pair at $\theta_{2}=0$ and one Kramers pair at $\theta_{2}=\pi$.  In the nested Wilson spectrum in SFig.~\ref{fig:nested}~$\bs{d}$ -- which corresponds to the four spinful $\tilde{P}_{out}$ Wilson bands in SFig.~\ref{fig:nested}~$\bs{b}$ -- the constraints of $\mathcal{I}$ and $\mathcal{T}$ symmetry conversely \emph{pin} the Kramers pairs in nested Wilson energy, and we cannot float the nested Wilson eigenstates away from their respectively values of $\theta_{2}$.  Furthermore, as shown in SRef.~\onlinecite{WiederAxion}, because of the constraints imposed by $\mathcal{I}$ symmetry on $W_{1}(k_{y},k_{z})$ [SEq.~(\ref{eq:invWilson1})], then Wilson gap closures unaccompanied by energy gap closures occur in pairs that do not change $\gamma_{2}(0)$ (or $\gamma_{2}(\pi)$, if we expand our focus beyond the $k_{z}=0$ plane).  Crucially, in the presence of an additional SU(2) or spinful $\mathcal{T}$ symmetry, which both enforce a twofold degeneracy in the eigenvalues of $W_{2}(0)$ [SEq.~(\ref{eq:TimeWilson2})], then SEq.~(\ref{eq:invWilson1}) implies that Wilson gap closures unaccompanied by energy gap closures \emph{also} do not change $\gamma_{2}^{\uparrow,\downarrow}(0)$, \emph{i.e.}, the nested Berry phase, \emph{per spin sector}.  As we will argue below, given orbitals originating from atoms at the $1a$ position (SFig.~\ref{fig:fragile}~$\bs{b}$), the nested Wilson spectrum in SFig.~\ref{fig:nested}~$\bs{c}$ (SFig.~\ref{fig:nested}~$\bs{d}$) corresponds to a trivial (filling-anomalous) 2D insulator without (with) Kramers pairs of corner modes that can only be removed by closing a bulk energy gap or breaking a symmetry.  Specifically, we will see that, even though both nested Wilson spectra in SFig.~\ref{fig:nested}~$\bs{c,d}$ exhibit $\gamma_{2}(0)=0$ [SEq.~(\ref{eq:nestedSum})], the two nested Wilson spectra can still be meaningfully distinguished by the presence of an even or odd number of Kramers pairs of nested Wilson eigenstates at $\theta_{2}=\pm \pi$.  Lastly, for completeness, we note that the results of SRefs.~\onlinecite{WiederAxion,TMDHOTI} imply that the $y$-directed nested Wilson spectrum of the two spinful $\tilde{P}_{in}$ Wilson bands in SFig.~\ref{fig:nested}~$\bs{b}$ would consist of a single Kramers pair of nested Wilson eigenstates pinned to $\theta_{2}=\pi$, and would therefore also contain the same number of Kramers pairs of nested Wilson eigenstates at $\theta_{2}=\pi$ (taken modulo $2$) as the four $\tilde{P}_{out}$ Wilson bands.

To demonstrate the robustness of the number of Kramers pairs at $\theta_{2}=\pm \pi$ in the nested Wilson spectrum, we will first consider the limit in which SU(2) spin rotation symmetry, and thus $s^{z}$ symmetry, is restored to the $\mathcal{I}$- and $\mathcal{T}$-symmetric 2D FTI discussed throughout this section.  In this case, the six valence bands remain separated from the conduction bands, and the six Wilson bands in SFig.~\ref{fig:nested}~$\bs{b}$ collapse into doubly degenerate pairs while preserving the Wilson gaps between the $\tilde{P}_{in/out}$ pairs of Wilson bands.  We can therefore still calculate the four-band nested Wilson spectrum of the $\tilde{P}_{\text{out}}$ Wilson bands.  As determined numerically in SRef.~\onlinecite{TMDHOTI}, in the SU(2)-symmetric limit, the trivialized FTI discussed in this section exhibits the nested Wilson spectrum shown in SFig.~\ref{fig:nested}~$\bs{d}$.  Specifically, $W_{2}(0)=W_{2}$ displays one pair of $s_{z}=\uparrow,\downarrow$ states at $\theta_{2}=0$ and another pair at $\theta_{2}=\pi$.  Hence, within each spin sector:
\begin{equation}
\gamma_{2}^{\uparrow,\downarrow}=\pi,
\end{equation}
where we have suppressed the explicit $k_{z}$ dependence $\gamma_{2}^{\uparrow,\downarrow}=\gamma_{2}^{\uparrow,\downarrow}(0)$.  We further recognize that, because $\mathcal{I}$ symmetry requires that nested Wilson eigenstates within each $s^{z}$ sector come in nested Wilson particle-hole-symmetric pairs [SEq.~(\ref{eq:invWilson2})], then the sum in SEq.~(\ref{eq:nestedSum}) simplifies:
\begin{equation}
\gamma_{2}^{\uparrow,\downarrow} = \pi\left(n_{\pi}^{\uparrow,\downarrow}\text{ mod }2\right),
\label{eq:parityPerSpin}
\end{equation}
where $n_{\pi}^{\uparrow,\downarrow}$ indicates the number of nested Wilson eigenvalues with $\theta_{2}=\pm \pi$ within each spin sector $s^{z}=\uparrow,\downarrow$.  For the 2D trivialized FTI discussed in this section, SEq.~(\ref{eq:parityPerSpin}) specifically indicates a filling anomaly when: 
\begin{equation}
n_{\pi}^{\uparrow,\downarrow} = L_{z,n}^{\mathrm{FTI}} = 1 + 2n,\ n\in\{\mathbb{Z}^{+},0\},
\end{equation}
where $L_{z,n}^{\mathrm{FTI}}$ also describes the number of spin-degenerate pairs of anomalous corner modes [SEq.~(\ref{eq:fragileAnomalyCounting})].  Because SU(2) symmetry requires that the eigenstates of $W_{2}$ appear in spin-degenerate pairs, then the total number of nested Wilson eigenstates at $\theta_{2}=\pm \pi$ is given by:
\begin{equation}
n_{\pi} = n_{\pi}^{\uparrow} + n_{\pi}^{\downarrow} = 2n_{\pi}^{\uparrow},
\end{equation}
and indicates a filling anomaly if:
\begin{equation}
n_{\pi}\text{ mod }4 = 2L_{z,n}^{\mathrm{FTI}}\text{ mod }4 = 2.
\label{eq:TruePartialParity}
\end{equation}

We next allow for weak, $\mathcal{T}$-symmetric SOC, which relaxes SU(2) and $s^{z}$ symmetry.  This splits the Wilson band degeneracies at generic crystal momenta, but does not close any additional bulk gaps (though it may close and reopen Wilson gaps).  Hence, the nested Wilson projector $\tilde{P}_{out}$ in SFig.~\ref{fig:nested}~$\bs{b}$, and thus $W_{2}$, remain well-defined throughout the introduction of SOC.  Crucially, as shown in SEq.~(\ref{eq:TimeWilson2}), because spinful $\mathcal{T}$ symmetry imposes \emph{the same} constraint on $W_{2}$ that SU(2) symmetry did previously, namely that all nested Wilson eigenstates are doubly degenerate, then $n_{\pi}\text{ mod }4$ \emph{cannot} change under the $\mathcal{T}$-symmetric relaxation of SU(2) symmetry.  Specifically, while $n_{\pi}$ itself can be changed by a Wilson gap closure unaccompanied by an energy gap closure~\cite{WiederAxion}, $\mathcal{I}$- and $\mathcal{T}$ symmetries require that Wilson gap closings at generic Wilson energies and crystal momenta $k_{y}$ in the $k_{z}=0$ plane come in sets of four (\emph{i.e.} at $\pm \theta_{1}$ and $\pm k_{y}$), such that $n_{\pi}\text{ mod }4$ cannot change without a bulk energy gap closure.  Additionally, as shown in SRefs.~\onlinecite{TMDHOTI,WiederAxion}, a filling anomaly in an $\mathcal{I}$-, $\mathcal{T}$-, and SU(2)-symmetric insulator \emph{remains preserved} under the $\mathcal{T}$-symmetric relaxation of SU(2) spin symmetry.  Therefore, because $n_{\pi}\text{ mod }4$ can only be changed through a bulk gap closure that removes the filling anomaly or by lowering the bulk symmetry, SEq.~(\ref{eq:TruePartialParity}) \emph{remains} an indicator of nontrivial (partial) polarization topology, \emph{even in the absence of $s^{z}$ symmetry}.

Finally, using $n_{\pi}$, we define a \emph{partial} or \emph{time-reversal} nested polarization:
\begin{equation}
\tilde{\gamma}_{2} = \pi\left[\left(\frac{n_{\pi}}{2}\right)\text{ mod }2\right],
\label{eq:nestedTRPolarization}
\end{equation}
that indicates the presence of Kramers pairs of corner states in an $\mathcal{I}$- and $\mathcal{T}$-symmetric 2D insulator (given orbitals originating from atoms at the $1a$ position in SFig.~\ref{fig:fragile}~$\bs{b}$).  SEq.~(\ref{eq:nestedTRPolarization}) therefore represents the higher-order analog of time-reversal polarization, which was introduced in SRefs.~\onlinecite{TRPolarization,KaneMeleZ2,FuKaneInversion} to predict Kramers pairs of end states in $\mathcal{I}$- and $\mathcal{T}$-symmetric 1D insulators (spinful SSH chains).

While it was shown in SRefs.~\onlinecite{TMDHOTI,WiederAxion,WladCorners} and SN~\ref{sec:FTIwithT} that $n_{\pi}\text{ mod }4$ [SEq.~(\ref{eq:TruePartialParity})] can be determined in an $\mathcal{I}$- and $\mathcal{T}$-symmetric 2D insulator using the parity eigenvalues of the occupied bands, the analysis in this section \emph{also} applies with only minor modification to spinful $C_{2z}$- (twofold-rotation-) and $\mathcal{T}$-symmetric 2D insulators~\cite{WiederAxion,KoreanAxion,NicoDavidAXI2,IvoAXI1,IvoAXI2}, whose topology cannot be inferred from symmetry eigenvalues~\cite{AshvinIndicators,ChenTCI,TMDHOTI}.  Therefore, our formulation of $\tilde{\gamma}_{2}$ [SEq.~(\ref{eq:nestedTRPolarization})] provides an essential bridge towards the identification of \emph{non-symmetry-indicated} higher-order topology in $\mathcal{T}$-symmetric 3D insulators, which has to date only been predicted in the noncentrosymmetric structural phases of XTe$_2$ (X$=$Mo,W)~\cite{TMDHOTI}.

\section{Numerical Calculation Details: Edge and Screw Dislocations}
\label{sec:numerics}

In this section, we will present numerical calculations for representative 2D and 3D insulators that confirm the dislocation responses discussed in this work.  First, in SN~\ref{sec:noTDislocation}, we will detail the dislocation response of 2D (SN~\ref{sec:2Dtoysubsec}) and 3D (SN~\ref{sec:3Dscrew}) insulators with broken $\mathcal{T}$ symmetry.  Then, in SN~\ref{sec:yesTDislocation}, we will detail analogous calculations demonstrating the dislocation response of $\mathcal{T}$-symmetric 2D (SN~\ref{sec:2DtoysubsecTRS}) and 3D (SN~\ref{sec:3DscrewTRSsec}) insulators.

\subsection{Dislocation Response without Time-Reversal Symmetry}
\label{sec:noTDislocation}

\subsubsection{2D Point Dislocations in the Absence of $\mathcal{T}$ Symmetry}
\label{sec:2Dtoysubsec}

\begin{figure}[t]
\begin{center}
\includegraphics[width=0.9\textwidth]{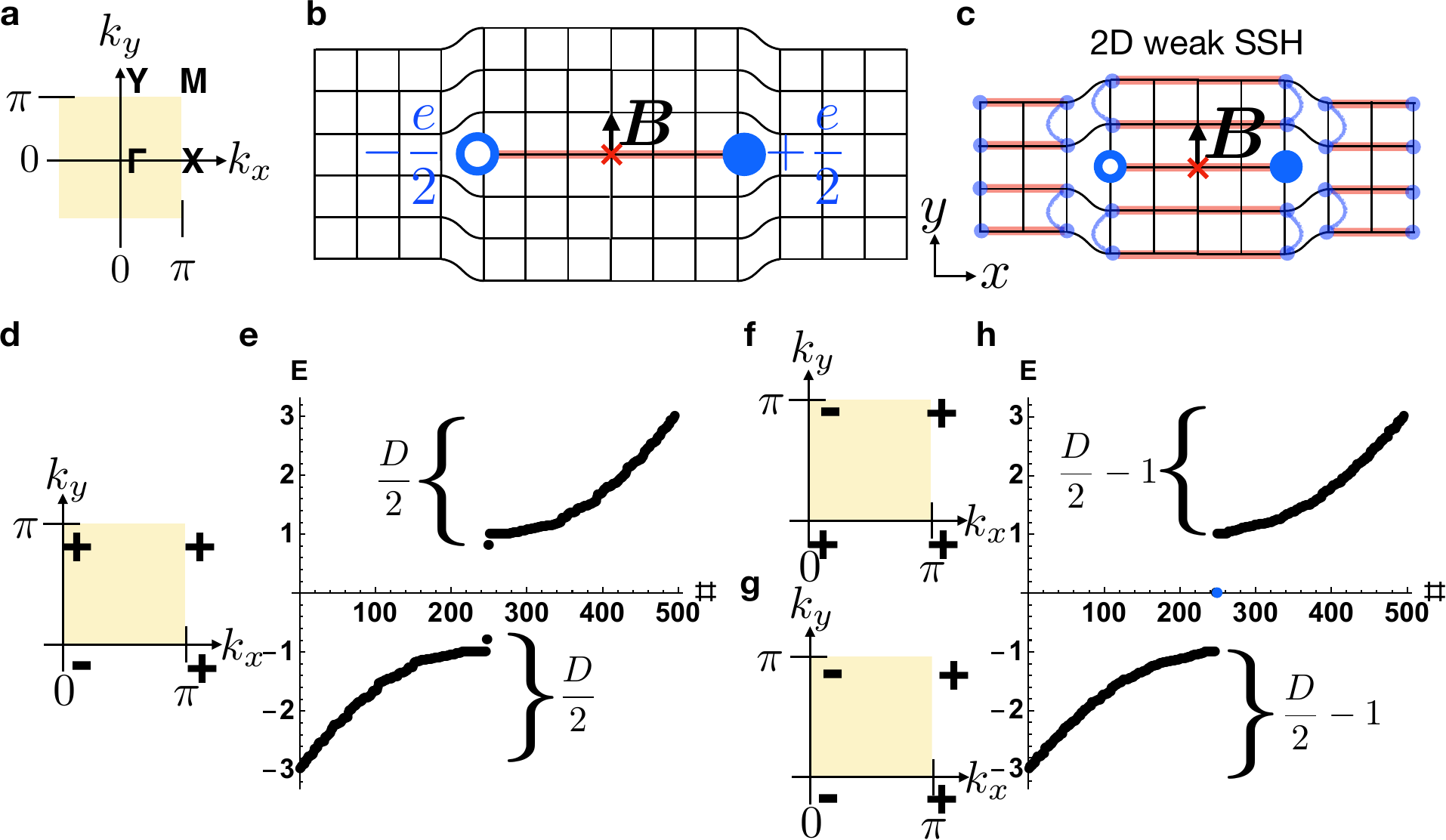}
\caption{{\bf 0D dislocation bound states in 2D insulators with inversion symmetry.} $\bs{a}$ The bulk BZ of a 2D rectangular magnetic crystal with only inversion ($\mathcal{I}$) symmetry.  $\bs{b}$ A pair of 0D dislocations with Burgers vector $\bs{B} =  \hat{y}$ that are related by a global $\mathcal{I}$ center (red $\times$) and bind anomalous fractional charges.  $\bs{c}$ A pair of fractionally charged 0D dislocation states can be obtained by coupling an array of 1D $x$-directed SSH chains along the $y$ direction in a manner in which one SSH chain remains decoupled from the others; two midgap states, equivalent to the end states of a nontrivial SSH chain, remain bound to the cores of the dislocations introduced by the interchain coupling.  $\bs{d}$ -- $\bs{h}$  Bulk parity eigenvalues and energy spectra with periodic boundary conditions (PBC) for a pair of dislocations with $\bs{B}=\hat{y}$ inserted into 2D insulators whose occupied bands carry the parity eigenvalues in $\bs{d}$, $\bs{f}$, and $\bs{g}$, respectively.  $\bs{d},\bs{e}$ characterize a $|C|=1$ Chern insulator with band inversion at $\Gamma$ [$\bs{M}^{\mathrm{SSH}}_{\nu} = (0,0)$ as defined in SN~\ref{sec:weakSSH}], $\bs{f},\bs{h}$ characterize a $|C|=1$ Chern insulator with band inversion at $Y$ [$\bs{M}^{\mathrm{SSH}}_{\nu} = (0,\pi)$], and $\bs{g},\bs{h}$ characterize a weak $y$-directed array of SSH chains~\cite{SSH,RiceMele} [$\bs{M}^{\mathrm{SSH}}_{\nu} = (0,\pi)$].  Anomalous 0D defect states with charge $\pm e/2$ are present in cases $\bs{f}$ and $\bs{g}$, but not in $\bs{d}$, and are equivalent to the end states of an $\mathcal{I}$-symmetric, $\mathcal{T}$-broken SSH chain~\cite{SSH,RiceMele} [SEq.~\eqref{eq:dislocsshchain}, red line in $\bs{b}$], and thus persist under the relaxation of particle-hole symmetry in the form of a filling anomaly~\cite{WilczekAxion,GoldstoneWilczek,NiemiSemenoff,WiederAxion,HingeSM}.  As derived in SEq.~(\ref{eq:dislocationtheory}), the dislocation response in $\bs{d}$ -- $\bs{h}$ is nontrivial when $\bs{B}\cdot \bs{M}^{\mathrm{SSH}}_{\nu}\text{ mod }2\pi = \pi$.}
\label{fig:2DAPPENDIX}
\end{center}
\end{figure}

In this section, we will demonstrate the dislocation response of 2D insulators with broken $\mathcal{T}$ symmetry.  We begin by comparing the minimal tight-binding model for an inversion- ($\mathcal{I}$-) symmetric Chern insulator with Chern number $|C| = 1$ to that of a stack (array) of Su-Schrieffer-Heeger (SSH) chains~\cite{SSH}.  Consider the Bloch Hamiltonian:
\begin{equation}
\begin{aligned}
\mathcal{H}(\bs{k}) =& \sin k_x \, \sigma^x + \xi \sin k_y \, \sigma^y \\&+ \left[\alpha \cos k_x + \xi \left(1 + \beta \cos k_y\right)\right] \sigma^z,
\label{eq:cherninsulatorsshmodel}
\end{aligned}
\end{equation}
defined on a square lattice, where $\sigma^i$, $i = 0,x,y,z$ are Pauli matrices indexing spinless valence $s$ and conduction $p$ orbitals on each site, and $\xi=0,1$ and $\alpha,\beta = \pm 1$ are parameters that can be tuned to realize different topological phases.  Inversion symmetry ($\mathcal{I}$) is represented by: 
\begin{equation}
\mathcal{I}:\ \mathcal{H}(\bs{k})\rightarrow\sigma^{z}\mathcal{H}(-\bs{k})\sigma^{z}.
\end{equation}
At half filling there is a single occupied band with the Bloch eigenstates $\ket{u(\bs{k})}$.  At the four time-reversal-invariant momenta (TRIM points) $\bs{\bar{k}}$ of the 2D BZ (SFig.~\ref{fig:2DAPPENDIX}~$\bs{a}$) -- here, because $\mathcal{T}$ symmetry is absent~\cite{BigBook,MagneticBook}, instead defined by $\mathcal{I}\bs{\bar{k}} = \bs{\bar{k}}\text{ mod }{\bf b}$, where ${\bf b}$ is a reciprocal lattice vector -- we can define the parity eigenvalue $\lambda_{\bs{\bar{k}}}$ of the occupied band as:
\begin{equation}
\mathcal{I} \ket{u(\bs{\bar{k}})} = \lambda_{\bs{\bar{k}}} \ket{u(\bs{\bar{k}})},
\end{equation}
where $\mathcal{I}^2 = 1$ implies that $\lambda_{\bs{\bar{k}}} = \pm 1$.  For $\alpha = \beta = 0$ and $\xi=0,1$, SEq.~(\ref{eq:cherninsulatorsshmodel}) is topologically trivial, and its occupied subspace is composed of a single band with the parity eigenvalues $\lambda_{\bs{\bar{k}}} = 1$ at all TRIM points.

For $\xi = 1$, $\mathcal{H}(\bs{k})$ describes a symmetry-indicated $|C|=1$ Chern insulator~\cite{QWZ,ChenBernevigTCI,ArisInversion,FuKaneMele,FuKaneInversion,AndreiInversion} that differs from a trivial (unobstructed) atomic limit by a single sign-change of the parity eigenvalue of the occupied band at $\bs{\bar{k}} = (1+\alpha,1+\beta)\pi/2$.  To determine the topology of $\mathcal{H}(\bs{k})$, we exploit that the Chern number modulo $2$ of a $\mathcal{T}$-broken 2D insulator is indicated by the product of the parity eigenvalues of the occupied bands~\cite{QWZ,ChenBernevigTCI,ArisInversion,FuKaneMele,FuKaneInversion,AndreiInversion}.  Here, and throughout this work, we consider the process of closing a gap and inverting bands to be a \emph{band inversion}, and restrict focus in this and the following section (SN~\ref{sec:numerics2}) to band inversions that exchange the parity eigenvalues of the valence and conduction bands.

When $\xi = 0$, $\mathcal{H}(\bs{k})$ is equivalent to an array of identical $x$-directed SSH chains~\cite{YoungkukWeakSSH} indexed by $k_{y}$ whose polarization (Berry phase) is indicated by the parity eigenvalues at $k_{x}=0,\pi$~\cite{ArisInversion}.  More generally, in this work, we define a stack or array as a \emph{weak insulating} phase whose Bloch Hamiltonian has the same symmetry eigenvalues along a specified momentum-space direction $k_{i}$ (reciprocal to a real-space direction $i$), and therefore the same $(d-1)$-D topological indices in BZ surfaces indexed by $k_{i}$.  For example, a 3D weak TI is equivalent to a stack of 2D TIs~\cite{FuKaneMele,FuKaneInversion,AdyWeak,MooreBalentsWeak}.

Having established the properties of the Hamiltonian of the pristine system [SEq.~(\ref{eq:cherninsulatorsshmodel})], we will now probe the dislocation response.  We introduce a pair of 0D dislocations with Burgers vector $\bs{B} = \hat{y}$.  Through SEq.~\eqref{eq:dislocationtheory}, we deduce that each dislocation realizes an interface between two insulators $S$ and $S'$ that ``differ'' by the Hamiltonian of the $k_{y}=\pi$ plane of SEq.~(\ref{eq:cherninsulatorsshmodel}):
\begin{equation}
\label{eq:dislocsshchain}
\mathcal{H}(k_x,\pi) = \sin k_x \, \sigma^x + [\alpha \cos k_x + \xi (1-\beta)] \sigma^z.
\end{equation}
Setting, $\alpha = -1$, $\mathcal{H}(k_x,\pi)$ describes an $\mathcal{I}$-symmetric SSH chain with a trivial (nontrivial) polarization for $\xi = 1$, $\beta = -1$ ($\xi = 1$, $\beta = + 1$ or $\xi = 0$) when measured with respect to the valence $s$ orbitals (SN~\ref{sec:weakSSH}).

To obtain the numerical results presented in SFig.~\ref{fig:2D} of the main text (reproduced in SFig.~\ref{fig:2DAPPENDIX}), we begin with a square lattice $\Lambda$ of size $|\Lambda| = L^2$ with $L = 16$ and periodic boundary conditions (PBC).  We then remove a line of $8$ sites to create a pair of dislocations with Burgers vector $\bs{B} = \hat{y}$.  We note that two is the minimal number of dislocations compatible with (untwisted) PBC: a single point dislocation introduces a fault line in the lattice that must terminate either at a boundary or at another dislocation.

Next, we analyze the system with two dislocations using the terminology established in SN~\ref{sec:defecttopology}.  We begin by denoting the pristine (dislocation-free) real-space Hamiltonian corresponding to the model in SEq.~\eqref{eq:cherninsulatorsshmodel} as $H$.  $H$ contains only nearest-neighbor hoppings, and has $(2 L^2 \times 2 L^2) = (512 \times 512)$ elements.  Let $H_{a,b}$ be the components of the $(2 |A| \times 2 |B|)$-dimensional submatrix that couples lattice sites taken from the sets $A$ and $B$.  $H_{a,b}$ is the matrix obtained by removing all of the rows from the $(2 L^2 \times 2 L^2)$ matrix $H$ that do not couple to lattice sites within the set $A$.  Similarly, we remove all of the columns in $H_{a,b}$ that do not couple to the set $B$.  We next introduce the shorthand notation:
\begin{equation}
\sum_{a,b} H_{a,b} \ketbra{a}{b} = \sum_{\bs{r} \in A} \sum_{\bs{r'} \in B} \sum_{\mu,\nu \in\{1,2\}}H_{\bs{r},\bs{r'}}^{\mu,\nu} \ketbra{\bs{r},\mu}{\bs{r'},\nu},
\label{eq:HabHab}
\end{equation}
where $\ket{\bs{r},\mu}$ denotes the basis state at position $\bs{r} \in \Lambda$ with an orbital indexed by $\mu$ [the Bloch Hamiltonian in SEq.~\eqref{eq:cherninsulatorsshmodel} is a $(2 \times 2)$ matrix in the basis of spinless $s$ and $p$ orbitals].

We will next detail how, in a tight-binding model with nearest-neighbor hoppings and orthogonal lattice vectors, we can numerically implement a pair of point dislocations by first removing a line of sites from a pristine lattice, and then subsequently reconnecting the two lines above and below the missing line.  We use the site labeling $U$, $M$, and $D$ established in SFig.~\ref{fig:2Dimplementation}~$\bs{a}$, and specialize to $\bs{B} = \hat{y}$ dislocations.  Defining $O = U \cup M \cup D$ and $\bar{O} = \Lambda \backslash O$, we form the expression: 
\begin{equation}
\begin{aligned}
H =& \sum_{\bar{o},\bar{o}'} H_{\bar{o},\bar{o}'} \ketbra{\bar{o}}{\bar{o}'} + \sum_{o,\bar{o}} H_{o,\bar{o}} \ketbra{o}{\bar{o}} \\&+ \sum_{\bar{o}, o} H_{\bar{o},o} \ketbra{\bar{o}}{o} + \sum_{o,o'} H_{o,o'} \ketbra{o}{o'}.
\label{eq:dislocalgo}
\end{aligned}
\end{equation}
Noting that $H$ only contains nearest-neighbor hoppings, we then expand SEq.~(\ref{eq:dislocalgo}):
\begin{equation}
\begin{aligned}
\sum_{o,o'} H_{o,o'} \ketbra{o}{o'} =& \sum_{u,u'} H_{u,u'} \ketbra{u}{u'} \\&+ \sum_{m,m'} H_{m,m'} \ketbra{m}{m'} + \sum_{d,d'} H_{d,d'} \ketbra{d}{d'} \\ &+\sum_{u,m} H_{u,m} \ketbra{u}{m}+\sum_{m,u} H_{m,u} \ketbra{m}{u}\\&+\sum_{m,d} H_{m,d} \ketbra{m}{d}+\sum_{d,m} H_{d,m} \ketbra{d}{m}.
\end{aligned}
\end{equation}
To introduce a pair of dislocations, we form the Hamiltonian $\tilde{H}$, which is defined on $\tilde{\Lambda} = \Lambda \backslash M$ -- \emph{i.e}., the lattice obtained by taking out the set of sites $M$.  Defining $\tilde{O} = U \cup D$ and $\bar{\tilde{O}} = \tilde{\Lambda} \backslash \tilde{O}$, $\tilde{H}$ can be expanded as:
\begin{equation}
\begin{aligned}
\tilde{H} = & \sum_{\bar{\tilde{o}},\bar{\tilde{o}}'} H_{\bar{\tilde{o}},\bar{\tilde{o}}'} \ketbra{\bar{\tilde{o}}}{\bar{\tilde{o}}'}+ \sum_{\tilde{o},\bar{\tilde{o}}} H_{\tilde{o},\bar{\tilde{o}}} \ketbra{\tilde{o}}{\bar{\tilde{o}}} \\&+ \sum_{\bar{\tilde{o}},\tilde{o}} H_{\bar{\tilde{o}},\tilde{o}} \ketbra{\bar{\tilde{o}}}{\tilde{o}}+ \sum_{\tilde{o},\tilde{o}'} H_{\tilde{o},\tilde{o}'} \ketbra{\tilde{o}}{\tilde{o}'},
\end{aligned}
\end{equation}
in which we can further expand:
\begin{equation}
\begin{aligned}
\sum_{\tilde{o},\tilde{o}'} H_{\tilde{o},\tilde{o}'} \ketbra{\tilde{o}}{\tilde{o}'} =& \sum_{u,u'} H_{u,u'} \ketbra{u}{u'} + \sum_{d,d'} H_{d,d'} \ketbra{d}{d'} \\&+ \sum_{u,d} \tilde{H}_{u,d} \ketbra{u}{d}+\sum_{d,u} \tilde{H}_{d,u} \ketbra{d}{u}.
\end{aligned}
\end{equation}
Finally, we conclude that:
\begin{equation}
\label{eq:substitutionsforedgedislocation}
\tilde{H}_{u,d} = H_{u,m}, \quad \tilde{H}_{d,u} = H_{d,m},
\end{equation}
such that the matrix elements of $\tilde{H}$ between sites in $U$ and sites in $D$ are given by those of $H$ between $U$ and $M$.

After following the prescription in SEqs.~(\ref{eq:dislocalgo}) through~(\ref{eq:substitutionsforedgedislocation}), we obtain a Hamiltonian $\tilde{H}$ that contains two $\bs{B} = \hat{y}$ point dislocations on the lattice $\tilde{\Lambda}$ and $[2 (L^2 - 8) \times 2 (L^2 - 8)] = (496 \times 496)$ elements.  We will now discuss the electronic structure of $\tilde{H}$ in detail for characteristic values of $\xi$ and $\beta$.

\paragraph{$\xi = 1$, $\beta = -1$: $|C|=1$ Chern insulator with band inversion at $\Gamma$.}
\label{subsec:ChernAtGamma}
This case corresponds to an inversion-symmetry-indicated $|C|=1$ Chern insulator~\cite{QWZ,ChenBernevigTCI,ArisInversion,FuKaneMele,FuKaneInversion,AndreiInversion} driven by a single band inversion at $\bs{k} = (0,0)$ in SEq.~(\ref{eq:cherninsulatorsshmodel}).  In this case, the bulk characterizes a strong topological phase.  Because the bands at $X$, $Y$, and $M$ are uninverted relative to an atomic insulator with orbitals at the $1a$ position, the weak SSH invariant $\bs{M}^{\mathrm{SSH}}_{\nu}=(0,0)$ (SN~\ref{sec:weakSSH}).  This implies that $\bs{B}\cdot\bs{M}^{\mathrm{SSH}}_{\nu}\text{ mod }{ 2\pi }=0$, and therefore, that the dislocations do not bind anomalous charges ($q\text{ mod }e = 0$).  In SFig.~\ref{fig:2DAPPENDIX}~$\bs{d},\bs{e}$, we respectively show the parity eigenvalues of the occupied bands and the dislocation spectrum, which does not exhibit a filling anomaly or midgap dislocation bound states.

\paragraph{$\xi = 1$, $\beta = +1$: $|C|=1$ Chern insulator with band inversion at $Y$.}
\label{subsec:ChernAtY}
This case describes a $|C|=1$ Chern insulator~\cite{QWZ,ChenBernevigTCI,ArisInversion,FuKaneMele,FuKaneInversion,AndreiInversion} driven by band inversion at the $Y$ point (SFig.~\ref{fig:2DAPPENDIX}~$\bs{a}$).  Relative to the initial $1a$ atomic insulator, the band inversion at $Y$ has changed not only the strong index (2D Chern number), but also the weak SSH indices [the parity eigenvalues of the occupied bands are shown in SFig.~\ref{fig:2DAPPENDIX}~$\bs{f}$].  Specifically, the bulk exhibits a weak SSH invariant $\bs{M}^{\mathrm{SSH}}_{\nu}=(0,\pi)$ as defined in SN~\ref{sec:weakSSH}.  This implies that $\bs{B}\cdot\bs{M}^{\mathrm{SSH}}_{\nu}\text{ mod }{ 2\pi }=\pi$, indicating that the dislocation response is nontrivial.  Correspondingly, in the dislocation spectrum (SFig.~\ref{fig:2DAPPENDIX}~$\bs{h}$), we observe one midgap-localized zero mode per dislocation.  The zero modes are protected by the chiral (\emph{i.e.} unitary particle-hole) symmetry $\Pi$, which is defined through the action:
\begin{equation}
\Pi:\ \mathcal{H}(k_x,\pi)\rightarrow \sigma^{y}\mathcal{H}(k_x,\pi)\sigma^{y},
\label{eq:chernchiralsym}
\end{equation}
such that $\Pi$ is a symmetry of $\mathcal{H}(k_{x},\pi)$ if $\Pi\mathcal{H}(k_{x},\pi)\Pi^{-1} =\sigma^{y}\mathcal{H}(k_x,\pi)\sigma^{y}= -\mathcal{H}(k_{x},\pi)$.  Crucially, if we were to relax $\Pi$ symmetry, then the midgap states could be pushed out of the gap.  However, if we preserve $\mathcal{I}$ symmetry while breaking $\Pi$, then $\pm e/2$ end charges would still remain bound to the dislocations~\cite{OrtixSSHoneHalfEndCharges}, in a generalization of the conclusions of Goldstone and Wilczek~\cite{WilczekAxion,GoldstoneWilczek,NiemiSemenoff}.

Equivalently, in the absence of chiral symmetry, we can also identify the nontrivial topology by counting the number of states that are occupied up to a given Fermi level in the gap.  Let $\tilde{H}$ be a $(D \times D)$ matrix [in our numerics, $D = 496$, as explained in the text following SEq.~(\ref{eq:substitutionsforedgedislocation})].  Comparing SFigs.~\ref{fig:2DAPPENDIX}~$\bs{e}$ and $\bs{h}$, we observe that the two spectra differ by the presence of two midgap states and the absence of one state from each of the valence and conduction manifolds.  When chiral symmetry is broken, the energy of the midgap states can be shifted, but only in a manner that preserves $\mathcal{I}$ symmetry, leaving the two states degenerate in the thermodynamic limit.  This implies that any gapped Fermi level that encloses either $(D/2-1)$ or $(D/2+1)$ occupied states indicates a nontrivial topology, whereas a Fermi level enclosing $D/2$ occupied states indicates trivial bulk topology (this is also detailed in SFigs.~\ref{fig:2Dimplementation}~$\bs{b},\bs{c}$).  More generally, the presence of anomalous 0D states in arbitrary dimensions can be diagnosed in a similar manner by constructing a 0D system with a boundary that preserves a global point group symmetry and observing an imbalance in the number of states above and below the gap that cannot be resolved without breaking a symmetry or closing the gap (\emph{i.e.} a filling anomaly)~\cite{HOTIChen,WiederAxion,HingeSM,WladCorners,AshvinFragile2,ZhidaFragileTwist1,ZhidaFragileTwist2}.

\paragraph{$\xi = 0$: Weak Array of $x$-Directed SSH Chains.}
\label{subsec:weakSSHstack}
In this case, $\mathcal{H}(\bs{k})$ is independent of $\beta$, and SEq.~\eqref{eq:cherninsulatorsshmodel} describes a $y$-directed array of identical $x$-directed SSH chains~\cite{YoungkukWeakSSH}.  Specifically, the Hamiltonian [SEq.~\eqref{eq:cherninsulatorsshmodel} with $\xi = 0$] is in this case completely independent of $k_{y}$.  For each SSH chain, the nontrivial polarization is indicated by the difference in the parity eigenvalues of the occupied band at $k_{x}=0,\pi$~\cite{SSH,SSHspinon,HeegerReview,RiceMele,ZakPhase,ArisInversion,TRPolarization,KaneMeleZ2}.  This case is an example of weak topology, because the Hamiltonian $\mathcal{H}(\bs{k})$ can be deformed into a real-space array of decoupled 1D chains without breaking a symmetry or closing the bulk gap.  Relative to the initial $1a$ atomic insulator, $\mathcal{H}(\bs{k})$ features band inversions at $\Gamma$ and $Y$ (the parity eigenvalues are shown in SFig.~\ref{fig:2DAPPENDIX}~$\bs{g}$), such that the bulk exhibits a trivial symmetry-indicated Chern number $C\text{ mod }2=0$ and nontrivial weak SSH indices, and is irrep-equivalent to an OAL from $1a$ [$(x,y)=(0,0)$] to $1c$ [($(x,y) = (1/2,0)$] (SFig.~\ref{fig:fragile}~$\bs{b}$).  Specifically, the bulk exhibits a weak SSH invariant $\bs{M}^{\mathrm{SSH}}_{\nu}=(0,\pi)$ as defined in SN~\ref{sec:weakSSH}.  This implies that, even though the bulk is an OAL (SN~\ref{sec:weakSSH}), the dislocation response is nontrivial.  Correspondingly, in the dislocation spectrum (SFig.~\ref{fig:2DAPPENDIX}~$\bs{h}$), we observe a filling anomaly.

For the present case of an array of SSH chains, we can also understand the existence of dislocation midgap states intuitively without invoking the more general theory used to derive SEq.~\eqref{eq:dislocsshchain}.  As depicted in SFig.~\ref{fig:2DAPPENDIX}~$\bs{c}$, the dislocations introduce an uncoupled SSH chain into the system whose end states become the dislocation bound states.  As long as the ``leftover'' SSH chain contains an inversion center, its end states also induce a system filling anomaly, consistent with the numerical results shown in SFig.~\ref{fig:2DAPPENDIX}~$\bs{h}$.

\subsubsection{3D Screw Dislocations in the Absence of $\mathcal{T}$ Symmetry}
\label{sec:3Dscrew}

In this section, we will demonstrate the screw dislocation response of 3D insulators with broken $\mathcal{T}$ symmetry.  We begin by comparing the minimal tight-binding model for an axion insulator (AXI)~\cite{HOTIBernevig,HOTIChen,WladTheory,AxionZhida,EslamInversion,FanHOTI,TMDHOTI,HarukiLayers,AshvinMagnetic,EzawaMagneticHOTI,VDBHOTI,WiederAxion,KoreanAxion,NicoDavidAXI2,IvoAXI1,IvoAXI2} to that of a weak stack of 2D fragile topological insulators (FTIs)~\cite{TMDHOTI,WiederAxion}, where we retain $\mathcal{I}$ symmetry in all cases.  Consider the Bloch Hamiltonian:
\begin{equation}
\begin{aligned}
\label{eq:axioninsulatorfragile2Dmodel}
\mathcal{H}(\bs{k}) =& \sin k_x \, \tau^z \sigma^x + \sin k_y \, \tau^z \sigma^y + \xi \sin k_z \, \tau^z \sigma^z \\&+ \left[\alpha \cos k_x + \beta \cos k_y + \xi \left(2 + \gamma \cos k_z\right)\right] \tau^x \sigma^0 \\
& + \xi \, \delta \, (\tau^x \sigma^x + \tau^x \sigma^y + \tau^x \sigma^z),
\end{aligned}
\end{equation}
defined on a square lattice.  In SEq.~(\ref{eq:axioninsulatorfragile2Dmodel}), $\tau^i$ and $\sigma^i$, $i = 0,x,y,z$ are respectively two sets of Pauli matrices indexing sublattice and orbital degrees of freedom, $\xi=0,1$ and $\alpha,\beta,\gamma = \pm 1$ are parameters that can be tuned to realize different topological phases, and $\delta$ is a small parameter that gaps out the bulk (we will use $\delta = 1/4$ in all calculations).  We note that when $\delta = 0$, SEq.~(\ref{eq:axioninsulatorfragile2Dmodel}) is instead the Hamiltonian of two uncoupled 3D Weyl semimetals~\cite{MTQC,BarryBenCDW}.  Throughout this work, we abbreviate the Kronecker product of Pauli matrices as $\tau^i \otimes \sigma^j \equiv \tau^i \sigma^j$.

\begin{figure}[t]
\begin{center}
\includegraphics[width=1 \textwidth]{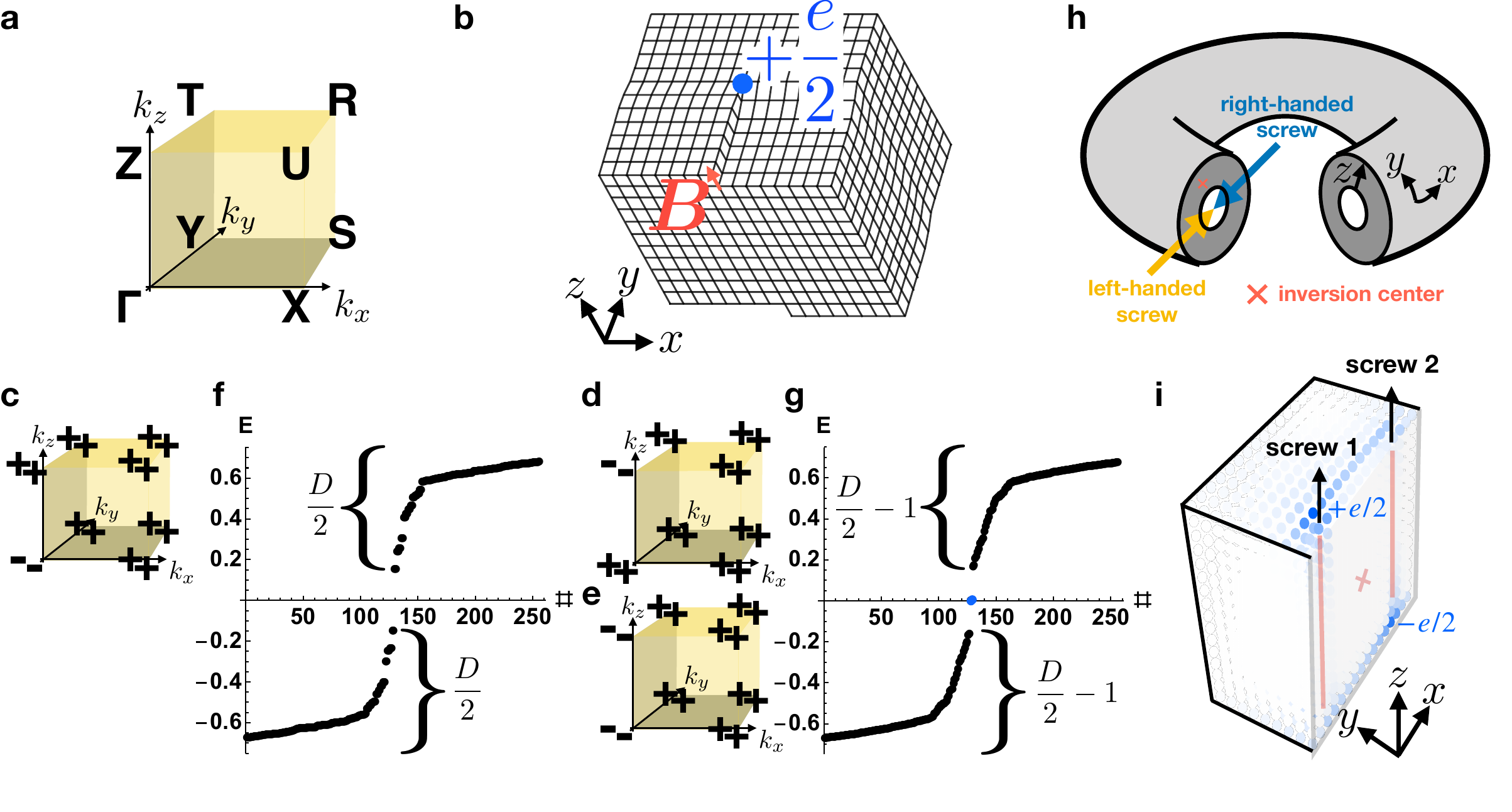}
\caption{{\bf 0D screw dislocation bound states in 3D insulators with inversion symmetry.} $\bs{a}$ The bulk BZ of a 3D orthorhombic crystal with only inversion ($\mathcal{I}$) symmetry.   $\bs{b}$ A screw dislocation in an $\mathcal{I}$-symmetric crystal with Burgers vector $\bs{B} =  \hat{z}$ in a geometry with open boundary conditions in all directions.  $\bs{c}$-$\bs{g}$ Bulk parity eigenvalues and hollow-doughnut-boundary-condition (HDBC) ($\bs{h}$) energy spectra for the defect in $\bs{b}$ when the crystal with the BZ in $\bs{a}$ is equivalent to $\bs{c},\bs{f}$ an $\mathcal{I}$-symmetric axion insulator (AXI)~\cite{HOTIBernevig,HOTIChen,WladTheory,AxionZhida,EslamInversion,FanHOTI,TMDHOTI,HarukiLayers,AshvinMagnetic,EzawaMagneticHOTI,VDBHOTI,WiederAxion,KoreanAxion,NicoDavidAXI2,IvoAXI1,IvoAXI2} with double band inversion at $\Gamma$ [$\bs{M}^{\mathrm{F}}_{\nu} = (0,0,0)$ as defined in SN~\ref{sec:weakFragile}], $\bs{d},\bs{g}$ an AXI with double band inversion at $Z$ [$\bs{M}^{\mathrm{F}}_{\nu} = (0,0,\pi)$], and $\bs{e},\bs{g}$ a weak $z$-directed stack of $\mathcal{I}$-symmetric 2D FTIs with $\pm e/2$ corner charges~\cite{TMDHOTI,WiederAxion} [$\bs{M}^{\mathrm{F}}_{\nu} = (0,0,\pi)$].  The HDBC geometry is closely related to the ``Corbino doughnut'' geometry employed in SRef.~\onlinecite{FuKaneInversion} to characterize 3D TIs; however, in this work, we insert screw dislocations (and later flux tubes) in a different arrangement than in SRef.~\onlinecite{FuKaneInversion}.  In the absence of dislocations, the HDBC geometry does not have any hinges. Hence, the system with HDBC, in the absence of dislocations, exhibits the same energy spectrum whether the bulk is topologically equivalent to a trivial insulator or an AXI.   In $\bs{f},\bs{g}$, only the $256$ lowest-lying states are shown.  Anomalous 0D (HEND) states with charge $\pm e/2$ are present in $\bs{d}$ and $\bs{e}$, but are absent in $\bs{c}$.  The HEND states in $\bs{d}$ and $\bs{e}$ are equivalent to the corner states of an $\mathcal{I}$-symmetric 2D FTI, and thus persist under the relaxation of particle-hole symmetry~\cite{TMDHOTI,WiederAxion,HingeSM}.  $\bs{i}$ Real-space-resolved absolute square of the midgap eigenstates in $\bs{g}$ for the model in SEq.~\eqref{eq:axioninsulatorfragile2Dmodel} on a $16 \times 16 \times 16$ lattice.  $\mathcal{I}$-related bound states appear on two of the four screw dislocation ends.  The inversion center in $\bs{i}$ is marked with a red $X$.  Consistent with the analytic calculations in the text preceding SEq.~(\ref{eq:dislocationtheory}), the dislocation response in $\bs{d}$ -- $\bs{h}$ is nontrivial when $\bs{B}\cdot \bs{M}^{\mathrm{F}}_{\nu}\text{ mod }2\pi = \pi$.}
\label{fig:3D}
\end{center}
\end{figure}

In SEq.~(\ref{eq:axioninsulatorfragile2Dmodel}), $\mathcal{I}$ symmetry is represented by: 
\begin{equation}
\mathcal{I}:\ \mathcal{H}(\bs{k})\rightarrow\tau^x\sigma^0\mathcal{H}(-\bs{k})\tau^x\sigma^0.
\end{equation}
At the eight time-reversal invariant momenta (TRIM points) $\bs{\bar{k}}$ of the 3D BZ (SFig.~\ref{fig:3D}~$\bs{a}$) -- here, because $\mathcal{T}$ symmetry is absent~\cite{BigBook,MagneticBook}, instead defined by $\mathcal{I}\bs{\bar{k}} = \bs{\bar{k}}\text{ mod }{\bf b}$, where ${\bf b}$ is a reciprocal lattice vector -- we can again compute the parity eigenvalues $\lambda^n_{\bs{\bar{k}}}$, $n=1,2$ of the two occupied bands, where $\lambda^n_{\bs{\bar{k}}} = \pm 1$.

For $\alpha = \beta = \gamma = 0$ and $\xi=0,1$, SEq.~\eqref{eq:axioninsulatorfragile2Dmodel} is topologically trivial, and the occupied subspace is composed of two bands with equal parity eigenvalues $\lambda_{\bs{\bar{k}}} = 1$ at all TRIM points.  For $\xi = 1$, $\mathcal{H}(\bs{k})$ describes an inversion-symmetry-indicated axion insulator (AXI)~\cite{AxionZhida,EslamInversion,MTQC,MTQCmaterials,AshvinMagnetic,EzawaMagneticHOTI,VDBHOTI,WiederAxion,KoreanAxion,NicoDavidAXI2,IvoAXI1,IvoAXI2} that differs from a trivial (unobstructed) atomic limit by a sign-change of the parity eigenvalues of the two occupied bands (\emph{i.e.} double band inversion) at $\bs{\bar{k}} = (1+\alpha,1+\beta,1+\gamma)\pi/2$.  For $\xi = 0$, $\mathcal{H}(\bs{k})$ describes a weak $z$-directed stack of identical inversion-symmetry-indicated~\cite{AxionZhida,EslamInversion,MTQC,MTQCmaterials,AshvinMagnetic,EzawaMagneticHOTI,VDBHOTI,WiederAxion,KoreanAxion,NicoDavidAXI2,IvoAXI1,IvoAXI2} 2D FTIs with $\pm e/2$ corner charges.

Having established the properties of the Hamiltonian of the pristine system [SEq.~(\ref{eq:axioninsulatorfragile2Dmodel})], we will now probe the dislocation response.  Previously, in SN~\ref{sec:2Dtoysubsec}, we discussed 0D states bound to point dislocations in 2D crystals.  In 3D, there are point, screw, and mixed~\cite{MerminReview} (as well as partial~\cite{JenDefect}) dislocations.  In this work, we focus on defects with integer Burgers vectors, for which the presence of topological bound states only depends on the direction and length of the Burgers vector, as opposed to the details of the dislocation (see SN~\ref{sec:proofs}).  Hence, in this section, we will restrict focus to 3D screw dislocations, with the understanding that our conclusions also apply to edge and mixed dislocations with integer Burgers vectors.  In particular in this work, for simplicity, we have not performed numerical analyses of systems with mixed dislocations, because mixed dislocations have position-dependent sense vectors, whereas the sense vectors of screw and edge dislocations are always parallel and perpendicular to the Burgers vector, respectively~\cite{MerminReview}.

We introduce a pair of 1D screw dislocations with Burgers vector $\bs{B} = \hat{z}$ and opposite chiralities [SEq.~(\ref{eq:chiralityScrew})] into the system described in the text surrounding SEq.~\eqref{eq:axioninsulatorfragile2Dmodel}.  Through SEq.~\eqref{eq:dislocationtheory}, we deduce that each screw realizes an interface between two insulators $S$ and $S'$ that ``differ" by the Hamiltonian of the $k_z = \pi$ plane of SEq.~\eqref{eq:axioninsulatorfragile2Dmodel}:
\begin{equation}
\label{eq:disloc2Dphase}
\begin{aligned}
\mathcal{H}(k_x,k_y,\pi) =& \sin k_x \, \tau^z \sigma^x + \sin k_y \, \tau^z \sigma^y \\&+ [\alpha \cos k_x + \beta \cos k_y + \xi (2-\gamma)] \tau^x \sigma^0 \\
& + \xi \, \delta \, (\tau^x \sigma^x + \tau^x \sigma^y + \tau^x \sigma^z).
\end{aligned}
\end{equation}
Setting $\alpha = \beta = -1$, $\delta=1/4$, SEq.~(\ref{eq:disloc2Dphase}) describes an $\mathcal{I}$-symmetric 2D FTI~\cite{TMDHOTI,WiederAxion} (trivial insulator) when $\xi = 1$, $\gamma = + 1$ ($\xi = 1$, $\gamma = -1$).

To obtain the numerical results presented in SFig.~\ref{fig:AXI} of the main text (reproduced in SFig.~\ref{fig:3D}), we begin with a square lattice $\Lambda$ of size $|\Lambda| = L^2$ with $L = 16$ and PBC to model the $(x,y)$-plane, while keeping $k_z$ as a quantum number that labels states by their momentum in the $z$ direction.  We then modify the hoppings across a plane of $8$ sites in $\Lambda$ with a fixed $y$ coordinate to create a pair of screw dislocations with Burgers vector $\bs{B} = \hat{z}$.  We note that two is the minimal number of screw dislocations compatible with untwisted PBC.

Next, we analyze the system with two dislocations using the terminology established in SN~\ref{sec:FrankScrew}.  We begin by denoting the pristine real-space Hamiltonian corresponding to the Fourier transform of the model in SEq.~\eqref{eq:axioninsulatorfragile2Dmodel} within the $(x,y)$-plane as $H(k_z)$.  $H(k_z)$ contains only nearest-neighbor hoppings, and has $(4 L^2 \times 4 L^2) = (1024 \times 1024)$ elements for each value of $k_z$.  As previously in SEq.~(\ref{eq:HabHab}), $H_{a,b}$ denotes the matrix obtained by removing all of the rows from $H$ that do not couple to lattice sites within the (freely specifiable) set $A$, and removing all of the columns in $H$ that do not couple to the set $B$, such that the only remaining couplings are between $A$ and $B$.  We next introduce the shorthand notation:
\begin{equation}
\begin{aligned}
\sum_{a,b} &H_{a,b} (k_z) \ketbra{a}{b} \\&= \sum_{\bs{r} \in A} \sum_{\bs{r'} \in B} \sum_{\mu,\nu=1 \dots 4}H_{\bs{r},\bs{r'}}^{\mu,\nu} (k_z) \ketbra{\bs{r},\mu}{\bs{r'},\nu},
\end{aligned}
\label{eq:temp3DnoTDefect}
\end{equation}
where $\ket{\bs{r},\mu}$ denotes the basis state at position $\bs{r} \in \Lambda$ with an orbital indexed by $\mu$, where the Bloch Hamiltonian in SEq.~\eqref{eq:axioninsulatorfragile2Dmodel} is a $(4 \times 4)$ matrix in a basis with four orbitals.  Additionally, in SEq.~(\ref{eq:temp3DnoTDefect}), $H_{\bs{r},\bs{r'}}^{\mu,\nu}$ are the real-space matrix elements that follow from Fourier transforming SEq.~\eqref{eq:axioninsulatorfragile2Dmodel}.

Using the site labeling $U$ and $M$ established in SFig.~\ref{fig:2Dimplementation}~$\bs{a}$, where here $M$ is the 8-site plane across which hoppings are modified, we will now detail how we numerically implement a pair of $\bs{B} = \hat{z}$ screw dislocations.
Defining $O = U \cup M$ and $\bar{O} = \Lambda \backslash O$, we form the expression: 
\begin{equation}
\label{eq:screwalgo}
\begin{aligned}
H (k_z) =& \sum_{\bar{o},\bar{o}'} H_{\bar{o},\bar{o}'} (k_z) \ketbra{\bar{o}}{\bar{o}'} + \sum_{o,\bar{o}} H_{o,\bar{o}} (k_z) \ketbra{o}{\bar{o}} \\&+ \sum_{\bar{o}, o} H_{\bar{o},o} (k_z) \ketbra{\bar{o}}{o} + \sum_{o,o'} H_{o,o'} (k_z) \ketbra{o}{o'}.
\end{aligned}
\end{equation}
We then expand SEq.~\eqref{eq:screwalgo}:
\begin{equation}
\begin{aligned}
\sum_{o,o'} &H_{o,o'} (k_z) \ketbra{o}{o'} \\=& \sum_{u,u'} H_{u,u'} (k_z) \ketbra{u}{u'} + \sum_{m,m'} H_{m,m'} (k_z) \ketbra{m}{m'} \\ &+\sum_{u,m} H_{u,m} (k_z) \ketbra{u}{m}+\sum_{m,u} H_{m,u} (k_z) \ketbra{m}{u}.
\end{aligned}
\end{equation}
To introduce a pair of screw dislocations, we form the Hamiltonian $\tilde{H} (k_z)$, which is defined on $\Lambda$ and labeled by $k_z$.  $\tilde{H} (k_z)$ can be expanded as:
\begin{equation}
\begin{aligned}
\tilde{H} (k_z) =& \sum_{\bar{o},\bar{o}'} H_{\bar{o},\bar{o}'} (k_z) \ketbra{\bar{o}}{\bar{o}'} + \sum_{o,\bar{o}} H_{o,\bar{o}} (k_z) \ketbra{o}{\bar{o}} \\&+ \sum_{\bar{o}, o} H_{\bar{o},o} (k_z) \ketbra{\bar{o}}{o} + \sum_{o,o'} \tilde{H}_{o,o'} (k_z) \ketbra{o}{o'},
\end{aligned}
\end{equation}
in which we can further expand:
\begin{equation}
\label{eq:screwalgo2}
\begin{aligned}
\sum_{o,o'} & \tilde{H}_{o,o'} (k_z) \ketbra{o}{o'} \\= &\sum_{u,u'} H_{u,u'} (k_z) \ketbra{u}{u'} + \sum_{m,m'} H_{m,m'} (k_z) \ketbra{m}{m'} \\ &+\sum_{u,m} e^{-\mathrm{i} k_z} H_{u,m} (k_z) \ketbra{u}{m}+\sum_{m,u} e^{\mathrm{i} k_z} H_{m,u} (k_z) \ketbra{m}{u},
\end{aligned}
\end{equation}
where the exponential factors $e^{\pm ik_{z}}$ implement the screw dislocations (we have set the lattice spacing to $a_{x,y,z}=1$).  To summarize, SEqs.~(\ref{eq:screwalgo}) through~(\ref{eq:screwalgo2}) describe how, in a tight-binding model with nearest-neighbor hoppings and orthogonal lattice vectors, we numerically implement a pair of screw dislocations by multiplying all hoppings across the plane $M$ in a pristine lattice by the momentum-dependent phase factor $e^{\mathrm{i} k_z}$ in one direction, and $e^{-\mathrm{i} k_z}$ in the other direction, to obtain a Hermitian Hamiltonian.  After following this prescription, we obtain a Hamiltonian $\tilde{H}(k_z)$ that contains two $\bs{B} = \hat{z}$ screw dislocations with opposite sense vectors [defined in the text surrounding SEq.~(\ref{eq:chiralityScrew})].  We will now discuss the electronic structure of $\tilde{H} (k_z)$ in detail for characteristic values of $\xi$ and $\gamma$.

\paragraph{$\xi = 1$, $\gamma = -1$: Axion insulator with double band inversion at $\Gamma$}
\label{subsec:AXIatGamma}
This case corresponds to an inversion-symmetry-indicated AXI driven by a pair of band inversions at $\bs{k} = (0,0,0)$ in SEq.~\eqref{eq:axioninsulatorfragile2Dmodel}.  In this case, the bulk characterizes a strong topological phase (specifically, an AXI).  Because the bands at $X$, $Y$, $M$, $Z$, $U$, $R$, and $T$ are uninverted relative to an atomic insulator with orbitals at the $1a$ position, then the weak fragile invariant $\bs{M}^{\mathrm{F}}_{\nu} = (0,0,0)$ (SN~\ref{sec:weakFragile}).  This implies that $\bs{B}\cdot\bs{M}^{\mathrm{F}}_{\nu}\text{ mod }{ 2\pi }=0$, and therefore, that the dislocations do not bind anomalous charges ($q\text{ mod }e = 0$).  In SFig.~\ref{fig:3D}~$\bs{c},\bs{f}$, we respectively show the parity eigenvalues of the occupied bands and the dislocation spectrum, which does not exhibit a filling anomaly or midgap dislocation bound states.

\paragraph{$\xi = 1$, $\gamma = +1$: Axion insulator with double band inversion at $Z$}
\label{subsec:AXIatZ}
This case describes an AXI driven by double band inversion at the $Z$ point (SFig.~\ref{fig:3D}~$\bs{a}$).  Relative to the initial trivial atomic insulator, the double band inversion at $Z$ has changed not only the strong AXI index, but also the weak fragile indices [the parity eigenvalues of the occupied bands are shown in SFig.~\ref{fig:3D}~$\bs{d}$].  Specifically, the bulk exhibits a weak fragile invariant $\bs{M}^{\mathrm{F}}_{\nu} = (0,0,\pi)$ as defined in SN~\ref{sec:weakFragile}.  This implies that $\bs{B}\cdot\bs{M}^{\mathrm{F}}_{\nu}\text{ mod }{ 2\pi }=\pi$, indicating that the dislocation response is nontrivial.  Correspondingly, in the dislocation spectrum (SFig.~\ref{fig:3D}~$\bs{g}$), we observe one midgap-localized zero mode per dislocation.  The zero modes are protected at zero energy by the chiral (\emph{i.e.} unitary particle-hole) symmetry $\Pi$, which is defined through the action:
\begin{equation}
\label{eq:axichiralsym}
\Pi:\ \mathcal{H}(k_x,k_y,\pi)\rightarrow \tau^{y}\sigma^{0}\mathcal{H}(k_x,k_y,\pi)\tau^{y}\sigma^{0},
\end{equation}
such that $\Pi$ is a symmetry of $\mathcal{H}(k_{x},k_y,\pi)$ if $\Pi\mathcal{H}(k_{x},k_y,\pi)\Pi^{-1} =\tau^{y}\sigma^{0}\mathcal{H}(k_x,k_y,\pi)\tau^{y}\sigma^{0}= -\mathcal{H}(k_{x},k_y,\pi)$.  Crucially, if we were to relax $\Pi$ symmetry, then the midgap states could be pushed out of the gap.  However, if we preserve $\mathcal{I}$ symmetry while breaking $\Pi$, then $\pm e/2$ end charges would still remain bound to the dislocations~\cite{OrtixSSHoneHalfEndCharges}, in a generalization of the conclusions of Goldstone and Wilczek~\cite{WilczekAxion,GoldstoneWilczek,NiemiSemenoff}.  Equivalently, in the absence of chiral symmetry, we can also identify the nontrivial topology by counting the number of states that are occupied up to a given Fermi level in the gap~\cite{WiederAxion,HingeSM,WladCorners,AshvinFragile2,ZhidaFragileTwist1,ZhidaFragileTwist2}, as we did previously in SN~\ref{subsec:ChernAtY}.  Let the 3D real-space Hamiltonian corresponding to the system with two screw dislocations be a $(D \times D)$ matrix.  Comparing SFigs.~\ref{fig:3D}~$\bs{f}$ and $\bs{g}$, we observe that the two spectra differ by the presence of two midgap states and the absence of one state from each of the valence and conduction manifolds.  When chiral symmetry is broken, the energy of the midgap states can be shifted in a manner that preserves $\mathcal{I}$ symmetry, leaving the two states degenerate in the thermodynamic limit.  This implies that any gapped Fermi level that encloses either $(D/2-1)$ or $(D/2+1)$ occupied states indicates a nontrivial topology, whereas a Fermi level enclosing $D/2$ occupied states indicates trivial bulk topology.  More generally, the presence of anomalous 0D states in arbitrary dimensions can be diagnosed in a similar manner by constructing a 0D system with a pair of generalized dislocations that preserve a global point group symmetry and observing an imbalance in the number of states above and below the gap that cannot be resolved without breaking a symmetry or closing the gap (\emph{i.e.}, a filling anomaly)~\cite{HOTIChen,WiederAxion,HingeSM,WladCorners,AshvinFragile2,ZhidaFragileTwist1,ZhidaFragileTwist2}.

\paragraph{$\xi = 0$: Weak FTI stack}
\label{subsec:weakFTIstack}
In this case, $\mathcal{H}(\bs{k})$ is independent of $\gamma$, and SEq.~\eqref{eq:axioninsulatorfragile2Dmodel} describes a $z$-directed stack of identical 2D FTI models. Specifically, the Hamiltonian [SEq.~\eqref{eq:axioninsulatorfragile2Dmodel} with $\xi = 0$] is in this case completely independent of $k_{z}$.  This is an example of weak topology, because the Hamiltonian $\mathcal{H}(\bs{k})$ can be deformed into a real-space, $z$-directed stack of decoupled 2D FTIs without breaking a symmetry or closing the bulk gap.  Relative to the initial trivial atomic insulator, $\mathcal{H}(\bs{k})$ features band inversions at $\Gamma$ and $Z$ (the parity eigenvalues are shown in SFig.~\ref{fig:3D}~$\bs{e}$), such that the bulk exhibits a nontrivial weak fragile invariant $\bs{M}^{\mathrm{F}}_{\nu} = (0,0,\pi)$ as defined in SN~\ref{sec:weakSSH}.  This implies that even though the bulk is an FTI (which is Wannierizable under the addition of trivial bands, see SN~\ref{sec:weakFragile}), the dislocation response is still nontrivial.  Correspondingly, in the dislocation spectrum (SFig.~\ref{fig:3D}~$\bs{g}$), we observe a filling anomaly.

For the present case of an array of FTI planes, we can also understand the existence of dislocation midgap states intuitively without invoking the more general theory used to derive SEq.~\eqref{eq:disloc2Dphase}.  Specifically, if we were to insert a loop of edge dislocations with Burgers vector $\bs{B}=\hat{z}$ (instead of screw dislocations), then we would effectively introduce an uncoupled FTI into the system whose corner states become dislocation end states.  As long as the ``leftover'' FTI contains an inversion center, then its corner states also induce a system filling anomaly, consistent with the numerical results shown in SFig.~\ref{fig:3D}~$\bs{g}$.

\begin{figure}[t]
\begin{center}
\includegraphics[width=0.7 \textwidth]{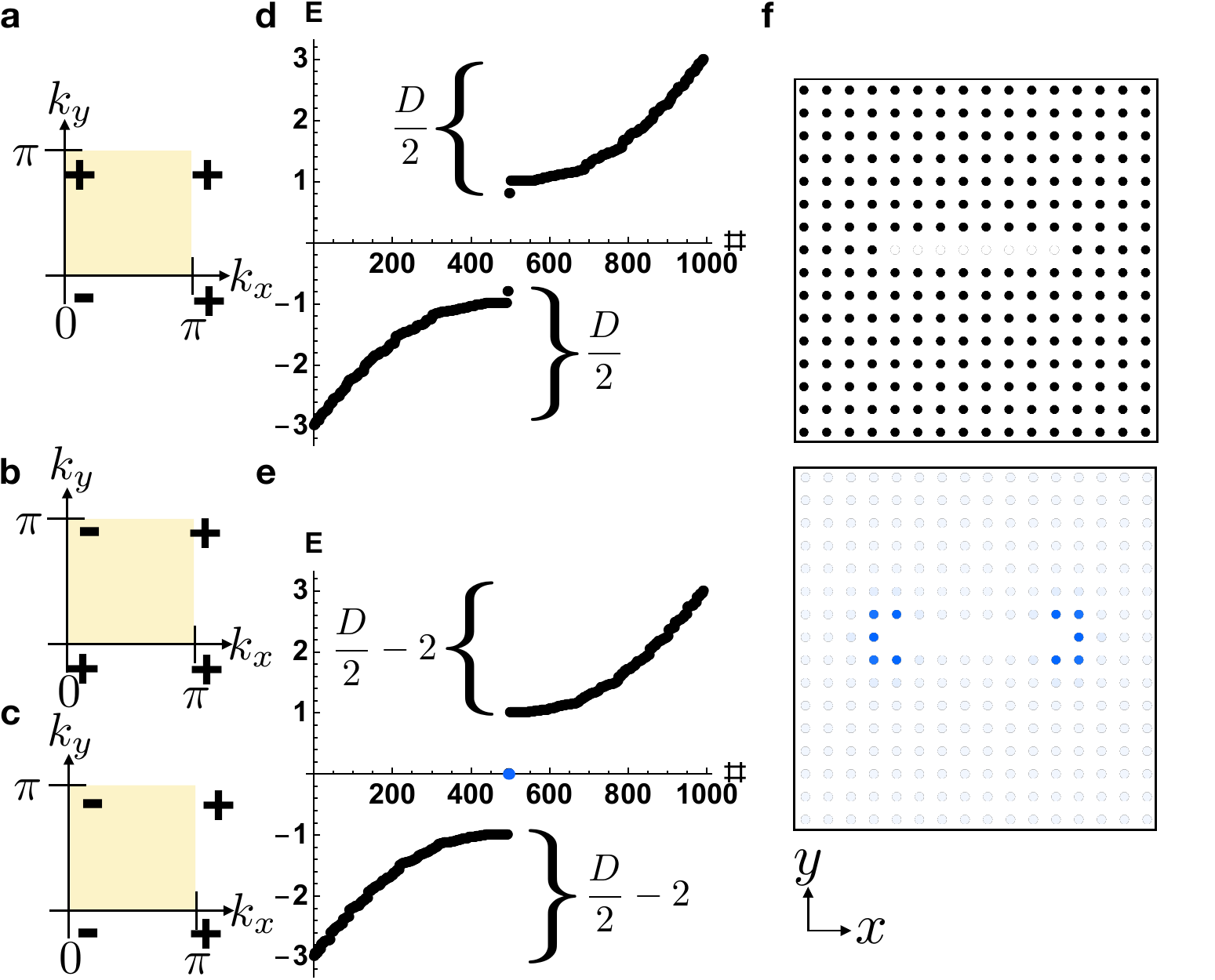}
\caption{{\bf 0D point dislocation bound states in 2D insulators with inversion and time-reversal symmetry.} $\bs{a}$-$\bs{e}$ Bulk parity [inversion ($\mathcal{I}$)] eigenvalues and PBC energy spectra for the defect in SFig.~\ref{fig:2DAPPENDIX}~$\bs{b}$ when the bulk is equivalent to $\bs{a},\bs{d}$ a 2D TI with a Kramers pair of band inversions at $\Gamma$ [$\bs{M}^{\mathrm{SSH}}_{\nu} = (0,0)$ as defined in SN~\ref{sec:weakSSH}], $\bs{b},\bs{e}$ a 2D TI with a Kramers pair of band inversions at $Y$ [$\bs{M}^{\mathrm{SSH}}_{\nu} = (0,\pi)$], $\bs{c},\bs{e}$ a weak $y$-directed array of time-reversal- ($\mathcal{T}$-) symmetric SSH chains [$\bs{M}^{\mathrm{SSH}}_{\nu} = (0,\pi)$].  Kramers pairs of parity eigenvalues are denoted by a single $\pm$ symbol.  Anomalous 0D defect states with spin-charge separation (SN~\ref{sec:kp}) are present in cases $\bs{b}$ and $\bs{c}$, but not in $\bs{a}$, and are equivalent to the end states of an $\mathcal{I}$- and $\mathcal{T}$-symmetric SSH chain~\cite{SSH,SSHspinon,HeegerReview,RiceMele,TRPolarization}, and thus persist under the relaxation of particle-hole symmetry in the form of a filling anomaly~\cite{WilczekAxion,GoldstoneWilczek,NiemiSemenoff,WiederAxion,HingeSM,WladCorners,AshvinFragile2,ZhidaFragileTwist1,ZhidaFragileTwist2}.  $\bs{f}$ Real-space geometry with a pair of point dislocations and absolute square of the wavefunction of the midgap states in $\bs{e}$ (indicated in blue in the lower panel) on a $16 \times 16$ lattice.  As derived in SEq.~(\ref{eq:dislocationtheory}), the dislocation response in $\bs{a}$ -- $\bs{e}$ is nontrivial when $\bs{B}\cdot \bs{M}^{\mathrm{SSH}}_{\nu}\text{ mod }2\pi = \pi$.}
\label{fig:2DTRS}
\end{center}
\end{figure}

\subsection{Dislocation Response with Time-Reversal Symmetry}
\label{sec:yesTDislocation}

\subsubsection{2D Point Dislocations in the Presence of $\mathcal{T}$ Symmetry}
\label{sec:2DtoysubsecTRS}

In this section, we will demonstrate the dislocation response of 2D insulators with $\mathcal{T}$ symmetry.  We begin by comparing the minimal tight-binding model of an inversion- ($\mathcal{I}$-) symmetric 2D topological insulator (TI) to that of a stack (array) of $\mathcal{I}$- and $\mathcal{T}$-symmetric Su-Schrieffer-Heeger (SSH) chains~\cite{SSH,SSHspinon,HeegerReview,RiceMele,TRPolarization}.  Consider the Bloch Hamiltonian:
\begin{equation}
\begin{aligned}
\label{eq:cherninsulatorsshmodelTRS}
\mathcal{H}(\bs{k}) =& \sin k_x \, \tau^0 \sigma^x + \xi \sin k_y \, \tau^0 \sigma^y \\&+ \left[\alpha \cos k_x + \xi \left(1 + \beta \cos k_y\right)\right] \tau^z \sigma^z,
\end{aligned}
\end{equation}
defined on a square lattice, where $\alpha,\beta = \pm 1$ are parameters that can be tuned to realize different topological phases.  Time-reversal and inversion symmetry are represented by: 
\begin{equation}
\begin{aligned}
\mathcal{T}&:\ \mathcal{H}(\bs{k})\rightarrow \tau^x\sigma^y\mathcal{H}^*(-\bs{k})\tau^x\sigma^y,\\
\mathcal{I}&:\ \mathcal{H}(\bs{k})\rightarrow\tau^z\sigma^z\mathcal{H}(-\bs{k})\tau^z\sigma^z.
\label{eq:matrixRep2DITdislocation}
\end{aligned}
\end{equation}
At half filling, there are two occupied bands with the Bloch eigenstates $\ket{u^n(\bs{k})}$, $n=1,2$.  At the four time-reversal-invariant momenta (TRIM points) of the 2D BZ shown in SFig.~\ref{fig:2DAPPENDIX}~$\bs{a}$ (\emph{i.e.} the $k$ points for which $\mathcal{T}\bs{\bar{k}}=\bs{\bar{k}}\text{ mod }{\bf b}$, where ${\bf b}$ is a reciprocal lattice vector), we can define the parity eigenvalue $\lambda_{\bs{\bar{k}}}$ of the occupied Kramers pair as:
\begin{equation}
\mathcal{I} \ket{u^1(\bs{\bar{k}})} = \lambda_{\bs{\bar{k}}} \ket{u^1(\bs{\bar{k}})},
\end{equation}
where $\mathcal{I}^2 = 1$ implies that $\lambda_{\bs{\bar{k}}} = \pm 1$.  At each TRIM point, states appear in Kramers pairs with the same parity ($\mathcal{I}$) eigenvalues, because the eigenvalues of $\mathcal{I}$ are real ($\lambda_{\bs{\bar{k}}} = \pm 1$), and because the matrix representatives of $\mathcal{I}$ and $\mathcal{T}$ commute at the TRIM points [SEq.~(\ref{eq:matrixRep2DITdislocation})].

For $\alpha = \beta = 0$ and $\xi=0,1$, SEq.~\eqref{eq:cherninsulatorsshmodelTRS} is topologically trivial, and its occupied subspace is composed of two bands with the same parity eigenvalues $\lambda_{\bs{\bar{k}}} = 1$ at all TRIM points.  For $\xi = 1$, $\mathcal{H}(\bs{k})$ describes a symmetry-indicated 2D TI that differs from a trivial (unobstructed) atomic limit by a single sign-change of the parity eigenvalue of the occupied Kramers pair of bands at $\bs{\bar{k}} = (1+\alpha,1+\beta)\pi/2$.  Specifically, the $\mathbb{Z}_2$ TI (Fu-Kane) invariant is indicated by the product of the parity eigenvalues of the occupied Kramers pairs~\cite{FuKaneInversion,AndreiInversion}.  For $\xi = 0$, $\mathcal{H}(\bs{k})$ becomes equivalent to an array of identical $x$-directed $\mathcal{T}$-symmetric SSH chains indexed by $k_y$ whose time-reversal polarization is indicated by the parity eigenvalues at $k_x = 0,\pi$~\cite{SSH,SSHspinon,HeegerReview,RiceMele,ZakPhase,ArisInversion,TRPolarization,KaneMeleZ2}.

Having established the properties of the Hamiltonian of the pristine system [SEq.~\eqref{eq:cherninsulatorsshmodelTRS}], we will now probe the dislocation response.  We introduce a pair of 0D dislocations with Burgers vector $\bs{B} = \hat{y}$.  Through SEq.~\eqref{eq:dislocationtheory}, we deduce that each dislocation realizes an interface between two insulators $S$ and $S'$ that ``differ" by the Hamiltonian of the $k_y = \pi$ plane of SEq.~\eqref{eq:cherninsulatorsshmodelTRS}:
\begin{equation}
\label{eq:dislocsshchainTRS}
\mathcal{H}(k_x,\pi) = \sin k_x \, \tau^0 \sigma^x + [\alpha \cos k_x + \xi (1-\beta)] \tau^z \sigma^z.
\end{equation}
Setting $\alpha = -1$, $\mathcal{H}(k_x,\pi)$ describes an $\mathcal{I}$- and $\mathcal{T}$-symmetric SSH chain with a trivial (nontrivial) time-reversal polarization for $\xi = 1$, $\beta = -1$ ($\xi = 1$, $\beta = + 1$).

To obtain the numerical results presented in SFig.~\ref{fig:2DTRS}, we use the same prescription as employed in SN~\ref{sec:2Dtoysubsec} to obtain a Hamiltonian $\tilde{H}$ that differs from the pristine 2D Hamiltonian $\mathcal{H}(\bs{k})$ by the presence of a pair of point dislocations.  With $L = 16$, $\tilde{H}$ has the dimensions $[4 (L^2 - 8) \times 4 (L^2 - 8)] = (992 \times 992)$.  We will now discuss the electronic structure of $\tilde{H}$ in detail for characteristic values of $\xi$ and $\beta$.

\paragraph{$\xi = 1$, $\beta = -1$: Topological insulator with band inversion at $\Gamma$}
\label{subsec:TIatGamma}
This case corresponds to an inversion-symmetry-indicated 2D TI driven by band inversion at $\bs{k} = (0,0)$ in SEq.~\eqref{eq:cherninsulatorsshmodelTRS}.  In this case, the bulk characterizes a strong topological phase.  Because the bands at $X$, $Y$, and $M$ are uninverted relative to an atomic insulator with orbitals at the $1a$ position, the time-reversal generalization of the weak SSH invariant $\bs{M}^{\mathrm{SSH}}_{\nu}=(0,0)$ (SN~\ref{sec:weakSSH}).  This implies that $\bs{B}\cdot\bs{M}^{\mathrm{SSH}}_{\nu}\text{ mod }{ 2\pi }=0$.  In SFig.~\ref{fig:2DTRS}~$\bs{a},\bs{d}$, we respectively show the parity eigenvalues of the occupied bands and the dislocation spectrum, which does not exhibit a filling anomaly or midgap dislocation bound states.

\paragraph{$\xi = 1$, $\beta = +1$: Topological insulator with band inversion at $Y$}
\label{subsec:TIatY}
This case corresponds to an inversion-symmetry-indicated 2D TI driven by band inversion at the $Y$ point (SFig.~\ref{fig:2DAPPENDIX}~$\bs{a}$).  Relative to the initial $1a$ atomic insulator, the band inversion at $Y$ has changed not only the strong index (the $\mathbb{Z}_2$ TI invariant), but also the weak SSH indices [the parity eigenvalues of the occupied bands are shown in SFig.~\ref{fig:2DTRS}~$\bs{b}$].  Specifically, the bulk exhibits a weak SSH invariant $\bs{M}^{\mathrm{SSH}}_{\nu}=(0,\pi)$ as defined in SN~\ref{sec:weakSSH}.  This implies that $\bs{B}\cdot\bs{M}^{\mathrm{SSH}}_{\nu}\text{ mod }{ 2\pi }=\pi$, indicating that the dislocation response is nontrivial.  Correspondingly, in the dislocation spectrum (SFig.~\ref{fig:2DTRS}~$\bs{e}$), we observe one midgap-localized Kramers pair of zero modes per dislocation.  The zero modes are protected by $\mathcal{T}$ symmetry and the chiral (\emph{i.e.}, unitary particle-hole) symmetry $\Pi$, which is defined through the action:
\begin{equation}
\label{eq:TIchiralsym}
\Pi:\ \mathcal{H}(k_x,\pi)\rightarrow \tau^{0}\sigma^{y}\mathcal{H}(k_x,\pi)\tau^{0}\sigma^{y},
\end{equation}
such that $\Pi$ is a symmetry of $\mathcal{H}(k_{x},\pi)$ if $\Pi\mathcal{H}(k_{x},\pi)\Pi^{-1} =\tau^{0}\sigma^{y}\mathcal{H}(k_x,\pi)\tau^{0}\sigma^{y}= -\mathcal{H}(k_{x},\pi)$.  Crucially, if we were to relax $\Pi$ symmetry, then the midgap states could be pushed out of the gap.  However, if we preserve $\mathcal{I}$ and $\mathcal{T}$ symmetries while breaking $\Pi$, then chargeless spin would still remain bound to the dislocations, in a generalization of the conclusions of Goldstone and Wilczek~\cite{WilczekAxion,GoldstoneWilczek,NiemiSemenoff} and SRefs.~\onlinecite{QiFlux,AshvinFlux,AdyFlux,MirlinFlux,CorrelatedFlux,OrtixSSHoneHalfEndCharges} (SN~\ref{sec:kp}).

\paragraph{$\xi = 0$: $\mathcal{T}$-symmetric weak SSH array}
\label{subsec:weakTRSSSHstack}
In this case, $\mathcal{H}(\bs{k})$ is independent of $\beta$, and SEq.~\eqref{eq:cherninsulatorsshmodelTRS} describes a $y$-directed array of identical $x$-directed $\mathcal{T}$-symmetric SSH chains~\cite{YoungkukWeakSSH}.  Specifically, the Hamiltonian [SEq.~\eqref{eq:cherninsulatorsshmodelTRS} with $\xi = 0$] is in this case completely independent of $k_{y}$.  For each SSH chain, the nontrivial time-reversal polarization is indicated by the difference in the parity eigenvalues of the occupied bands at $k_{x}=0,\pi$~\cite{SSH,SSHspinon,HeegerReview,RiceMele,ZakPhase,ArisInversion,TRPolarization,KaneMeleZ2}.  This case is an example of weak topology, because the Hamiltonian $\mathcal{H}(\bs{k})$ can be deformed into a real-space array of decoupled 1D chains without breaking a symmetry or closing the bulk gap.  Relative to the initial $1a$ atomic insulator, $\mathcal{H}(\bs{k})$ features band inversions at $\Gamma$ and $Y$ (the parity eigenvalues are shown in SFig.~\ref{fig:2DTRS}~$\bs{c}$), such that the bulk exhibits a trivial symmetry-indicated $\mathbb{Z}_2$ TI invariant and nontrivial weak SSH indices, and is irrep-equivalent to an OAL from $1a$ [$(x,y)=(0,0)$] to $1c$ [($(x,y) = (1/2,0)$] (SFig.~\ref{fig:fragile}~$\bs{b}$).  Specifically, the bulk exhibits a weak SSH invariant $\bs{M}^{\mathrm{SSH}}_{\nu}=(0,\pi)$ as defined in SN~\ref{sec:weakSSH}.  This implies that, even though the bulk is an OAL (SN~\ref{sec:weakSSH}), the dislocation response is nontrivial.  Correspondingly, in the dislocation spectrum (SFig.~\ref{fig:2DTRS}~$\bs{e}$), we observe a filling anomaly.

For the present case of an array of SSH chains, we can also understand the existence of dislocation midgap states intuitively without invoking the more general theory used to derive SEq.~\eqref{eq:dislocsshchainTRS}.  The dislocations introduce an uncoupled SSH chain into the system whose end states become the dislocation bound states.  As long as the ``leftover'' SSH chain is $\mathcal{T}$-symmetric and contains an inversion center, its end states also induce a system filling anomaly, consistent with the numerical results shown in SFig.~\ref{fig:2DTRS}~$\bs{e}$.

\subsubsection{3D Screw Dislocations in the Presence of $\mathcal{T}$ Symmetry}
\label{sec:3DscrewTRSsec}

In this section, we will demonstrate the screw dislocation response of 3D insulators with $\mathcal{T}$ symmetry.  We will here focus on higher-order TIs and weak FTIs, which, as we will numerically demonstrate, exhibit 0D dislocation bound states.  However, we have also verified that our numerical models exhibit helical modes bound to screw dislocations inserted into insulators with nontrivial weak TI indices, in agreement with the results of SRefs.~\onlinecite{AshvinScrewTI,TeoKaneDefect,QiDefect2,Vlad2D,TanakaDefect,VladScrewTI}.  Specifically, we observe that when the vector of weak TI indices in a $\mathcal{T}$-invariant 3D insulator $\bs{M}_\nu$ satisfies $\bs{B} \cdot \bs{M}_\nu\text{ mod }2\pi = \pi$, the plane spanning the screw dislocations in SEq.~\eqref{eq:dislocationtheory} corresponds to 2D TI with 1D helical edge states that coincide with the bulk screw dislocations and boundary step edges.  For the remainder of this section and work, we will restrict focus to $\mathcal{T}$-invariant 3D insulators for which line-like dislocations bind 0D (HEND) states (\emph{i.e.} non-axionic HOTIs and weak FTIs/OALs).

While we will focus in this section on the topological response of 3D insulators to screw dislocations, we note that we can also consider $\mathcal{I}$-symmetric arrangements of edge dislocations.  As shown in SN~\ref{sec:kpEdge}, when the HEND-state weak index vector $\bs{M}_{\nu}^{\mathrm{F}}$ introduced in this work is nontrivial, an edge dislocation with a Burgers vector $\bs{B}$ will also bind an anomalous number of 0D states if $\bs{B}\cdot \bs{M}_{\nu}^{\mathrm{F}}\text{ mod }2\pi = \pi$.  Later, in SN~\ref{sec:DFTSnTe}, we will use edge dislocations to demonstrate the nontrivial HEND-state defect response of 3D SnTe crystals (SFig.~\ref{fig:SnTeTBresults}~$\bs{a}$).

\begin{figure}[t]
\begin{center}
\includegraphics[width=0.65 \textwidth]{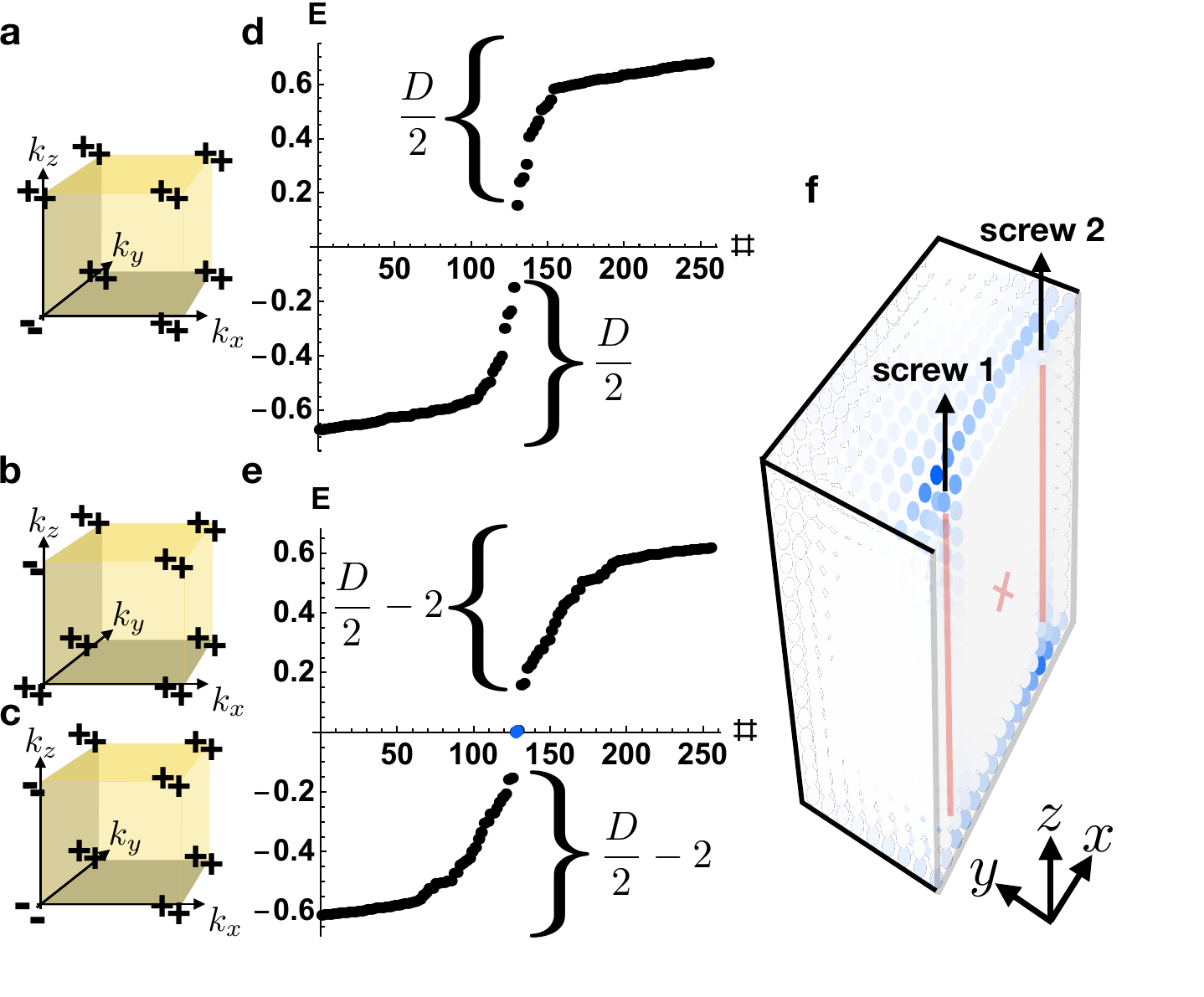}
\caption{{\bf 0D screw dislocation bound states in 3D insulators with time-reversal and inversion symmetry.} $\bs{a}$-$\bs{e}$ Bulk parity [inversion ($\mathcal{I}$)] eigenvalues and hollow-doughnut-boundary-condition (HDBC) energy spectra for a pair of screw dislocations (SFig.~\ref{fig:3D}~$\bs{b}$) inserted into $\bs{a},\bs{d}$ an $\mathcal{I}$- and time-reversal- ($\mathcal{T}$-) symmetric HOTI with double band inversion at $\Gamma$ [$\bs{M}^{\mathrm{F}}_{\nu} = (0,0,0)$ as defined in SN~\ref{sec:weakFragile}], $\bs{b},\bs{e}$ a HOTI with double band inversion at $Z$ [$\bs{M}^{\mathrm{F}}_{\nu} = (0,0,\pi)$], and $\bs{c},\bs{e}$ a weak stack of $\mathcal{I}$- and $\mathcal{T}$-symmetric 2D FTIs [$\bs{M}^{\mathrm{F}}_{\nu} = (0,0,\pi)$].  In $\bs{a},\bs{b},\bs{c}$, Kramers pairs of parity eigenvalues are denoted with a single $\pm$ symbol.  Anomalous 0D HEND states with spin-charge separation (SN~\ref{sec:kp}) are present on the dislocation ends in $\bs{b}$ and $\bs{c}$, but not in $\bs{a}$.  The 0D (HEND) states in $\bs{b}$ and $\bs{c}$ are equivalent to the corner states of an $\mathcal{I}$- and $\mathcal{T}$-symmetric 2D FTI~\cite{TMDHOTI,WiederAxion}, and thus persist under the relaxation of particle-hole symmetry~\cite{TMDHOTI,WiederAxion,HingeSM}.  In $\bs{d},\bs{e}$, we only show the $256$ lowest-lying states.  In the presence of an additional chiral symmetry, the two midgap Kramers pairs of dislocation end states in $\bs{b}$ and $\bs{c}$ are pinned to zero energy.  In the absence of chiral symmetry, a nontrivial dislocation response can still be diagnosed by observing filling anomaly~\cite{WiederAxion,HingeSM,WladCorners,AshvinFragile2,ZhidaFragileTwist1,ZhidaFragileTwist2}.  $\bs{f}$ Real-space-resolved absolute square of the midgap eigenstates shown in $\bs{e}$ on a $16 \times 16 \times 16$ lattice (cut in half along the transparent gray plane to expose the bound states).  $\mathcal{I}$-related Kramers pairs of bound states appear on two of the four screw dislocation ends.  The inversion center in $\bs{f}$ is marked with a red $X$.  As derived in SEq.~(\ref{eq:dislocationtheory}), the dislocation response in $\bs{a}$ -- $\bs{c}$ is nontrivial when $\bs{B}\cdot \bs{M}^{\mathrm{F}}_{\nu}\text{ mod }2\pi = \pi$.}
\label{fig:3DTRS}
\end{center}
\end{figure}

We begin by comparing the minimal tight-binding model for a non-axionic higher-order topological insulator (HOTI)~\cite{HOTIBernevig,HOTIChen,WladTheory,AxionZhida,EslamInversion,FanHOTI,TMDHOTI,HarukiLayers,AshvinMagnetic,EzawaMagneticHOTI,VDBHOTI,WiederAxion,KoreanAxion,NicoDavidAXI2,IvoAXI1,IvoAXI2} to that of a stack of 2D $\mathcal{T}$-symmetric fragile topological insulators ($\mathcal{T}$-symmetric FTIs)~\cite{TMDHOTI,WiederAxion}, where we retain $\mathcal{I}$ and $\mathcal{T}$ symmetries for all models.  Consider the Bloch Hamiltonian:
\begin{equation}
\begin{aligned}
\label{eq:axioninsulatorfragile2DmodelTRS}
\mathcal{H}(\bs{k}) =& \sin k_x \, \rho^z \tau^z \sigma^x + \sin k_y \, \rho^z \tau^z \sigma^y + \xi \sin k_z \, \rho^z \tau^z \sigma^z \\&+ \left[\alpha \cos k_x + \beta \cos k_y + \xi \left(2 + \gamma \cos k_z\right)\right] \rho^z \tau^x \sigma^0 \\
& + \xi \, \delta \, (\rho^y \tau^z \sigma^x + \rho^y \tau^z \sigma^y + \rho^y \tau^z \sigma^z),
\end{aligned}
\end{equation}
defined on a square lattice, where $\rho^i$, $\tau^i$ and $\sigma^i$, $i = 0,x,y,z$ are three sets of Pauli matrices indexing sublattice and orbital degrees of freedom, respectively, and $\xi=0,1$ and $\alpha,\beta,\gamma = \pm 1$ are parameters that can be tuned to realize different topological phases, while $\delta$ is a small parameter that gaps out the helical states along dislocations (we will use $\delta = 1/4$ in all calculations, for $\delta = 0$ the Hamiltonian is a model of two uncoupled 3D TIs~\cite{AshvinIndicators}).  We abbreviate the Kronecker product by $\rho^i \otimes \tau^j \otimes \sigma^k \equiv \rho^i \tau^j \sigma^k$.  Time-reversal ($\mathcal{T}$) and inversion ($\mathcal{I}$) symmetries are respectively represented by: 
\begin{equation}
\begin{aligned}
\mathcal{T}&:\ \mathcal{H}(\bs{k})\rightarrow \rho^0\tau^0\sigma^y\mathcal{H}^*(-\bs{k})\rho^0\tau^0\sigma^y,\\
\mathcal{I}&:\ \mathcal{H}(\bs{k})\rightarrow \rho^z\tau^x\sigma^0\mathcal{H}(-\bs{k})\rho^z\tau^x\sigma^0.
\end{aligned}
\end{equation}
At the eight TRIM points of the 3D BZ shown in SFig.~\ref{fig:3D}~$\bs{a}$, we can again compute the parity eigenvalues $\lambda^n_{\bs{\bar{k}}}$, $n=1,2$ of the two occupied Kramers pairs of bands, where $\lambda^n_{\bs{\bar{k}}} = \pm 1$.  For $\alpha = \beta = \gamma = 0$ and $\xi=0,1$, SEq.~\eqref{eq:axioninsulatorfragile2DmodelTRS} is topologically trivial and its occupied subspace is composed of four bands with the parity eigenvalues $\lambda_{\bs{\bar{k}}} = 1$ at all TRIM points.  For $\xi = 1$, $\mathcal{H}(\bs{k})$ describes a symmetry-indicated HOTI~\cite{AshvinIndicators,AshvinTCI,ChenTCI,TMDHOTI} driven by a sign-change of the parity eigenvalues of the two occupied Kramers pairs of bands at $\bs{\bar{k}} = (1+\alpha,1+\beta,1+\gamma)\pi/2$.  For $\xi = 0$, $\mathcal{H}(\bs{k})$ becomes equivalent to a $z$-directed stack of identical $\mathcal{I}$-symmetry-indicated~\cite{AxionZhida,EslamInversion,MTQC,MTQCmaterials,AshvinMagnetic,EzawaMagneticHOTI,VDBHOTI,WiederAxion,KoreanAxion,NicoDavidAXI2,IvoAXI1,IvoAXI2} $\mathcal{T}$-symmetric FTIs indexed by $k_z$.

Having established the properties of the Hamiltonian of the pristine system [SEq.~\eqref{eq:axioninsulatorfragile2DmodelTRS}], we will now probe its dislocation response.  We introduce a pair of 1D screw dislocations with Burgers vector $\bs{B} = \hat{z}$ and opposite chiralities [SEq.~(\ref{eq:chiralityScrew})].  Through SEq.~\eqref{eq:dislocationtheory}, we deduce that each dislocation realizes an interface between two insulators $S$ and $S'$ that ``differ" by the Hamiltonian of the $k_z = \pi$ plane of SEq.~\eqref{eq:axioninsulatorfragile2DmodelTRS}:
\begin{equation}
\label{eq:disloc2DphaseTRS}
\begin{aligned}
\mathcal{H}(k_x,k_y,\pi) =& \sin k_x \, \rho^0 \tau^z \sigma^x + \sin k_y \, \rho^0 \tau^z \sigma^y \\&+ [\alpha \cos k_x + \beta \cos k_y + \xi (2-\gamma)] \rho^z \tau^x \sigma^0 \\
& + \xi \, \delta \, (\rho^z \tau^x \sigma^x + \rho^z \tau^x \sigma^y + \rho^z \tau^x \sigma^z).
\end{aligned}
\end{equation}
Setting $\alpha = \beta = -1$, $\mathcal{H}(k_x,k_y,\pi)$ describes a 2D trivial insulator ($\mathcal{T}$-symmetric FTI~\cite{TMDHOTI,WiederAxion}) for $\xi = 1$, $\gamma = -1$ ($\xi = 1$, $\gamma = + 1,\delta=1/4$).  We will now discuss the electronic structure of $\tilde{H} (k_z)$ in detail for characteristic values of $\xi$ and $\gamma$ (SFig.~\ref{fig:3DTRS}).  To obtain the numerical results presented in SFig.~\ref{fig:3DTRS}, we use the same prescription as employed in SN~\ref{sec:3Dscrew} [specifically SEqs.~\eqref{eq:screwalgo}-\eqref{eq:screwalgo2}] to obtain a Hamiltonian $\tilde{H}(k_z)$ that differs from a pristine crystal by a pair of screw dislocations.

\paragraph{$\xi = 1$, $\gamma = -1$: Higher-order topological insulator with double band inversion at $\Gamma$}
\label{subsec:HOTIatGamma}
This case corresponds to an inversion-symmetry-indicated HOTI driven by two Kramers pairs of band inversions at $\bs{k} = (0,0,0)$ in SEq.~\eqref{eq:axioninsulatorfragile2DmodelTRS}.  In this case, the bulk characterizes a strong topological phase.  Because the bands at $X$, $Y$, $M$, $Z$, $U$, $R$, and $T$ are uninverted relative to an atomic insulator with orbitals at the $1a$ position, the weak fragile invariant $\bs{M}^{\mathrm{F}}_{\nu} = (0,0,0)$ (SN~\ref{sec:weakFragile}).  This implies that $\bs{B}\cdot\bs{M}^{\mathrm{F}}_{\nu}\text{ mod }{ 2\pi }=0$, and therefore, that the dislocations do not bind anomalous Kramers pairs of spin-charge-separated HEND states.  In SFig.~\ref{fig:3DTRS}~$\bs{a},\bs{d}$, we respectively show the parity eigenvalues of the occupied bands and the dislocation spectrum, which does not exhibit a filling anomaly or midgap dislocation bound states.

\paragraph{$\xi = 1$, $\gamma = +1$: Higher-order topological insulator with double band inversion at $Z$}
\label{subsec:HOTIatZ}
This case describes a HOTI driven by double band inversion at the Z point (SFig.~\ref{fig:3D}~$\bs{a}$).  Relative to a $1a$ trivial (unobstructed) atomic limit, the double band inversion at $Z$ has changed not only the strong $\mathbb{Z}_{4}$ HOTI index~\cite{AshvinIndicators,AshvinTCI,ChenTCI,TMDHOTI}, but also the weak fragile indices [the parity eigenvalues of the occupied bands are shown in SFig.~\ref{fig:3DTRS}~$\bs{b}$].  Specifically, the bulk exhibits a nontrivial weak fragile index $\bs{M}^{\mathrm{F}}_{\nu} = (0,0,\pi)$ as defined in SN~\ref{sec:weakFragile}.  This implies that $\bs{B}\cdot\bs{M}^{\mathrm{F}}_{\nu}\text{ mod }{ 2\pi }=\pi$, indicating that the dislocation response is nontrivial.  Correspondingly, in the dislocation spectrum (SFig.~\ref{fig:3DTRS}~$\bs{e}$), we observe one midgap-localized Kramers pair of modes per dislocation.  The zero modes are protected by the chiral (\emph{i.e.} unitary particle-hole) symmetry $\Pi$, which is defined through the action:
\begin{equation}
\label{eq:HOTIchiralsym}
\Pi:\ \mathcal{H}(k_x,k_y,\pi)\rightarrow \rho^0\tau^{y}\sigma^{0}\mathcal{H}(k_x,k_y,\pi)\rho^0\tau^{y}\sigma^{0},
\end{equation}
such that $\Pi$ is a symmetry of $\mathcal{H}(k_{x},k_y,\pi)$ if $\Pi\mathcal{H}(k_{x},k_y,\pi)\Pi^{-1} =\rho^0\tau^{y}\sigma^{0}\mathcal{H}(k_x,k_y,\pi)\rho^0\tau^{y}\sigma^{0}= -\mathcal{H}(k_{x},k_y,\pi)$.  Crucially, if we were to relax $\Pi$ symmetry, the midgap states could be pushed out of the gap.
Equivalently, in the absence of chiral symmetry, we can also identify the nontrivial topology by counting the number of states that are occupied up to a given Fermi level in the gap~\cite{WiederAxion,HingeSM,WladCorners,AshvinFragile2,ZhidaFragileTwist1,ZhidaFragileTwist2}, just as we did in SN~\ref{subsec:TIatY}.  Let the 3D real-space Hamiltonian corresponding to the system with two screw dislocations be a $(D \times D)$ matrix.  Comparing SFigs.~\ref{fig:3DTRS}~$\bs{d}$ and $\bs{e}$, we observe that the spectra differ by the presence of two Kramers pairs of midgap states and the absence of two states from each of the valence and conduction manifolds.  When chiral symmetry is broken while preserving $\mathcal{I}$ and $\mathcal{T}$ symmetries, then the energy of the midgap states can be shifted, but only in a manner that preserves $\mathcal{I}$ and $\mathcal{T}$, leaving the four states degenerate in the thermodynamic limit.  This implies that any gapped Fermi level that encloses either $(D/2-2)$ or $(D/2+2)$ occupied states indicates a nontrivial dislocation response, whereas a Fermi level enclosing $D/2$ occupied states indicates a trivial screw dislocation response.

\paragraph{$\xi = 0$: Weak $\mathcal{T}$-symmetric FTI stack}
\label{subsec:weakTRSFTIstack}
In this case, $\mathcal{H}(\bs{k})$ is independent of $\gamma$, and SEq.~\eqref{eq:axioninsulatorfragile2DmodelTRS} describes a $z$-directed stack of identical 2D $\mathcal{T}$-symmetric FTIs. Specifically, the Hamiltonian [SEq.~\eqref{eq:axioninsulatorfragile2DmodelTRS} with $\xi = 0$] is in this case completely independent of $k_{z}$.  This is an example of weak topology, because the Hamiltonian $\mathcal{H}(\bs{k})$ can be deformed into a real-space stack of decoupled $\mathcal{I}$- and $\mathcal{T}$-symmetric 2D FTIs without breaking a symmetry or closing the bulk gap.  Relative to the initial trivial atomic insulator, $\mathcal{H}(\bs{k})$ features double band inversions at $\Gamma$ and $Z$ (the parity eigenvalues are shown in SFig.~\ref{fig:3DTRS}~$\bs{c}$), such that the bulk exhibits a nontrivial weak fragile invariant $\bs{M}^{\mathrm{F}}_{\nu} = (0,0,\pi)$ as defined in SN~\ref{sec:weakSSH}.  This implies that, even though the bulk is Wannierizable (after the addition of trivial bands, see SN~\ref{sec:weakFragile}), the dislocation response is nontrivial.  Correspondingly, in the dislocation spectrum (SFig.~\ref{fig:3DTRS}~$\bs{e}$), we observe a filling anomaly.

For the present case of an array of FTI planes, we can also understand the existence of dislocation midgap states intuitively without invoking the more general theory used to derive SEq.~\eqref{eq:disloc2DphaseTRS}.  Specifically, if we were to instead insert a loop of edge dislocations with Burgers vector $\bs{B}=\hat{z}$, then we would effectively introduce an uncoupled $\mathcal{I}$- and $\mathcal{T}$-symmetric FTI into the system whose corner states become dislocation end states.  As long as the ``leftover'' FTI plane contains an inversion center, then its corner states will also induce a system filling anomaly, consistent with the numerical results shown in SFig.~\ref{fig:3DTRS}~$\bs{e}$.

\section{Numerical Calculation Details: Fluxes and Flux Tubes}
\label{sec:numerics2}
\subsection{Flux Response without Time-Reversal Symmetry}
\subsubsection{2D Fluxes in the Absence of $\mathcal{T}$ Symmetry}
\label{sec:2DVORTEXtoysubsec}

In this section, we will demonstrate the $\pi$-flux response of $\mathcal{I}$-symmetric 2D insulators with broken $\mathcal{T}$ symmetry~\cite{TeoKaneDefect,QiDefect2,Vlad2D,QiFlux,AshvinFlux,AdyFlux,MirlinFlux,CorrelatedFlux,Radzihovsky2019}.  We begin by considering a pristine 2D insulator described by the Bloch Hamiltonian $\mathcal{H}(\bs{k})$ in SEq.~\eqref{eq:cherninsulatorsshmodel} of SN~\ref{sec:2Dtoysubsec}.  We then numerically introduce a pair of $\pi$-fluxes related by a bulk $\mathcal{I}$ center as discussed in SN~\ref{sec:fluxfluxtubetopologymapping}.

\begin{figure}[t]
\begin{center}
\includegraphics[width=0.9 \textwidth]{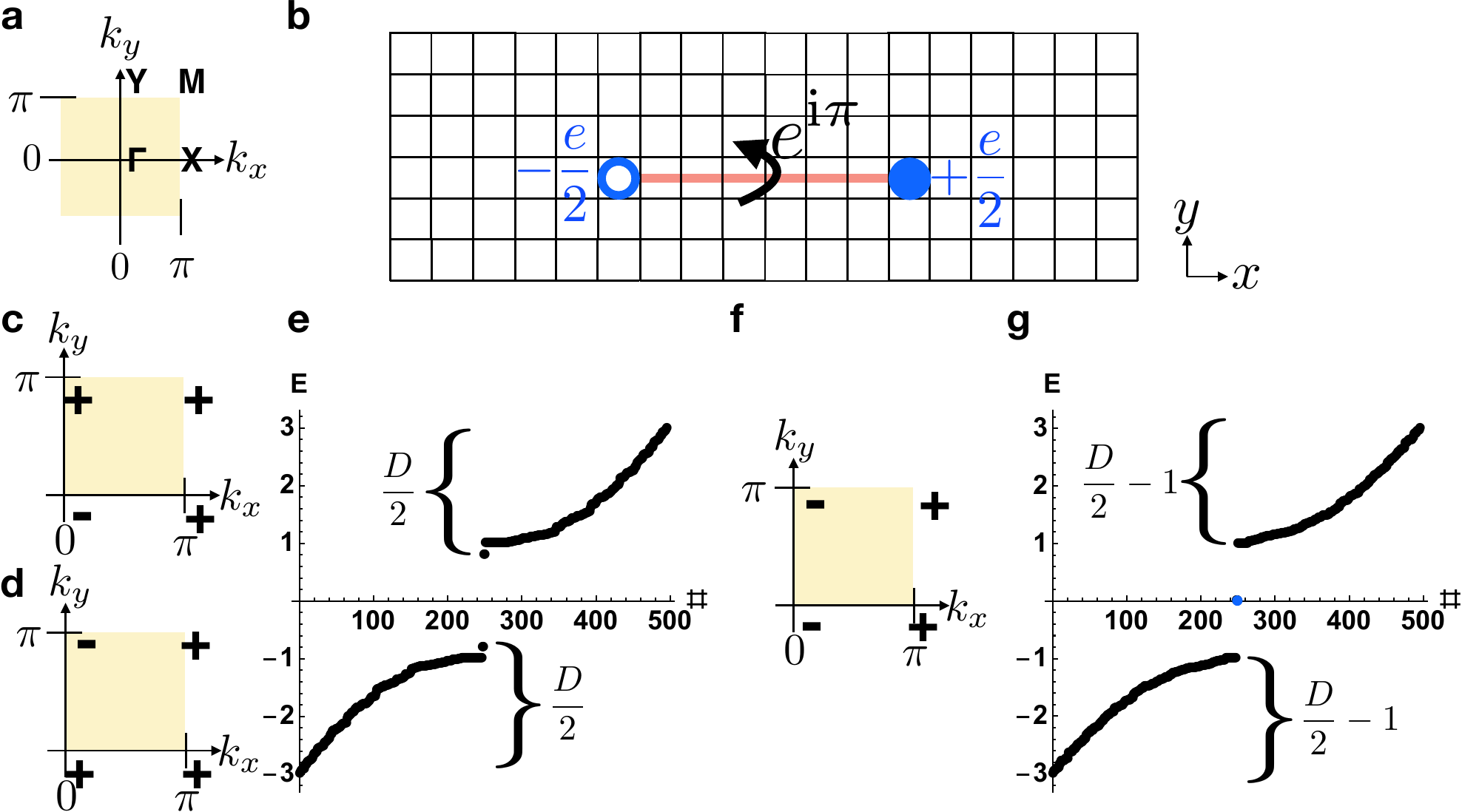}
\caption{{\bf 0D $\pi$-flux bound states in 2D insulators with inversion symmetry.} $\bs{a}$~The bulk BZ of a 2D rectangular magnetic crystal with only inversion ($\mathcal{I}$) symmetry.   $\bs{b}$~We implement a pair of 0D $\pi$-fluxes related by $\mathcal{I}$ by multiplying all of the hoppings along the red line of sites by $e^{\mathrm{i} \pi}$ in the positive $y$ direction, and by multiplying all of the hoppings along the same line by $e^{-\mathrm{i} \pi}$ in the negative $y$ direction.  $\bs{c}$ -- $\bs{g}$ Bulk parity ($\mathcal{I}$) eigenvalues and PBC energy spectra for the flux in $\bs{b}$ when the crystal with the BZ in $\bs{a}$ is equivalent to $\bs{c},\bs{g}$ a $|C|=1$ Chern insulator with band inversion at $\Gamma$, $\bs{d},\bs{g}$ a $|C|=1$ Chern insulator with band inversion at Y, and $\bs{f},\bs{e}$ a weak $y$-directed array of SSH chains~\cite{SSH,RiceMele}.  Anomalous 0D flux states with charge $\pm e/2$ are present in cases $\bs{c}$ and $\bs{d}$, but not in $\bs{f}$, and are equivalent to the end states of an $\mathcal{I}$-symmetric, $\mathcal{T}$-broken SSH chain~\cite{SSH,RiceMele} [SEq.~\eqref{eq:dislocsshchain}, red line in $\bs{b}$], and thus persist under the relaxation of particle-hole symmetry~\cite{WilczekAxion,GoldstoneWilczek,NiemiSemenoff,WiederAxion,HingeSM}.}
\label{fig:2Dvortex}
\end{center}
\end{figure}

To obtain the numerical results presented in SFig.~\ref{fig:2Dvortex}, we begin with a square lattice $\Lambda$ of size $|\Lambda| = L^2$ with $L = 16$ and PBC to model the $(x,y)$-plane.  We then modify the $y$-directed hoppings crossing a line of length $8$ at fixed (but necessarily fractional) $y$ coordinate to create a pair of $\pi$-fluxes, where two is the minimal number of fluxes compatible with (untwisted) PBC.  We define $H$ and $H_{A,B} \ketbra{A}{B}$ using the same notation employed in SN~\ref{sec:2Dtoysubsec}.  Using the site labeling $U$ and $M$ established in SFig.~\ref{fig:2Dimplementation}~$\bs{a}$, we will now detail how we numerically implement a pair of $\pi$-fluxes.

Defining $S = U \cup M$ and $\bar{S} = \Lambda \backslash S$, we form the expression:
\begin{equation}
\label{eq:fluxalgo}
\begin{aligned}
H= H_{\Lambda,\Lambda} \ketbra{\Lambda}{\Lambda} =& H_{\bar{S},\bar{S}} \ketbra{\bar{S}}{\bar{S}} + H_{S,\bar{S}} \ketbra{S}{\bar{S}} \\&+ H_{\bar{S},S} \ketbra{\bar{S}}{S} + H_{S,S} \ketbra{S}{S}.
\end{aligned}
\end{equation}
We then expand SEq.~\eqref{eq:fluxalgo}:
\begin{equation}
\begin{aligned}
H_{S,S} \ketbra{S}{S} =& H_{U,U} \ketbra{U}{U} + H_{M,M} \ketbra{M}{M} \\ &+H_{U,M} \ketbra{U}{M}+H_{M,U}\ketbra{M}{U}.
\end{aligned}
\end{equation}

To introduce a pair of $\pi$-fluxes, we form the Hamiltonian $\tilde{H}$, which is defined on $\Lambda$, and can be expanded as:
\begin{equation}
\begin{aligned}
\tilde{H} = \tilde{H}_{\Lambda,\Lambda} \ketbra{\Lambda}{\Lambda} =& H_{\bar{S},\bar{S}} \ketbra{\bar{S}}{\bar{S}} + H_{S,\bar{S}} \ketbra{S}{\bar{S}} \\&+ H_{\bar{S},S} \ketbra{\bar{S}}{S} + \tilde{H}_{S,S} \ketbra{S}{S},
\end{aligned}
\end{equation}
in which we can further expand:
\begin{equation}
\label{eq:fluxalgo2}
\begin{aligned}
\tilde{H}_{S,S} \ketbra{S}{S} =& H_{U,U} \ketbra{U}{U} + H_{M,M} \ketbra{M}{M} \\ &+e^{-\mathrm{i} \pi} H_{U,M} \ketbra{U}{M}+ e^{\mathrm{i} \pi} H_{M,U}\ketbra{M}{U},
\end{aligned}
\end{equation}
where the exponential factors $e^{\pm\mathrm{i}\pi}$ implement the $\pi$-fluxes.  To summarize, SEqs.~\ref{eq:fluxalgo} through~\ref{eq:fluxalgo2} describe how, in a tight-binding model with nearest-neighbor hoppings and orthogonal lattice vectors, we numerically implement a pair of $\pi$-fluxes by multiplying all of the hoppings across a line between sites in a pristine lattice by the phase factors $e^{\pm\mathrm{i} \pi}$.

We will now discuss the electronic structure of $\tilde{H}$ in detail for characteristic values of $\xi$ and $\beta$, setting $\alpha = -1$ for all of the cases discussed in this section.

\paragraph{$\xi = 1$, $\beta = -1$: Chern insulator with band inversion at $\Gamma$}
In this case, we consider the system response to a pair of $\pi$-fluxes spatially separated in the $x$ direction.  As shown in SN~\ref{sec:fluxfluxtubetopologymapping}, the $\pi$-flux response is given in this case by the summed topologies of the $k_y = 0,\pi$ lines in the 2D BZ of $\mathcal{H}(\bs{k})$ (which can be summarized as the bulk Chern number $C\text{ mod }2$).  In the pristine insulating bulk, the $k_{y}=0$ ($k_{y}=\pi$) line characterizes a nontrivial (trivial) SSH chain [SEq.~\eqref{eq:dislocsshchain}], and the bulk correspondingly exhibits an inversion-symmetry-indicated Chern number $|C|=1$~\cite{QHZ}.  We observe a single midgap state at each $\pi$-flux core, which can be pinned to zero energy by the chiral symmetry in SEq.~\eqref{eq:chernchiralsym} or diagnosed via a filling anomaly~\cite{WiederAxion,HingeSM,WladCorners,AshvinFragile2,ZhidaFragileTwist1,ZhidaFragileTwist2}.  The bulk parity eigenvalues and the flux spectrum are respectively shown in SFig.~\ref{fig:2Dvortex}~$\bs{c},\bs{g}$.  The dislocation response of a $|C|=1$ Chern insulator driven by band inversion at $\Gamma$ was previously discussed in detail in SN~\ref{subsec:ChernAtGamma} -- further details of our numerical implementation are provided in that section.

\paragraph{$\xi = 1$, $\beta = +1$: Chern insulator with band inversion at $Y$}
As in the previous case of a $|C|=1$ Chern insulator driven by band inversion at $\Gamma$, because the summed topology of the $k_y = 0,\pi$ lines in the 2D BZ of $\mathcal{H}(\bs{k})$ is again nontrivial (and, equivalently, because the bulk Chern number is odd), then we observe a nontrivial flux response.  This underlines the fact that, in contrast to the dislocation response, the $\pi$-flux response of 2D and 3D insulators only depends on the strong topological index (SN~\ref{sec:kpFlux} and~\ref{sec:fluxfluxtubetopologymapping}), and in particular does not depend on where in the BZ bands are inverted.  The bulk parity eigenvalues and the flux spectrum are respectively shown in SFig.~\ref{fig:2Dvortex}~$\bs{d},\bs{g}$.  The dislocation response of a $|C|=1$ Chern insulator driven by band inversion at $Y$ was previously discussed in detail in SN~\ref{subsec:ChernAtY} -- further details of our numerical implementation are provided in that section.

\paragraph{$\xi = 0$: Weak SSH array}
The bulk parity eigenvalues and the flux spectrum are shown in SFig.~\ref{fig:2Dvortex}~$\bs{f},\bs{e}$.  Because the bulk carries a trivial Chern number, then the $\pi$-flux response is trivial.  The dislocation response of a weak array of $\mathcal{I}$-symmetric, $\mathcal{T}$-broken SSH chains was previously discussed in detail in SN~\ref{subsec:weakSSHstack} -- further details of our numerical implementation are provided in that section.

\subsubsection{3D Flux Tubes in the Absence of $\mathcal{T}$ Symmetry}
\label{sec:3Dvortex}

\begin{figure}[t]
\begin{center}
\includegraphics[width=0.7 \textwidth]{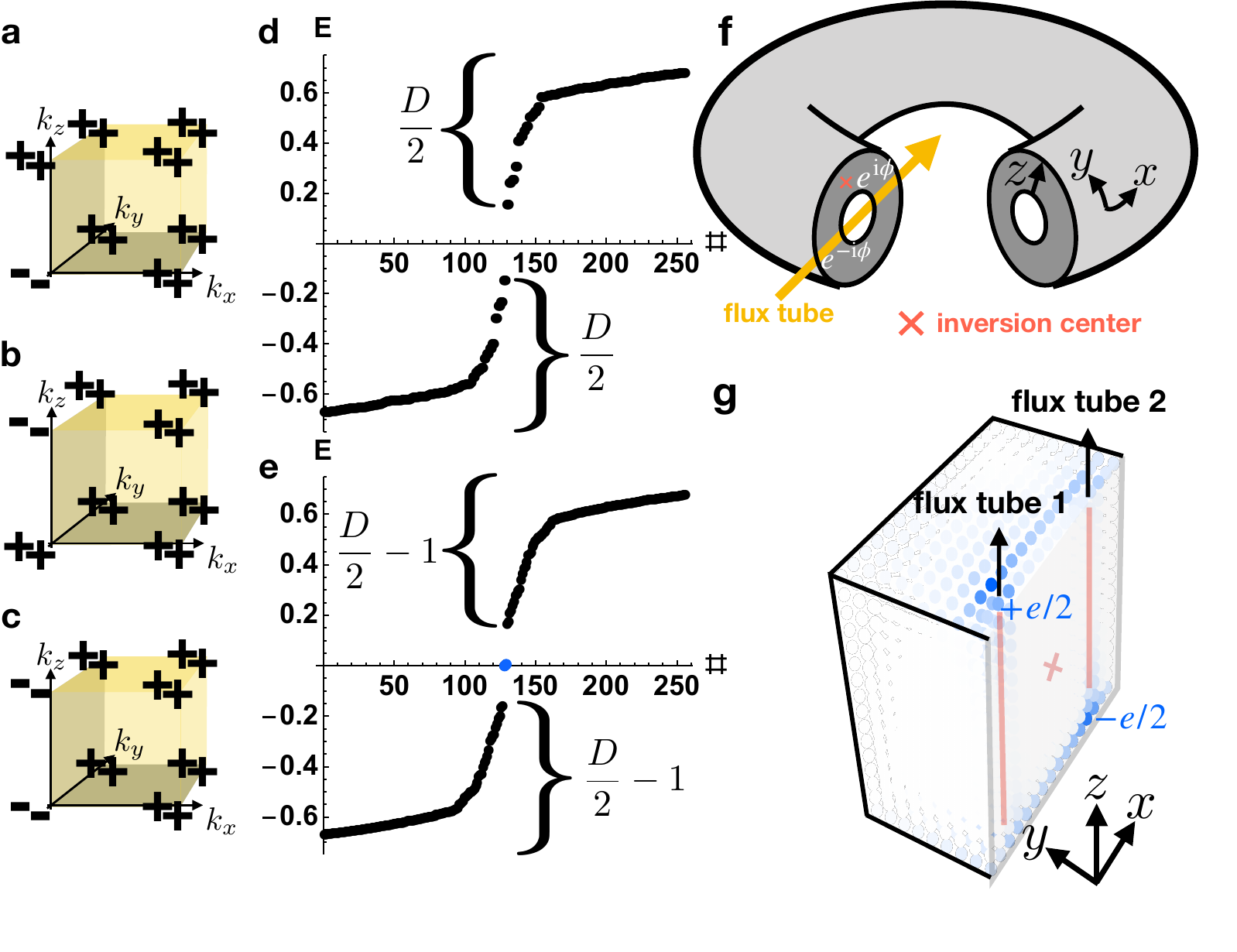}
\caption{{\bf 0D $\pi$-flux bound states in 3D insulators with inversion symmetry.} $\bs{a}$ -- $\bs{e}$ Bulk parity eigenvalues and HDBC energy spectra for the flux in SFig.~\ref{fig:2Dvortex}~$\bs{b}$, extended along the $z$ direction, when the crystal with  the BZ in SFig.~\ref{fig:3D}~$\bs{a}$ is equivalent to $\bs{a},\bs{e}$ an $\mathcal{I}$-symmetric axion insulator (AXI)~\cite{HOTIBernevig,HOTIChen,WladTheory,AxionZhida,EslamInversion,FanHOTI,TMDHOTI,HarukiLayers,AshvinMagnetic,EzawaMagneticHOTI,VDBHOTI,WiederAxion,KoreanAxion,NicoDavidAXI2,IvoAXI1,IvoAXI2} with double band inversion at $\Gamma$, $\bs{b},\bs{e}$ an AXI with double band inversion at $Z$, $\bs{c},\bs{d}$ a weak stack of $\mathcal{I}$-symmetric 2D FTIs with $\pm e/2$ corner charges~\cite{TMDHOTI,WiederAxion}.  Only the $256$ lowest-lying states are shown in $\bs{d},\bs{e}$.  Anomalous 0D (HEND) states with charge $\pm e/2$ are present on $\mathcal{I}$-related flux tube ends in $\bs{a}$ and $\bs{b}$, but not in $\bs{c}$.  The HEND states in $\bs{a}$ and $\bs{b}$ are equivalent to the corner states of an $\mathcal{I}$-symmetric, $\mathcal{T}$-broken 2D FTI~\cite{TMDHOTI,WiederAxion}, and thus persist under the relaxation of particle-hole symmetry~\cite{TMDHOTI,WiederAxion,HingeSM}.  This is consistent with the conclusions of SN~\ref{sec:kpFlux} and~\ref{sec:fluxfluxtubetopologymapping}, in which it was determined that only topological phases with nontrivial strong indices exhibit a nontrivial $\pi$-flux response.  $\bs{f}$ Hollow doughnut boundary conditions (HDBC) used to obtain the flux tube end states, which correspond to periodic boundary conditions in the $x$, $y$ directions and open boundary conditions in $z$ direction.  Crucially, the HDBC geometry does not have any hinges, such that, absent $\pi$-flux tubes, trivial insulators and AXIs exhibit the same gapped energy spectra when terminated in an HDBC geometry.  $\bs{g}$ Absolute square of the wavefunction of the midgap states in $\bs{e}$ on a $16 \times 16 \times 16$ lattice (cut in half along the transparent gray plane to expose the bound states, which appear on two of the four flux tube ends).  The inversion center in $\bs{g}$ is marked with a red $X$.  In $\bs{g}$, the bound state wavefunctions exhibit some residual localization on all four flux tube ends.  However, there is no symmetry that relates the two flux tube ends on each surface (as opposed to flux tube ends on opposing tubes and surfaces, which are conversely related by bulk $\mathcal{I}$).  Hence, in the presence of symmetry-allowed terms that break all artificial mirror reflection symmetries~\cite{TMDHOTI,KoreanFragile}, we expect that as the system size is increased, the two anomalous midgap states will more strongly localize on only two of the four flux tube ends.}
\label{fig:3Dvortex}
\end{center}
\end{figure}

In this section, we will demonstrate the $\pi$-flux response of $\mathcal{I}$-symmetric 3D insulators with broken $\mathcal{T}$ symmetry~\cite{TeoKaneDefect,QiDefect2,Vlad2D,QiFlux,AshvinFlux,AdyFlux,MirlinFlux,CorrelatedFlux,Radzihovsky2019}.  We begin by considering a pristine 3D insulator described by the Bloch Hamiltonian $\mathcal{H}(\bs{k})$ in SEq.~\eqref{eq:axioninsulatorfragile2Dmodel} of SN~\ref{sec:3Dscrew}.  We then numerically introduce a pair of $\pi$-flux tubes related by a bulk $\mathcal{I}$ center as discussed in SN~\ref{sec:fluxfluxtubetopologymapping}.  Each tube corresponds to a $\pi$-flux in each real-space plane indexed by $z$, or equivalently, at each momentum $k_z$.  As in SN~\ref{sec:3Dscrew}, we set $\alpha = \beta = -1$ in SEq.~\eqref{eq:axioninsulatorfragile2Dmodel}.  In our numerics, the flux tubes are implemented by following the procedure employed for the 2D insulators in SN~\ref{sec:2DVORTEXtoysubsec} in each $z$-indexed, real-space plane of the 3D insulators examined in this section.

We will now discuss the electronic structure of the Hamiltonian with two $\pi$-flux tubes $\tilde{H} (k_z)$ in detail for representative values of $\xi$ and $\gamma$.  The numerical results are summarized in SFig.~\ref{fig:3Dvortex}.

\paragraph{$\xi = 1$, $\gamma = -1$: Axion insulator with double band inversion at $\Gamma$}
In this case, we consider the system response to a pair of $z$-directed $\pi$-flux tubes that are spatially separated in the $x$ direction.  As shown in SN~\ref{sec:fluxfluxtubetopologymapping}, the $\pi$-flux response is given in this case by the summed topologies of the $k_y = 0,\pi$ planes in the 3D BZ of $\mathcal{H}(\bs{k})$ (which can be summarized by the value of the strong AXI index~\cite{WiederAxion}).  In the pristine insulating bulk, the $k_{y}=0$ ($k_{y}=\pi$) plane characterizes an $\mathcal{I}$-symmetric 2D FTI (trivial insulator), and the bulk correspondingly exhibits an inversion-symmetry-indicated nontrivial AXI index~\cite{HOTIBernevig,HOTIChen,WladTheory,AxionZhida,EslamInversion,FanHOTI,TMDHOTI,HarukiLayers,MTQC,MTQCmaterials,AshvinMagnetic,EzawaMagneticHOTI,VDBHOTI,WiederAxion,KoreanAxion,NicoDavidAXI2,IvoAXI1,IvoAXI2}.  We observe a single midgap state on two of four flux tube ends, which can be pinned to zero energy by the chiral symmetry in SEq.~\eqref{eq:axichiralsym} or diagnosed via a filling anomaly in the absence of chiral symmetry~\cite{WiederAxion,HingeSM,WladCorners,AshvinFragile2,ZhidaFragileTwist1,ZhidaFragileTwist2}.  The bulk parity eigenvalues and the flux spectrum are respectively shown in SFig.~\ref{fig:3Dvortex}~$\bs{a},\bs{e}$. The dislocation response of an AXI with double band inversion at $\Gamma$ was previously discussed in detail in SN~\ref{subsec:AXIatGamma} -- further details of our numerical implementation are provided in that section.

\paragraph{$\xi = 1$, $\gamma = +1$: Axion insulator with double band inversion at $Z$}
As in the previous case of an AXI driven by double band inversion at $\Gamma$, because the summed topology of the $k_y = 0,\pi$ planes in the 3D BZ of $\mathcal{H}(\bs{k})$ is again nontrivial (and, equivalently, because the bulk strong AXI index is nontrivial), then we observe a nontrivial flux response.  This further underlines the fact that, in contrast to the dislocation response, the $\pi$-flux response of 2D and 3D insulators only depends on the strong topological index (SN~\ref{sec:kpFlux} and~\ref{sec:fluxfluxtubetopologymapping}), and in particular does not depend on where bands are inverted.  The bulk parity eigenvalues and the flux spectrum are respectively shown in SFig.~\ref{fig:3Dvortex}~$\bs{b},\bs{e}$.  The dislocation response of an AXI with double band inversion at $Z$ was previously discussed in detail in SN~\ref{subsec:AXIatZ} -- further details of our numerical implementation are provided in that section.

\paragraph{$\xi = 0$: Weak FTI stack}
The bulk parity eigenvalues and the flux spectrum are respectively shown in SFig.~\ref{fig:3Dvortex}~$\bs{c},\bs{d}$.  Because the bulk carries a trivial strong (AXI) index, then the $\pi$-flux response is trivial.  The dislocation response of a weak stack of $\mathcal{I}$-symmetric, $\mathcal{T}$-broken 2D FTIs was previously discussed in detail in SN~\ref{subsec:weakFTIstack} -- further details of our numerical implementation are provided in that section.

\subsection{Flux Response with Time-Reversal Symmetry}
\subsubsection{2D Fluxes in the Presence of $\mathcal{T}$ Symmetry}
\label{sec:2DVORTEXtoysubsecTRS}

\begin{figure}[t]
\begin{center}
\includegraphics[width=0.7 \textwidth]{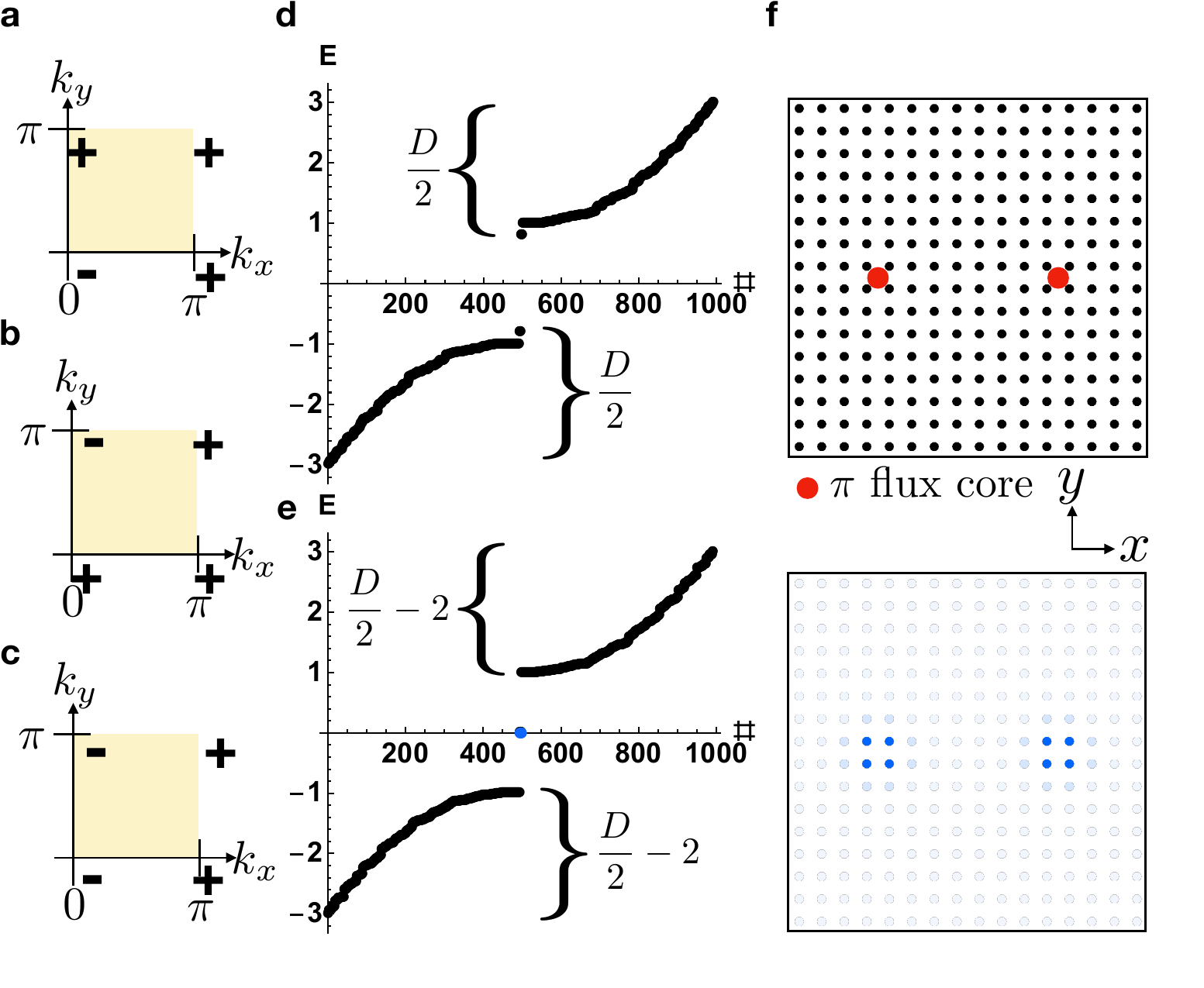}
\caption{{\bf 0D $\pi$-flux bound states in 2D insulators with time-reversal and inversion symmetry.} $\bs{a}$ -- $\bs{e}$ Bulk parity eigenvalues and PBC energy spectra for the flux in SFig.~\ref{fig:2Dvortex}~$\bs{b}$ when the crystal with the BZ in SFig.~\ref{fig:2Dvortex}~$\bs{a}$ is equivalent to $\bs{a},\bs{e}$ a 2D TI driven by band inversion at $\Gamma$, $\bs{b},\bs{e}$ a 2D TI driven by band inversion at Y, and $\bs{c},\bs{d}$ a weak $y$-directed array of $\mathcal{I}$- and $\mathcal{T}$-symmetric SSH chains~\cite{SSH,RiceMele,SSHspinon,HeegerReview,TRPolarization,YoungkukWeakSSH}.  In $\bs{a}$ -- $\bs{c}$, we denote Kramers pairs of parity eigenvalues with a single $\pm$ sign.  Anomalous 0D flux states with spin-charge separation~\cite{QiFlux,AshvinFlux,AdyFlux,MirlinFlux,CorrelatedFlux} are present in cases $\bs{a}$ and $\bs{b}$, but not in $\bs{c}$, and are equivalent to the end states of an $\mathcal{I}$- and $\mathcal{T}$-symmetric SSH chain~\cite{SSH,RiceMele,SSHspinon,HeegerReview,TRPolarization}.  With chiral symmetry, the two midgap Kramers pairs that are present in $\bs{a}$ and $\bs{b}$ are pinned to zero energy.  In the absence of chiral symmetry, a nontrivial $\pi$-flux response can still be diagnosed by observing a filling anomaly~\cite{WiederAxion,HingeSM,WladCorners,AshvinFragile2,ZhidaFragileTwist1,ZhidaFragileTwist2}.  $\bs{f}$ Real-space geometry with a pair of $\pi$-flux cores and absolute square of the wavefunction of the midgap states in $\bs{e}$ on a $16 \times 16$ lattice.}
\label{fig:2DvortexTRS}
\end{center}
\end{figure}

In this section, we will demonstrate the $\pi$-flux response of $\mathcal{I}$- and $\mathcal{T}$-symmetric 2D insulators~\cite{TeoKaneDefect,QiDefect2,Vlad2D,QiFlux,AshvinFlux,AdyFlux,MirlinFlux,CorrelatedFlux,Radzihovsky2019}.  We begin by considering a pristine 2D insulator described by the Bloch Hamiltonian $\mathcal{H}(\bs{k})$ in SEq.~\eqref{eq:cherninsulatorsshmodelTRS} of SN~\ref{sec:2DtoysubsecTRS}.  We then numerically introduce a pair of $\pi$-fluxes related by a bulk $\mathcal{I}$ center as discussed in SN~\ref{sec:fluxfluxtubetopologymapping} (SFig.~\ref{fig:2DvortexTRS}).  To obtain the numerical results presented in SFig.~\ref{fig:2DvortexTRS}, we use the same prescription as employed in SN~\ref{sec:2DVORTEXtoysubsec} to create a pair of $\pi$-fluxes.  We will now discuss the electronic structure of the resulting model $\tilde{H}$ in detail for $\alpha=-1$ and for representative values of $\xi$ and $\beta$.

\paragraph{$\xi = 1$, $\beta = -1$: 2D topological insulator with band inversion at $\Gamma$}
In this case, we consider the system response to a pair of $\pi$-fluxes spatially separated in the $x$ direction.  As shown in SN~\ref{sec:fluxfluxtubetopologymapping}, the $\pi$-flux response is given in this case by the summed topologies of the $k_y = 0,\pi$ lines in the 2D BZ of $\mathcal{H}(\bs{k})$ (which can be summarized as the bulk $\mathbb{Z}_{2}$ invariant~\cite{AndreiTI,CharlieTI,KaneMeleZ2}).  In the pristine insulating bulk, the $k_{y}=0$ ($k_{y}=\pi$) line characterizes a nontrivial (trivial) $\mathcal{T}$-symmetric SSH chain [SEq.~(\ref{eq:dislocsshchainTRS})], and the bulk correspondingly exhibits an inversion-symmetry-indicated nontrivial 2D TI $\mathbb{Z}_{2}$ Fu-Kane index~\cite{FuKaneMele,FuKaneInversion}.  We observe a Kramers pair of midgap states at each $\pi$-flux core, which can be pinned to zero energy by the chiral symmetry in SEq.~\eqref{eq:TIchiralsym} or diagnosed via a filling anomaly~\cite{WiederAxion,HingeSM,WladCorners,AshvinFragile2,ZhidaFragileTwist1,ZhidaFragileTwist2}.  The bulk parity eigenvalues and the flux spectrum are respectively shown in SFig.~\ref{fig:2DvortexTRS}~$\bs{a},\bs{e}$.  The dislocation response of a 2D TI with band inversion at $\Gamma$ was previously discussed in detail in SN~\ref{subsec:TIatGamma} -- further details of our numerical implementation are provided in that section.

\paragraph{$\xi = 1$, $\beta = +1$: 2D topological insulator with band inversion at $Y$}
 As in the previous case of a 2D TI driven by band inversion at $\Gamma$, because the summed topology of the $k_y = 0,\pi$ lines in the 2D BZ of $\mathcal{H}(\bs{k})$ is again nontrivial (and, equivalently, because the bulk 2D TI index is nontrivial), then we observe a nontrivial flux response.  This further underlines the fact that, in contrast to the dislocation response, the $\pi$-flux response of 2D and 3D insulators only depends on the strong topological index (SN~\ref{sec:kpFlux} and~\ref{sec:fluxfluxtubetopologymapping}), and in particular does not depend on where bands are inverted.  The bulk parity eigenvalues and the flux spectrum are respectively shown in SFig.~\ref{fig:2DvortexTRS}~$\bs{b},\bs{e}$.  The dislocation response of a 2D TI with band inversion at $Y$ was previously discussed in detail in SN~\ref{subsec:TIatY} -- further details of our numerical implementation are provided in that section.

\paragraph{$\xi = 0$: Weak $\mathcal{T}$-symmetric SSH array}
The bulk parity eigenvalues and the flux spectrum are respectively shown in SFig.~\ref{fig:2DvortexTRS}~$\bs{c},\bs{d}$.  Because the bulk carries a trivial 2D TI index, then the $\pi$-flux response is trivial.  The dislocation response of a weak array of $\mathcal{I}$- and $\mathcal{T}$-symmetric SSH chains was previously discussed in detail in SN~\ref{subsec:weakTRSSSHstack} -- further details of our numerical implementation are provided in that section.

\subsubsection{3D Flux Tubes in the Presence of $\mathcal{T}$ Symmetry}
\label{sec:3DvortexTRS}

\begin{figure}[t]
\begin{center}
\includegraphics[width=0.75 \textwidth]{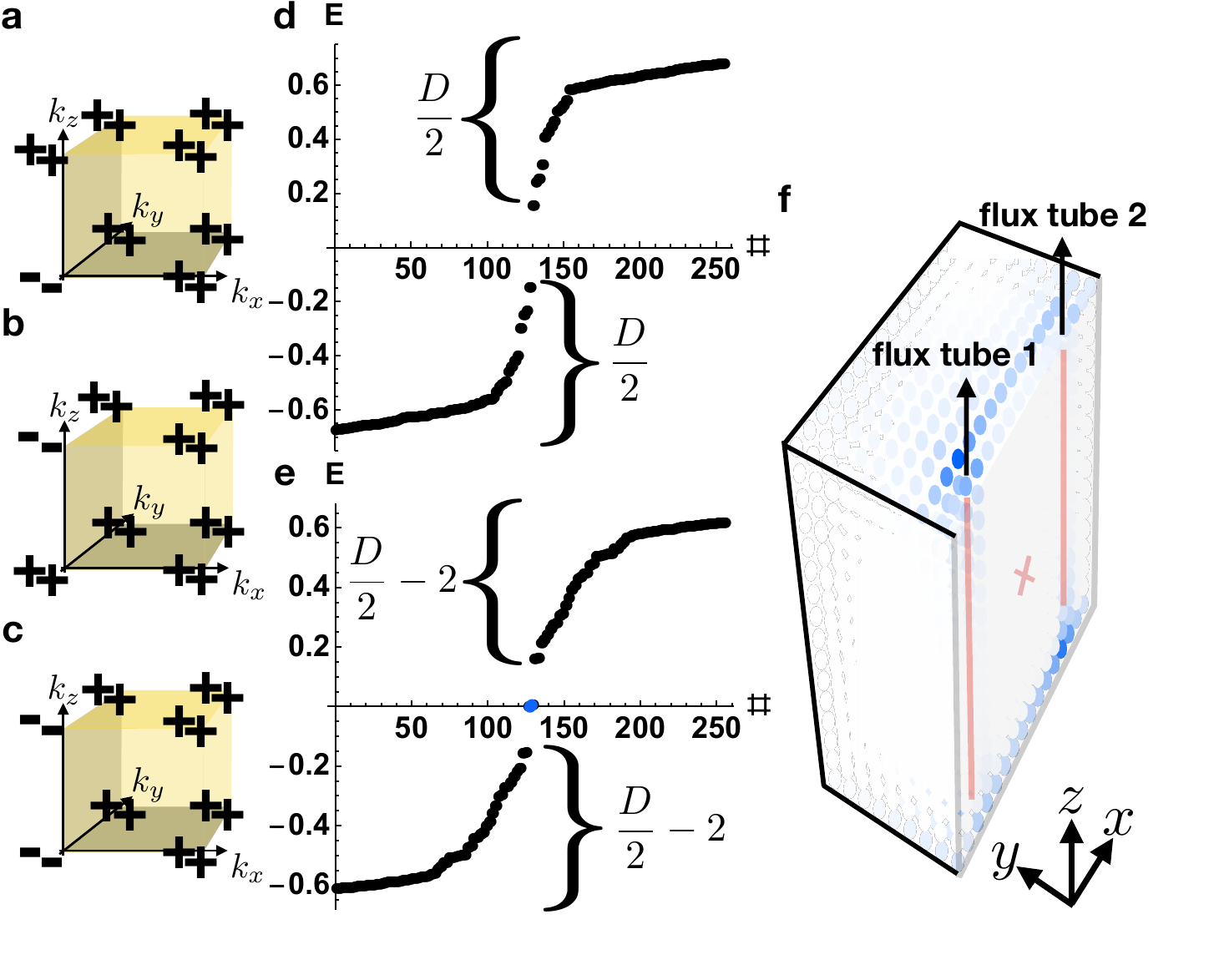}
\caption{{\bf 0D $\pi$-flux bound states in 3D insulators with time-reversal and inversion symmetry.} $\bs{a}$ -- $\bs{e}$ Bulk parity eigenvalues and HDBC energy spectra for the flux in SFig.~\ref{fig:2Dvortex}~$\bs{b}$, extended along the $z$ direction, when the crystal with the BZ shown in SFig.~\ref{fig:3D}~$\bs{a}$ is equivalent to $\bs{a},\bs{e}$ an $\mathcal{I}$- and $\mathcal{T}$-symmetric HOTI driven by double band inversion at $\Gamma$, $\bs{b},\bs{e}$ a HOTI with double band inversion at $Z$, and $\bs{c},\bs{d}$ a weak stack of $\mathcal{I}$- and $\mathcal{T}$-symmetric 2D FTIs~\cite{TMDHOTI} with corner midgap Kramers pairs.  In $\bs{a}$ -- $\bs{c}$, we denote Kramers pairs of parity eigenvalues with a single $\pm$ sign.  Only the $256$ lowest-lying states are shown in $\bs{d},\bs{e}$.  Anomalous 0D (HEND) states with spin-charge separation~\cite{QiFlux,AshvinFlux,AdyFlux,MirlinFlux,CorrelatedFlux} are present on $\mathcal{I}$-related flux tube ends in $\bs{a}$ and $\bs{b}$, but not in $\bs{c}$.  The HEND states in $\bs{a}$ and $\bs{b}$ are equivalent to the corner states of an $\mathcal{I}$- and $\mathcal{T}$-symmetric 2D FTI~\cite{TMDHOTI,WiederAxion}, and thus persist under the relaxation of particle-hole symmetry~\cite{TMDHOTI,WiederAxion,HingeSM}.  This is consistent with the conclusions of SN~\ref{sec:kpFlux} and~\ref{sec:fluxfluxtubetopologymapping}, in which it was determined that only topological phases with nontrivial strong indices exhibit a nontrivial $\pi$-flux response.  With chiral symmetry, the two midgap Kramers pairs that are present in cases $\bs{a}$ and $\bs{b}$ are pinned to zero energy.  In the absence of chiral symmetry, the $\pi$-flux-tube response can still be diagnosed as topologically nontrivial by observing a filling anomaly~\cite{WiederAxion,HingeSM,WladCorners,AshvinFragile2,ZhidaFragileTwist1,ZhidaFragileTwist2}.  $\bs{f}$ Absolute square of the midgap states in $\bs{e}$ on a $16 \times 16 \times 16$ lattice (cut in half along the transparent gray plane to expose the bound states, which appear on two of the four flux tube ends).  The inversion center in $\bs{f}$ is marked with a red $X$.  In $\bs{f}$, the bound state wavefunctions exhibit some residual localization on all four flux tube ends.  However, there is no symmetry that relates the two flux tube ends on each surface (as opposed to flux tube ends on opposing tubes and surfaces, which are conversely related by bulk $\mathcal{I}$).  Hence, in the presence of symmetry-allowed terms that break all artificial mirror reflection symmetries~\cite{TMDHOTI,KoreanFragile}, we expect that as the system size is increased, the two anomalous Kramers pairs of midgap states will more strongly localize on only two of the four flux tube ends.}
\label{fig:3DvortexTRS}
\end{center}
\end{figure}

In this section, we will demonstrate the $\pi$-flux response of 3D insulators with $\mathcal{I}$ and $\mathcal{T}$ symmetry~\cite{TeoKaneDefect,QiDefect2,Vlad2D,QiFlux,AshvinFlux,AdyFlux,MirlinFlux,CorrelatedFlux,Radzihovsky2019}.  We begin by considering a pristine 3D insulator described by the Bloch Hamiltonian $\mathcal{H}(\bs{k})$ in SEq.~\eqref{eq:axioninsulatorfragile2DmodelTRS} of SN~\ref{sec:3DscrewTRSsec}.  We then numerically introduce a pair of $\pi$-flux tubes related by a bulk $\mathcal{I}$ center as detailed in SN~\ref{sec:3Dvortex}.  As in SN~\ref{sec:3DscrewTRSsec}, we set $\alpha = \beta = -1$ in SEq.~\eqref{eq:axioninsulatorfragile2DmodelTRS}.  We will now discuss the electronic structure of the Hamiltonian with two $\pi$-flux tubes $\tilde{H} (k_z)$ in detail for representative values of $\xi$ and $\gamma$.  The numerical results are summarized in SFig.~\ref{fig:3DvortexTRS}.

\paragraph{$\xi = 1$, $\gamma = -1$: Higher-order topological insulator with double band inversion at $\Gamma$}
In this case, we consider the system response to a pair of $z$-directed $\pi$-flux tubes that are spatially separated in the $x$ direction.  As shown in SN~\ref{sec:fluxfluxtubetopologymapping}, the $\pi$-flux response is given in this case by the summed topologies of the $k_y = 0,\pi$ planes in the 3D BZ of $\mathcal{H}(\bs{k})$ (which can be summarized by the value of the strong $\mathbb{Z}_{4}$ HOTI index~\cite{WiederAxion,TMDHOTI}).  In the pristine insulating bulk, the $k_{y}=0$ ($k_{y}=\pi$) plane characterizes an $\mathcal{I}$- and $\mathcal{T}$-symmetric 2D FTI (trivial insulator) [SEq.~(\ref{eq:disloc2DphaseTRS})], and the bulk correspondingly exhibits an inversion-symmetry-indicated nontrivial $\mathbb{Z}_{4}$ HOTI index~\cite{AshvinIndicators,AshvinTCI,ChenTCI,TMDHOTI}.  We observe a Kramers pair of midgap states on two of four flux tube ends, which can be pinned to zero energy by the chiral symmetry in SEq.~\eqref{eq:HOTIchiralsym} or diagnosed via a filling anomaly in the absence of chiral symmetry~\cite{WiederAxion,HingeSM,WladCorners,AshvinFragile2,ZhidaFragileTwist1,ZhidaFragileTwist2}.  The bulk parity eigenvalues and the flux spectrum are respectively shown in SFig.~\ref{fig:3DvortexTRS}~$\bs{a},\bs{e}$. The dislocation response of an $\mathcal{I}$- and $\mathcal{T}$-symmetric HOTI with double band inversion at $\Gamma$ was previously discussed in detail in SN~\ref{subsec:HOTIatGamma} -- further details of our numerical implementation are provided in that section.

\paragraph{$\xi = 1$, $\gamma = +1$: HOTI with double band inversion at $Z$}
As in the previous case of a HOTI driven by double band inversion at $\Gamma$, because the summed topology of the $k_y = 0,\pi$ planes in the 3D BZ of $\mathcal{H}(\bs{k})$ is again nontrivial (and, equivalently, because the bulk is a symmetry-indicated HOTI~\cite{WiederAxion,TMDHOTI}), then we observe a nontrivial flux response.  This further underlines the fact that, in contrast to the dislocation response, the $\pi$-flux response of 2D and 3D insulators only depends on the strong topological index (SN~\ref{sec:kpFlux} and~\ref{sec:fluxfluxtubetopologymapping}), and in particular does not depend on where bands are inverted.  The bulk parity eigenvalues and the flux spectrum are respectively shown in SFig.~\ref{fig:3DvortexTRS}~$\bs{b},\bs{e}$.  The dislocation response of an $\mathcal{I}$- and $\mathcal{T}$-symmetric HOTI with double band inversion at $Z$ was previously discussed in detail in SN~\ref{subsec:HOTIatZ} -- further details of our numerical implementation are provided in that section.

\paragraph{$\xi = 0$: Weak $\mathcal{T}$-symmetric FTI stack}
The bulk parity eigenvalues and the flux spectrum are respectively shown in SFig.~\ref{fig:3DvortexTRS}~$\bs{c},\bs{d}$.  Because the bulk carries trivial strong TI and HOTI indices, then the $\pi$-flux response is trivial.  The dislocation response of a weak stack of $\mathcal{I}$- and $\mathcal{T}$-symmetric 2D FTIs was previously discussed in detail in SN~\ref{subsec:weakTRSFTIstack} -- further details of our numerical implementation are provided in that section.

\section{First-Principles Calculation Details}
\label{sec:DFT}

\subsection{2D PbTe Monolayers}
\label{sec:DFTPbTe}

\begin{figure}[b]
\includegraphics[width=0.95\textwidth]{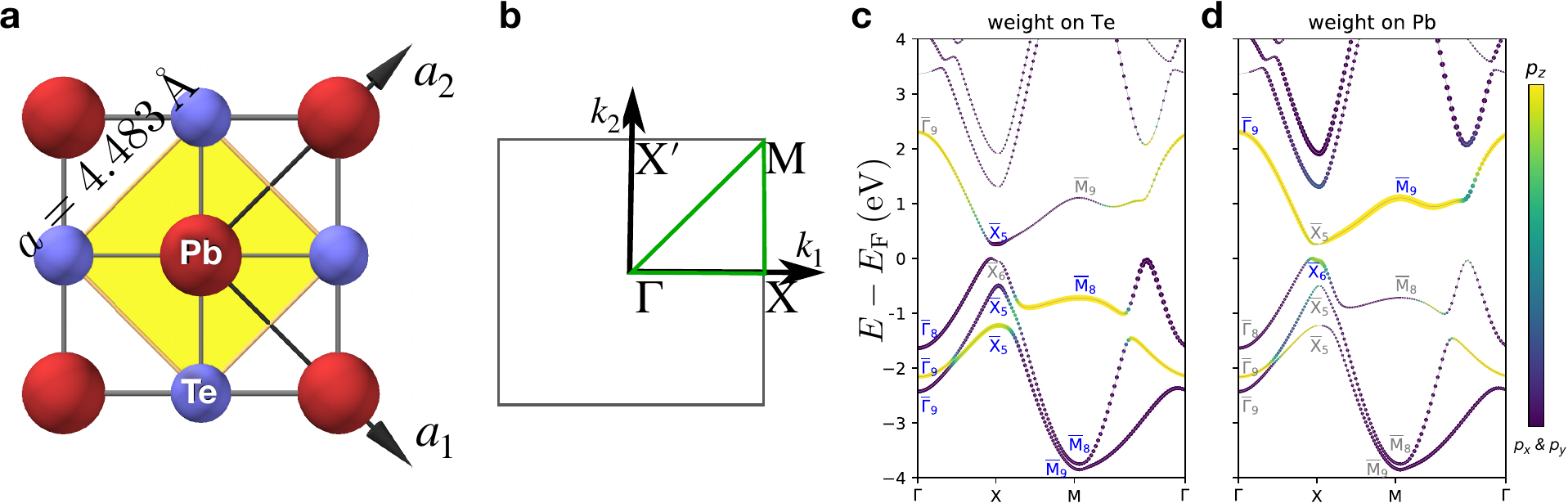}
\caption{{\bf Bulk electronic structure of monolayer PbTe.} $\bs{a}$~The crystal structure of monolayer PbTe in layer group (LG) $p4/mmm1'$~\cite{BigBook,MagneticBook,subperiodicTables,WiederLayers,SteveMagnet,DiracInsulator,HingeSM}.  The primitive (unit) cell is highlighted in yellow.  $\bs{b}$~The BZ of LG $p4/mmm1'$.  $\bs{c,d}$~The band structure of monolayer PbTe calculated from first principles; bands at the TRIM points are labeled with the irreducible small corepresentations of $p4/mmm1'$ using the convention established in the~\href{http://www.cryst.ehu.es/cgi-bin/cryst/programs/representations.pl?tipogrupo=dbg}{REPRESENTATIONS DSG} tool on the BCS~\cite{QuantumChemistry,Bandrep1} for SG 123 $P4/mmm1'$, the index-2 space supergroup of LG $p4/mmm1'$ generated by adding lattice translations in the $z$ direction.  Bands at each $k$ point are doubly degenerate due to the presence of $\mathcal{I}\times\mathcal{T}$ symmetry~\cite{WiederLayers}.  The relative projected spectral weight of the Te (Pb) $p_{z}$ and $p_{x,y}$ orbitals nearest the Fermi energy is respectively indicated with yellow and blue, as shown in the color bar to the right of panel $\bs{d}$.  The band structures and spectral weights in $\bs{c}$ and $\bs{d}$ indicate that PbTe monolayers differ from an unobstructed atomic limit by a band inversion between Bloch states with oppositely-signed Kramers pairs of parity eigenvalues [SEq.~(\ref{eq:parityCharacter})] at the $X$ [${\bs k}_{X}={\bs b}_{1}/2$] and $X'$ [${\bs k}_{X'}=C_{4z}{\bs k}_{X}\text{ mod }{\bs b}_{1}\text{ mod }{\bs b}_{2}= {\bs b}_{2}/2$] points.}
\label{fig:PbTe-dft} 
\end{figure}

In this section, we will detail our first-principles and tight-binding calculations confirming a nontrivial dislocation response in 2D lead telluride (PbTe) monolayers.  PbTe monolayers (SFig.~\ref{fig:PbTe-dft}~$\bs{a}$) have theoretically been predicted to be 2D TCIs~\cite{PbTeMonolayer1,PbTeMonolayer2,LiangTCIMonolayer} 
with mirror Chern number~\cite{TeoFuKaneTCI} $C_{M_{z}} = 2$.  Previous works~\cite{PbTeMonolayer1,PbTeMonolayer2,LiangTCIMonolayer} have specifically determined that the nontrivial crystalline topology of PbTe monolayers is driven by band inversion at two ${\bs k}$ points along the 2D BZ boundary (SFig.~\ref{fig:PbTe-dft}~$\bs{b}$), suggesting the possibility of a nontrivial point dislocation response (see SN~\ref{sec:weakSSH}).  In this section, we will use first-principles calculations to reproduce the electronic structure and topology of PbTe monolayers, which we will then use to determine the electronic response to point dislocations.

We begin by calculating the electronic structure of a PbTe monolayer.  To obtain the crystal structure of a single, pristine monolayer, we start with a 3D crystal of rock-salt-structure PbTe [SG 225 $Fm\bar{3}m1'$, Inorganic Crystal Structure Database (ICSD)~\cite{ICSD} No. 194220, further details available at~\url{https://topologicalquantumchemistry.com/#/detail/194220}~\cite{QuantumChemistry,AndreiMaterials,AndreiMaterials2,BCS1,BCS2}], increase the lattice spacing in the $z$ ($c$-axis) direction to isolate a single plane of Pb and Te atoms (SFig.~\ref{fig:PbTe-dft}~$\bs{a}$), and then restrict the system symmetry to layer group (LG)~\cite{BigBook,MagneticBook,subperiodicTables,WiederLayers,SteveMagnet,DiracInsulator,HingeSM} $p4/mmm1'$.  Notably, 3D PbTe crystals, depending on the choice of simulation parameters and initial structure (indexed by ICSD number), have been shown to be  ``rotation-anomaly'' TCIs with helical hinge states~\cite{ChenRotation,HOTIChen,HOTIBernevig,AndreiMaterials,AndreiMaterials2,BarryPbTe}.  We next perform fully relativistic density functional theory (DFT) calculations of the electronic structure using the Vienna Ab initio Simulation Package (VASP)~\cite{VASP1,VASP2} employing the projector-augmented wave (PAW) method~\cite{PAW1,PAW2} and the Perdew, Burke, and Ernzerhof generalized-gradient approximation (GGA-PBE)~\cite{PBE-1996} for the exchange-correlation functional.  In our first-principles calculations, we have used the primitive unit cell shown in SFig.~\ref{fig:PbTe-dft}~$\bs{a}$, which contains one Pb atom at $(x,y) = (0,0)$ and one Te atom at $(1/2,0)$.  The lattice vectors of the primitive cell (see SFig.~\ref{fig:PbTe-dft}~$\bs{a}$) are given by:
\begin{equation}
\bs{a}_1 = (1/2,-1/2),\ \bs{a}_2 = (1/2,1/2),
\end{equation} 
and the reciprocal lattice vectors are given by:
\begin{equation}
\bs{b}_1 = 2\pi(1,-1),\ \bs{b}_2 = 2\pi(1,1). 
\label{eq:monolayerGvecs}
\end{equation}
Lastly, we have allowed the in-plane lattice spacing $a_{1}=a_{2}=a$ to relax from its experimental value to an equilibrium length of $a=4.483$ \AA.

In SFig.~\ref{fig:PbTe-dft}~$\bs{c,d}$, we plot the electronic structure of a PbTe monolayer.  The bands at each $k$ point are doubly degenerate due to the presence of spinful $\mathcal{I}\times\mathcal{T}$ symmetry~\cite{WiederLayers}.  We find that the $12$ bands closest to the Fermi energy ($E_{F}$) are induced from the $5p$ orbitals of Te and the $6p$ orbitals of Pb.  To determine the location in momentum space and associated irreducible small corepresentations (coreps) of the bulk band inversion, we construct maximally-localized, symmetric Wannier functions (MLWFs)~\cite{MarzariPRB,MarzariReview} using~\textsc{Wannier90}~\cite{Mostofi2008,Mostofi2014}.  Specifically, to capture the six valence bands and two conduction bands closest to $E_{F}$, we construct MLWFs using Kramers pairs of $5p_{x,y,z}$ orbitals from Te and Kramers pairs of $6p_{z}$ orbitals from Pb.  In SFig.~\ref{fig:PbTe-dft}~$\bs{c}$ ($\bs{d}$) we show the electronic structure of a PbTe monolayer with the projected weight of the states at each $k$ point onto the twelve total $5p$ Te and $6p$ Pb orbitals.  We observe that the bands closest to $E_{F}$ are well captured by the eight orbitals (Kramers pairs of $5p_{x,y,z}$ Te orbitals and $6p_{z}$ Pb orbitals) used to construct the MLWFs, and that the bulk differs from an obstructed atomic limit by band inversion at the $X$ point between states from the $6p_{z}$ orbitals of Pb and the $5p_{x,y}$ orbitals of Te.

To determine the topology of the PbTe monolayer, we use the~\href{https://github.com/stepan-tsirkin/irrep}{IrRep} program~\cite{StepanIrRepCode} to deduce the small coreps of the six highest valence and the two lowest conduction bands (\emph{i.e.} the bands previously selected to construct MLWFs), which are shown in SFig.~\ref{fig:PbTe-dft}~$\bs{c,d}$ and labeled employing the convention of the~\href{http://www.cryst.ehu.es/cgi-bin/cryst/programs/representations.pl?tipogrupo=dbg}{REPRESENTATIONS DSG} tool on the BCS~\cite{QuantumChemistry,Bandrep1} for the $k_{z}=0$ plane of SG 123 $P4/mmm1'$, the index-2 supergroup of LG $p4/mmm1'$ generated by adding lattice translations in the $z$ direction.  Layer group $p4/mmm1'$, because of the presence of bulk $C_{4z}$ and $M_{z}$ symmetries, supports a symmetry-indicated mirror Chern number~\cite{ChenBernevigTCI}:
\begin{equation}
\begin{aligned}
C_{M_{z}}\text{ mod } 4 = 
& \frac{3}{2}n\left(\bar{\Gamma}_{6}\right)
-\frac{3}{2}n\left(\bar{\Gamma}_{8}\right)
-\frac{1}{2}n\left(\bar{\Gamma}_{7}\right)
+\frac{1}{2}n\left(\bar{\Gamma}_{9}\right)\\
&+\frac{3}{2}n\left(\bar{M}_{6}\right)
-\frac{3}{2}n\left(\bar{M}_{8}\right)
-\frac{1}{2}n\left(\bar{M}_{7}\right)
+\frac{1}{2}n\left(\bar{M}_{9}\right)\\
&+n\left(\bar{X}_{5}\right)
-n\left(\bar{X}_{6}\right)\text{ mod }4.
\label{eq:c4LG}
\end{aligned}
\end{equation}
The six highest valence bands in energy in SFig.~\ref{fig:PbTe-dft}~$\bs{c,d}$ are separated from the other valence bands by a large energy gap, and exhibit the small corep multiplicities:
\begin{eqnarray}
n\left(\bar{\Gamma}_{6}\right)&=&n\left(\bar{\Gamma}_{7}\right)=0,\ n\left(\bar{\Gamma}_{8}\right)=1,\ n\left(\bar{\Gamma}_{9}\right)=2, \nonumber \\
n\left(\bar{M}_{6}\right)&=&n\left(\bar{M}_{7}\right)=0,\ n\left(\bar{M}_{8}\right)=2,\ n\left(\bar{M}_{9}\right)=1, \nonumber \\
n\left(\bar{X}_{5}\right) &=& 2,\ n\left(\bar{X}_{6}\right) = 1.
\label{eq:numCoreps2D}
\end{eqnarray}
Substituting SEq.~(\ref{eq:numCoreps2D}) into SEq.~(\ref{eq:c4LG}), we determine that PbTe monolayers exhibit:
\begin{equation}
C_{M_{z}}\text{ mod }4 = 2,
\end{equation}
in agreement with the results of SRefs.~\onlinecite{PbTeMonolayer1,PbTeMonolayer2,LiangTCIMonolayer}.

Next, to determine the dislocation response of PbTe monolayers, we calculate the weak (partial) SSH invariant vector $\bs{M}_{\nu}^{\mathrm{SSH}}$, which is defined in the text surrounding SEq.~(\ref{eq:SSHinvariant}).  $\bs{M}_{\nu}^{\mathrm{SSH}}$ can be obtained by counting the number of parity-eigenvalue-exchanging band inversions by which a set of bands differs from an unobstructed (trivial) atomic limit with a trivial dislocation response.  As shown in SFig.~\ref{fig:PbTe-dft}~$\bs{c,d}$, PbTe monolayers differ from an unobstructed atomic limit through band inversion at the $X$ point [${\bs k}_{X}={\bs b}_{1}/2 = (\pi,-\pi)$] between bands labeled by the small coreps $\bar{X}_{5,6}$ of the little group at $X$.  The small coreps $\bar{X}_{5,6}$ correspond to doubly-degenerate pairs of states with the same parity ($\mathcal{I}$) eigenvalues within each pair, such that:
\begin{equation}
\chi_{\bar{X}_{5}}(\mathcal{I})=2,\ \chi_{\bar{X}_{6}}(\mathcal{I})=-2,
\label{eq:parityCharacter}
\end{equation}
where $\chi_{\rho}(h)$ is the character of the unitary symmetry $h$ in the corep $\rho$, and is equal to the sum of the eigenvalues of $h$ in $\rho$.  Because the $X$ and symmetry-equivalent $X'$ [${\bs k}_{X'}=C_{4z}{\bs k}_{X}\text{ mod }{\bs b}_{1}\text{ mod }{\bs b}_{2}={\bs b}_{2}/2 = (\pi,\pi)$] points lie along the BZ-edge $XM$ and $X'M$ lines, then, following the discussion in SN~\ref{sec:weakSSH}, we conclude that PbTe monolayers exhibit a nontrivial weak partial (time-reversal) SSH invariant vector:
\begin{equation}
\bs{M}_{\nu}^{\mathrm{SSH}} = \frac{1}{2}(\bs{b}_1 + \bs{b}_2) = (2\pi,0).
\label{eq:PbTeWeakIndex}
\end{equation}
We emphasize that, despite $\nu^{\mathrm{SSH}}_{x}\text{ mod }2\pi = \nu^{\mathrm{SSH}}_{y}\text{ mod }2\pi = 0$ in SEq.~(\ref{eq:PbTeWeakIndex}), $\bs{M}_{\nu}^{\mathrm{SSH}}$ is still nontrivial, because $(2\pi,0)$ and $(0,2\pi)$ are \emph{not} reciprocal lattice vectors [SEq.~(\ref{eq:monolayerGvecs})] in the rotated coordinates used in this section (which we have chosen because, as will be discussed in SN~\ref{sec:DFTSnTe}, the Bravais lattice of 3D PbTe is face-centered cubic, see SFig.~\ref{fig:PbTe-dft}~$\bs{a,b}$).

\begin{figure}[t]
\begin{center}
\includegraphics[width=\textwidth]{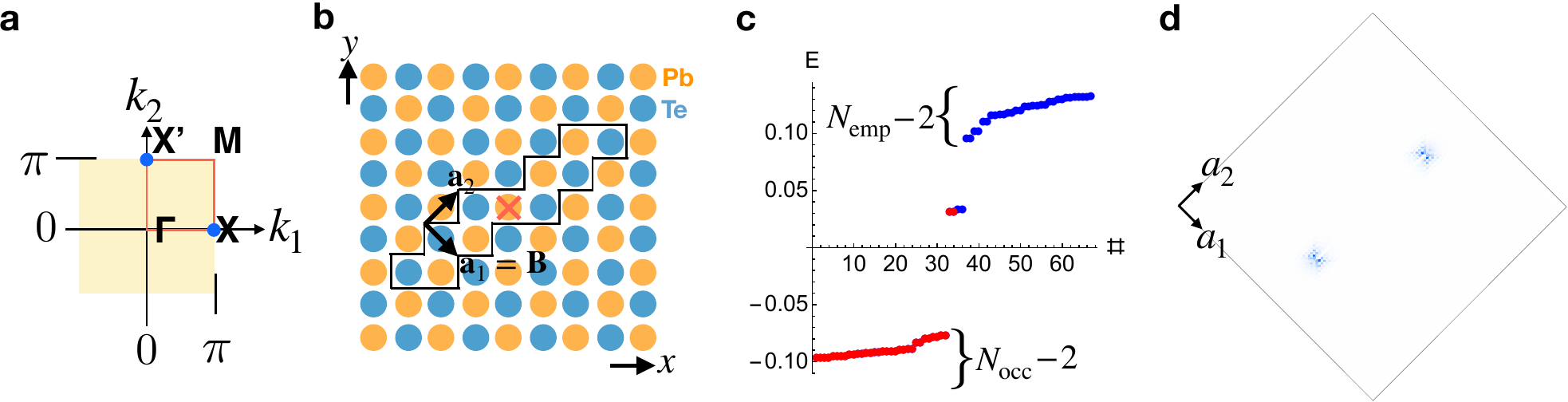}
\caption{{\bf Dislocation response of a PbTe monolayer.} $\bs{a}$~Brillouin zone (BZ) of a pristine monolayer of PbTe, which respects layer group (LG) $p4/mmm1'$~\cite{BigBook,MagneticBook,subperiodicTables,WiederLayers,SteveMagnet,DiracInsulator,HingeSM}.  Focusing on the six highest valence bands and the two lowest conduction bands in SFig.~\ref{fig:PbTe-dft}~$\bs{c,d}$, the highest (lowest) doubly degenerate pair of valence (conduction) bands is inverted at the $X$ and symmetry-equivalent $X'$ points (blue dots in~$\bs{a}$).  The band inversion exchanges Kramers pairs of states with opposite parity eigenvalues, driving the monolayer to exhibit nontrivial weak (partial) SSH indices $\bs{M}_{\nu}^{\mathrm{SSH}} = ({\bs b}_{1} + {\bs b}_{2})/2$ [see SN~\ref{sec:weakSSH} and the text surrounding SEq.~(\ref{eq:PbTeWeakIndex})].  $\bs{b}$~Schematic of the real-space implementation of an $\mathcal{I}$-related pair of $\bs{B} = \bs{a}_1$ point dislocations in a tight-binding model of a PbTe monolayer obtained from first-principles calculations, where the $\mathcal{I}$ center is marked with a red $\times$ symbol.  The black line in $\bs{b}$ encloses atoms with a fixed average value of the $a_1$ coordinate that were removed to create the pair of dislocations.  $\bs{c}$~Energy spectrum of a tight-binding model of PbTe with the $\mathcal{I}$-related pair of $\bs{B} = \bs{a}_1$ point dislocations shown in $\bs{b}$ and periodic boundary conditions.  The filling $N_\text{occ}$ is set by assigning six occupied electrons to each Te atom in the system, corresponding to the six occupied Te $p$ orbitals used to construct our Wannier-based tight-binding model [see the text following SEq.~(\ref{eq:monolayerGvecs})].
$\bs{d}$~The real-space localization of the four midgap states.  One Kramers pair of states is localized on each dislocation core, and, when half-filled, corresponds to a chargeless, spin-1/2 quasiparticle (\emph{i.e.} a spinon) that is equivalent to the end state of a spinful SSH chain (SN~\ref{sec:kp}).  The white dislocation line in $\bs{d}$ does not carry any spectral weight because it represents a line of removed sites (\emph{i.e.}, the atoms surrounded by black lines in panel $\bs{b}$).  Because the dislocation geometry in $\bs{b}$ preserves $\mathcal{I}$ and spinful $\mathcal{T}$ symmetries, then the spectrum in $\bs{c}$ cannot be deformed into that of a trivial insulator with dislocations, and is thus \emph{filling-anomalous}~\cite{TMDHOTI,WiederAxion,WladCorners,HingeSM}.  The presence of anomalous chargeless spin bound to dislocations is consistent with SEq.~(\ref{eq:dislocationtheory}) and with the discussion in SN~\ref{sec:weakSSH}, in which it was determined that 2D insulators with nontrivial weak (partial) SSH indices $\bs{M}^{\mathrm{SSH}}_{\nu}$ exhibit a filling-anomalous energy spectrum in the presence of point dislocations that satisfy $\bs{B}\cdot \bs{M}^{\mathrm{SSH}}_{\nu}\text{ mod }2\pi = \pi$.}
\label{fig:PbTe}
\end{center}
\end{figure}

To confirm the nontrivial dislocation response of a PbTe monolayer, we will now insert a pair of 0D dislocations into the previous eight-band tight-binding model obtained from MLWFs through~\textsc{Wannier90}~\cite{Mostofi2008,Mostofi2014} [detailed in the text preceding SEq.~(\ref{eq:c4LG})].  In practice, when mapping a DFT calculation to a tight-binding model, one must choose a cutoff distance for hopping interactions.  Surprisingly, even though the band inversion in PbTe monolayers is relatively strong (the negative band gap at the $X$ and $X'$ points is roughly $\sim 260$ meV)~\cite{PbTeMonolayer1,PbTeMonolayer2,LiangTCIMonolayer}, we find that the strong and weak partial-polarization topology of a PbTe monolayer is only reproduced in a tight-binding model that is truncated to a minimum range of \emph{sixth-nearest-neighbor} hopping.

We next place the tight-binding model on periodic boundary conditions (PBC).  Because the tight-binding model has six valence states and two conduction states per unit cell, then the PBC energy spectrum exhibits a valence (conduction) manifold with $6N/8$ ($2N/8$) states (\emph{i.e.}, the spectrum is gapped at an electronic filling fraction $\nu/N=3/4$).  To probe the dislocation response, we insert a pair of 0D dislocations with Burgers vector $\bs{B} = \bs{a}_1$, which satisfies $\bs{B} \cdot \bs{M}^{\mathrm{SSH}}_{\nu} \text{ mod }2\pi = \pi$ [SEq.~(\ref{eq:PbTeWeakIndex})].  We take the filling fraction $\nu/N=3/4$ to be derived from three occupied Kramers pairs of electrons per Te atom, corresponding to the six occupied Te $p$ orbitals used to construct our Wannier-based tight-binding model [see the text following SEq.~(\ref{eq:monolayerGvecs})].  We then denote the corresponding number of filled states in the defect geometry as $N_{\text{occ}}$.  In terms of the tight-binding model used to demonstrate the dislocation response, prior to inserting the dislocations, we have employed a PBC geometry with $71 \times 71$ unit cells.  We next implement the dislocations by removing the atoms surrounding a line $M$ (defined in SN~\ref{sec:2Dtoysubsec}) of a linear extent of $40$ cells and reconnecting the sites across $M$ (schematically shown in SFig.~\ref{fig:PbTe}~$\bs{b}$ for a smaller system size).  Hence, in our numerical implementation, $N_{\text{occ}} = 29994$.

In SFig.~\ref{fig:PbTe}~$\bs{c}$, we plot the PBC spectrum of the $\mathcal{I}$- and spinful $\mathcal{T}$-symmetric Hamiltonian $\tilde{H}$ with two $\bs{B} = \bs{a}_1$ dislocations.  The PBC spectrum in SFig.~\ref{fig:PbTe}~$\bs{c}$ exhibits a filling anomaly~\cite{TMDHOTI,WiederAxion,WladCorners,HingeSM} at $N_{\text{occ}}$ filled states, with the valence and conduction manifolds each specifically missing one Kramers pair of states.  The four anomalous states appear as midgap Kramers pairs in the PBC defect spectrum at filling $\nu = N_{\text{occ}}$ in SFig.~\ref{fig:PbTe}~$\bs{c}$, and are localized on the dislocation cores (SFig.~\ref{fig:PbTe}~$\bs{d}$).  When the system is three-quarters-filled, we expect the dislocation bound states to each be half filled in the limit that the dislocation separation is thermodynamically large.  Analogous to the domain-wall (end) states of a spinful SSH chain~\cite{SSH,SSHspinon,HeegerReview,RiceMele,TRPolarization}, each Kramers pair of dislocation bound states in SFig.~\ref{fig:PbTe}~$\bs{c,d}$ corresponds to a chargeless, spin-1/2 quasiparticle (\emph{i.e.} a spinon) when half filled (see SN~\ref{sec:kp}).

To conclude, in this section, we have demonstrated that PbTe monolayers exhibit an anomalous dislocation response that is consistent with SEq.~(\ref{eq:dislocationtheory}) and with the discussion in SN~\ref{sec:weakSSH}, in which it was determined that 2D insulators with nontrivial weak (partial) SSH indices $\bs{M}^{\mathrm{SSH}}_{\nu}$ [SEq.~(\ref{eq:PbTeWeakIndex})] exhibit an anomalous energy spectrum in the presence of point dislocations that satisfy $\bs{B}\cdot \bs{M}^{\mathrm{SSH}}_{\nu}\text{ mod }2\pi = \pi$.

\subsection{3D SnTe}
\label{sec:DFTSnTe}

\begin{figure}[t]
\begin{center}
\includegraphics[width=0.9\textwidth]{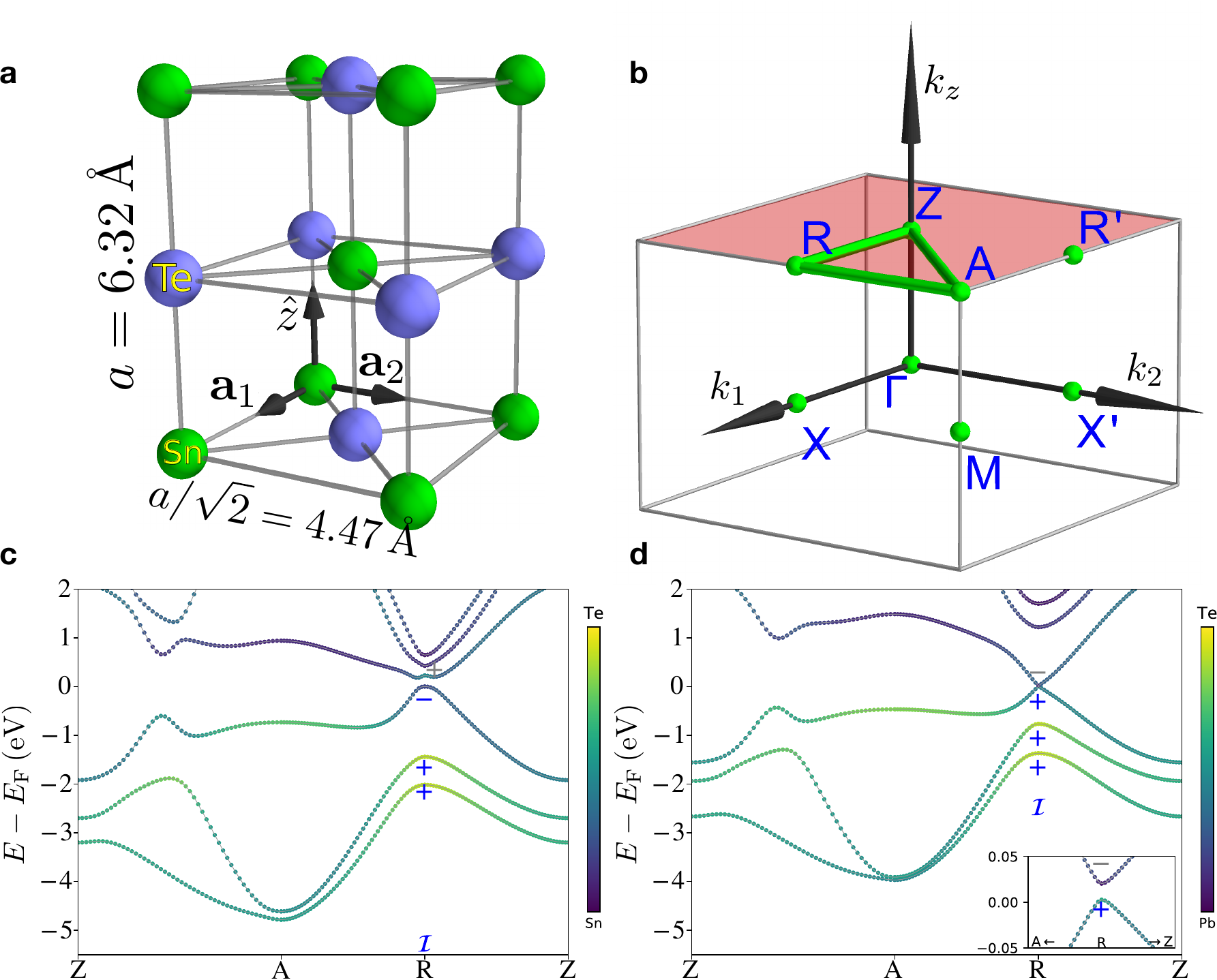}
\caption{{\bf Bulk electronic structures of 3D SnTe and PbTe.} $\bs{a}$~Crystal structure of 3D SnTe in a tetragonal supercell.  The supercell has twice as many atoms (four in total) as the primitive (unit) cell, and respects the symmetries of SG 123 $P4/mmm1'$.  In 3D PbTe, the crystal structure is the same as in~$\bs{a}$, with the Sn atoms replacing the Pb atoms.  $\bs{b}$~The BZ of SnTe and PbTe in the tetragonal supercell in $\bs{a}$.  
$\bs{c},\bs{d}$~The electronic structures of $\bs{c}$ SnTe and $\bs{d}$ PbTe calculated from first-principles and plotted along the path indicated in $\bs{b}$ with a green line.  The bands in $\bs{c}$ and $\bs{d}$ exhibit an artificial fourfold degeneracy at all ${\bs k}$ points, due to the combined effects of spinful $\mathcal{I} \times \mathcal{T}$ symmetry and supercell BZ folding.  The $+$ and $-$ signs in $\bs{c,d}$ denote the parity eigenvalues per Kramers pair of the Bloch states at the TRIM point $R$ [${\bs k}_{R} = {\bs b}_{1}/2$].  The band structures in $\bs{c,d}$ indicate that 3D SnTe differs from 3D PbTe by a double band inversion at the $R$ point and at the symmetry-related point $R'$ [${\bs k}_{R'} = C_{4z}{\bs k}_{R}\text{ mod }{\bs b}_{1}\text{ mod }{\bs b}_{2} = {\bs b}_{2}/2$] between two pairs of Kramers pairs of states with opposite parity eigenvalues [four valence bands and four conduction bands become inverted at $R$ and at $R'$].  Because PbTe in $\bs{d}$ is an unobstructed atomic limit, then the double band inversion in SnTe leads to a nontrivial weak (partial) fragile index vector $\bs{M}^{\mathrm{F}}_{\nu} = ({\bs b}_{1} + {\bs b}_{2})/2$ [see SN~\ref{sec:weakFragile} and the text surrounding SEq.~(\ref{eq:SnTe3DHENDresponse})].
\label{fig:DFTSnTe3D}
}
\end{center}
\end{figure}

Three-dimensional tin telluride (SnTe) has been predicted to be a rotation-anomaly TCI with helical hinge states~\cite{ChenRotation,HOTIChen,HOTIBernevig}.  In this section, we will demonstrate that 3D SnTe crystals exhibit a nontrivial HEND-state dislocation response.  First, in SN~\ref{sec:SnTeDFTSPECIFIC}, we will use first-principles calculations to reproduce the electronic structure and diagnose the topology of 3D SnTe -- as well as its topologically trivial counterpart 3D PbTe.  We will then specifically demonstrate in SN~\ref{sec:SnTeTB} that 3D SnTe (PbTe) hosts a nontrivial (trivial) $\mathcal{I}$-symmetry-indicated response to 1D dislocations.

\subsubsection{First-Principles Calculation of the Nontrivial Dislocation Response Vector in SnTe}
\label{sec:SnTeDFTSPECIFIC}

In this section, we will use first principles to calculate the weak fragile index $\bs{M}^{\mathrm{F}}_{\nu}$ [defined in SEqs.~\eqref{eq:fragileMComponents},~(\ref{eq:weakTIIndex}), and in the surrounding text in SN~\ref{sec:FTIwithT}] for 3D PbTe and SnTe.  We begin by performing fully-relativistic DFT calculations of the electronic structure of 3D SnTe and PbTe using VASP~\cite{VASP1,VASP2} employing the PAW method~\cite{PAW1,PAW2} and GGA-PBE~\cite{PBE-1996} for the exchange-correlation functional~\cite{AndreiMaterials,TQCwebpage}.  The lattice parameters of the rock-salt structure [SG 225 $Fm\bar{3}m1'$] were fixed to their experimental values~\cite{latt-SnTePbTe} $a=6.32$ {\AA} for SnTe and $a=6.46$ {\AA} for PbTe.

Below, we will specifically compute the dislocation response for the shortest possible dislocation Burgers vectors -- \emph{i.e.} dislocations for which the Burgers vector ${\bs B}$ is equal to one of the primitive, face-centered-cubic lattice vectors of SnTe or PbTe.  We are particularly interested in the case in which ${\bs B}$ is equal to a primitive lattice vector, as the Frank energy criterion~\cite{FrankReadEnergy} implies that dislocations become increasingly energetically unfavorable with increasing $|{\bs B}|$.  Hence, for geometric simplicity and because 3D SnTe and PbTe are cubic, we will, without loss of generality, form a tetragonal supercell in which the ${\bs a}_{1}$ and ${\bs a}_{2}$ primitive lattice vectors are also lattice vectors in the face-centered cubic cell, but in which ${\bs a}_{3}$ is $\sqrt{2}$ times the length of a face-centered-cubic primitive lattice vector [see SFig.~\ref{fig:DFTSnTe3D}~$\bs{a}$].  The tetragonal cell specifically contains two Sn/Pb atoms at $(x,y,z) = (0,0,0)$ and $(1/2,1/2,1/2)$ and two Te atoms at $(0,0,1/2)$ and $(1/2,1/2,0)$, and respects the symmetries of SG 123 $P4/mmm1'$.  As shown in SFig.~\ref{fig:DFTSnTe3D}~$\bs{a}$, the lattice vectors of the tetragonal cell are given by:
\begin{equation}
\bs{a}_1 = (1/2,-1/2,0),\  \bs{a}_2 = (1/2,1/2,0),\ \bs{a}_3 = (0,0,1),
\end{equation}
in units in which the lattice spacing $a=1$, and the reciprocal lattice vectors are given by:
\begin{equation}
\bs{b}_1 = 2\pi(1,-1,0),\ \bs{b}_2 = 2\pi(1,1,0),\ \bs{b}_3 = 2\pi(0,0,1).  
\label{eq:3DSnTeReciprocalVector}
\end{equation}
In our first-principles calculations, we only incorporate valence-shell states -- hence, our calculations only include the $5p$ orbitals of Te and $5p$ ($6p$) orbitals of Sn (Pb), as well as twelve total empty conduction bands from higher-shell (empty) valence orbitals.    Therefore, at each TRIM point in SFig.~\ref{fig:DFTSnTe3D}~$\bs{c,d}$, the lower twelve (upper twelve) bands are occupied (unoccupied) [the bands in SFig.~\ref{fig:DFTSnTe3D}~$\bs{c,d}$ are fourfold degenerate due to the combined effects of spinful $\mathcal{I}\times\mathcal{T}$ symmetry and supercell BZ folding].

As shown in SN~\ref{sec:proofs} and~\ref{sec:indices}, $\bs{M}^{\mathrm{F}}_{\nu}$ can be obtained by counting the number of parity-eigenvalue-exchanging band inversions by which a set of bands differs from an unobstructed atomic limit with a trivial dislocation response.  We first establish that 3D PbTe is an unobstructed atomic limit~\cite{HsiehTCI} in which three Kramers pairs of states are located on each of the four Te atoms in the tetragonal supercell -- this can be seen from the absence of band inversions in SFig.~\ref{fig:DFTSnTe3D}~$\bs{c}$, and has been shown in previous works~\cite{HsiehTCI}.  The band structures in SFig.~\ref{fig:DFTSnTe3D}~$\bs{c,d}$ indicate that 3D SnTe differs from 3D PbTe by double band inversions at the $R$ point [${\bs k}_{R} = {\bs b}_{1}$] and at the symmetry-related point $R'$ [${\bs k}_{R'} = C_{4z}{\bs k}_{R}\text{ mod }{\bs b}_{1}\text{ mod }{\bs b}_{2} = {\bs b}_{2}/2$] between two pairs of Kramers pairs of states with opposite parity eigenvalues [four valence states become inverted with four conduction states at $R$ and at $R'$].

Before calculating the dislocation response index vector $\bs{M}^{\mathrm{F}}_{\nu}$, we will next calculate the stable topological indices of 3D SnTe (as we have established that 3D PbTe is topologically trivial when its electronic structure is calculated as detailed above).  First, SRefs.~\onlinecite{ChenTCI,HOTIChen,AshvinTCI} previously established the existence of a strong $\mathbb{Z}_{8}$-valued topological index $z_{8}$ in SG 123 $P4/mmm1'$.  Below, we reproduce the specific formula for $z_{8}$ provided in Supplementary Table~2 in SRef.~\onlinecite{ChenTCI}:
\begin{equation}
z_8 = \frac{1}{2} \left[3 n^+_{\frac{3}{2}} - 3 n^-_{\frac{3}{2}} - n^+_{\frac{1}{2}} + n^-_{\frac{1}{2}}\right] \text{ mod } 8.
\label{eq:temp3Dz8}
\end{equation}
Applying the~\href{https://github.com/stepan-tsirkin/irrep}{IrRep} program~\cite{StepanIrRepCode} to the output of our first-principles calculations to obtain the small coreps that correspond to the occupied Bloch states of 3D SnTe, we find that, using the definitions for the small corep multiplicities $n^{\pm}_{1/2,3/2}$ established in SRef.~\onlinecite{ChenTCI} [\emph{c.f.} Supplementary Table~1 in SRef.~\onlinecite{ChenTCI}]:
\begin{equation}
n^+_{\frac{1}{2}} = 11, \quad n^-_{\frac{1}{2}} = 15, \quad n^+_{\frac{3}{2}} = 9, \quad n^-_{\frac{3}{2}} = 13.
\label{eq:SnTeSmallCorep}
\end{equation}
SEqs.~(\ref{eq:temp3Dz8}) and~(\ref{eq:SnTeSmallCorep}) together indicate that:
\begin{equation}
z_8 = 4,
\end{equation}
confirming that SnTe realizes an $\mathcal{I}$- and $C_{4z}$-indicated fourfold-rotation-anomaly TCI.

In SG 123 $P4/mmm1'$ there are also weak mirror Chern numbers and $\mathbb{Z}_{2}$ 2D TI indices~\cite{ChenTCI,HOTIChen,AshvinTCI}.  First, through the Fu-Kane weak-index parity indices~\cite{FuKaneInversion}, the small corep multiplicities in SEq.~(\ref{eq:SnTeSmallCorep}) imply that the weak $\mathbb{Z}_{2}$ indices in the ${\bs b}_{1,2}/2$ planes are trivial.  Next, we calculate the $\mathbb{Z}_{4}$-valued index $z_{4m,\pi}$ established in SRefs.~\onlinecite{ChenTCI,ChenBernevigTCI,HOTIChen,AshvinTCI}:
\begin{equation}
\begin{aligned}
z_{4m,\pi}= 
& \frac{3}{2}n\left(E_{\frac{3}{2}g}^Z  \right)
-\frac{3}{2}n\left(E_{\frac{3}{2}u}^Z  \right)
-\frac{1}{2}n\left(E_{\frac{1}{2}g}^Z  \right)
+\frac{1}{2}n\left(E_{\frac{1}{2}u}^Z  \right)\\
&+\frac{3}{2}n\left(E_{\frac{3}{2}g}^A  \right)
-\frac{3}{2}n\left(E_{\frac{3}{2}u}^A  \right)
-\frac{1}{2}n\left(E_{\frac{1}{2}g}^A  \right)
+\frac{1}{2}n\left(E_{\frac{1}{2}u}^A  \right)\\
&+n\left(E_{\frac{1}{2}g}^R  \right)
-n\left(E_{\frac{1}{2}u}^R  \right)
\;\mathrm{mod}\; 4,
\label{eq:z4mpi}
\end{aligned}
\end{equation}
where $z_{4m,\pi}$ indicates the mirror Chern number in the $k_{z}=\pi$ plane $C_{M_{z}}\text{ mod }4$.  Restricting the small corep multiplicities obtained in our first-principles calculations to the $k_{z}=\pi$ plane, we determine that 
\begin{equation}
\begin{aligned}
&n\left(E_{\frac{3}{2}g}^Z\right) = 1, \quad n\left(E_{\frac{3}{2}u}^Z\right) = 1, \quad n\left(E_{\frac{1}{2}g}^Z\right) = 2, \quad n\left(E_{\frac{1}{2}u}^Z\right) = 2, \\
&n\left(E_{\frac{3}{2}g}^A\right) = 1, \quad n\left(E_{\frac{3}{2}u}^A\right) = 2, \quad n\left(E_{\frac{1}{2}g}^A\right) = 2, \quad n\left(E_{\frac{1}{2}u}^A\right) = 1, \\
&n\left(E_{\frac{1}{2}g}^R\right) = 4, \quad n\left(E_{\frac{1}{2}u}^R\right) = 2,
\end{aligned}
\end{equation}
which implies that $z_{4m,\pi}=0$, such that $C_{M_{z}}(k_{z}=\pi)\text{ mod }4=0$.  In agreement with previous works~\cite{HsiehTCI,ChenTCI,HOTIBernevig}, we have additionally numerically confirmed through Wilson-loop calculations that $C_{M_{z}}(k_{z}=\pi)=0$ in the tetragonal supercell of SnTe.

In this work, we have focused on SnTe specifically because the $R$ and $R'$ points lie on the BZ boundary, such that the previously-established band inversions in SnTe~\cite{HsiehTCI} imply a nontrivial dislocation response (see SN~\ref{sec:proofs}).  To determine the dislocation response of SnTe, we first establish that $\bs{M}^{\mathrm{F}}_{\nu} = {\bs 0}$ in PbTe, because PbTe is an unobstructed atomic limit.  Hence, because SnTe differs from PbTe by double band inversions at the $R$ and $R'$ points in the tetragonal cell in SFig.~\ref{fig:DFTSnTe3D}~$\bs{a}$ between two pairs of Kramers pairs of states with opposite parity eigenvalues [four valence states become inverted with four conduction states at $R$ and at $R'$], then the HEND-state response of SnTe is \emph{nontrivial}:
\begin{equation}
\bs{M}^{\mathrm{F}}_{\nu} = (\bs{b}_1 + \bs{b}_2)/2 = (2 \pi,0,0).
\label{eq:SnTe3DHENDresponse}
\end{equation}
We emphasize that, despite $\nu^{\mathrm{F}}_{x}\text{ mod }2\pi = \nu^{\mathrm{F}}_{y}\text{ mod }2\pi = 0$ in SEq.~(\ref{eq:SnTe3DHENDresponse}), $\bs{M}_{\nu}^{\mathrm{F}}$ is still nontrivial, because $(2\pi,0,0)$ and $(0,2\pi,0)$ are \emph{not} reciprocal lattice vectors in the tetragonal supercell of SnTe [SEq.~(\ref{eq:3DSnTeReciprocalVector})].

\subsubsection{Tight-Binding Calculation of Topological HEND States in 3D SnTe}
\label{sec:SnTeTB}

In this section, we will use tight-binding calculations to demonstrate a nontrivial $\mathcal{I}$-symmetry-indicated HEND-state response in 3D SnTe.  Specifically, in the text surrounding SEq.~(\ref{eq:SnTe3DHENDresponse}), we previously demonstrated that the bulk of 3D SnTe carries a nontrivial HEND-state dislocation response vector $\bs{M}^{\mathrm{F}}_{\nu}= ({\bs b}_{1} + {\bs b}_{2})/2$, where ${\bs b}_{1,2}$ are defined in SEq.~(\ref{eq:3DSnTeReciprocalVector}).  Below, we will demonstrate that, as predicted by SEq.~(\ref{eq:BurgersFragileAppendix}) in SN~\ref{sec:kpEdge}, an $\mathcal{I}$-symmetric configuration of edge dislocations in SnTe with Burgers vector $\bs{B}$ satisfying $\bs{B} \cdot \bs{M}^{\mathrm{F}}_{\nu} \text{ mod } 2\pi = \pi$ will bind anomalous 0D states at $\mathcal{I}$-related locations along the set of dislocations.  In fact, the shortest dislocations with integer Burgers vectors $\bs{B} = \bs{a}_{1,2}$ satisfy the requirement that $\bs{B} \cdot \bs{M}^{\mathrm{F}}_{\nu} \text{ mod } 2\pi = \pi$.  Furthermore, because the Frank energy criterion~\cite{FrankReadEnergy} indicates that dislocations with larger values of $|{\bs B}|$ are energetically unfavorable, then dislocations with the smallest possible integer Burgers vectors $\bs{B} = \bs{a}_{1,2}$ may be energetically favorable and present in SnTe samples.  We note that screw dislocations with $\bs{B} = \bs{a}_{1,2}$, unlike edge dislocations, must reach a crystal surface of SnTe, and cannot form internal loops with a spatially constant Burgers vector (see SN~\ref{sec:kpEdge} and~\ref{sec:kpScrew}).  Additionally, because SnTe is a (rotation-anomaly) TCI with gapless surface states~\cite{HsiehTCI,ChenTCI,HOTIBernevig}, then the HEND states bound to screw dislocations in SnTe may be obscured due to their coexistence with gapless surface states.

We will now explicitly confirm the nontrivial defect response of 3D SnTe predicted by our first-principles calculations in SN~\ref{sec:SnTeDFTSPECIFIC}.  To model an edge dislocation in SnTe, we use the tight-binding model from SRef.~\onlinecite{HsiehTCI}, with the parameters listed in SRef.~\onlinecite{FulgaCoupledLayers}.  We first enlarge the model unit cell to obtain the tetragonal supercell shown in SFig.~\ref{fig:DFTSnTe3D}~$\bs{a}$.  We then determine the locations of the $\mathcal{I}$ centers in the supercell from the mirror symmetry representations given in SRef.~\onlinecite{FulgaCoupledLayers} -- in real space, the Sn and Te atoms in the model in SRef.~\onlinecite{HsiehTCI} occupy inversion centers that coincide with lines of $C_{4z}$ symmetry in the tetragonal supercell (SFig.~\ref{fig:DFTSnTe3D}~$\bs{a}$).

\begin{figure}[t]
\begin{center}
\includegraphics[width=\textwidth]{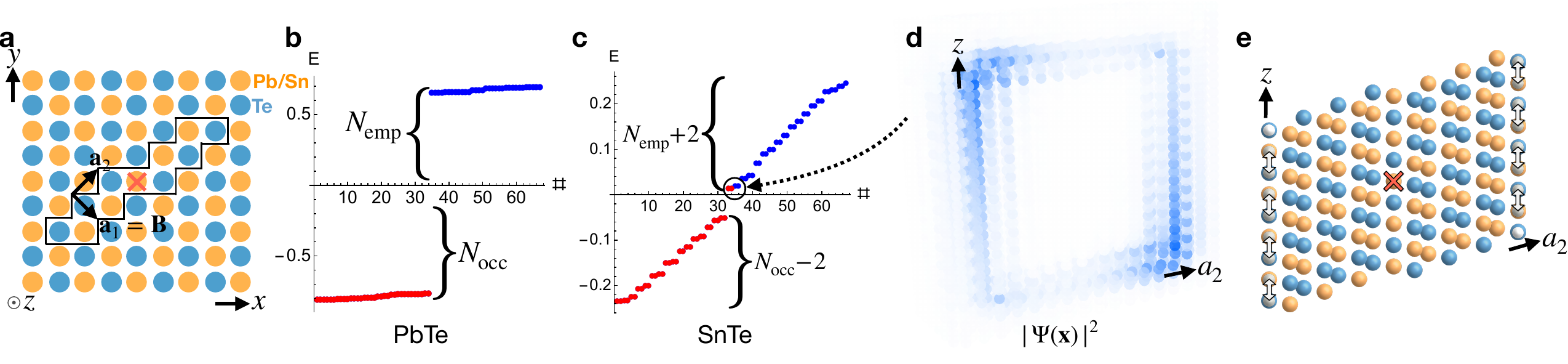}
\caption{{\bf Defect spectra of 3D PbTe and SnTe for edge dislocations with Burgers vector $\bs{B} = \bs{a}_1$.} $\bs{a}$~Defect geometry for an $\mathcal{I}$-related pair of edge dislocations, where the $\mathcal{I}$ center is marked with a red $\times$ symbol.  The black line in $\bs{a}$ encloses the cross section of the sites surrounding a plane with fixed $\bs{a}_1$ that were removed to create the internal dislocation pair [our numerical calculations were performed employing a PBC grid of size $11$ (along $\bs{a}_1$) $\times\ 17$ (along $\bs{a}_2$) $\times\ 17$ (along $\bs{a}_3 = \hat{z}$) unit cells and a $23$ (in the $\bs{a}_1,\bs{a}_2$ plane) $\times\ 25$ (along $\hat{z}$) site defect plane (where a single unit cell is taken to contain four sites, see SFig.~\ref{fig:DFTSnTe3D}~$\bs{a}$)].  The calculations of the dislocation spectrum were performed using the tight-binding model for SnTe introduced in SRef.~\onlinecite{HsiehTCI}.  $\bs{b,c}$ The PBC dislocation spectrum of $\bs{b}$ 3D PbTe and $\bs{c}$ 3D SnTe using the edge dislocation geometry in $\bs{a}$.  As previously for 2D monolayer PbTe (see SN~\ref{sec:DFTPbTe}), the electronic filling of $\nu=N_\text{occ}$ states is set by the unobstructed atomic limit of six electrons per Te atom, in addition to the tightly-bound core-shell electrons (which coincides with the bulk trivial topology of PbTe, see SN~\ref{sec:DFTPbTe}).  The defect spectrum for PbTe in $\bs{b}$ exhibits a large, trivial gap.  Conversely, the dislocation spectrum in ${\bs c}$ for SnTe exhibits midgap states.  Specifically, because the dislocation geometry in $\bs{a}$ preserves $\mathcal{I}$ and spinful $\mathcal{T}$ symmetries, then the spectrum in $\bs{c}$ for SnTe cannot be deformed into that of a trivial insulator with dislocations, and is thus filling-anomalous~\cite{TMDHOTI,WiederAxion,WladCorners,HingeSM}.  $\bs{d}$ The real-space profile of the four (circled) states in $\bs{c}$ that contribute to the filling anomaly of the SnTe spectrum.  In $\bs{d}$, two total Kramers pairs of states are localized on $\mathcal{I}$-related dislocation corners (one Kramers pair of states is bound to every other corner, for a total of four, filling-anomalous HEND states).  When the HEND states in ${\bs d}$ are half-filled, each Kramers pair corresponds to a chargeless, spin-1/2 quasiparticle (\emph{i.e.} a spinon) that is equivalent to the corner state of an $\mathcal{I}$- and $\mathcal{T}$-symmetric 2D FTI (see SN~\ref{sec:3DscrewTRSsec} and SRefs.~\onlinecite{TMDHOTI,WiederAxion}). $\bs{e}$~The SnTe defect plane (black cross-section) from $\bs{a}$ schematically depicted as a stack of PbTe monolayer defect lines (SFig.~\ref{fig:PbTe}~$\bs{b,d}$).  In $\bs{e}$, each defect line has two 0D dislocations on its end, which each bind first-order 0D topological dislocation states.  In an $\mathcal{I}$-symmetric stack of dislocation lines, the 0D dislocations evolve into 1D edge dislocations, and the 0D states pairwise annihilate, leaving two, filling-anomalous HEND states.  We choose PbTe for the monolayers -- rather than SnTe -- because the interlayer coupling in realistic 3D PbTe drives additional band inversions, whereas a tetragonal supercell of 3D SnTe has the same $x,y$ components of the $\bs{M}_{\nu}^{\mathrm{F}}$ vector as a decoupled stack of PbTe monolayers (see SN~\ref{sec:DFTPbTe}).  Hence, in the same sense that a helical (non-axionic) HOTI is equivalent to an $\mathcal{I}$-symmetric stack of 2D TIs (with an odd total number of layers)~\cite{AdyWeak,ChenTCI,HOTIBernevig,HOTIBismuth,TMDHOTI,MTQC}, HEND dislocation states can be considered the result of stacking and symmetrically coupling (gray arrows in $\bs{e}$) an odd number of 2D monolayers that each contain first-order dislocation bound states.}
\label{fig:SnTeTBresults}
\end{center}
\end{figure}

We next implement an internal edge dislocation with $\bs{B} = \bs{a}_1$ as detailed in SN~\ref{sec:2Dtoysubsec}.  First, because 3D PbTe is isostructural to 3D SnTe, then the tetragonal supercell in SFig.~\ref{fig:DFTSnTe3D}~$\bs{a}$ can be considered a bilayer of the 2D PbTe crystal structure in SFig.~\ref{fig:PbTe-dft}~$\bs{a}$.  Hence, a finite 2D plane bounded on two sides by $\bs{B} = \bs{a}_1$, 1D dislocations in 3D SnTe can be constructed by stacking alternating bilayers of the 1D line bounded by two $\bs{B}=\bs{a}_{1}$, 0D dislocations in a monolayer of SnTe with the same crystal structure as PbTe in SFig.~\ref{fig:PbTe}~$\bs{b}$.  Importantly, in order to use filling anomalies to diagnose the nontrivial HEND-state dislocation response, we must implement the defect plane in an $\mathcal{I}$-symmetric manner, which we accomplish with the alternating pattern of site removal depicted in SFig.~\ref{fig:SnTeTBresults}~$\bs{a}$.

To provide a reference for our numerical analysis of the defect response in 3D SnTe, we have also implemented a $\bs{B} = \bs{a}_1$ pair of edge dislocations in a tight-binding model of 3D PbTe.  To construct the tight-binding model, we have increased the on-site energy difference between the two inequivalent atoms in the primitive unit cell [specifically, in the notation of SRef.~\onlinecite{FulgaCoupledLayers}, we have changed the parameter $m$ from $1.65$ to $3$ in Eq.~(16) in SRef.~\onlinecite{FulgaCoupledLayers}].  Increasing the on-site energies reverses the pair of double band inversions at $R$ and $R'$ (see SN~\ref{sec:SnTeDFTSPECIFIC}), and reproduces the first-principles-derived parity eigenvalues and electronic structure of PbTe (SFig.~\ref{fig:DFTSnTe3D}~$\bs{b}$).  The on-site potential can also be understood as a chemical potential that localizes all of the electrons on the Te atoms of PbTe, in agreement with our diagnosis in SN~\ref{sec:SnTeDFTSPECIFIC} of 3D PbTe as an unobstructed (trivial) atomic limit.  Because PbTe is isostructural to SnTe, then the real-space defect geometry for our tight-binding model of PbTe is identical to the defect geometry previously employed in SnTe (depicted in SFig.~\ref{fig:SnTeTBresults}~$\bs{a}$).

In SFig.~\ref{fig:SnTeTBresults}~$\bs{b}$ (SFig.~\ref{fig:SnTeTBresults}~$\bs{c}$), we plot the PBC defect spectrum for PbTe (SnTe) using the tight-binding models and $\bs{B} = \bs{a}_1$ internal edge dislocations described earlier in this section.  For both SnTe and PbTe, we consider the valence electrons to originate from the Te $p$ orbitals, as described in SN~\ref{sec:SnTeDFTSPECIFIC} -- hence, $N_\text{occ}$ in the defect spectra in SFig.~\ref{fig:SnTeTBresults}~$\bs{b,c}$ is equal to six times the number of Te atoms in the PBC defect geometry (\emph{i.e.}, excluding the Te atoms removed to implement the dislocations).  The dislocation spectrum of PbTe exhibits a large gap and is trivial, whereas the defect spectrum of SnTe is conversely filling-anomalous, specifically exhibiting four midgap states (two Kramers pairs corresponding to the circled states in SFig.~\ref{fig:SnTeTBresults}~$\bs{c}$).  As shown in SFig.~\ref{fig:SnTeTBresults}~$\bs{d}$, the four filling-anomalous defect states in SnTe are localized on two $\mathcal{I}$-related dislocation corners, and are thus HEND states.  When the system has an electronic filling of $\nu=N_\text{occ}$, the dislocation bound states will each be half filled in the limit that the dislocation separation is thermodynamically large.  Analogous to the corner states of $\mathcal{I}$- and $\mathcal{T}$-symmetric 2D FTIs, each Kramers pair of dislocation bound states corresponds to a chargeless, spin-1/2 quasiparticle (\emph{i.e.} a spinon) when half filled (see SN~\ref{sec:kp}).  To conclude, we have demonstrated a nontrivial HEND-state dislocation response in 3D SnTe that is consistent with SEq.~(\ref{eq:dislocationtheory}) and with the discussion in SN~\ref{sec:weakFragile}.

\bibliographystyle{naturemag}
\bibliography{../defectBibs}